\documentclass[12pt]{report}      
\pdfoutput=1

\usepackage{textcomp}
\usepackage{enumerate}
\usepackage{amsmath, amssymb,mathtools}
\usepackage{bbold}
\usepackage{bbm}
\usepackage{mathbbol}
\usepackage{slashed}
\usepackage{graphics}
\usepackage{shadow}
\usepackage{float}
\usepackage{graphicx}
\usepackage{color}

\usepackage{slashed}
\usepackage{epsfig}
\usepackage{epsf}
\usepackage{epstopdf}

\usepackage{pdfsync}
\usepackage{mathrsfs}

  \usepackage[english]{babel} 
   \usepackage{graphicx}        
      \RequirePackage{hyperref} 

\newcommand{\tr}{{\rm tr}}

\newcommand{\braket}[2]{ \langle #1 \lvert #2 \rangle }  
\newcommand{\ket}[1]{\lvert #1 \rangle} 
\newcommand{\bra}[1]{\langle #1 \rvert} 

\def\3i{\int\!\!\!\int\!\!\!\int}
\def\2i{\int\!\!\!\int}

\def\del{\Delta}
\def\ddel{{}^\bullet\! \Delta}
\def\deld{\Delta^{\hskip -.5mm \bullet}}
\def\dddel{{}^{\bullet \bullet} \! \Delta}
\def\ddeld{{}^{\bullet}\! \Delta^{\hskip -.5mm \bullet}}

\begin{document}

           

\def\no{\noindent}
\def\non{\nonumber\\}
\def\green{\color{green}}
\def\blue{\color{blue}}
\def\red{\color{red}}
\def\black{\color{black}}
%
\def\cosech{\rm cosech}
\def\sech{\rm sech}
\def\coth{\rm coth}
\def\tanh{\rm tanh}
\def\half{{1\over 2}}
\def\third{{1\over3}}
\def\fourth{{1\over4}}
\def\fifth{{1\over5}}
\def\sixth{{1\over6}}
\def\seventh{{1\over7}}
\def\eigth{{1\over8}}
\def\ninth{{1\over9}}
\def\tenth{{1\over10}}
\def\conj{{{\rm c.c.}}}
\def\bN{\mathop{\bf N}}
\def\R{{\rm I\!R}}
\def\Eins{{\mathchoice {\rm 1\mskip-4mu l} {\rm 1\mskip-4mu l}
{\rm 1\mskip-4.5mu l} {\rm 1\mskip-5mu l}}}
\def\Z{{\mathchoice {\hbox{$\sf\textstyle Z\kern-0.4em Z$}}
{\hbox{$\sf\textstyle Z\kern-0.4em Z$}}
{\hbox{$\sf\scriptstyle Z\kern-0.3em Z$}}
{\hbox{$\sf\scriptscriptstyle Z\kern-0.2em Z$}}}}
\def\abs#1{\left| #1\right|}
\def\com#1#2{
        \left[#1, #2\right]}
\def\square{\kern1pt\vbox{\hrule height 1.2pt\hbox{\vrule width 1.2pt
   \hskip 3pt\vbox{\vskip 6pt}\hskip 3pt\vrule width 0.6pt}
   \hrule height 0.6pt}\kern1pt}
      \def\boxop{{\raise-.25ex\hbox{\square}}}
\def\contract{\makebox[1.2em][c]{
        \mbox{\rule{.6em}{.01truein}\rule{.01truein}{.6em}}}}
\def\ltap{\ \raisebox{-.4ex}{\rlap{$\sim$}} \raisebox{.4ex}{$<$}\ }
\def\gtap{\ \raisebox{-.4ex}{\rlap{$\sim$}} \raisebox{.4ex}{$>$}\ }
\def\ab{{\alpha\beta}}
\def\kl{{\kappa\lambda}}
\def\mn{{\mu\nu}}
\def\rs{{\rho\sigma}}
\newcommand{\Det}{{\rm Det}}
\def\Tr{{\rm Tr}\,}
\def\tr{{\rm tr}\,}
\def\sumij{\sum_{i<j}}
\def\e{\,{\rm e}}
\def\br{{\bf r}}
\def\bp{{\bf p}}
\def\bq{{\bf q}}
\def\bx{{\bf x}}
\def\by{{\bf y}}
\def\brhat{{\bf \hat r}}
\def\bv{{\bf v}}
\def\ba{{\bf a}}
\def\bE{{\bf E}}
\def\bB{{\bf B}}
\def\bA{{\bf A}}
\def\b0{{\bf 0}}
\def\pa{\partial}
\def\dA{\partial^2}
\def\ddx{{d\over dx}}
\def\ddt{{d\over dt}}
\def\der#1#2{{d #1\over d#2}}
\def\lie{\hbox{\it \$}} 
\def\partder#1#2{{\partial #1\over\partial #2}}
\def\secder#1#2#3{{\partial^2 #1\over\partial #2 \partial #3}}
%
\def\be{\begin{equation}}
\def\ee{\end{equation}\noindent}
\def\bear{\begin{eqnarray}}
\def\ear{\end{eqnarray}\noindent}
\def\bec{\blue\begin{equation}}
\def\eec{\end{equation}\black\noindent}
\def\bearc{\blue\begin{eqnarray}}
\def\earc{\end{eqnarray}\black\noindent}
\def\benn{\begin{enumerate}}
\def\enn{\end{enumerate}}
\def\ee{&=&}
\def\veject{\vfill\eject}
\def\ven{\vfill\eject\noindent}
%
\def\eq#1{{eq. (\ref{#1})}}
\def\eqs#1#2{{eqs. (\ref{#1}) -- (\ref{#2})}}
\def\la{\langle}
\def\ra{\rangle}
%
\def\sumninf{\sum_{n=0}^{\infty}}
%
\def\totint{\int_{-\infty}^{\infty}}
\def\posint{\int_0^{\infty}}
\def\negint{\int_{-\infty}^0}
\def\pint{{\dps\int}{dp_i\over {(2\pi)}^d}}
\def\intdp3{\int\frac{d^3p}{(2\pi)^3}}
\def\intdp4{\int\frac{d^4p}{(2\pi)^4}}
\def\scalprop#1{\frac{-i}{#1^2+m^2-i\epsilon}}
%
\newcommand{\GeV}{\mbox{GeV}}
\def\FFdual{F\cdot\tilde F}
\def\bra#1{\langle #1 |}
\def\ket#1{| #1 \rangle}
\def\braket#1#2{\langle {#1} \mid {#2} \rangle}
\def\vev#1{\langle #1 \rangle}
\def\matel#1#2#3{\langle {#1} \mid #2 \mid{#3} \rangle}
\def\rightvac{\mid0\rangle}
\def\leftvac{\langle0\mid}
\def\ihbar{{i\over\hbar}}
\def\lagr{{\cal L}}
\def\sigmabar{{\bar\sigma}}
\def\ge{\hbox{$\gamma_1$}}
\def\gz{\hbox{$\gamma_2$}}
\def\gd{\hbox{$\gamma_3$}}
\def\go{\hbox{$\gamma_1$}}
\def\gt{\hbox{$\gamma_2$}}
\def\gth{\hbox{$\gamma_3$}} 
\def\gf{\hbox{$\gamma_5\;$}}
\def\slash#1{#1\!\!\!\raise.15ex\hbox {/}}
\newcommand{\slD}{\,\raise.15ex\hbox{$/$}\kern-.27em\hbox{$\!\!\!D$}}
\newcommand{\slpartial}{\raise.15ex\hbox{$/$}\kern-.57em\hbox{$\partial$}}
\newcommand{\PP}{\cal P}
\newcommand{\G}{{\cal G}}
\newcommand{\nc}{\newcommand}
\nc{\spa}[3]{\left\langle#1\,#3\right\rangle}
\nc{\spb}[3]{\left[#1\,#3\right]}
\nc{\ksl}{\not{\hbox{\kern-2.3pt $k$}}}
\nc{\hf}{\textstyle{1\over2}}
\nc{\pol}{\varepsilon}
\nc{\tq}{{\tilde q}}
\nc{\esl}{\not{\hbox{\kern-2.3pt $\pol$}}}
\newcommand{\cL}{\cal L}
\newcommand{\D}{\cal D}
\newcommand{\Dhalf}{{D\over 2}}
\def\eps{\epsilon}
\def\epshalf{{\epsilon\over 2}}
\def\lag{( -\partial^2 + V)}
\def\veps{\varepsilon}
\def\freeexp{{\rm e}^{-\int_0^Td\tau {1\over 4}\dot x^2}}
\def\kinb{\frac{1}{4}\dot x^2}
\def\kinf{{1\over 2}\psi\dot\psi}
\def\expk{{\rm exp}\biggl[\,\sum_{i<j=1}^4 G_{Bij}k_i\cdot k_j\biggr]}
\def\expp{{\rm exp}\biggl[\,\sum_{i<j=1}^4 G_{Bij}p_i\cdot p_j\biggr]}
\def\expshort{{\e}^{\half G_{Bij}k_i\cdot k_j}}
\def\expabb{{\e}^{(\cdot )}}
\def\epseps#1#2{\varepsilon_{#1}\cdot \varepsilon_{#2}}
\def\epsk#1#2{\varepsilon_{#1}\cdot k_{#2}}
\def\kk#1#2{k_{#1}\cdot k_{#2}}
\def\G#1#2{G_{B#1#2}}
\def\Gp#1#2{{\dot G_{B#1#2}}}
\def\GF#1#2{G_{F#1#2}}
\def\Dab{{(x_a-x_b)}}
\def\Dsq{{({(x_a-x_b)}^2)}}
\def\PITD{{(4\pi T)}^{-{D\over 2}}}
\def\4piTD{{(4\pi T)}^{-{D\over 2}}}
\def\4piT4{{(4\pi T)}^{-2}}
\def\TintmD{{\dps\int_{0}^{\infty}}
\frac{dT}{T}
\,e^{-m^2T}
    {(4\pi T)}^{-\frac{D}{2}}}
\def\Tintm4{{\dps\int_{0}^{\infty}}{dT\over T}\,e^{-m^2T}
    {(4\pi T)}^{-2}}
\def\Tintm{{\dps\int_{0}^{\infty}}\frac{dT}{T}\,e^{-m^2T}}
\def\Tint{{\dps\int_{0}^{\infty}}{dT\over T}}
\def\np{n_{+}}
\def\nm{n_{-}}
\def\Np{N_{+}}
\def\Nm{N_{-}}
\newcommand{\slG}{{{\dot G}\!\!\!\! \raise.15ex\hbox {/}}}
\newcommand{\Gd}{{\dot G}}
\newcommand{\Gund}{{\underline{\dot G}}}
\newcommand{\Gdd}{{\ddot G}}
\def\GBd12{{\dot G}_{B12}}
\def\Dx{\dps\int{\cal D}x}
\def\Dy{\dps\int{\cal D}y}
\def\Dpsi{\dps\int{\cal D}\psi}
\def\dint#1{\int\!\!\!\!\!\int\limits_{\!\!#1}}
\def\ddtau{{d\over d\tau}}
\def\ie{\hbox{$\textstyle{\int_1}$}}
\def\iz{\hbox{$\textstyle{\int_2}$}}
\def\id{\hbox{$\textstyle{\int_3}$}}
\def\ldop{\hbox{$\lbrace\mskip -4.5mu\mid$}}
\def\rdop{\hbox{$\mid\mskip -4.3mu\rbrace$}}
%
\newcommand{\1}{{\'\i}}
\def\dps{\displaystyle}
\def\sy{\scriptscriptstyle}
\def\sy{\scriptscriptstyle}
\newcommand{\ogf}{{^{\circ}\gf}}
\newcommand{\gfo}{{\gf^{\circ}}}
\newcommand{\ogfo}{{^{\circ}\gf^{\circ}}}
\newcommand{\ep}{\varepsilon}
\newcommand{\linep}{\textrm{lin } \ep_{1}\ldots\ep_{N}}
\newcommand{\linn}{\textrm{lin } n_{1}\ldots n_{N}}

\newcommand{\hint}{{\overset{\wedge\!\!\!\!\hspace{0.04em}\wedge}{\phantom{\int}\,}\hspace{-1.5em}}\int}
\newcommand{\scal}{\textrm{scal}}
\newcommand{\spin}{\textrm{spin}}
\newcommand{\DBC}{\textrm{DBC}}
\newcommand{\xm}{x_{-}}
\newcommand{\xp}{x_{+}}

\newcommand{\dup}{{^{\bullet}\!\Upsilon}}
\newcommand{\upd}{{\Upsilon\!^{\bullet}}}
\newcommand{\dupd}{{^{\bullet}\!\Upsilon\!^{\bullet}}}

\newcommand{\delC}{\underset{\smile}{\Delta}}
\newcommand{\ddelC}{{^{\bullet}\!\delC}}
\newcommand{\delCd}{{\delC\!^{\bullet}}}
\newcommand{\ddelCd}{{^{\bullet}\!\delC\!^{\bullet}}}
\newcommand{\odelC}{{^{\circ}\!\delC}}
\newcommand{\delCo}{{\delC\!^{\circ}}}
\newcommand{\odelCo}{{^{\circ}\!\delC\!^{\circ}}}
\newcommand{\odelCd}{{^{\circ}\!\delC\!^{\bullet}}}
\newcommand{\ddelCo}{{^{\bullet}\!\delC\!^{\circ}}}

\newcommand{\bone}{1\!\!1}

\def\e{\,{\rm e}}

\newcommand{\symb}{\textrm{symb}}
\newcommand{\symbi}{\symb^{-1}}
\newcommand{\sep}{\slashed{\epsilon}}
\newcommand{\ps}{\slashed{p}}
\newcommand{\pps}{\slashed{p}^{\prime}}
\newcommand{\ks}{\slashed{k}}
\newcommand{\As}{\slashed{A}}
\newcommand{\hgamma}{\hat{\gamma}}

\def\targ{(\tau,\tau')}
\def\argN{(k_1,\varepsilon_1;\ldots ; k_N,\varepsilon_N)}
\def\argS{(k_1,\varepsilon_1;\ldots ; k_S,\varepsilon_S)}
\newcommand{\perms}{\textrm{ permutations }}
\newcommand{\s}{\slash}

\newcommand{\Zz}{\mathcal{Z}}
\newcommand{\Zzp}{\mathcal{Z}^{\prime}}
\newcommand{\detZ}{\textrm{det}^{-\frac{1}{2}}\left[\frac{\sin(\Zz)}{\mathcal{\Zz}}\right]}
\newcommand{\detZp}{\textrm{det}^{-\frac{1}{2}}\left[\frac{\sin (\Zzp)}{\Zzp}\right]}
\newcommand{\detZs}{\textrm{det}^{-\frac{1}{2}}\left[\frac{\tan(\Zz)}{\mathcal{\Zz}}\right]}
\newcommand{\detZps}{\textrm{det}^{-\frac{1}{2}}\left[\frac{\tan(\Zzp)}{\Zzp}\right]}
\newcommand{\pdetZ}{\textrm{det}^{-\frac{1}{2}}\left[\cos( \Zz) \right]}
\newcommand{\pdetZp}{\textrm{det}^{-\frac{1}{2}}\left[\cos(\Zzp)\right]}

\newcommand{\tZz}{\frac{\tan\Zz}{\Zz}}
\newcommand{\tZzp}{\frac{\tan\Zzp}{\Zzp}}

\begin{titlepage}
\begin{figure}[h]
\begin{center}
\vspace{-6em}
\includegraphics[scale=0.7]{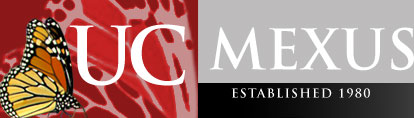}
\end{center}
\end{figure}
\vspace{-2em}
\begin{center}
{\Large {\bf Spinning Particles in Quantum Mechanics and Quantum Field Theory}\\}
\vspace*{0.2cm}
{\large Olindo Corradini$^{a,b,c}$ {\it and} Christian Schubert$^d$ \\
 James P. Edwards$^{d}$ {\it and} Naser Ahmadiniaz$^{e}$}
\vspace*{.1cm}

\item  [$^a$] 
{\it Dipartimento di Scienze Fisiche, Informatiche e Matematiche\\ Universit\`a degli Studi di Modena e Reggio Emilia\\ Via Campi 213/A, I-41125 Modena, Italy\\
olindo.corradini@unimore.it}

\item [$^b$]
{\it Istituto Nazionale di Fisica Nucleare, Sezione di Bologna\\ Via Irnerio 46, I-40126 Bologna, Italy}

\item  [$^c$]
{\it Facultad de Ciencias en F\'isica y Matem\'aticas\\
Universidad Aut\'onoma de Chiapas, Ciudad Universitaria\\ Tuxtla Guti\'errez 29050, Chiapas, M\'exico}

\item [$^d$]
{\it 
Instituto de F\'{\i}sica y Matem\'aticas,
\\
Universidad Michoacana de San Nicol\'as de Hidalgo,\\
Edificio C-3, Apdo. Postal 2-82,\\
C.P. 58040, Morelia, Michoac\'an, M\'exico\\
schubert@ifm.umich.mx\\
james.p.edwards@umich.mx
}

\item [$^e$]
{\it 
Helmholtz-Zentrum Dresden-Rossendorf, Bautzner Landstra\ss e 400, 01328 Dresden, Germany,
\\
n.ahmadiniaz@hzdr.de

}

\vspace*{0.4cm}
{\large School on 
{\it Spinning Particles in Quantum Field Theory: 
Worldline Formalism, Higher Spins, and Conformal Geometry}\\}
\vspace*{0.4cm}
{Morelia, Michoac\'an, M\'exico, November 19--23, 2012}

\end{center}

\end{titlepage}                                       
          


 \no
\textit{ABSTRACT:}\\
 These are notes of lectures on spinning particles and the worldline formalism 
 originally given by Olindo Corradini and Christian Schubert at the School on
{\it Spinning Particles in Quantum Field Theory: 
Worldline Formalism, Higher Spins, and Conformal Geometry},
held at Morelia, Mexico, from November 19 through November 23, 2012.
The lectures were addressed to graduate level students with a background in relativistic quantum mechanics
and at least a rudimentary knowledge of field theory. They have since been updated to include a further set of
lecture notes on tree level processes from a worldline perspective
based on a mini-course by James P. Edwards at the Instituto de Fisica y Matematicas 
in Morelia, Mexico given to graduates and visiting professors during July 2017 and in various later classes, complemented by a series of three lectures titled {\it New techniques for amplitude calculation in QED} given by Naser Ahmadiniaz at the {\it Center for Relativistic Laser Science (CoReLS), Institute for Basic Science (IBS), November 2015, Gwangju, South Korea} and as an invited lecturer at the {\it Helmholtz International Summer School (HISS) - Dubna International Advanced School of Theoretical Physics (DIAS-TH): ``Quantum Field Theory at the Limits: from Strong Fields to Heavy Quarks'', July-August 2019, Dubna, Russia}.  
These additional notes complete the picture of first quantised techniques and bring the worldline description up to date.

The first part of the lectures
 covers introductory material on quantum-mechanical Feynman path integrals, which are here derived and applied to several particle models. We start considering the nonrelativistic bosonic particle, for which we compute the exact path integrals for the case of the free particle and for the harmonic oscillator, and then describe perturbation theory for an arbitrary potential. We then move to relativistic particles, considering both bosonic and fermionic (spinning) particles. We first investigate them from the classical view-point, studying the symmetries of their actions, then consider their canonical quantization and path integrals, and underline the role these models have in the study of space-time quantum field theories (QFT), by introducing the ``worldline'' path integral representation of propagators and effective actions. We also describe a special class of spinning particles that constitute a first-quantized approach to higher-spin fields. 

The second section gives a more specialist treatment of applications of the worldline approach. Since the fifties the quantum-mechanical path integral representation of QFT propagators and effective actions has been known. However it is  only in recent years
that they have gained some popularity for certain types of calculations in quantum field theory. 
The second part of the lectures describes two of the main techniques for the evaluation of such path integrals, which are
gaussian integration (also called ``string-inspired formalism'') and semiclassical approximation (``worldline instantons''). 
We first give a detailed discussion of photonic amplitudes in Scalar and Spinor QED, in vacuum as well as in a constant external field.
Then we shortly discuss the main issues encountered in the generalization to the nonabelian case,
and finally the case of gravity, where subtle mathematical issues arise. 

In the final part of these notes we detail recent results
applying first quantised techniques to tree level processes in
QED.  Historically the worldline formalism was successfully applied at one- and multi-loop order, but a complete treatment
requires extension of these techniques to encompass external scalar or spinor legs. As discussed in section 3,
this will require ``opening up'' the closed worldlines met in the first
two sections so as to analyse relativistic point particle trajectories
between fixed endpoints. The applications include the 
calculation of (Compton) scattering amplitudes and determination of
the scalar and spinor self-energies to one-loop order. Furthermore, in
the fermionic case we show that the worldline approach leads to
a convenient decomposition of amplitudes into ``leading'' and
``sub-leading'' contributions, closely related to
gauge invariance. In the on-shell case only the leading part
contributes to the cross section, which allows immediate
identification of the physical information in the amplitude, that would be extremely difficult to uncover
using standard (second quantised) techniques. There is also a useful
decomposition of the amplitudes into spin and orbital contributions
explained briefly in this section.

\vspace{2em}

\no
{\bf Acknowledgements:} \\
We thank A. Raya, the late V.M. Villanueva and A. Waldron
for the coorganization of this School,  and UCMEXUS for providing the funding through Collaborative Grant
CN-12-564. Special thanks to Misha Arturo Lopez-Lopez, Victor Banda Guzmán, José Nicasio and Guopeng Xu for informing us of a number of errors in an earlier version of these lectures notes.

\tableofcontents                         
 
 \vfill\eject

 \chapter{Quantum Mechanical Path Integrals: from Transition Amplitudes to the Worldline Formalism}
 \label{ch:ch1}

\section{Introduction}
\label{sec:intro}
The notion of {\it path integral} as integral over trajectories was first introduced by Wiener in the 1920's to solve problems related to the Brownian motion.
Later, in 1940's, it was reintroduced by Feynman as an alternative to operatorial methods to compute transition amplitudes in quantum mechanics~\cite{Feynman:1965}. Feynman path integrals use a lagrangian formulation instead of a hamiltonian one and can be seen as a quantum-mechanical generalization of the least-action principle (see e.g.~\cite{Chaichian:2001cz}). 

\begin{figure}
\begin{center}
\includegraphics[scale=0.65]{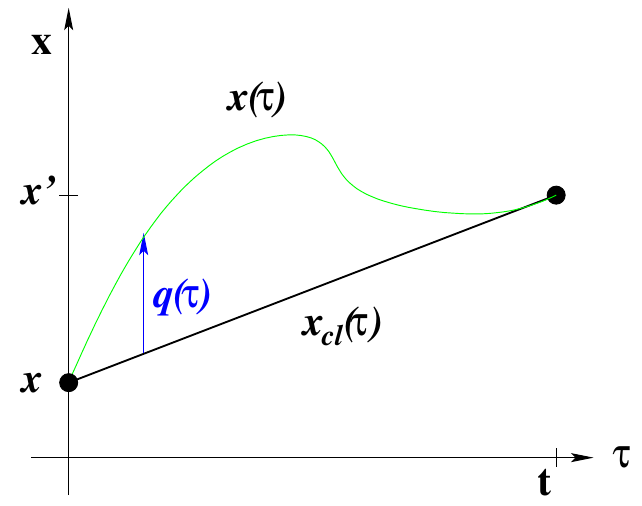}
\end{center}
\label{fig:paths}
\caption{A useful parametrization of paths}
\end{figure}

In electromagnetism the linearity of the Maxwell equations in vacuum allows to formulate the Huygens-Fresnel principle that in turn allows to write the wave scattered by a multiple slit as a sum of waves generated by each slit, where each single wave is characterized by the optical length $I(x_i,x')$ between the i-th slit and the field point $x'$, and the final amplitude is thus given by $A =\sum_i e^{iI(x_i,x')}$, whose squared modulus describes patterns of interference between single waves. In quantum mechanics, a superposition principle can be formulated in strict analogy to electromagnetism and a linear equation of motion, the Schr\"odinger equation, can be correspondingly postulated. Therefore, the analogy can be carried on further replacing the electromagnetic wave amplitude by a transition amplitude between an initial point $x$ at time $0$ and a final point $x'$ at time $t$, whereas the optical length is replaced by the classical action for going from $x$ and $x'$ in time $t$ (divided by $\hbar$, that has dimensions of action). The full transition amplitude will thus be a ``sum" over all paths connecting $x$ and $x'$ in time~$t$:
\begin{equation}
K(x',x;t) \sim \sum_{\{x(\tau)\}} e^{iS[x(\tau)]/\hbar}~.
\end{equation}         
In the rest of the chapter we will derive the previous expression for a few simple cases, and introduce a set of modern topics where it can be applied.  However, to begin let us simply justify the presence of the action in the expression above by recalling that the action principle applied to the classical action function $S_c(x',t)$---that corresponds to $S[x_{cl}(\tau)]$ evaluated on the classical path, with only the initial point $x=x(0)$ fixed---implies that $S_c$ satisfies the classical Hamilton-Jacobi equation. Hence the Schr\"odinger equation imposed on the amplitude $K(x',x;t) \sim e^{iS(x',t)/\hbar}$ yields an equation for $S(x',t)$ that deviates from the Hamilton-Jacobi equation by a linear term in $\hbar$. So that  $\hbar$ measures the departure from classical mechanics, which thus corresponds to the limit $\hbar \to 0$, and in turn the classical action function determines the transition amplitude to leading order in $\hbar$. It will be often useful to parametrize an arbitrary path $x(\tau)$ between $x$ and $x'$ as $x(\tau)= x_{cl}(\tau)+q(\tau)$ with $x_{cl}(\tau)$ being the ``classical path" i.e. the path that satisfies the equations of motion with the above boundary conditions, and $q(\tau)$ is an arbitrary deviation (see Figure~\ref{fig:paths} for graphical description.) 
With such parametrization the transition amplitude reads
\begin{equation}
K(x',x;t) \sim \sum_{\{q(\tau)\}} e^{iS[x_{cl}(\tau)+q(\tau)]/\hbar}~,
\end{equation}  
and $x_{cl}(\tau)$ represents as a sort of ``origin" in the space of all paths connecting $x$ and $x'$ in time $t$.

\section{Path integral representation of quantum mechanical transition amplitude: non relativistic bosonic particle}
\label{sec:nonrel-bos}
The quantum-mechanical transition amplitude for a time-independent hamiltonian operator is given by (here and henceforth we use natural units and thus set $\hbar=c=1$; see Appendix~\ref{app:NU} for a brief review on the argument)\footnote{A generalization to time-dependent hamiltonian ${\mathbb H}(t)$ can be obtained with the replacement $e^{-it {\mathbb H}} \ \to\ {\cal T} e^{-i\int_0^t d\tau {\mathbb H}(\tau)}$ and ${\cal T} e$ being a time-ordered exponential.\label{footenote1}}
\begin{align}\label{eq:evol}
K(x',x;t) &=\langle x'| e^{-it {\mathbb H}} |x\rangle = \langle x',t| x,0\rangle\\
 K(x',x;0) &=\delta(x'-x) 
\label{eq:evol-bc}
\end{align}
and describes the evolution of the wave function from time $0$ to time $t$

\begin{equation}
\psi(x',t) = \int dx~K(x',x;t) \psi(x,0) ~.
\label{eq:evolint}
\end{equation}
It satisfies the Schr\"odinger equation
\begin{equation}
i\partial_t K(x',x;t) = H  K(x',x;t)
\label{eq:Schrody}
\end{equation}
with $H$ being the hamiltonian operator in the coordinate representation; for a non-relativistic particle on a line we have $H=-\frac{\hbar^2}{2m}\frac{d^2}{dx'^2} + V(x')$.  
In particular, for a free particle $V=0$, it is easy to solve the Schr\"odinger equation~\eqref{eq:Schrody} with boundary conditions~\eqref{eq:evol-bc}. One obtains  
\begin{align}
K_f(x',x;t) = N_f(t)~ e^{iS_{cl}(x',x;t)}~,
\label{eq:Kf}
\end{align}
with
\begin{align}
\label{eq:Nf}
N_f(t) &= \sqrt{\frac{m}{i2\pi t}}\\
S_{cl}(x',x;t) &\equiv \frac{m(x'-x)^2}{2t} ~.
\label{eq:Scl}
\end{align}
The latter as we will soon see is the action for the classical path of the free particle.  

In order to introduce the path integral we ``slice" the evolution operator in~\eqref{eq:evol} by defining $\epsilon = t/N$ and insert $N-1$ decompositions of unity in terms of position eigenstates ${\mathbb 1} = \int dx |x\rangle \langle x| $; namely
\begin{align}
K(x',x;t) &=   \langle x'| e^{-i\epsilon {\mathbb H}} e^{-i\epsilon {\mathbb H}} \cdots e^{-i\epsilon {\mathbb H}}|x\rangle\\
&=\int \left(\prod_{i=1}^{N-1} dx_i\right) ~\prod_{j=1}^{N}  \langle x_j|e^{-i\epsilon {\mathbb H}} |x_{j-1}\rangle 
\end{align}
with $x_0\equiv x$ and $x_N\equiv x'$. We now insert $N$ spectral decomposition of unity in terms of momentum eigenstates ${\mathbb 1} = \int \frac{dp}{2\pi} |p\rangle \langle p| $
and get
\begin{align}
K(x',x;t) &= \int \left(\prod_{i=1}^{N-1} dx_i \right) \left(\prod_{k=1}^{N} \frac{dp_k}{2\pi}\right)~\prod_{j=1}^{N}  \langle x_j|e^{-i\epsilon {\mathbb H}}|p_j\rangle \langle p_j |x_{j-1}\rangle. 
\label{eq:PI1} 
\end{align}
For large $N$,  assuming ${\mathbb H} ={\mathbb T}+{\mathbb V}= \frac{1}{2m}{\mathbb p}^2 +V({\mathbb q})$, we can use the ``Trotter formula" (see e.g.~\cite{Schulman:1981vu}) 
\begin{align}
\Big(e^{-i\epsilon {\mathbb H}}\Big)^N = \Big( e^{-i\epsilon{\mathbb V}} e^{-i\epsilon{\mathbb T}}+O(1/N^2)\Big)^N \approx \Big( e^{-i\epsilon{\mathbb V}} e^{-i\epsilon{\mathbb T}}\Big)^N
\label{eq:trotter}
\end{align}
that allows us to replace $e^{-i\epsilon {\mathbb H}}$ with $e^{-i\epsilon{\mathbb V}} e^{-i\epsilon{\mathbb T}}$ in~\eqref{eq:PI1}, so that  one gets\\  $ \langle x_j|e^{-i\epsilon {\mathbb H}}|p_j\rangle \approx e^{i(x_jp_j-\epsilon H(x_j,p_j))}$, with $\langle x|p\rangle = e^{ixp}$. Hence,
\begin{align}
K(x',x;t) &= \int \left(\prod_{i=1}^{N-1} dx_i \right) \left(\prod_{k=1}^{N} \frac{dp_k}{2\pi}\right) \exp\Biggl[ i\sum_{j=1}^N \epsilon \Big(p_j \frac{x_j-x_{j-1}}{\epsilon} -H(x_j,p_j)\Big)\Biggr]
\label{eq:Kxx'}
\end{align}
with
\begin{align}
H(x_j,p_j) &= \frac{1}{2m}p_j^2+V(x_j)
\end{align}
the hamiltonian function. In the large $N$ limit we can formally write the above expression~\eqref{eq:Kxx'} as 
\begin{align}
K(x',x;t) &= \int_{x(0)=x}^{x(t)=x'} Dx Dp \exp\Biggl[ i\int_0^td\tau \Big(p \dot x -H(x,p)\Big)\Biggr]\\ 
& Dx\equiv \prod_{0<\tau <t} dx(\tau)\,,\ Dp\equiv \prod_{0<\tau <t} dp(\tau)
\end{align}
that is referred to as the ``phase-space path integral." Alternatively, the momenta can be integrated out in~\eqref{eq:PI1} as the integrals are (analytic continuation of) gaussian ones. 
Completing the square one gets

\begin{align}
K(x',x;t) &= \int \left(\prod_{i=1}^{N-1} dx_i \right) \left(\frac{m}{2\pi i \epsilon}\right)^{N/2}\nonumber\\
&\times \exp\Biggl[ i\sum_{j=1}^N \epsilon \Biggl(\frac{m}{2} \left(\frac{x_j-x_{j-1}}{\epsilon}\right)^2 -V(x_j)\Biggr)\Biggr]
\label{eq:PI-CF}
\end{align}
that can be formally written as 
\begin{align}
K(x',x;t) &= \int_{x(0)=x}^{x(t)=x'} Dx~e^{iS[x(\tau)]}
\label{eq:PI2} 
\end{align}
with
\begin{align}
S[x(\tau)] = \int_0^t d\tau \Big( \frac{m}{2}\dot x^2 -V(x(\tau))\Big)~.
\label{eq:SV}
\end{align}
The expression~\eqref{eq:PI2} is referred to as ``configuration space path integral" and is interpreted  as a functional integral over trajectories with boundary condition $x(0)=x$ and $x(t)=x'$. 

 To date no mathematically precise definition of the path integral measure $Dx$ is known and one has to rely on  some regularization methods. For example one may expand paths on a suitable basis to turn the functional integral into a countably infinite-dimensional Riemann integral. Thus one may regularize taking a large (though finite) number of vectors in the basis, perform the integrals and take the limit at the very end. This regularization procedure, as we will see, fixes the path integral up to an overall normalization constant that must be fixed using some consistency conditions. Nevertheless, ratios of  path integrals are well-defined objects and turn out to be quite convenient tools in several areas of modern physics.  Moreover one may fix---as we do in the next section---the above constant using the simplest possible model (the free particle) and compute other path integrals via their ratios with the free particle path integral.     

\subsection{Wick rotation to euclidean time: from quantum mechanics to statistical mechanics}
As already mentioned path integrals were born in statistical physics. In fact we can easily obtain the ``heat kernel" from~\eqref{eq:PI2}  by ``Wick rotating" to imaginary time  $it \equiv \sigma$
and get 
\begin{align}
K(x',x;-i\sigma) =\langle x'| e^{-\sigma {\mathbb H}}|x\rangle = \int_{x(0)=x}^{x(\sigma)=x'} Dx ~e^{-S_E[x(\tau)]}
\label{eq:H1}
\end{align}    
where the euclidean action
\begin{align}
S_E[x(\tau)] =\int_0^\sigma d\tau~\Big( \frac{m}{2} \dot x^2+V(x(\tau))\Big) 
\label{eq:EA}
\end{align}
has been obtained by Wick rotating the worldline time $i\tau \to \tau$. The heat kernel~\eqref{eq:H1} satisfies the heat equation 
\begin{align}
(\partial_\sigma + H) K(x',x;-i\sigma) =0~.
\label{eq:EE}
\end{align}
In particular, by setting $m=1/2\alpha$ and $V=0$, one gets  $H = -\alpha \partial_x'^2$ with $\alpha$ the ``thermal diffusivity." Hence, if $T(x,0)$ is the temperature profile on a one-dimensional rod at time $\sigma=0$, then
\begin{align}
T(x',\sigma) = \int dx K_f(x',x;-i\sigma) T(x,0)
\end{align} 
will be the temperature profile at time $\sigma$. Here
\begin{align}
K_f(x',x;-i\sigma) = \frac{1}{\sqrt{4\pi \alpha\sigma}} e^{-\frac{(x'-x)^2}{4\sigma\alpha}} ~.
\end{align}
 Equation~\eqref{eq:EE} has a number of applications in mathematical physics ranging from the Fokker-Planck equation in statistical physics to the Black-Scholes model in financial mathematics. 

The particle partition function can also be easily obtained from~\eqref{eq:PI2} by,  (a) ``Wick rotating" time to imaginary time, namely $it \equiv \beta =1/k \theta$ (where $\theta$ is the temperature) and (b) taking the trace
\begin{align}
Z(\beta) &= {\rm tr} ~e^{-\beta {\mathbb H}} = \int dx ~\langle x| e^{-\beta {\mathbb H}}|x\rangle = \int dx ~K(x,x;-i\beta) \nonumber\\
&= \int_{PBC} Dx ~e^{-S_E[x(\tau)]}
\end{align}    
where  $PBC$ stands for {\it periodic boundary conditions} and means that the path integral is taken over all closed paths. Here the euclidean action is the same as in Eq.~\eqref{eq:EA} with $\sigma$ replaced by $\beta$. 

\subsection{The free particle path integral}
\label{sec:free-path}
We consider the path integral~\eqref{eq:PI2} for the special case of a free particle, i.e. $V=0$. For simplicity we consider a particle confined on a line and rescale the worldline time $\tau \to t \tau$ in such a way that free action and boundary conditions turn into  
\begin{align}
S_f[x(\tau)] =\frac{m}{2t} \int_0^1 d\tau ~\dot x^2\,,\quad x(0)=x\,,\ x(1)=x'
\end{align}
where $x$ and $x'$ are points on a straight line. The free equation of motion is obviously $\ddot x=0$ so that the aforementioned parametrization yields
\begin{align}
x(\tau) &=x_{cl}(\tau) +q(\tau)\\
x_{cl}(\tau) &= x +(x'-x)\tau = x'+(x-x')(1-\tau)\,,\quad q(0)=q(1)=0~
\end{align}
and the above action reads
\begin{align}
S_f[x_{cl}+q] &=\frac{m}{2t} \int_0^1 d\tau ~(\dot x_{cl}^2 +\dot q^2 +2 \dot x_{cl} \dot q) = \frac{m}{2t} (x'-x)^2 + \frac{m}{2t} \int_0^1 d\tau ~\dot q^2 \\
&= S_f[x_{cl}]+S_f[q]\nonumber
\end{align}
and the mixed term identically vanishes due to equation of motion and boundary conditions. The path integral can thus be written as 
\begin{align}
K_f(x',x;t) &= e^{iS_f[x_{cl}]}\int_{q(0)=0}^{q(1)=0} Dq~e^{iS_f[q(\tau)]} = e^{i\frac{m}{2t} (x'-x)^2} \int_{q(0)=0}^{q(1)=0} Dq~e^{i\frac{m}{2t}\int_0^1 \dot q^2} 
\end{align}
 so that by comparison with~(\ref{eq:Kf},\ref{eq:Nf},\ref{eq:Scl})  one can infer that, for the free particle,
 \begin{equation}
 \int_{q(0)=0}^{q(1)=0} Dq~e^{i\frac{m}{2t}\int_0^1 \dot q^2}  = \sqrt{\frac{m}{i2\pi t}}~. 
 \label{freepiqm}
 \end{equation}
The latter results easily generalize to $d$ space dimensions, where one obtains 
\begin{align}
\int_{q(0)=0}^{q(1)=0} Dq~e^{i\frac{m}{2t}\int_0^1 \dot q^2}=\left(\frac{m}{i2\pi t}\right)^{d/2}~.
\end{align}

\subsubsection{Direct evaluation of path integral normalization}
In order to directly compute the path integral normalization one must rely on a regularization scheme that allows one to handle the otherwise ill-defined measure $Dq$.  
One may expand $q(\tau)$ on a basis of functions with Dirichlet boundary condition on the line
\begin{align}
q(\tau) = \sum_{n=1}^\infty q_n \sin(n\pi \tau)
\end{align} 
with $q_n$ arbitrary real coefficients. The measure can be parametrized as
\begin{align}
Dq\equiv A \prod_{n=1}^\infty dq_n  a_n
\label{eq:measure}
\end{align} 
with $A$ and $a_n$ numerical coefficients that we fix shortly. One possibility, quite popular among string theorists (see e.g.~\cite{Polchinski:1985zf}), is to use a gaussian definition for the measure, namely
\begin{align}
\int \prod_n dq_n a_n~e^{-||q||^2} =1, \quad ||q||^2\equiv \int_0^t d\tau' q^2(\tau') = t \int_0^1 d\tau q^2(\tau) 
\end{align}
so that $a_n=\sqrt{\frac{t}{2\pi}}$.  For the path integral normalization one gets
\begin{align}
\int_{q(0)=0}^{q(1)=0} Dq~e^{i\frac{m}{2t}\int_0^1 \dot q^2}  = A \int \prod_{n=1}^\infty dq_n  \sqrt{\frac{t}{2\pi}} e^{i\frac{m}{4t}\sum_n (\pi n q_n)^2} = A \prod_{n=1}^\infty 
\sqrt{\frac{2i}{m\pi^2}}t~\frac{1}{n}
\label{eq:free-PI}
\end{align}
that is thus expressed in terms of an ill-defined infinite product. In general, the class of infinite products given by $\prod_n a n^b$, may be computed using the analytical continuation of the Riemann zeta-function. For any complex number $k$, with ${\rm Re}(k)>1$, such function is defined by 
\begin{align}\label{eq:zeta}
\zeta(k)= \sum_{n=1}^\infty \frac{1}{n^k}~,
\end{align}
whose derivative reads
\begin{align}
\zeta'(k)= -\sum_{n=1}^\infty \frac{\log n}{n^k}~.
\end{align}
The expression~\eqref{eq:zeta} can be analytically continued to any complex number, except $k=1$, where it diverges. 
One can thus use this analytical property to make sense of the above, otherwise ambiguous, infinite product. One thus gets

\begin{align}
\prod_{n=1}^\infty a n^b =\exp\Big\{ \log a\sum_n 1+b\sum_n \log n\Big\} =a^{\zeta(0)}e^{-b\zeta'(0)}=\sqrt{\frac{(2\pi)^b}{a}}~,
\end{align}
where we have used $\zeta(0) = -1/2, \zeta'(0) = -\half \ln (2\pi)$. Applied to the free path integral~\eqref{eq:free-PI}, this identity gives  

\begin{align}
\int_{q(0)=0}^{q(1)=0} Dq~e^{i\frac{m}{2t}\int_0^1 \dot q^2}  = A \left(\frac{m}{2i}\right)^{1/4} \frac{1}{\sqrt{2t}} = \sqrt{\frac{m}{i2\pi t}}
\end{align}
provided $A=\left(\frac{2m}{i\pi^2}\right)^{1/4}$. The Riemann zeta-function ``regularization'' thus provides the correct functional form (in terms of $t$) of the path integral normalization,
although it still does not fix the absolute normalization. 

A different normalization can be obtained by asking instead that each mode be normalized with respect to the free kinetic action, namely

 \begin{align}
\int  dq_n  a_n  e^{i\frac{m}{4t}\sum_n (\pi n q_n)^2} = 1\ \Rightarrow\ a_n = \sqrt{\frac{m n^2\pi}{4it}}
\end{align}
and therefore the overall normalization must be fixed as $A = \sqrt{\frac{m}{i2\pi t}}$.  The latter normalization is quite useful if used with {\it mode regularization} where the product~\eqref{eq:measure} is truncated to a large finite mode $M$. This method can be employed to compute more generic particle path integrals where interaction terms may introduce computational ambiguities. Namely: whenever an ambiguity appears one can always rely on the mode expansion, truncated at $M$, and then take the large limit at the very end, after having resolved the ambiguities. Other regularization schemes that have been adopted to such purpose are: {\it time slicing} that rely on the well-defined expression for the path integral as multiple time slices (cfr. Eq.~\eqref{eq:PI-CF}) and {\it dimensional regularization} that regulates ambiguities by dimensionally extending the worldline (see~\cite{Bastianelli:2006rx} for a review on such issues).

\subsubsection{The free particle partition function}
The partition function for a free particle in $d$-dimensional space can be obtained as
\begin{align}
Z_f (\beta)= \int d^dx ~K(x,x;-i\beta) =  \left(\frac{m}{2\pi \beta}\right)^{d/2} \int d^dx = {\cal V}\left(\frac{m}{2\pi \beta}\right)^{d/2}~,
\end{align}
${\cal V}$ being the spatial volume. It is easy to check that this is the correct result:
\begin{align}
Z_f (\beta)= \sum_p~e^{-\beta p^2/2m} = \frac{\cal V}{(2\pi)^d} \int d^d p  ~e^{-\beta p^2/2m} = {\cal V}\left(\frac{m}{2\pi \beta}\right)^{d/2}~,
\label{eq:Zfb}
\end{align}
with $\frac{\cal V}{(2\pi)^d} $ being the ``density of states". 

\subsubsection{Perturbation theory about the free particle solution: Feynman diagrams}
\label{sec:free-PT}
In the presence of an arbitrary potential the path integral for the transition amplitude is not exactly solvable. However if the potential is ``small" compared to the kinetic term one can rely on perturbation theory about the free particle solution. The significance of being ``small" will be clarified a posteriori.

Let us then obtain a perturbative expansion for the transition amplitude~\eqref{eq:PI2} with action~\eqref{eq:SV}. As done above we split the arbitrary path in terms of the classical path (with respect to the free action) and a deviation $q(\tau)$ and, again making use of the rescaled time, we can rewrite the amplitude as 
\begin{align}
K(x',x;t) = K_f(x',x;t) ~\frac{{\displaystyle \int}_{q(0)=0}^{q(1)=0} Dq ~e^{i\int_0^1 (\frac{m}{2t}  \dot q^2-t V(x_{cl}+q))}}{{\displaystyle \int}_{q(0)=0}^{q(1)=0} Dq ~e^{i\frac{m}{2t} \int_0^1 \dot q^2}}~.
\end{align}
We then Taylor expand the potential in the exponent about the classical free solution: this gives rise to a infinite set of interaction terms (the $\tau$ dependence in $x_{cl}$ and $q$ is left implied) 
\begin{align}
S_{int} = -t \int_0^1 d\tau \Big( &V(x_{cl})+V'(x_{cl}) q +\frac{1}{2!} V^{(2)}(x_{cl})q^2+ \frac{1}{3!} V^{(3)}(x_{cl}) q^3+\cdots\Big)\nonumber\\
& \raisebox{-.25cm}{ \includegraphics[scale=.5]{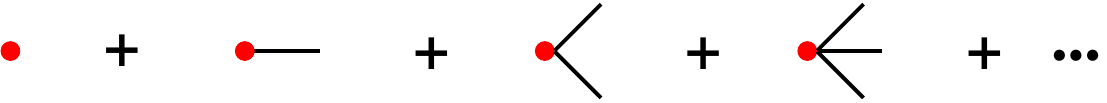}} 
\end{align}
Next we expand the exponent $e^{iS_{int}}$ so that we only have polynomials in $q$ to integrate with the path integral weight; in other words  we only need to compute expressions like
\begin{align}
\frac{{\displaystyle \int}_{q(0)=0}^{q(1)=0} Dq ~e^{i\frac{m}{2t} \int_0^1 \dot q^2} q(\tau_1) q(\tau_2)\cdots q(\tau_n)}{{\displaystyle \int}_{q(0)=0}^{q(1)=0} Dq ~e^{i\frac{m}{2t} \int_0^1 \dot q^2}} \equiv \Big\langle q(\tau_1) q(\tau_2)\cdots q(\tau_n) \Big\rangle
\end{align}
and the full (perturbative) path integral can be compactly written as
\begin{align}
K(x',x;t) = K_f(x',x;t)\Big\langle e^{-it\int_0^1 V(x_{cl}+q)}\Big\rangle
\end{align}
and the expressions $\Big\langle f(q) \Big\rangle$ are referred to as ``correlation functions". In order to compute the above correlations functions we define and compute the so-called ``generating functional"
\begin{align}
{\cal Z}[j] \equiv \int_{q(0)=0}^{q(1)=0} Dq ~e^{i\frac{m}{2t} \int_0^1 \dot q^2 + i\int_0^1 q j} = N_f(t) \Big\langle e^{i\int_0^1 q j} \Big \rangle 
\end{align} 
in terms of which 
\begin{align}
\Big\langle q(\tau_1) q(\tau_2)\cdots q(\tau_n) \Big\rangle = (-i)^n \frac{1}{{\cal Z}[0]}\frac{\delta^n}{\delta j(\tau_1)\delta j(\tau_2)\cdots \delta j(\tau_n) } {\cal Z}[j]\Big|_{j=0}~.
\end{align}
By partially integrating the kinetic term and completing the square we get

\begin{align}
{\cal Z}[j] = e^{\frac{i}{2}\int\!\! \int j  {\cal D}^{-1}  j}  \int_{q(0)=0}^{q(1)=0} D\tilde q ~e^{i\frac{m}{2t} \int_0^1 \dot {\tilde q}^2}
\label{eq:Zj}
\end{align}
where ${\cal D}^{-1}(\tau,\tau')$, ``the propagator",  is the Green's function of the kinetic operator ${\cal D}\equiv \frac{m}{t}\partial_\tau^2$, 

\begin{align}
{\cal D} {\cal D}^{-1}(\tau,\tau') =\delta(\tau - \tau')~,
\end{align} 
in the basis of functions with Dirichlet boundary conditions. Above we indicated $\tilde q(\tau) \equiv q(\tau)-\int_0^1 j(\tau') {\cal D}^{-1}(\tau',\tau)$. By defining ${\cal D}^{-1}(\tau,\tau') = \frac{t}{m}\Delta(\tau,\tau')$ we get
\begin{align}
& \dddel(\tau,\tau') = \delta(\tau - \tau') \label{eq:dddel}\\
\Rightarrow\quad  & \Delta(\tau,\tau') = \frac12 |\tau -\tau'|-\frac{\tau+\tau'}{2} +\tau\tau'
\label{eq:del} 
\end{align}
where ``dot" on the left (right) means derivative with respect to $\tau$ ($\tau'$). The propagator thus satisfies the following properties

\begin{align}
\Delta(\tau,\tau') &= \Delta(\tau',\tau)\\
\Delta(\tau,0) &= \Delta(\tau,1) =0
\end{align}
from which $\tilde q(0) = \tilde q(1)=0$. Therefore we can shift the integration variable in~\eqref{eq:Zj} from $q$ to $\tilde q$ and get
\begin{align}
{\cal Z}[j] = N_f(t)~e^{\frac{it}{2m}\int\!\! \int j  \Delta  j} 
\label{eq:Zj2}
\end{align}
and finally obtain
\begin{align}
\Big\langle q(\tau_1) q(\tau_2)\cdots q(\tau_n) \Big\rangle = (-i)^n \frac{\delta^n}{\delta j(\tau_1)\delta j(\tau_2)\cdots \delta j(\tau_n) } e^{\frac{it}{2m}\int\!\! \int j  \Delta  j} \Big|_{j=0}~.
\end{align}
In particular:
\begin{enumerate}
\item correlation functions of an odd number of ``fields"  vanish;
\item the 2-point function is nothing but the propagator 
\begin{align}
\Big\langle q(\tau_1) q(\tau_2)\Big\rangle =-i\frac{t}{m}\Delta(\tau_1,\tau_2) = \includegraphics[scale=.5]{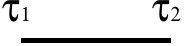}
\label{eq:prop}
\end{align} 
\item correlation functions of an even number of fields are obtained by all possible contractions of pairs of fields. For example, for $n=4$ we have 
$\langle q_1 q_2 q_3 q_4 \rangle = \frac{1}{2}(-i\frac{t}{m})^2 (\Delta_{12} \Delta_{34}+\Delta_{13} \Delta_{24}+\Delta_{14} \Delta_{23})$ where an obvious shortcut notation has been used. The latter statement is known as the ``Wick's theorem".  Diagrammatically  $$ \Big\langle q_1 q_2 q_3 q_4 \Big\rangle =\quad \raisebox{-0.8cm}{\includegraphics[scale=.5]{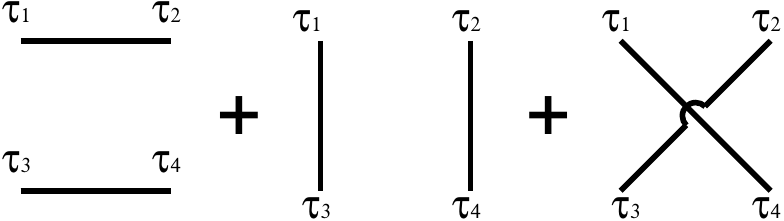}} $$
\end{enumerate} 
Noting that each propagator carries a power of $t/m$ we can write the perturbative expansion as a short-time expansion. It is thus not difficult to convince oneself that the expansion reorganizes as
\begin{align}
K(x',x\begin{scriptsize}
{\footnotesize •}
\end{scriptsize};t) &= N_f(t)~ e^{i\frac{m}{2t}(x'-x)^2} \exp\Big\{ {\rm connected\ diagrams} \Big\}\nonumber\\
&=N_f(t)~ e^{i\frac{m}{2t}(x'-x)^2} \exp\left\lbrace \raisebox{-1.55cm}{\includegraphics[scale=0.4]{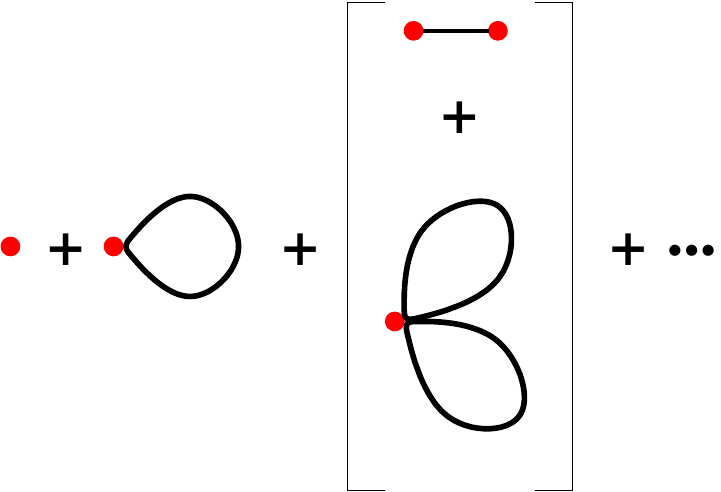}}\right\rbrace
\end{align}
where the diagrammatic expansion in the exponent (that is ordered by increasing powers of $t$) only involves connected diagrams, i.e. diagrams whose vertices are connected by at least one propagator. We recall that $N_f(t) =\sqrt{\frac{m}{i2\pi t}}$, and we can also give yet another representation for the transition amplitude, the so-called ``heat-kernel expansion"~\footnote{Strictly speaking what is referred to as the ``heat-kernel expansion", is the analytical continuation of Eq.~\eqref{eq:HK} obtained by setting $it=\sigma$, which thus satisfies the heat equation.}
 \begin{align}
 K(x',x;t) &=\sqrt{\frac{m}{i2\pi t}}~ e^{i\frac{m}{2t}(x-x')^2} \sum_{n=0}^\infty a_n(x',x) t^n
 \label{eq:HK}
 \end{align}
 where the terms $a_n(x,x')$ are known as Seeley-DeWitt coefficients. We can thus express such coefficients in terms of  Feynman diagrams. For $n\leq 2$ we thus get
 \begin{align}
 a_0(x',x) &=1\\
 a_1(x',x) &= \includegraphics[scale=0.5]{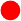}= -i\int_0^1 d\tau ~V(x_{cl}(\tau))\label{eq:a1}\\
 a_2(x',x) &=\raisebox{-0.4cm}{\includegraphics[scale=0.5]{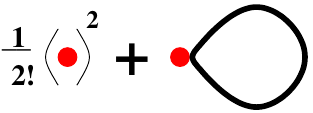}}\nonumber\\&=\frac{1}{2!}\left(-i\int_0^1 d\tau ~V(x_{cl}(\tau))\right)^2 -\frac{1}{2!\, m}\int_0^1d\tau~V^{(2)}(x_{cl}(\tau)) \tau(\tau-1) \label{eq:a2}
 \end{align}
where in~\eqref{eq:a2} we have used that $\langle q(\tau) q(\tau) \rangle = -i\frac{t}{m} \Delta(\tau,\tau) = -i\frac{t}{m} \tau(\tau -1)$. Let us emphasize that, for a generic potential, the $\tau$ integrals appearing in~(\ref{eq:a1},\ref{eq:a2}) are not analytically solvable. However those expressions are helpful shortcuts to represent the infinite series of terms one obtains expanding the integrands about, for instance, the final point $x'$. For example, the expansion of the first Seeley-DeWitt coefficient reads:
\begin{align}
a_1(x',x) &=-i\int_0^1 d\tau ~V(x_{cl}(\tau)) =-i\sum_{k=0}^\infty \frac{V^{(k)}(x')}{k!}(x-x')^k \int_0^1d\tau (1-\tau)^k\nonumber\\
&=-i\sum_{k=0}^\infty \frac{V^{(k)}(x')}{(k+1)!}(x-x')^k~,
\end{align} 
and similarly for the other coefficients. Note that, inserting the heat-kernel expansion~\eqref{eq:HK} into the expression~\eqref{eq:evolint}, all the terms $(x-x')^k$ are integrated with a Gaussian weight: those which have odd powers give vanishing contributions, whereas those which have even powers give contributions of order $(\frac{t}{m})^{\frac{k}{2}}$, to the wave function.  

Let us now conclude by commenting on the validity of the expansion~\eqref{eq:HK}: each propagator  inserts a power of $\frac{t}{m}$. Therefore for a fixed potential $V$, the larger the mass, the larger the time for which the expansion is accurate. In other words for a very massive particle it results quite costly to move away from the classical path.

\subsection{The harmonic oscillator path integral}
\label{sec:harmonic-path}
If the particle is subject to a harmonic potential $V(x)=\frac12 m\omega^2 x^2$ the path integral is again exactly solvable. In the rescaled time adopted above the path integral can be formally written as in~\eqref{eq:PI2} with action
\begin{align}
S_h[x(\tau)] =\frac{m}{2t}\int_0^1d\tau\Big[ \dot x^2-(\omega t)^2 x^2\Big] ~. 
\end{align}  
Again we parametrize $x(\tau)=x_{cl}(\tau) +q(\tau)$ with 
\begin{align}\left.
\begin{array}{l}
\ddot{x}_{cl}+(\omega t)^2 x_{cl} =0\\
x_{cl}(0)=x\,,\ x_{cl}(1)=x'\\
(q(0)=q(1)=0) 
\end{array}\right\rbrace
\ \Rightarrow\ x_{cl}(\tau) =x\cos(\omega t \tau) + \frac{x'-x\cos(\omega t)}{\sin(\omega t)} \sin(\omega t \tau)
\end{align}
and, since the action is quadratic,  similarly to the free particle case there is no mixed term between $x_{cl}(\tau) $ and $q(\tau)$, i.e. $S_h[x(\tau)] = S_h[x_{cl}(\tau)] + S_h[q(\tau)]$ and the path integral reads
\begin{align}
K_h(x',x;t) &= N_h(t) ~e^{iS_h[x_{cl}]}
\label{eq:PI-CH}
\end{align}
with
\begin{align}\label{eq:Scl-AH}
S_h[x_{cl}] &= \frac{m\omega}{2\sin(\omega t)} \Big((x^2+x'^2)\cos(\omega t) -2xx' \Big)\\
N_h(t) &= \int_{q(0)=0}^{q(1)=0} Dq ~e^{i\frac{m}{2t} \int_0^1d\tau\Big( \dot q^2-(\omega t)^2 q^2\Big)}~.
\end{align} 
We now use the above mode expansion to compute the latter: 
\begin{align}
N_h(t) &= N_f(t)~\frac{N_h(t)}{N_f(t)} = \sqrt{\frac{m}{i2\pi t}} ~\frac{{\displaystyle \int}_{q(0)=0}^{q(1)=0} Dq ~e^{i\frac{m}{2t} \int_0^1 (\dot q^2-(\omega t)^2 q^2)}}{{\displaystyle \int}_{q(0)=0}^{q(1)=0} Dq ~e^{i\frac{m}{2t} \int_0^1 \dot q^2}}\nonumber\\
&= \sqrt{\frac{m}{i2\pi t}}  \prod_{n=1}^\infty \frac{{\displaystyle \int} dq_n~e^{i\frac{imt}{4} \sum_n (\omega_n^2-\omega^2)q_{n}^2}}{{\displaystyle \int}dq_n~e^{i\frac{imt}{4} \sum_n \omega_n^2 q_{n}^2} }=  \sqrt{\frac{m}{i2\pi t}}  \prod_{n=1}^\infty \left(1-\frac{\omega^2}{\omega_n^2}\right)^{-1/2}\nonumber\\
&= \sqrt{\frac{m}{i2\pi t}} \left( \frac{\omega t}{\sin(\omega t)} \right)^{1/2}~.
\label{eq:rtpi}
\end{align}
Above we defined $\omega_n\equiv \frac{\pi n}{t}$ and used the infinite product formula 
\begin{align}
\frac{\sin x}{x} =\prod_{n=1}^\infty\left( 1-\left(\frac{x}{\pi n}\right)^2\right)~.
\end{align}
 It is not difficult to check that, with the previous expression for $N_h(t)$, the path integral~\eqref{eq:PI-CH} satisfies the Schr\"odinger equation with hamiltonian $H=-\frac{1}{2m}\frac{d^2}{dx'^2}+\frac12 m\omega^2 x^2$ and boundary condition $K(x',x;0)=\delta(x'-x)$. The propagator can also be easily generalized 
\begin{align}
\Delta_\omega(\tau,\tau') = \frac{1}{\omega t \sin(\omega t)}\Big\lbrace &\theta(\tau-\tau') \sin(\omega t(\tau-1)) \sin(\omega t \tau') \nonumber\\
+& \theta(\tau'-\tau) \sin(\omega t(\tau'-1)) \sin(\omega t \tau)\Big\rbrace
\label{eq:prop-HOM}
\end{align}
and can be used in the perturbative approach.

In the limit $t \to 0$ (or $\omega \to 0$), all previous expressions reduce to their free particle counterparts.  

\subsubsection{The harmonic oscillator partition function}
\label{sec:PF-HO}
Similarly to the free particle case we can switch to statistical mechanics by Wick rotating the time, $it=\beta$ and get
\begin{align}
K_h(x',x;-i\beta) = e^{-S_h[x_{cl}]} \int_{q(0)=0}^{q(1)=0}Dq~e^{-S_h[q] }~,
\end{align}  
where now
\begin{align}
S_h[x] = \frac{m}{2\beta} \int_0^1 d\tau \Big(\dot x^2 +(\omega\beta)^2 x^2 \Big)
\end{align}
is the euclidean action, and
\begin{align}
S_h[x_{cl}] &= \frac{m\omega}{2\sinh(\beta\omega)} \Big[ (x^2 +x'^2)\cosh(\beta\omega)-2xx'\Big]\\
\int_{q(0)=0}^{q(1)=0} Dq~e^{-S_h[q]} &= \left(\frac{m\omega}{2\pi \sinh(\beta\omega)}\right)^{1/2}
\end{align}
that, unlike the real-time path integral~\eqref{eq:rtpi}, is always regular. Taking the trace of the amplitude (heat kernel) one gets the partition function
\begin{align}
Z_h(\beta) &= \int dx~K_h(x,x;-i\beta) = \int_{PBC} Dx~e^{-S_h[x]} =\sum\raisebox{-.5cm}{ \includegraphics[scale=.2]{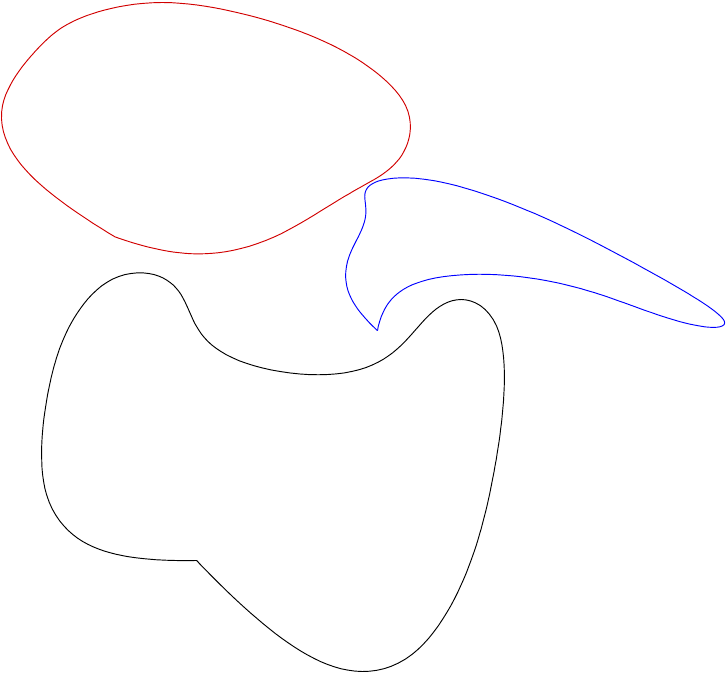}}\nonumber\\
&= \frac{1}{\sqrt2} \frac{1}{\left(\cosh(\beta\omega) -1\right)^{1/2}} = \frac{e^{-\frac{\beta\omega}{2}}}{1-e^{-\beta\omega}} 
\label{eq:HO-PF}
\end{align}
where $PBC$ stands for ``periodic boundary conditions" $x(0)=x(1)$ and implies that the path integral is a ``sum" over all closed trajectories, as suggested by the pictorial representation. Above we used the fact that the partition function for the harmonic oscillator can be obtained either using a transition amplitude computed with Dirichlet boundary conditions $x(0)=x(1)=x$ and integrating  over the initial=final point $x$, or with periodic boundary conditions  $x(0)=x(1)$, integrating over the ``center of mass" of the path $x_0 \equiv \int_0^1 d\tau x(\tau)$.

It is easy to check that Eq.~\eqref{eq:HO-PF} matches the result one obtains from
the (geometric) series ${\sum_{n=0}^\infty}e^{-\beta E_n}={\sum_{n=0}^\infty} e^{-\beta\omega (n+\frac12)}$. In particular, taking the zero-temperature limit ($\beta\to\infty$), the latter 
singles out the vacuum energy $Z_h(\beta) \approx e^{-\beta \frac{\omega}{2}}$ of the harmonic oscillator. Thus, for generic particle models, the computation of the above path integral can be seen as a method to obtain an estimate of the vacuum energy. For example, in the case of an anharmonic oscillator, if the deviation from harmonicity is small, perturbation theory can be used to compute corrections to the vacuum energy, as we do below. 

\subsubsection{Perturbation theory about the harmonic oscillator partition function solution}
\label{sec:PT-HO}
\begin{figure}
\begin{center}
\includegraphics[scale=0.4]{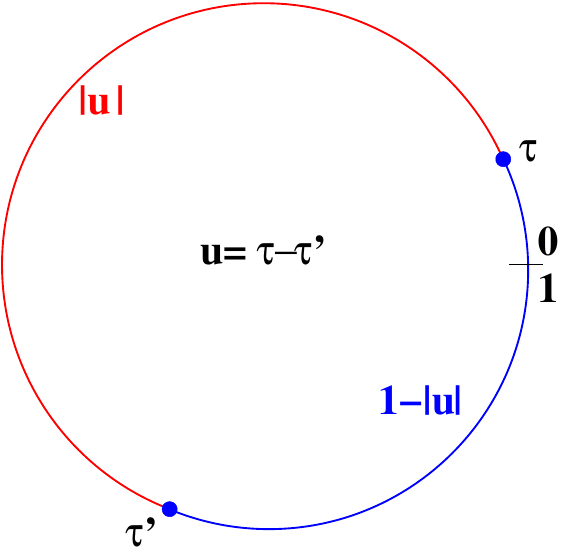}
\end{center}
\caption{Shortest distance on the circle between $\tau$ and $\tau'$\label{fig:HOcircle}}
\end{figure}

Perturbation theory about the harmonic oscillator partition function solution goes essentially the same way as  for the free particle transition amplitude considered above, except that now we may use  periodic boundary conditions for the quantum fields rather than Dirichlet boundary conditions. Of course one can keep using DBC, factor out the classical solution, with $x_{cl}(0)=x_{cl}(1)=x$ and integrate over $x$. However for completeness let us choose the former parametrization and let us focus on the case where the interacting action is polynomial $S_{int}[x] = \beta \int_0^1 d\tau \sum_{n>2} \frac{g_n}{n!} x^n$; we can define the generating functional
\begin{align}
{\cal Z}[j] = \int_{PBC} Dx~e^{-S_h[x]+\int_0^1 j x}  
\end{align}      
that similarly to the free particle case yields
\begin{align}
{\cal Z}[j] = Z_h(\beta)~e^{\frac12\int\!\!\int j {\cal D}_h^{-1} j}
\end{align}
and the propagator results
\begin{align}
& \Big\langle x(\tau') x(\tau) \Big\rangle ={\cal D}_h^{-1}(\tau-\tau') \equiv G_\omega(\tau-\tau')\nonumber\\
&\frac{m}{\beta} (-\partial_\tau^2 +(\omega\beta)^2) G_\omega(\tau-\tau') =\delta(\tau-\tau')~,
\label{eq:Green:HO}
\end{align}
on the space of functions with periodicity $\tau\cong \tau+1$, i.e. on the circle of unit circumference. On the infinite line the latter equation can be easily inverted using the Fourier transformation that yields
\begin{align}
G^l_\omega(u)=
\frac{1}{2m\omega} e^{-\beta\omega|u|}~, \quad u\equiv \tau -\tau'~.
\label{eq:prop-HO}
\end{align}
In order to get to Green's function on the circumference-one circle  we need to render~\eqref{eq:prop-HO} periodic~\cite{18}. Using Fourier analysis in~\eqref{eq:Green:HO} we get
\begin{align}
G_\omega(u) &= \frac{\beta}{m} \sum_{k\in {\mathbb Z}} \frac{1}{(\beta\omega)^2+(2\pi k)^2} e^{i2\pi k u} 
\nonumber \\&= \frac{\beta}{m}\int_{-\infty}^{\infty}d\lambda \sum_{k\in {\mathbb Z}} \delta(\lambda+k)\frac{1}{(\beta\omega)^2+(2\pi \lambda)^2} e^{i2\pi\lambda u}
\end{align}
where in the second passage we inserted an auxiliary integral. Now we can use the Poisson resummation formula $\sum_{k} f(k) = \sum_{n} \hat f(n)$ where $\hat f(\nu)$ is the Fourier transform of $f(x)$, with $\nu$ being the frequency, i.e. $\hat f(\nu) = \int dx f(x) e^{-i2\pi \nu x}$. In the above case $\hat \delta (n) = e^{i2\pi \lambda n}$. Hence the leftover integral over $\lambda$ yields 
\begin{align}
G_\omega(u) = \sum_{n\in {\mathbb Z}} G^l_\omega(u+n) 
\label{eq:Green-periodic}
\end{align}
that is explicitly periodic. The latter expression involves simple geometric series that can be summed to give
\begin{align}
G_\omega(u)=\frac{1}{2m\omega} \frac{\cosh\big( \omega\beta\big( \frac12 -|u|\big)\big)}{\sinh\big(\frac{\omega\beta}{2} \big)}~.
\label{eq:prop-HOcircle}
\end{align}
This is the Green's function for the harmonic oscillator with periodic boundary conditions (on the circle). Notice that in the large $\beta$ limit one gets an expression that is slightly different from the Green's function on the line (cfr. Eq.\eqref{eq:prop-HO}), namely:
\begin{align}
G^\infty_\omega(u)=\frac{1}{2m\omega} 
\left\lbrace
\begin{array}{ll}
e^{-\beta \omega|u|}~, & |u|<\frac12\\[1mm]
e^{-\beta \omega(1-|u|)}~, \quad &  \frac12 < |u|<1~.
\end{array}\right.
\label{eq:prop-HOinfty}
\end{align} 
Basically the Green's function is the exponential of the shortest distance (on the circle) between $\tau$ and $\tau'$, see Figure~\ref{fig:HOcircle}.

 Finally, the partition function for the anharmonic oscillator can be formally written as
\begin{align}
Z_{ah}(\beta) &= Z_h(\beta) ~e^{- S_{int}[\delta/\delta j]} ~e^{\frac{1}{2}\int\!\!\int j G_\omega j}\Big|_{j=0}\nonumber\\
&=Z_h(\beta) ~e^{\{\rm connected\ diagrams\}} ~.
\end{align}
As an example let us consider the case where a  cubic and a quartic interaction terms are present, namely 
\begin{align}
S_{int}[x] = \beta \int_0^1 d\tau \Big( \frac{g}{3!} x^3 +\frac{\lambda}{4!}x^4\Big) = \beta\left( \raisebox{-0.5cm}{\includegraphics[scale=.5]{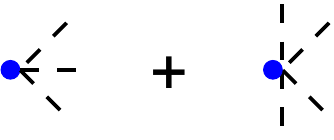}} \right)
\label{eq:ao-example}
\end{align}
so that, to lowest order in perturbation theory, the finite temperature partition function reads
\begin{align}\label{eq:partition-ah}
Z_{ah}(\beta) &=Z_h(\beta) \exp\left\lbrace \beta\left( \raisebox{-0.72cm}{\includegraphics[scale=.35]{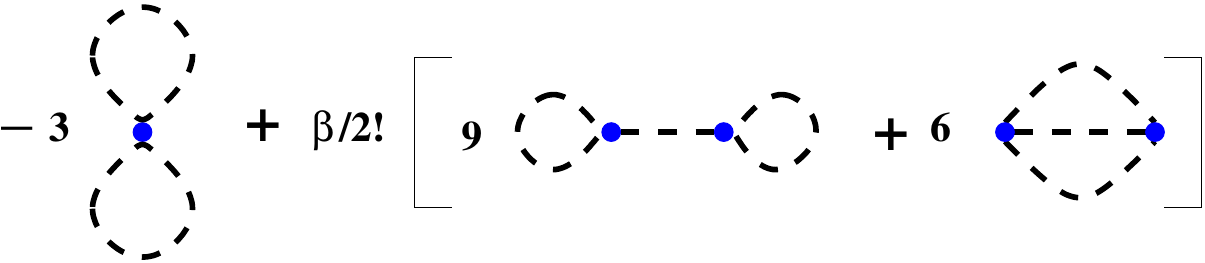}} +\cdots\right)\right\rbrace
\end{align}
and the single diagrams are given by
\begin{align}\label{eq:otto}
\raisebox{-0.72cm}{\includegraphics[scale=.35]{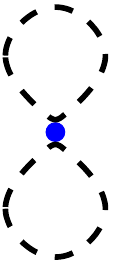}} = \frac{\lambda}{4!} \int_0^1d\tau~\big(G_\omega(0)\big)^2 \stackrel{\beta\to\infty }{\longrightarrow} &=\frac{\lambda}{4\cdot 4! (m\omega)^2}  \\ \label{eq:glasses}
\raisebox{-0.17cm}{\includegraphics[scale=.35]{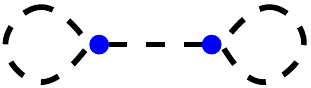}} = \left(\frac{g}{3!}\right)^2 \int_0^1\!\!\int_0^1~\big(G_\omega(0)\big)^2 G_\omega(\tau-\tau')\stackrel{\beta\to\infty }{\longrightarrow} &=\frac{g^2}{4 (3!)^2 \beta m^3\omega^4} 
\\ \label{eq:melon}
\raisebox{-0.17cm}{\includegraphics[scale=.35]{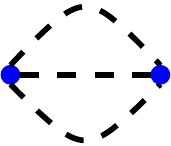}} = \left(\frac{g}{3!}\right)^2 \int_0^1\!\!\int_0^1~
\big(G_\omega(\tau-\tau')\big)^3\stackrel{\beta\to\infty }{\longrightarrow} &=\frac{g^2}{12 (3!)^2 \beta m^3\omega^4} ~.
\end{align}
Above we took the zero-temperature limit of the diagrams, which leads to 
\begin{align}
Z_{ah}(\beta) &\approx e^{-\beta E_0'}\\
E_0' &= \frac{\omega}{2}\left(1+\frac{\lambda}{16 m^2 \omega^3}-\frac{11g^2}{144 m^3 \omega^5} \right)
\label{eq:vacuum}
\end{align}
and yields the lowest perturbative order of the vacuum energy for the anharmonic oscillator with cubic and quartic interaction terms.

\subsection{Problems for Section~\ref{sec:nonrel-bos}}
\label{sec:pb-sec2}
\begin{enumerate}[(1)]
\item Starting from ${\mathbb q}|x\rangle = x |x\rangle$, show that ${\mathbb q}(t)|x,t\rangle = x |x,t\rangle$, where ${\mathbb q}(t)$ is the position operator in the Heisenberg picture and $ |x,t\rangle = e^{i{\mathbb H}t}  |x\rangle$ thus is the eigenstate of ${\mathbb q}(t)$ with eigenvalue $x$.
\item Show that the classical action for the free particle on the line is $S_f[x_{cl}]=\frac{m(x-x')^2}{2t}$.
\item Show that expression~\eqref{eq:Kf} satisfies the Schr\"odinger equation and, setting $K_f(x',x;t) = e^{i\Gamma(x',x;t)}$, show that $\Gamma$ satisfies the modified Hamilton-Jacobi equation $\frac{\partial \Gamma}{\partial t} +\frac{1}{2m}\left( \frac{\partial \Gamma}{\partial x'}\right)^2+V(x') = \frac{i}{2m} \frac{\partial^2\Gamma}{\partial x'^2}$.
\item Show that $\int_{R} dy K_f(x',y;t) K_f(y,x;t) = K_f(x',x;2t)$.  
\item Show that~\eqref{eq:del} satisfies~\eqref{eq:dddel}. 
\item Compute the 6-point correlation function. 
\item Compute the Seeley-DeWitt coefficient $a_3(x',x)$ both diagrammatically and in terms of vertex functions.
\item Compute the classical action~\eqref{eq:Scl-AH} for the harmonic oscillator on the line.
\item Show that the transition amplitude~\eqref{eq:PI-CH} satisfies the Schr\"odinger equation for the harmonic oscillator. 
\item Show that the propagator~\eqref{eq:prop-HOM} satisfies the Green equation $(\partial_\tau^2 +(\omega t)^2)\Delta_{\omega} (\tau,\tau') = \delta(\tau,\tau')$.
\item Show that the propagator~\eqref{eq:prop-HO} satisfies eq.~\eqref{eq:Green:HO}.
\item Using Fourier transformation, derive~\eqref{eq:prop-HO} from~\eqref{eq:Green:HO}. 
\item Using the geometric series obtain \eqref{eq:prop-HOcircle} from~\eqref{eq:Green-periodic} .
\item Check that, in the large $\beta$ limit, up to exponentially decreasing terms, expressions (\ref{eq:otto},\ref{eq:glasses},\ref{eq:melon}) give the same results, both 
 with the Green's function~\eqref{eq:prop-HO} and with~\eqref{eq:prop-HOinfty}.      
\item Compute the next-order correction to the vacuum energy~\eqref{eq:vacuum}.
\item Compute the finite temperature partition function for the anharmonic oscillator given by~\eqref{eq:ao-example} to leading order in perturbation theory, i.e. only consider the eight-shaped diagram.
\end{enumerate}

\section{Path integral representation of quantum mechanical transition amplitude: fermionic degrees of freedom}
\label{sec:fermi}
We employ the coherent state approach to generalize the path integral to transition amplitude of models with fermionic degrees of freedom. The simplest fermionic system is a two-dimensional Hilbert space representation of the anticommutator algebra of a pair of fermionic ladder operators,
\begin{align}
[ a, a^\dagger ]_{_+} = 1\,,\quad a^2=(a^\dagger)^2=0~.
\label{eq:ladder-f}
\end{align} 
The spin basis for such algebra is given by $(|-\rangle, |+\rangle)$ where
\begin{align}
a|-\rangle =0\,,\quad |+\rangle = a^\dagger|-\rangle\,,\quad |-\rangle = a |+\rangle
\end{align}
and a spin state is thus a two-dimensional object (a spinor) in such a basis. An alternative, overcomplete, basis for spin states is the so-called ``coherent state basis" that, for the previous simple system, is simply given by the following bra's and ket's (below we define the ``vacuum'' as the state with zero fermionic excitations, i.e. $|0\rangle \equiv|-\rangle$) 
\begin{align}
&|\xi\rangle = e^{a^\dagger \xi} |0\rangle=(1+a^\dagger \xi) |0\rangle \quad \to \quad a|\xi \rangle = \xi |\xi \rangle\nonumber \\
&\langle \bar \eta|= \langle 0| e^{\bar\eta a}= \langle 0| (1+\bar\eta a)\quad \to \quad  \langle \bar \eta| a^\dagger = \langle \bar \eta|  \bar \eta 
\label{eq:CS}
\end{align}
where $\xi$ and $\bar\eta$ are Grassmann numbers. The latter can be generalized to an arbitrary set of pairs of fermionic generators; see Appendix~\ref{app:FCS} for further details. Coherent states~\eqref{eq:CS} satisfy the following properties

\begin{align}
&\langle \bar \eta|\xi\rangle = e^{\bar \eta \xi} \label{eq:p1}\\
&\int d\bar \eta d \xi ~e^{-\bar \eta \xi} |\xi\rangle\langle \bar \eta| = {\mathbb 1}\label{eq:p2}\\
&\int d \xi ~e^{-(\bar\lambda-\bar \eta) \xi} = \bar \eta-\bar\lambda=\delta(\bar \eta-\bar\lambda)\label{eq:p3}\\
&\int d \bar \eta ~e^{\bar \eta(\rho -\xi)} = \rho-\xi=\delta(\rho-\xi)\label{eq:p4}\\
& {\rm tr} A = \int d\bar \eta d \xi ~e^{-\bar \eta \xi} \langle -\bar\eta |A|\xi\rangle = \int  d \xi d\bar \eta~e^{\bar \eta \xi} \langle \bar\eta |A|\xi\rangle\label{eq:p5}~.
\end{align} 
Let us take $|\phi\rangle$ as initial state, then the evolved state will be
\begin{align}
|\phi(t)\rangle = e^{-it{\mathbb H}} |\phi\rangle~,
\end{align} 
that in the coherent state representation becomes
\begin{align}
\phi(\bar\lambda;t) \equiv \langle\bar\lambda |\phi(t)\rangle = \langle\bar\lambda |e^{-it{\mathbb H}} |\phi\rangle = \int d\bar\eta d\eta e^{-\bar \eta \eta}\langle\bar\lambda |e^{-it{\mathbb H}} |\eta\rangle \phi(\bar\eta;0) ~.
\label{eq:fermi-Schr}
\end{align}
In the last equality we have used property~\eqref{eq:p2}. The integrand $\langle\bar\lambda |e^{-it{\mathbb H}} |\eta\rangle$ in~\eqref{eq:fermi-Schr} assumes the form of a transition amplitude as in the bosonic case.  It is thus possible to represent it with a fermionic path integral. In order to do that let us first take the trivial case ${\mathbb H}=0$  
and insert $N$  decompositions of identity $\int d\bar \xi_i d \xi_i ~e^{-\bar \xi_i \xi_i} |\xi_i\rangle\langle \bar \xi_i| = {\mathbb 1}$.  We thus get
\begin{align}
\langle\bar\lambda |\eta\rangle = \int \prod_{i=1}^N d\bar\xi_i d\xi_i ~\exp\Biggl[{\bar\lambda \xi_N -{\displaystyle \sum_{i=1}^N} \bar\xi_i(\xi_i-\xi_{i-1})}\Biggr]\,,\quad \bar\xi_N\equiv \bar\lambda\,,\ \xi_0 \equiv \eta
\end{align}  
that in the large $N$ limit can be written as 
\begin{align}
\langle\bar\lambda |\eta\rangle = \int_{\xi(0)=\eta}^{\bar\xi(1)=\bar\lambda} D\bar\xi D\xi~e^{iS[\xi,\bar\xi]}
\end{align}
with
\begin{align}
S[\xi,\bar\xi] = i\left( \int_0^1d\tau \bar\xi \dot\xi(\tau) -\bar\xi \xi(1)\right)~. 
\end{align}
In the presence of a nontrivial hamiltonian ${\mathbb H}$ the latter becomes
\begin{align}
&\langle\bar\lambda |e^{-it{\mathbb H}} |\eta\rangle  = \int_{\xi(0)=\eta}^{\bar\xi(1)=\bar\lambda} D\bar\xi D\xi~e^{iS[\xi,\bar\xi]}\label{eq:fermi-H}\\
&S[\xi,\bar\xi] = \int_0^1d\tau \left( i\bar\xi \dot\xi(\tau) -H(\xi,\bar\xi)\right)-i\bar\xi \xi(1)\label{eq:fermi-action}
\end{align}
that is the path integral representation of the fermionic transition amplitude. Here a few comments are in order: (a) the fermionic path integral resembles more a bosonic phase space path integral than a configuration space one. (b) The boundary term $\bar\xi \xi(1)$, that naturally comes out from the previous construction, plays a role when extremizing the action to get the equations of motion; namely, it cancels another boundary term that comes out from partial integration. It also plays a role when computing the trace of an operator: see below. (c) The generalization from the above naive case to~\eqref{eq:fermi-H} is a priori not  trivial, because of ordering problems. In fact ${\mathbb H}$ may involve mixing terms between $a$ and $a^\dagger$. However result~\eqref{eq:fermi-H} is guaranteed in that form (i.e. the quantum ${\mathbb H}(a,a^\dagger)$ is replaced by $H(\xi,\bar\xi)$ without the addition of counterterms) if the hamiltonian operator ${\mathbb H}(a,a^\dagger)$ is written in (anti-)symmetric form: for the present simple model, composed of a single pair of fermionic ladder operators, it simply means ${\mathbb H}_S(a,a^\dagger) = c_0+c_1 a +c_2 a^\dagger + c_3(aa^\dagger -a^\dagger a)$. In general the hamiltonian will not have such form and it is necessary to order it as ${\mathbb H} ={\mathbb H}_S+ {\rm ``counterterms"}$, where ``counterterms" come from anticommuting $a$ and $a^\dagger$ in order to put  $ {\mathbb H}$ in symmetrized form. The present ordering is called Weyl-ordering. (For details about Weyl ordering in bosonic and fermionic path integrals see~\cite{Bastianelli:2006rx}.)      

Let us now compute the trace of the evolution operator. It yields
\begin{align}
{\rm tr} ~e^{-it{\mathbb H}} = \int  d \eta d\bar \lambda~e^{\bar \lambda \eta} \langle\bar\lambda | e^{-it{\mathbb H}}| \eta \rangle = 
\int  d \eta  \int_{\xi(0)=\eta}^{\bar\xi(1)=\bar\lambda} D\bar\xi D\xi d\bar\lambda ~e^{\bar\lambda (\eta+\xi(1))} e^{i\int_0^1(i\bar\xi\dot \xi-H)}
\end{align}
then the integral over $\bar\lambda$ gives a Dirac delta that can be integrated with respect to $\eta$. Hence,
\begin{align}
{\rm tr} ~e^{-it{\mathbb H}} = \int_{\xi(0)=-\xi(1)}  D\bar\xi D\xi~ e^{i\int_0^1(i\bar\xi\dot \xi-H(\xi,\bar\xi))}
\end{align}
where we notice that the trace in the fermionic variables corresponds to a path integral with anti-periodic boundary conditions (ABC), as opposed to the periodic boundary conditions of the bosonic case. Finally, we can rewrite the latter by using real (Majorana) fermions defined as
\begin{align}
\xi =\frac{1}{\sqrt2} \big(\psi^1 +i\psi^2 \big)\,,\quad \bar\xi =\frac{1}{\sqrt2} \big(\psi^1 -i\psi^2 \big)
\end{align}
and
\begin{align}
{\rm tr} ~e^{-it{\mathbb H}} = \int_{ABC}D\psi ~e^{i\int_0^1(\frac{i}{2} \psi_a\dot \psi^a -H(\psi))}~.
\end{align}
In particular, for ${\mathbb H}=0$, we have
\begin{align}
2= {\rm tr}{\mathbb 1}  = \int_{ABC}D\psi ~e^{i\int_0^1 \frac{i}{2} \psi_a\dot \psi^a}\,,\quad a=1,2
\end{align}
that is the dimension of the Hilbert space.
For an arbitrary number of pairs of fermionic ladder operators  $a_i\,,\ a^\dagger_i\,,\ \ i=1,...,l$, the latter of course generalizes to
\begin{align}
2^{D/2}= {\rm tr}{\mathbb 1} = \int_{ABC}D\psi ~e^{i\int_0^1 \frac{i}{2} \psi_a\dot \psi^a}\,,\quad a=1,...,D=2l
\label{eq:f-DoF}
\end{align}
that sets the normalization of the fermionic path integral with anti-periodic boundary conditions. 

The fermionic action that appears in Eq.~\eqref{eq:f-DoF} plays a fundamental role in the description of relativistic spinning particles, that is the subject of the next section.

\subsection{Problems for Section~\ref{sec:fermi}}
\label{sec:pb-secfermi}
\begin{enumerate}[(1)]
\item Show that the ket and bra defined in~\eqref{eq:CS} are eigenstates of $a$ and $a^\dagger$ respectively. 
\item Demonstrate properties~\eqref{eq:p1}-\eqref{eq:p4}.
\item Show that $\langle \bar \lambda| a |\phi\rangle =\frac{\partial}{\partial \bar \eta} \phi(\bar \lambda)$.
\item Test property~\eqref{eq:p5} using $A= {\mathbb 1}$. 
\item Obtain the equations of motion from the action~\eqref{eq:fermi-action} and check that the boundary terms cancel. 
\end{enumerate}

\section{Non-relativistic particle in curved space}
\label{sec:nonrel-curved}
The generalization of non-relativistic particle path integrals to curved space was a source of many controversies in the past and several erroneous statements are present in the literature. It was only around the year 2000 that all the doubts were dispelled. The main source of controversy was, as we shall see, the appearance of a non-covariant potential. The main issue is that the transition amplitude satisfies a Schr\"odinger equation with an operator ${\mathbb H}$ that must be invariant under coordinate reparametrization (for a complete discussion about particle path integrals in curved space see the book~\cite{Bastianelli:2006rx} and references therein.)

For an infinitesimal reparametrization $x^{i'} = x^{i} +\xi^{i}(x)$ the coordinate operator must transform as  ${\mathbb x}^{i'} = {\mathbb x}^{i} +\xi^{i}({\mathbb x})$ so that ${\mathbb x}^{i'} |x\rangle = (x^{i} +\xi^{i}(x))|x\rangle$. One can easily check that 
\begin{align}
\delta {\mathbb x}^{i} = [ {\mathbb x}^{i}, G_\xi] 
\label{eq:delx}
\end{align}  
where
\begin{align}
G_\xi = \frac{1}{2i}\Big( {\mathbb p}_k \xi^k({\mathbb x})+\xi^k({\mathbb x}) {\mathbb p}_k\Big) 
\end{align}
is the generator of reparametrizations; hence
\begin{align}
\delta {\mathbb p}_i = [ {\mathbb p}_i, G_\xi] = -\frac12\Big({\mathbb p}_k \partial_i\xi^k({\mathbb x}) + \partial_i\xi^k({\mathbb x}) {\mathbb p}_k\Big) ~.
\label{eq:delp}
\end{align}
For a finite transformation we have (we remove operator symbols for simplicity)
\begin{align}
p_{i'}=\frac{\partial x^i}{\partial x^{i'}}\Big( p_i -\frac{i}{2} \partial_i \log \left| \frac{\partial x}{\partial x'}\right|\Big)	~.	 
\label{eq:p'p}
\end{align} 
It is thus easy to check that 
\begin{align}
(g')^{1/4} p_{i'} (g')^{-1/4} &= \frac{\partial x^i}{\partial x^{i'}} g^{1/4} p_{i} g^{-1/4}\label{eq:gpg-}\\
(g')^{-1/4} p_{i'} (g')^{1/4} &=  g^{-1/4} p_{i} g^{1/4}\frac{\partial x^i}{\partial x^{i'}}\label{eq:g-pg}
\end{align} 
so that the hamiltonian (we set m=1 for simplicity and for later convenience)
\begin{align}
H = \frac12 g^{-1/4} p_{i} g^{1/2} g^{ij}  p_{i} g^{-1/4}
\label{eq:H-c}
\end{align}
is Einstein invariant. Using the conventions
\begin{align}
&{\mathbb 1} = \int dx |x\rangle \sqrt{g(x)} \langle x| = \int \frac{dp}{(2\pi)^d} |p\rangle \langle p| \\
& \langle x |x'\rangle = \frac{\delta(x-x')}{\sqrt{g(x)}}\,,\quad  \langle p |p'\rangle = \delta(p-p')
\end{align}
that are consistent with
\begin{align}
\langle x |p\rangle =  \frac{e^{ix\cdot p}}{g^{1/4}(x)}
\end{align}
it is easy to convince oneself that
\begin{align}
\langle x |p_i|p\rangle = -ig^{-1/4} \partial_i g^{1/4}\frac{e^{ix\cdot p}}{g^{1/4}(x)}
\end{align}
i.e. the momentum operator in the coordinate representation is given by $p_i = -ig^{-1/4} \partial_i g^{1/4}$ and the hamiltonian above reduces to
\begin{align}
H(x) = -\frac{1}{2\sqrt{g}} \partial_i \sqrt{g} g^{ij} \partial_j 
\end{align} 
that is the known expression for the laplacian in curved space.

We want now to use the hamiltonian~\eqref{eq:H-c} to obtain a particle path integral in curved space as we did in the previous sections. The tricky point now is that here---unlike in the flat case---the metric is space-dependent and the hamiltonian cannot be written in the form $T(p) +V(x)$. Hence the Trotter formula cannot be applied straightforwardly. One possible way out to this problem is to re-write $H$ with a particular ordering that allows to apply a Trotter-like formula. One possibility relies on the Weyl ordering that amounts to re-write an arbitrary phase-space operator as a sum of a symmetric (in the canonical variables) operator plus a remainder
\begin{align}
{\cal O} (x,p) = {\cal O}_s(x,p) +\Delta {\cal O} \equiv {\cal O}_w (x,p)~. 
\end{align}      
For example
\begin{align}
x^i p_j = \frac12 \big( x^i p_j + p_j x^i\big) +\frac12 [x^i,p_j] = \big( x^i p_j\big)_s+\frac{i}{2}\delta^i_j \equiv \big( x^i p_j\big)_w~.
\end{align}
The simple reason why this is helpful is that (to avoid ambiguities we reinstate operator symbols here)
\begin{align}
\langle x'|  {\cal O}_w ({\mathbb x},{\mathbb p}) |x\rangle= \int dp ~\langle x'|  {\cal O}_w ({\mathbb x},{\mathbb p}) |p\rangle \langle p|x\rangle =  \int dp~
 {\cal O}_w \big( \frac{x'+x}{2}, p\big)\langle x' |p\rangle \langle p|x\rangle ~,
\end{align}
that is the so-called ``mid-point rule''.
For the path integral we can thus slice the evolution operator, as we did in flat space, insert decompositions of unity in terms of position and momentum eigenstates,  and get
\begin{align}
K(x',x;t) &= \int \left(\prod_{l=1}^{N-1} dx_l \sqrt{g(x_l)}\right) \left(\prod_{k=1}^{N} \frac{dp_k}{2\pi}\right)~\prod_{k=1}^{N}  \langle x_k|\left(e^{-i\epsilon {\mathbb H}}\right)_w|p_k\rangle \langle p_k |x_{k-1}\rangle
\label{eq:PI1c} 
\end{align}
where sliced operators are taken to be Weyl-ordered. The Trotter-like formula now consists of the statement $\left(e^{-i\epsilon {\mathbb H}}\right)_w \approx  e^{-i\epsilon {\mathbb H}_w}$, that allows one to use the above expression to get
 \begin{align}
K(x',x;t) &= [g(x')g(x)]^{-1/4}\int \left(\prod_{l=1}^{N-1} dx_l \right) \left(\prod_{k=1}^{N} \frac{dp_k}{2\pi}\right)\nonumber\\
&\times \exp\left\{i\sum_{k=1}^N \epsilon\left[p_k\cdot \frac{x_k-x_{k-1}}{\epsilon} -H_w(\bar x_k,p_k)\right] \right\}
\label{eq:PI1c'} 
\end{align}
where $\bar x_k = \frac12(x_k+x_{k-1})$. The Einstein-invariant hamiltonian given above, when Weyl-ordered, gives rise to the expression
\begin{align}
H_w({\mathbb x}, {\mathbb p}) = \frac12 \big( g^{ij} ({\mathbb x}) {\mathbb p}_i{\mathbb p}_j\big)_s +V_{TS}({\mathbb x}) 
\end{align}
where
\begin{align}
V_{TS} = \frac18 \Big( -R+g^{ij} \Gamma_{ia}^b \Gamma_{jb}^a\Big)~.  
\end{align}
Therefore we note that the slicing of the evolution operator in curved space produces a non-covariant potential. This might be puzzling at first. However, let us point out that a discretized particle action cannot be Einstein-invariant, so the non-covariant potential is precisely there to ``compensate" the breaking of Einstein invariance due to discretization. In other words discretization works as a regularization  and $V_{TS}$ can be interpreted as the counterterm necessary to ensure covariance of the final result. 

The continuous (formal) limit of the above integral yields the phase space path integral. One may also integrate out momenta as they appear at most quadratic and get
   \begin{align}
K(x',x;t) &= [g(x')g(x)]^{-1/4}\int \left(\prod_{l=1}^{N-1} dx_l \right) \frac{1}{(i2\pi \epsilon)^{Nd/2}} \prod_{k=1}^N \sqrt{g(\bar x_k)}\nonumber\\&
 \times\exp\left\{i\sum_{k} \epsilon\left[\frac12 g_{ij}(\bar x_k) \frac{x^i_k-x^i_{k-1}}{\epsilon}\frac{x^j_k-x^j_{k-1}}{\epsilon} -V_{TS}(\bar x_k)\right] \right\}~.
\label{eq:PI1c''} 
\end{align}
This is the configuration space path integral that, in the continuum limit, we indicate with
\begin{align}
K(x',x;t) = \int_{x(0)=x}^{x(t)=x'} {\cal D}x~e^{iS_{TS}[x]}
\end{align}
where
\begin{align}
S_{TS}[x] = \int_0^t d\tau \Big[ \frac12 g_{ij}(x(\tau)) \dot x^i \dot x^j-V_{TS}(x(\tau))\Big]  
\end{align}  
is the particle action in curved space, and 
\begin{align}
{\cal D}x \sim  \prod_{0<\tau< t} \sqrt{g(x(\tau))}  d^dx(\tau)
\end{align}  
  is the Einstein invariant formal measure. The action above corresponds to what is called a ``non-linear sigma model" as the metric, in general, is $x$-dependent. For such a reason the above path integral, in general, is not analytically solvable. One must thus rely on approximation methods such as perturbation theory, or numerical implementations. In a short-time perturbative approach we can expand the metric about a fixed point, for example the final point of the path
  \begin{align}
  g_{ij}(x(\tau)) = g_{ij} + \partial_l g_{ij} (x^l(\tau) -x'^{l}) +\frac12 \partial_k \partial_l g_{ij} (x^k(\tau) -x'^{k})(x^l(\tau) -x'^{l}) + \cdots
\end{align}    
where the tensors are evaluated at $x'$. The latter expansion leads to a free kinetic action and an infinite set of vertices. These vertices, unlike the potential vertices of the flat case, involve derivatives, i.e. they give rise to derivative interactions. These interaction vertices lead to divergences in some Feynman diagrams and a regularization procedure, that allows to deal with it, is thus needed. The Time Slicing procedure decribed above provides one possible regularization. However other regularization schemes can be employed and are caracterized by different rules than those provided by Time Slicing. They will thus be accompanied by counterterms that differ from those associated to Time Slicing. One possibility, already mentioned in Sec.~\ref{sec:free-path}, is to expand the field $x(\tau)$ in a function basis: taking the number of basis vectors---i.e. the number of modes---large but finite, provides a regularization, as the divergences are bounded by the finite number of modes. Of course this regularization, named Mode Regularization (MR), breaks Einstein invariance too, and a suitable non-covariant counterterm must be provided. Another regularization scheme used in this context is Dimensional Regularization (DR) where one dimensionally extends the one-dimensional $\tau$ space and computes dimensionally-extended Feynman diagrams.  Unlike TS and MR, DR only needs a covariant counterterm that is nothing but $-\frac18 R$. On the other hand DR is an only-perturbative regularization meaning that it allows to regularize diagrams, whereas DR and MR can be used at the non-perturbative (numerical) level as they provide a specific representation of the path integral measure. Let us conclude by mentioning a technical issue that can be faced in a rather neat way. The above formal measure involves a $\sqrt{g(x(\tau))}$ for each point of the path. In other words the measure is not Poincar\`e invariant. One may however exponentiate such measure using auxiliary fields, the so-called ``Lee-Yang" ghosts, namely
\begin{align}
\prod_{0 <\tau < t} \sqrt{g(x(\tau))} = \int Da Db Dc ~e^{i\int_0^td\tau\frac12 g_{ij}(x(\tau))(a^i a^j +b^ic^j)}
\label{eq:LY}
\end{align}     
with $a$'s being commuting fields, and $b$'s and $c$'s anticommuting fields; $Da \equiv \prod_{\tau} d^d a(\tau)$ and the same for $b$  and $c$. 
Thus the full measure $Dx Da Db Dc$ is now Poincar\`e invariant. In the perturbative approach where $g_{ij}$ is expanded in power series, the ghost action that appears in~\eqref{eq:LY} provides new vertices involving ghosts and $x$ fields, without derivatives. These new vertices collaborate in cancelling divergences in Feynman diagrams: the ghost propagator is proportional to $\delta(\tau-\tau')$ and can thus cancel divergences coming from derivatives of the $x$ propagator that involve Dirac deltas as well.   

\subsection{Problems for Section~\ref{sec:nonrel-curved}}
\label{sec:pb-secnonrel-curved}
\begin{enumerate}[(1)]
\item Compute $\delta {\mathbb x}^i$ and $\delta {\mathbb p}_i$ using the rules~\eqref{eq:delx} and~\eqref{eq:delp}.
\item Show that $[p_i, x^j ] = -i[\Omega^{-1}\partial_i \Omega, x^j ] = -i\delta_i^j$
\item Show that $[p_i, \left| \frac{\partial x}{\partial x'}\right|^a] =-ia \left| \frac{\partial x}{\partial x'}\right|^a \partial_i\log \left| \frac{\partial x}{\partial x'}\right|$, and in turn demostrate expressions~(\ref{eq:gpg-},\ref{eq:g-pg}). 
\item Use the ``symmetrized tensor law"~\cite{Bastianelli:2006rx}, $p'_i=\frac12 \{ \frac{\partial x^j}{\partial x'^i},p_j\}$,   and the result from the previous problem to demostrate expression~\eqref{eq:p'p}.
\item Show that  $(x^2 p)_s \equiv \frac13 (x^2p + xpx+ p x^2) = \frac14(x^2p + 2xpx + p x^2) = \frac12 (x^2 p +px^2)$
\item Show that $(x^2 p^2)_s \equiv \frac16 (x^2p^2 + xpxp+ xp^2 x + pxpx+p^2x^2 +px^2p) = \frac14(x^2p^2 + 2xp^2x + p^2 x^2)$
\end{enumerate}

\section{Relativistic particles: bosonic particles and $O(N)$ spinning particles}
\label{sec:rel-bosefermi}
We consider a generalization of the previous results to relativistic particles in flat space. In order to do that we start analyzing particle models at the classical level, then consider their quantization, in terms of canonical quantization and path integrals.

\subsection{Bosonic particles: symmetries and quantization. The worldline formalism}
\label{sec:bose-WF}
The dynamics of  a nonrelativistic free particle moving in a $d$-dimensional space, is described by the action
\begin{align}
S[{\bf x}] = \frac{m}{2} \int_0^td\tau~ \dot{\bf x}^2\,,\quad {\bf x} =(x^i)\,,\ i=1,...,d  
\end{align}  
which is invariant under a set of continuous global symmetries that correspond to an equal set of conserved charges. 
\begin{itemize}
\item time translation $\delta x^i = \xi \dot x^i\ \longrightarrow\ E= \frac{m}{2} \dot{\bf x}^2$, the energy
\item space translations  $\delta x^i = a^i\ \longrightarrow\ P^i= m \dot x^i$, linear momentum
\item spatial rotations $\delta x^i = \theta^{ij} x^j\ \longrightarrow\ L^{ij}= m (x^i\dot x^j-x^j\dot x^i)$, angular momentum
\item Galilean boosts $\delta x^i = v^i t\ \longrightarrow\ x^i_0:\ x^i=x_0^i +P^i t $, center of mass motion.
\end{itemize}
These symmetries are isometries of a one-dimensional euclidean space (the time) and a three-dimensional euclidean space (the space).  However the latter action is not, of course, Lorentz invariant.  

A Lorentz-invariant generalization of the free-particle action can be simply obtained by starting from the Minkowski line element $ds^2=-dt^2+d{\bf x}^2$. For a particle described by ${\bf x}(t)$ we have $ds^2=- (1-\dot{\bf x}^2 )dt^2$ that is Lorentz-invariant and measures the (squared) proper time of the particle along its path. Hence the Lorentz-invariant action for the massive free particle, referred to as ``the geometric action",  reads
\begin{align}
S[{\bf x}] = -m\int_0^t dt \sqrt{1-\dot{\bf x}^2}~.
\label{eq:geo}
\end{align}  
It is, by construction, invariant under the Poincar\'e group of transformations
\begin{itemize}
\item $x'^\mu = \Lambda^\mu{}_\nu x^\nu +a^\mu\,, \quad x^\mu= (t, x^i)\,,\quad (\Lambda^\mu{}_\nu,a^\mu) \in ISO(1,3)$,
\end{itemize}
which is thus the isometry group of Minkowski space.  Conserved charges are four-momentum, angular momentum and center of mass motion (from Lorentz boosts). The above action can be reformulated by making $x^0$ a dynamical field as well, in order to render the action explicitly Lorentz-invariant. It can be achieved by introducing a gauge symmetry. Hence
\begin{align}
S[x]= -m \int_0 ^1 d\tau \sqrt{-\eta_{\mu\nu} \dot x^\mu \dot x^\nu}
\label{eq:geo-cov}
\end{align} 
 where now $\dot{} = \frac{d}{d\tau}$, and $\eta_{\mu\nu}$ is the Minkowski metric. The latter is indeed explicitly Lorentz-invariant as it is written in the four-tensor notation and it also gauge invariant upon the reparametrization $\tau \to \tau'(\tau)$. Action~\eqref{eq:geo} can be recovered upon the gauge choice $x^0=t \tau$.  Yet another action for the relativistic particle can be obtained by introducing a gauge field, the einbein $e$, that renders explicit the above gauge invariance. 
 \begin{align}
 S[x,e] = \int_0^1 d\tau \left(\frac{1}{2e} \dot x^2- \frac{m^2 e}{2} \right)~.
 \label{eq:BdVH}
 \end{align}
 For an infinitesimal time reparametrization
 \begin{align}
 \delta \tau=-\xi(\tau)\,,\quad \delta x^\mu = \xi \dot x^\mu\,,\quad \delta e = \big( e\xi \big)^{\bullet}    
 \end{align}
 we have $\delta S[x,e] = \int d\tau \big(\xi L \big)^\bullet =0$.
 Now a few comments are in order:  (a) action~\eqref{eq:geo-cov} can be recovered from~\eqref{eq:BdVH} by replacing $e$ with its on-shell expression; namely,
 \begin{align}
 0 = m^2 + \frac{1}{e^2} \dot x^2 \quad \Rightarrow \quad e = \frac{\sqrt{-\dot x^2}}{m}~;
\end{align}   
 (b) unlike the above geometric actions, expression~\eqref{eq:BdVH}, that is known as the Brink-di Vecchia-Howe action, is also suitable for massless particles; (c) equation~\eqref{eq:BdVH} is quadratic in $x$ and therefore is more easily quantizable. In fact we can switch to the phase-space action by taking $p_\mu = \frac{\partial L}{\partial \dot x^\mu} = \dot x_\mu /e$ ($e$ has vanishing conjugate momentum, it yields a constraint) 
 \begin{align}
 S[x,p,e] = \int_0^1 d\tau \Big[p_\mu \dot x^\mu -e \frac12\big(p^2+ m^2\big)  \Big] 
 \end{align}
which is like a standard (nonrelativistic) hamiltonian action, with hamiltonian ${\cal H} =e \frac12\big(p^2+ m^2\big)\equiv e H_0$ and phase space constraint $H_0=0$. The constraint $H_0$ works also as gauge symmetry generator $\delta x^\mu = \{x^\mu, \xi H_0\} = \xi p^\mu$ and, by requiring that $\delta S =0$, one gets $\delta e =\dot \xi$.  Here $\{\ ,\ \}$ are Poisson brackets.

Upon canonical quantization  the dynamics is governed by a Schr\"odinger equation with hamiltonian operator ${\cal H}$ and the constraint becomes an operatorial constraint that must be imposed on physical states, i.e.
\begin{align}
&i\partial_\tau |\phi (\tau)\rangle = {\cal H} |\phi (\tau)\rangle = eH_0 |\phi (\tau)\rangle =0 \\
&\quad \Rightarrow\quad (p^2 +m^2)  |\phi\rangle \label{eq:KG}
\end{align}
with $|\phi\rangle$ being $\tau$ independent. In the coordinate representation Eq.~\eqref{eq:KG} is nothing but the Klein-Gordon equation. In conclusion, the canonical quantization of the relativistic, 1d-reparametrization invariant particle action~\eqref{eq:BdVH} yields a wave function that satisfies the Klein-Gordon equation. This is the essence of the ``worldline formalism" that uses (the quantization) of particle models to obtain results in quantum field theory---see~\cite{41} for a review of the method. Another important comment here  is that the local symmetry (1d reparametrization) ensures the propagation of physical degrees of freedom; i.e. it guarantees unitarity.  Before switching to path integrals let us consider the coupling to external fields: in order to achieve that, one needs to covariantize the particle momentum in $H_0$. 

For a coupling to an abelian vector field one obtains
\begin{align}
p_\mu \quad \to \quad &\pi_\mu = p_\mu -q A_\mu\quad \Rightarrow\quad \{\pi_\mu, \pi_\nu\} = q F_{\mu\nu}\label{eq:cov-mom}\\
& H_0= \frac{1}{2}\Big(\eta^{\mu\nu} \pi_\mu \pi_\nu+m^2\Big)
\end{align} 
and
\begin{align}
S[x,p,e; A_\mu] = \int_0^1 d\tau \Big[p_\mu \dot x^\mu -e \frac12\big(\pi^2+ m^2\big)  \Big] 
\label{eq:bosA-PS}
 \end{align}
with  $q$ being the charge of the particle and $F_{\mu\nu}$ the vector field strength. 
In order to switch to configuration space we just solve for $\pi_\mu$ in~\eqref{eq:bosA-PS}, $\pi_\mu= \eta_{\mu\nu}\dot x^\nu/e$ and get 
\begin{align}
S[x,e; A_\mu] = \int_0^1 d\tau \Big[\frac{1}{2e} \dot x^2 -\frac{e}{2} m^2 +q \dot x^\mu A_\mu \Big] 
\label{eq:bosA-CS}
\end{align}
so, although the hamiltonian involves a term quadratic in $A_\mu$, in configuration space the coupling between the particle worldline and the external vector field  is linear. 
The action~\eqref{eq:bosA-CS} is obviously gauge invariant upon the abelian gauge transformation $A_\mu\to A_\mu +\partial_\mu \alpha$. A naive extension of~\eqref{eq:bosA-CS} to a nonabelian vector field $A_\mu = A_\mu^a T_a$, with $T_a\in$ Lie algebra of a gauge group $G$, would not be gauge-invariant since the transformation of the  gauge potential is $A_\mu \to U^{-1}(A_\mu -i\partial_\mu) U$, with $U=e^{i\alpha^a T_a}$. However in the path integral the action enters in the exponent so it is possible to give the following gauge-covariant prescription 
\begin{align}
e^{iS[x,e;A_\mu]}  \quad \longrightarrow \quad  {\cal P} e^{iS[x,e;A_\mu]}
\end{align}   
i.e. one replaces the simple exponential with a Wilson line. Here ${\cal P}$ defines the ``path ordering" that, for the worldline integral $e^{iq\int_0^1 \dot x^\mu A_\mu}$, is nothing but the ``time-ordering" mentioned in footnote~\ref{footenote1}; namely
\begin{align}
{\cal P} e^{iq\int_0^1 \dot x^\mu A_\mu} = 1+iq\int_0^1 d\tau~\dot x^\mu A_\mu+(iq)^2\int_0^1d\tau_1 ~\dot x^{\mu_1} A_{\mu_1}  \int_0^{\tau_1}d\tau_2 ~\dot x^{\mu_2} A_{\mu_2}  +\cdots~,
\label{eq:Wilson}
\end{align}
that transforms covariantly ${\cal P} e^{iq\int_0^1 \dot x^\mu A_\mu}\to U^{-1} {\cal P} e^{iq\int_0^1 \dot x^\mu A_\mu} U$ under a gauge transformation; its trace is therefore gauge-invariant. Obviously, for abelian fields, the expression~\eqref{eq:Wilson} reduces to the conventional expansion for the exponential.  For completeness let us mention that,  for such bosonic particle,  the (minimal) coupling to gravity is immediate to achieve, and amounts to the replacement $\eta_{\mu\nu} \to  g_{\mu\nu}(x)$; for a spinning particle it would be rather more involved (see e.g.~\cite{Bastianelli:2008nm}), and it will not be treated here. 

Let us now consider a path integral for the action~\eqref{eq:bosA-CS}. For convenience we consider its Wick rotated ($i\tau \to \tau$) version (we also change $q\to -q$)
 \begin{align}
S[x,e; A_\mu] = \int_0^1 d\tau \Big[\frac{1}{2e} \dot x^2 +\frac{e}{2} m^2 +iq \dot x^\mu A_\mu \Big] ~,
\label{eq:bosA-CSE}
\end{align}
for which the path integral formally reads
\begin{align}
\int \frac{DxDe}{{\rm Vol\ (Gauge)}}~e^{-S[x,e;A_\mu]} ~,
\end{align}
where ``Vol (Gauge)" refers to the fact that we have to divide out all configurations that are equivalent upon gauge symmetry, that in this case reduces to 1d reparametrization. The previous path integral can be taken over two possible topologies: on the {\it line} where $x(0)=x$ and $x(1)=x'$, and on the {\it circle} for which bosonic fields have periodic boundary conditions.  However, such path integrals can be used to compute more generic tree-level and (multi-)loop graphs~\cite{15,Dai:2008bh}. 

\subsubsection{QM Path integral on the line: QFT propagator}
\label{sec:boseWF-line}
Worldline path integrals on the line are linked to quantum field theory propagators. In particular, from the previous 1d-reparametrization invariant bosonic model coupled to external abelian vector field,  one obtains the full propagator of scalar quantum electrodynamics (QED), i.e. a scalar propagator {\it dressed} with the insertion of an arbitrary number of couplings to $A_\mu$. 

On the line we keep fixed the extrema of $x(\tau)$ and the worldline gauge parameter is thus constrained to satisfy $\xi(0)=\xi(1)=0$, and the einbein can be gauge-fixed to an arbitrary positive constant  $\hat e \equiv 2T$ where
\begin{align}
2T \equiv \int_0^1d\tau~e \,,\quad \delta (2T) = \int_0^1d\tau~\big(e\xi\big)^{\bullet} =0~,
\end{align} 
and therefore
\begin{align}
De = dT D\xi~,
\end{align}
where $D\xi$ is the measure of  the gauge group. Moreover, there are no Killing vectors---i.e. transformations that leave $e$ invariant---since $(\hat e \xi)^{\bullet} =0$ on the line has only a trivial solution $\xi =0$.  
Hence the gauge-fixed path integral reads
\begin{align}
\Big\langle \phi(x') \bar\phi(x)\Big\rangle_{A} &= \int_0^\infty dT \int_{x(0)=x}^{x(1)=x} Dx ~e^{-S[x,2T;A_\mu]} 
\label{eq:propA} 
\end{align}
and 
\begin{align}
S[x,2T;A_\mu] = \int_0^1d\tau \Big( \frac{1}{4T} \dot x^2 + T m^2+iq \dot x^\mu A_\mu\Big)
\label{SA}
\end{align}
is the gauge-fixed action. For $A_\mu=0$ it is easy to convince oneself that~\eqref{eq:propA} reproduces the free bosonic propagator; namely
\begin{align}
\int_0^\infty dT \int_{x(0)=x}^{x(1)=x'} Dx ~e^{-\int_0^1(\frac{\dot x^2}{4T}+Tm^2)} &=  \int_0^\infty dT ~\langle x'| e^{-T({\mathbb p}^2 +m^2)} |x\rangle 
\nonumber\\&= \langle x' \big|\frac{1}{{\mathbb p}^2 +m^2} \big|x\rangle~. 
\label{eq:propS}
\end{align}  
A complete proof of~\eqref{eq:propS} will be given in Chapter~\ref{ch:ch2}.
 
In perturbation theory about the trivial vector field background, with  $A_\mu(x) =\sum_i \varepsilon^{i}_\mu e^{ip_i\cdot x}$, i.e. sum of external photons, expression~\eqref{eq:propA}  becomes nothing but the sum of the following Feynman diagrams 
\begin{align}
\int_0^\infty dT \int_{x(0)=x}^{x(1)=x'} Dx ~e^{-S[x,2T;A_\mu]}  =& \
\raisebox{-1.8cm}{
\includegraphics[scale=.45]{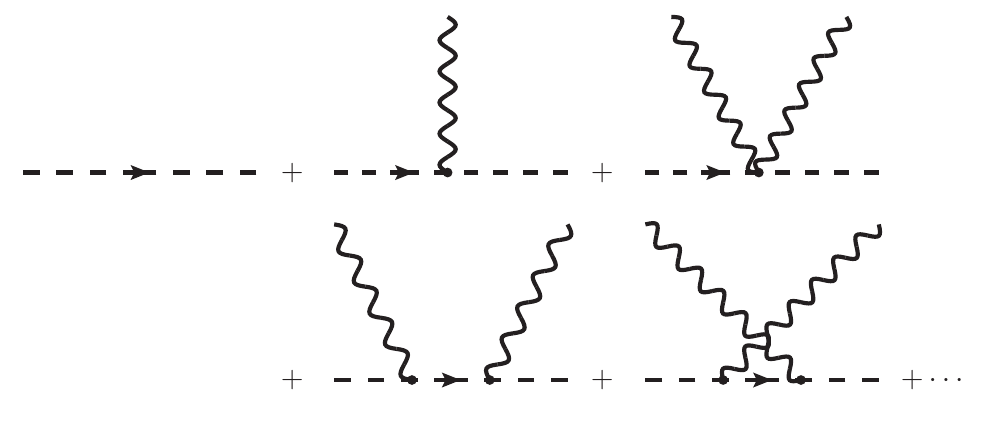} }
 \end{align}
as two types of vertices appear in scalar QED
\begin{align}
\label{ScalarQEDV}
iqA_\mu (\bar \phi \partial_\mu \phi-\phi \partial_\mu \bar \phi) \quad \longrightarrow 
\raisebox{-1cm}{
\includegraphics[scale=.4]{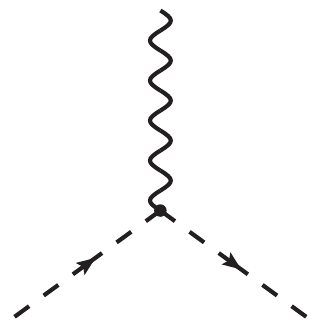} }
   \,,\qquad q^2 A_\mu A^\mu \bar \phi \phi \quad \longrightarrow 
   \raisebox{-1cm}{
\includegraphics[scale=.4]{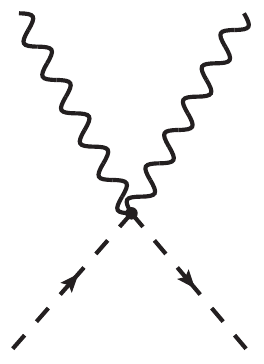} }
\end{align}
i.e. the linear vertex and the so-called ``seagull" vertex.
It is interesting to note that the previous expression for the propagator of scalar QED was already proposed by Feynman in his famous ``Mathematical formulation of the quantum theory of electromagnetic interaction,''~\cite{Feynman:1950ir} where he also included the interaction with an arbitrary number of virtual photons emitted and re-absorbed along the trajectory of the scalar particle.

\subsubsection{QM Path integral on the circle: one-loop QFT effective action}
\label{sec:boseWF-circle}
Worldline path integrals on the circle are linked to quantum field theory one-loop effective actions.  With the particle model of~\eqref{eq:bosA-CSE} the circle path integral yields the one-loop effective action of scalar QED. The gauge fixing goes similarly to previous case, except that on the circle we have periodic conditions $\xi(0)=\xi(1) $. This leaves a non-trivial solution, $\xi={\rm constant}$, for the Killing equation that corresponds to the arbitrariness on the choice of origin of the circle. One takes care of this further symmetry, dividing by the length of the circle itself. Therefore
\begin{align}
\Gamma[A_\mu] = \int_0^\infty \frac{dT}{T}  \int_{PBC} Dx ~e^{-S[x,2T;A_\mu]} 
\end{align}     
yields the worldline representation for the one loop effective action of scalar QED. Perturbatively the latter corresponds to the following sum of one-particle irreducible Feynman diagrams
\begin{align}\label{eq:QED-effaction}
\Gamma[A_\mu] =\sum 
\raisebox{-1.25cm}{
\includegraphics[scale=.45]{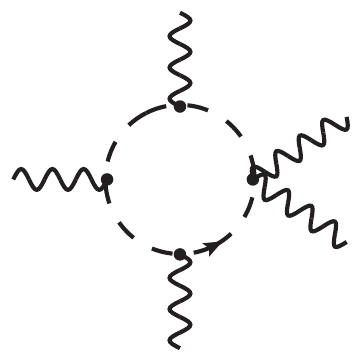}}
\end{align}
i.e. it corresponds to the sum of one-loop photon amplitudes. The figure above is meant to schematically convey the information that the scalar QED effective action involves both types of vertices.
  Further details about the many applications of the previous effective action representation will be given in Chapter~\ref{ch:ch2}.

\subsection{Spinning particles: symmetries and quantization. The worldline formalism}
\label{sec:spin-WF}
We can extend the phase space bosonic form by adding fermionic degrees of freedom. For example we can add Majorana worldline fermions that carry a space-time index $\mu$ and an internal index $i$ and get
\begin{align}
I_{sf}[x,p,\psi]=\int_0^1 d\tau \Big( p_\mu \dot x^\mu +\frac{i}{2} \psi_{\mu i}\dot \psi_i^\mu\Big) ~\,,\quad i=1,\dots, N~. 
\end{align} 
The latter expression is invariant under the following set of continuous global worldline symmetries, with their associated conserved Noether charges
\begin{itemize}
\item time translation: $\delta x^\mu = \xi p^\mu\,,\ \delta p_\mu = \delta \psi^\mu_i =0\ \longrightarrow\ H_0=\frac12 p_\mu p^\mu$
\item supersymmetries: $\delta x^\mu = i\epsilon_i \psi^\mu_i\,,\ \delta p^\mu=0\,,\ \delta \psi^\mu_i =-\epsilon_i p^\mu\ \longrightarrow\ Q_i = p_\mu \psi^\mu_i$
\item $O(N)$ rotations:  $\delta x^\mu = \delta p_\mu =0\,,\ \delta \psi^\mu_i =\alpha_{ij} \psi^\mu_j\ \longrightarrow\ J_{ij} = i \psi_{\mu i} \psi^\mu_j $
\end{itemize}
with arbitrary constant parameters $\xi,\ \epsilon_i,\ \alpha_{ij}$, and $\alpha_{ij}=-\alpha_{ij}$. Conserved charged also work as symmetry generators $\delta z = \{ z, G\}$ with $z= (x,p,\psi)$ and $G=\Xi^A G_A\equiv \xi H_0 + i\epsilon_i Q_i +\frac{1}{2} \alpha_{ij} J_{ij}$, and $\{\ ,\ \}$ being graded Poisson brackets; in flat space the generators $G_A$ satisfies a first-class algebra $\{ G_A,G_B \} = C_{AB}^CG_C$, see~\cite{Bastianelli:2008nm} for details. Taking the parameters to be time-dependent we have that the previous symplectic form transforms as
\begin{align}
\delta I_{sf}[x,p,\psi]=\int_0^1 d\tau \Big( \dot\xi H_0 +  i\dot\epsilon_i Q_i +\frac{1}{2}\dot\alpha_{ij} J_{ij} \Big) 
\label{eq:ST-local}
\end{align} 
so that we can add gauge fields $E=(e,\chi_i,a_{ij})$ and get the following locally-symmetric particle action
\begin{align}
S[x,p,\psi,E] = \int_0^1 d\tau \Big( p_\mu \dot x^\mu +\frac{i}{2} \psi_{\mu i}\dot \psi_i^\mu -eH_0 -i\chi_i Q_i -\frac12 a_{ij} J_{ij}\Big)~.
\label{eq:spinningN}
\end{align}
This is a spinning particle model with gauged $O(N)$-extended supersymmetry. The fact that the symmetry algebra is first class ensures that~\eqref{eq:spinningN} is invariant under the local symmetry generated by  $G=\Xi^A(\tau) G_A$, provided the fields $E$ transform as
\begin{align}
\begin{split}
\delta e &= \dot \xi  + 2 i \chi_i \epsilon_i\\
\delta \chi_i &= \dot \epsilon_i - a_{ij} \epsilon_j + \alpha_{ij} \chi_j\\
\delta a_{ij} &= \dot \alpha_{ij} + \alpha_{im} a_{mj} - a_{im} \alpha_{mj} 
\end{split}\label{eq:gauge-transf}
\end{align}
from which it is clear that they are gauge fields.

Upon canonical quantization Poisson brackets turn into (anti-)commutators $[p_\mu,x^\nu] = -i\delta^\nu_\mu\,,\ [ \psi_i^\mu,\psi_j^\nu]_{_+} = \delta_{ij}\eta^{\mu\nu}$. One possible representation of the previous  fermionic algebra, that is nothing but a multi-Clifford algebra, is the spin-basis, where $\psi_i^\mu$ are represented as Gamma-matrices. So, in the spin-basis and in bosonic coordinate representation the wave function is a multispinor $\phi_{\alpha_1\cdots \alpha_N} (x)$ where $\psi_i^\mu$ acts as Gamma-matrix on the $i-$th $\alpha-$index.  First class constraints again act {\it \`a la  Dirac-Gupta-Bleuler} on the wave function. In particular the susy constraints   
\begin{align}
Q_i |\phi\rangle =0\ \longrightarrow\ \big(\gamma^\mu\big)_{\alpha_i\tilde\alpha_i} \partial_\mu \phi_{\alpha_1\cdots \tilde\alpha_i \cdots \alpha_N}(x) =0
\end{align}
amount to $N$ massless Dirac equations, whereas the $O(N)$ constraints
\begin{align}
J_{ij} |\phi\rangle =0\ \longrightarrow\ \big(\gamma^\mu\big)_{\alpha_i\tilde\alpha_i} \big(\gamma_\mu\big)_{\alpha_j\tilde\alpha_j}\phi_{\alpha_1\cdots \tilde\alpha_i \cdots \tilde \alpha_j\cdots \alpha_N}(x)=0
\end{align}
are ``irreducibility" constraints, i.e. they impose the propagation of a field that is described by a single Young tableau of $SO(1,D-1)$, with $N/2$ columns and $D/2$ rows. The previous set of constraints yields Bargmann-Wigner equations for spin-$N/2$ fields in flat space. For generic $N$ only particle models in even dimensions are non-empty, whereas for $N\leq 2$ the $O(N)$ constraints are either trivial or abelian and the corresponding spinning particle models can be extended to odd-dimensional spaces. 

Coupling to external fields is now much less trivial. Coupling to gravity can be achieved by covariantizing momenta, and thus susy generators; however, for $N>2$, in a generically curved background the constraints algebra ceases to be first class. For conformally flat spaces the algebra turns into a first-class non-linear algebra that thus describes the propagation of spin-$N/2$ fields in such spaces~\cite{Bastianelli:2008nm}.    

\subsubsection{$N=1$ spinning particle: coupling to vector fields}
We consider the spinning particle model with $N=1$ that describes the first quantization of a Dirac field.  For the free model, at the classical level, the constraint algebra is simply 
\begin{align}
\{Q,Q\} = -2i H_0\,,\quad  \{Q,H_0\} =0
\end{align}
that is indeed first class. To couple the particle model to an external vector field we covariantize the momentum as in~\eqref{eq:cov-mom}, and consequently
\begin{align}
Q \equiv \pi_\mu \psi^\mu\,,\quad  \{ Q,Q\} = -2iH~,   
\end{align} 
with
\begin{align}
H= \frac12 \eta^{\mu\nu} \pi_\mu \pi_\nu +\frac{i}{2}q F_{\mu\nu}\psi^\mu\psi^\nu~,
\end{align}
and the phase-space locally symmetric action reads
\begin{align}
S[x,p,\psi,e,\chi;A_\mu] = \int_0^1d\tau \Big[ p_\mu \dot x^\mu +\frac{i}{2}\psi_\mu \dot \psi^\mu -eH -i\chi Q\Big]~,
\end{align}
whereas
\begin{align}
S[x,\psi,e,\chi;A_\mu] = \int_0^1d\tau \Big[ &\frac{1}{2e}\eta_{\mu\nu}(\dot x^\mu-i\chi\psi^\mu)(\dot x^\nu-i\chi\psi^\nu) +\frac{i}{2}\psi_\mu \dot \psi^\mu  \nonumber \\
&+q\dot x^\mu A_\mu -eq\frac{i}{2} F_{\mu\nu}\psi^\mu\psi^\nu\Big]
\end{align}
is the locally symmetric configuration space action where, along with the bosonic coupling found previously, a Pauli-type coupling between the field strength and the spinorial coordinates appears.
\subsubsection{QM Path integral on the circle: one loop QFT effective action}
\label{sec:fermiWF-circle}
We now consider the above spinning particle models on a path integral on the circle, i.e. we consider the one loop effective actions produced by the spin-$N/2$ fields whose first quantization is described by the spinning particle models. On the circle (fermionic) bosonic fields have (anti-)periodic boundary conditions. It is thus not difficult to convince oneself that gravitini $\chi_i$ can be gauged-away completely. For the $N=1$ model of the previous section this yields the spinor QED effective action
\begin{align}
\Gamma[A_\mu] = \int_0^\infty \frac{dT}{2T}  \int_{PBC} Dx \int_{ABC} D\psi ~e^{-S[x,\psi,2T,0;A_\mu]} ~,
\end{align} 
with
\begin{align}
S[x,\psi,2T,0;A_\mu] = \int_0^1d\tau \Big[ \frac{1}{4T}\dot x^2 +\frac12 \psi_\mu \dot \psi^\mu +i q \dot x^\mu A_\mu -iTq F_{\mu\nu}\psi^\mu\psi^\nu\Big]
\label{eq:N1global}
\end{align}
being the (euclidean)  gauge-fixed spinning particle action, that is globally supersymmetric. Perturbatively the previous path integral is the sum of one particle irreducible diagrams with external photons and a Dirac fermion in the loop. 

For arbitrary $N$ we will not consider the coupling to external fields as it is a too complicated topic to be covered here. The interested reader may consult the manuscript~\cite{Bastianelli:2012bn} and references therein. Let us consider instead the circle path integral for the free $O(N)-$extended spinning particle. The euclidean configuration space action can be obtained from~\eqref{eq:spinningN} by solving for the particle momenta and Wick rotating. We thus get
\begin{align}
S[x,\psi,E] = \int_0^1d\tau \Big[ \frac{1}{2e}\eta_{\mu\nu}(\dot x^\mu-\chi_i\psi_i^\mu)(\dot x^\nu-\chi_i\psi_i^\nu) &+\frac{1}{2}\psi_\mu \dot \psi^\mu  -\frac12a_{ij} \psi_{\mu i}\psi^\mu_j\Big]~,
\end{align}  
that yields the circle path integral
\begin{align}
\Gamma = \frac{1}{\rm Vol\ (Gauge)} \int_{PBC} Dx De Da \int_{ABC} D\psi D\chi ~e^{-S[x,\psi,E]}~.
\end{align}
Using~\eqref{eq:gauge-transf},  with antiperiodic boundary conditions for fermions, gravitini can be gauged away completely, $\chi_i=0$. On the other hand $O(N)$ gauge fields enter with periodic boundary conditions and cannot be gauged away completely. In fact, as shown in~\cite{Bastianelli:2007pv}, $a_{ij}$ can be gauged to a skew-diagonal constant matrix parametrized by $n=[N/2]$ angular variables, $\theta_k$. The whole effective action is proportional to the number of degrees of freedom of  fields described by a Young tableau with $n$ columns and $D/2$ rows. Such Young tableaux correspond to the field strengths of higher-spin fields. For $D=4$  this
involves all possible massless representations of the Poincar\'e group, that at the level of gauge potentials are given by 
totally symmetric (spinor-) tensors, whereas for $D>4$ it corresponds to
conformal multiplets only.

\subsection{Problems for Section~\ref{sec:rel-bosefermi}}
\label{sec:pb-secrel}
\begin{enumerate}[(1)]
\item Use the Noether trick to obtain the conserved charges for the free particle described by the geometric action~\eqref{eq:geo}.
\item Repeat the previous problem with action~\eqref{eq:geo-cov}. 
\item Show that the interaction term $L_{int} = q\dot x^\mu A_\mu$, with $A^\mu =(\phi, {\bf A})$, yields the Lorentz force. 
\item Show that, with time-dependent symmetry parameters, the symplectic form transforms as~\eqref{eq:ST-local}. 
\item Show that action~\eqref{eq:N1global} is invariant under global susy $\delta x^\mu = \epsilon \psi^\mu\,,\ \delta \psi^\mu = -\frac{1}{2T} \epsilon \dot x^\mu$.
\end{enumerate}

\section{Final Comments}
\label{eq:comments}
 In the previous lectures we reviewed some elementary material on particle path integrals and introduced a list of more or less recent topics where particle path integrals can be efficiently applied.  In particular we pointed out the role that some locally-symmetric relativistic particle models have in the first-quantized description of higher spin fields~\cite{Sorokin:2004ie}. 
However the list of topics reviewed here is by no means complete.  Firstly, coupling to external gravity has been overlooked (it will be partly covered below, in Chapter~\ref{ch:ch2}.) In fact, path integrals in curved spaces involve nonlinear sigma models and perturbative calculations have to be done carefully as superficial divergences appear. Although this issue is well studied and understood by now, it had been source of controversy (and errors) in the past (see~\cite{Bastianelli:2005rc} for a review.) Moreover, for the $O(N)$ spinning particle models discussed above, coupling to gravity is not straightforward as (for generic $N$) the background appears to be constrained~\cite{Bastianelli:2008nm} and the symmetry algebra is not linear:  the topic is an on-going research argument---for a review see~\cite{Bastianelli:2015tha} and references therein.    

Some other modern applications of particle path integrals that have not been covered here, include: the numerical worldline approach to the Casimir effect~\cite{Gies:2003cv}, AdS/CFT correspondence and string dualities~\cite{Gopakumar:2003ns}, photon-graviton amplitudes computations~\cite{61}, nonperturbative worldline methods in QFT~\cite{18,Gies:2005sb}, QFT in curved space time~\cite{Hollowood:2007kt}, the worldgraph approach to QFT~\cite{Dai:2008bh}, as well as the worldline approach to QFT on manifolds with boundary~\cite{Bastianelli:2008vh} and to noncommutative QFT~\cite{Bonezzi:2012vr}.    

\section{Appendix A: Natural Units}
\label{app:NU}
In quantum field theory it is often convenient to use so called ``natural units": for a generic physical quantity $X$, its physical units can always be express in terms of energy, angular-momentum $\times$ velocity and angular momentum as
\begin{equation}
[X] = E^a (Lc)^b L^c = m^\alpha l^\beta t^{\gamma} 
\end{equation} 
with 
\begin{equation}
\begin{split}
a &= \alpha-\beta-\gamma\\ 
b &= \beta- 2\alpha\\
c &=\gamma+ 2\alpha~.
\end{split}
\end{equation}
Therefore, if velocities are measured in units of the speed of light $c$ and angular momenta are measured in units of the Planck constant $\hbar$, and thus are numbers, we have the natural units for $X$ given by
\begin{equation}
[X]\big|_{n.u.} = E^{\alpha-\beta-\gamma}\,,\quad (\hbar c)\big|_{n.u.}  = 1\,,\quad \hbar\big|_{n.u.} =1  
\end{equation} 
and energy is normally given in MeV. Conversion to standard units is then easily obtained by using
\begin{equation}
\hbar c   = 1.97\times 10^{-11}\ {\rm MeV\cdot cm}  \,,\quad \hbar =6.58\times 10^{-22}\ {\rm MeV\cdot s} ~.
\end{equation} 
For example a distance expressed in natural units by $d = 1~( {\rm MeV} )^{-1}$, corresponds to $d = 1.97\times 10^{-11}~{\rm 
cm}$. 

Another set of units, very useful in general relativity, makes use of the Newton constant $G$ (divided by $c^2$) and the speed of light. The first in fact has dimension of inverse mass times length. For a generic quantity we thus have  
\begin{equation}
[X] = l^a v^b (m^{-1} l)^c = m^\alpha l^\beta t^{\gamma} 
\end{equation} 
with 
\begin{equation}
\begin{split}
a &= \alpha+\beta+\gamma\\ 
b &= -\gamma\\
c &=-\alpha~.
\end{split}
\end{equation}
Therefore, if velocities are measured in units of the speed of light $c$ and $m^{-1} l$ measured in units of $ G/c^2$, and thus are numbers, we have the geometric units for $X$ given by
\begin{equation}
[X]\big|_{g.u.} = l^{\alpha+\beta+\gamma}\,,\quad ( c)\big|_{g.u.}  = 1\,,\quad \frac{G}{c^2}\big|_{g.u.} =1  
\end{equation} 
and length is normally given in (centi)meters. Conversion to standard units is then easily obtained by using
\begin{equation}
c\big|_{s.i.}    = 3\times 10^{8}\ {\rm m\ s^{-1}}  \,,\quad \frac{G}{c^2}\big|_{s.i.}  =7.425\times 10^{-28}\ {\rm m\ kg^{-1}} ~.
\end{equation} 
Then for an arbitrary quantity we have
\begin{align}
X\big|_{s.i.} =\underbrace{x|_{g.u.} c^{-\gamma} \left(\frac{G}{c^2}\right)^{-\alpha}}_{x|_{s.i.}}\ {\rm kg^\alpha\ m^\beta\ s^\gamma }  
\end{align}
where $x$'s are numbers. For example, for the electron mass we have
\begin{align}
m_e|_{s.i.} = 9.11\times 10^{-31}\ {\rm kg}\,,\quad\underbrace{\longrightarrow}_{\alpha=1,\, \beta=\gamma=0} \quad  
m_e|_{g.u.} = 6.76\times 10^{-58}\ {\rm m}~.
\end{align}
The mass in geometric units represents (half of) the Schwarzschild radius of the particle, $r_S$. For astronomical objects it is a quite important quantity as it represents the event horizon in the Schwarzschild metric (for the sun we have $2M \approx 3\ {\rm km}$). Such horizon does not depend on the mass density, but only on the mass; on the other hand the physical radius does depend on the mass density.  Then, if $r_S\geq R_{phys}$ the object is very dense (the mass of the sun condensed in a sphere smaller than $3\ {\rm km}$!) and the horizon appears outside the physical radius where the  Schwarzschild solution is valid (for $r< R_{phys}$ the Schwarzchild solution is not valid). Only in that case does the spherical object give rise to a horizon and the object is a ``black hole".

To conclude let us just mention another set of units, the Planck units, where $G=c= \hbar =(4\pi\epsilon_0)^{-1} = k_B =1$, that is, all physical quantities are represented by numbers.

\section{Appendix B: Action principle and functional derivative}
\label{app:AP}
We briefly review the action principle for a particle as a way to introduce functional derivatives. A particle on a line is described by the action
\begin{align}
S[x] = \int_0^t d\tau~L(x,\dot x)~,
\end{align} 
with $x(\tau)$ describing the ``path" of the particle from time $0$ to time $t$. The classical path with fixed boundary  conditions $x(0)=x$ and $x(t)=x'$ is the one that stabilizes the functional. Considering a small variation $\delta x(\tau)$, centered on an arbitrary path  $x(\tau)$ with the above b.c.'s, so that $\delta x(0) = \delta x(t)=0$, we have
\begin{align}
\delta S[x] = \int_0^t d\tau~\delta x(\tau) \Biggl( \frac{\partial L}{\partial x} -\frac{d}{d\tau} \frac{\partial L}{\partial \dot x} \Biggr)
\end{align} 
as linear variation of the action, that can be interpreted as functional differential of the action functional, in strict analogy with multivariable differentials of a function. The functional derivative of the action functional  can thus be defined as 
\begin{align}
\frac{\delta S[x]}{\delta x(\tau)} = \frac{\partial L}{\partial x} -\frac{d}{d\tau} \frac{\partial L}{\partial \dot x} 
\end{align}
and the Euler-Lagrange equation thus follows from the vanishing of the functional derivative. That is we can treat $x(\tau)$ as a set of independent real variables, with a continuum index $\tau$ as use the analogy with multivariable calculus to define the rules for the functional derivative, that are as follows
\begin{enumerate}
\item $\frac{\delta x(\tau^i)}{\delta x(\tau^j)} = \delta(\tau^i-\tau^j)$, generalization of $\frac{\partial x^i}{\partial x^j} = \delta^i_j$;
\item $\frac{\delta}{\delta x(\tau)} \Big( S T\Big)  =\frac{\delta S}{\delta x(\tau)} T + S\frac{\delta T}{\delta x(\tau)}  $, the Leibnitz rule;
\item viewing $y(\tau)$ as a functional of  $\tau$, we can use the ``functional chain rule", $\frac{\delta S[y]}{\delta x(\tau)} = \int d\tau' \frac{\delta S[y]}{\delta y(\tau')} \frac{\delta y(\tau')}{\delta x(\tau)} $. 
\end{enumerate} 
The above analogy also allows to use indexed variables $x^i$ as shortcuts for functions $x(\tau^i)$ and (implied) sums $\sum_i$ as shortcut for integrals $\int d\tau^i$. 
 
\section{Appendix C: Fermionic coherent states}
\label{app:FCS}
This is a compendium of properties of fermionic coherent states. Further details can be found for instance in~\cite{Negele:1988vy}.

The even-dimensional Clifford algebra
\begin{eqnarray}
  [\psi^M,\psi^N]_{_+}=\delta^{MN}\,,\quad M,N=1,\dots,2l
\end{eqnarray}
can be written as a set of $l$ fermionic harmonic oscillators (the
index $M$ may collectively denote a set of indices that may involve
internal indices as well as a space-time index), by
simply taking complex combinations of the previous operators
\begin{eqnarray}
  && a^m =\frac1{\sqrt{2}} \left(\psi^{m}+i\psi^{m+l}\right)\\
  && a^\dagger_m =\frac1{\sqrt2}
  \left(\psi^{m}-i\psi^{m+l}\right)\,,\quad\quad m=1,\dots,l\\
  && [a^m,a^{\dagger}_{n}]_{_+} = \delta_{n}^m
\end{eqnarray}
and it can be thus represented in the vector space spanned by the $2^l$ orthonormal states $|{\bf
  k}\rangle = $ $\prod_m (a_m^\dagger)^{k_m} |0\rangle$ with
$a_m|0\rangle =0$ and the vector $\bf k$ has elements taking only two possible values,
$k_m=0,1$. This basis (often called spin-basis)
  yields a standard representation of the Clifford algebra,
  i.e. of the Dirac gamma matrices.

An alternative overcomplete basis is given by the coherent states
that are eigenstates of creation or annihilation operators
\begin{eqnarray}
  && |\xi\rangle= e^{a_m^\dagger \xi^m} |0\rangle\quad\rightarrow\quad
  a^m|\xi\rangle = \xi^m |\xi\rangle = |\xi\rangle \xi^m \\
  && \langle \bar\eta| = \langle 0| e^{\bar \eta_m a^m}  \quad\rightarrow\quad
  \langle \bar\eta|a^\dagger_m = \langle \bar \eta|\bar\eta_m =
  \bar\eta_m \langle \bar \eta| ~,
\end{eqnarray}
where $\xi^m$ and $\bar \eta_m$ are real anticommuting (Grassmann) numbers.
Below we list some of the useful properties satisfied by these
states. Using the Baker-Campbell-Hausdorff formula $e^X e^Y = e^Y e^X
e^{[X,Y]}$, valid if $[X,Y] = c$-number, one finds
\begin{eqnarray}
  \langle \bar \eta | \xi\rangle = e^{\bar\eta\cdot \xi}
\end{eqnarray}
that in turn implies
\begin{eqnarray}
  && \langle \bar \eta | a^m | \xi \rangle = \xi^m \langle \bar \eta |
  \xi\rangle= \frac\partial{\partial \bar\eta_m}\langle \bar \eta |
  \xi\rangle\\
  &&\langle \bar \eta | a^\dagger_m | \xi\rangle = \bar\eta_m \langle \bar \eta |
  \xi\rangle
\end{eqnarray}
so that $[\frac\partial{\partial
  \bar\eta_m},\bar\eta_{n}]_{_+}=\delta_{n}^m$.
 Defining
\begin{eqnarray}
  \label{bar-eta}
  d \bar\eta = d\bar\eta_l\cdots d\bar\eta_1\;, \qquad
  d\xi = d\xi^1\cdots d\xi^l
\end{eqnarray}
so that $d\bar\eta d\xi 
=d\bar\eta_1 d\xi^1d\bar\eta_2 d\xi^2\cdots d\bar\eta_l d\xi^l$, one finds the following relations
\begin{eqnarray}
  &&\int d\bar\eta d\xi\ e^{-\bar\eta\cdot \xi} =1\label{10}\\
  &&\int d\bar\eta d\xi\ e^{-\bar\eta\cdot \xi}\ |\xi\rangle\langle
  \bar\eta|={\mathbb 1}\label{11}
\end{eqnarray}
where ${\mathbb 1}$ is the identity in the Fock space.
One can also define fermionic delta functions
with respect to the measures (\ref{bar-eta}) by

\begin{eqnarray}
    \delta(\bar\eta-\bar\lambda) &\equiv & 
(\bar\eta_1-\bar\lambda_1)\cdots(\bar\eta_l-\bar\lambda_l) =  
\int d\xi\ e^{(\bar\lambda-\bar\eta)\cdot\xi}\\
    \delta(\rho-\xi) &\equiv & 
(\rho^l-\xi^l)\cdots(\rho^1-\xi^1) =  
\int d\bar\eta\ e^{\bar \eta\cdot (\rho-\xi)}  \;,
\end{eqnarray}
which act as

\begin{align}
& \int d\bar \eta\, \delta(\bar\eta-\bar\lambda) F(\bar \eta) =  F(\bar \lambda)\\
& \int d\rho\, \delta(\rho-\xi) G(\rho) =  G(\xi)
\end{align}
on functions that are totally arbitrary, in that the addenda that appear in the expansions of $F$ and $G$ can be of arbitrary Grassmann parity. 
Finally, the trace of an arbitrary, Grassmann-even, operator can be written  as
\begin{eqnarray}
  {\rm Tr}\, A =
    \int d\bar\eta d\xi\ e^{-\bar\eta\cdot \xi}  \langle-\bar\eta| A |\xi\rangle
    = \int     d\xi d\bar\eta
    \ e^{\bar\eta\cdot \xi}  \langle\bar\eta| A |\xi\rangle \;.
\end{eqnarray}
As a check one may compute the trace of the identity
\begin{eqnarray}
  {\rm Tr}\, {\mathbb 1} &=& \int d\xi d\bar\eta\ e^{\bar\eta\cdot \xi} \langle \bar\eta|\xi\rangle
  = \int d\xi d\bar\eta\ e^{2\bar\eta\cdot \xi} =2^l~.
\end{eqnarray}

\section{Appendix D: Noether theorem}
\label{sec:noether}
The Noether theorem is a powerful tool that relates continuous global symmetries to conserved charges. Let us consider a particle action $S[x]$: dynamics associated to this action is the same as that obtained from the action $S'[x]$ that differs from $S[x]$ by a boundary term, i.e. $S'[x]\cong S[x]$. Then, if a continuous transformation of fields $\delta x^i = \alpha \Delta x^i$, parametrized by a continuous parameter $\alpha$, yields $S[x+\alpha \Delta x] = S'[x]\,,\ \forall x^i(\tau)$, the action is said to be symmetric upon that transformation. Moreover, taking the parameter to be time-dependent the variation of the action will be given by
\begin{align}
\delta S \equiv S[x+\alpha \Delta x]- S'[x] = \int d\tau ~\dot \alpha Q~.
\label{eq:NT}
\end{align} 
The latter tells us that if we have a continuous symmetry then
\begin{align}
\dot \alpha =0 \quad\Rightarrow\quad \delta S=0~.
\end{align}
Moreover, for $\alpha(\tau)$ and  for an arbitrary trajectory $x(\tau)$, the variation of the action does not vanish. However, it does vanish ($\forall \alpha(\tau)$) on the mass shell of the particle, i.e. imposing equations of motion. Hence 
\begin{align}
\frac{d}{d\tau} Q\big|_{EoM} =0
\end{align}
in other words, $Q\big|_{EoM} $ is a conserved quantity. The formal construction that lead to~\eqref{eq:NT}, i.e. the use of a time-dependent parameter, is often called the ``Noether trick". The just-described Noether theorem holds at the classical level. Below we describe how it gets modified if we consider a quantum version of the classical particle system considered here. However let us first consider an example at the classical level; let us take the non-relativistic free particle, described by
\begin{align}
S[x] =\frac m2 \int_0^t d\tau ~\dot x^2\,,\quad x=(x^1,\dots,x^d) 
\end{align}
for which the on-shell condition is obviously $\ddot x^i=0$. A set of continuous symmetry for which the latter is invariant is:
\begin{enumerate}
\item time translation: $\delta \tau =-a\,,\quad \delta x^i = a\dot x^i$;
\item spatial translations: $\delta x^i = a^i$;
\item rotations: $\delta x^i = \theta^{ij} x^j\,,\ {\rm with}\ \theta^{ij}=-\theta^{ji}$;
\item Galilean boosts: $\delta x^i=-v^i \tau$;
\end{enumerate}
with time-independent infinitesimal parameters $a,a^i,\theta^{ij},v^i$. In order to check the invariance and obtain the conserved charges with turn the parameters into time-dependent ones $a(\tau),a^i(\tau),\theta^{ij}(\tau),v^i(\tau)$ and consider the variation of the above action. We get
\begin{enumerate}
\item $\delta S = \int_0^t d\tau~\dot a E\,,\ \longrightarrow$ conservation of energy $E=\frac m2 \dot x^2$;
\item $\delta S = \int_0^t d\tau~\dot a^i P^i\,,\ \longrightarrow$ conservation of momentum $P^i= m \dot x^i$;
\item $\delta S = \int_0^t d\tau~\dot \theta^{ij} J^{ij}\,,\ \longrightarrow$ conservation of angular momentum $J^{ij} = \frac m2 (\dot x^i x^j-x^i\dot x^j)$;
\item $\delta S = \int_0^t d\tau~\dot v^i X^i\,,\ \longrightarrow$ conservation of center of mass motion $X^i=x^i-\frac 1m P^i\tau$.
\end{enumerate}
The result is thus two-fold. On the one hand, for time-independent parameter the action is invariant off-shell, and secondly, for time-dependent parameter its variation is proportional to the conserved charge. In fact on-shell $\delta S =0$ and by partial integration one gets the conservation laws 
$\dot E|_{EoM} = \dot P^i|_{EoM}=\dot J^{ij}|_{EoM}=\dot X^i|_{EoM}=0$. The latter can be easily checked to hold.    

The quantum version of the Noether theorem can be easily obtained  in the path integral approach. In such a case we deal with sourceful functionals like
\begin{align}
F[j] = \int Dx~e^{iS[x] +i\int j x} = \int D\bar x~e^{iS[\bar x] +i\int j \bar x}
\end{align} 
where in the second equality we just renamed the dummy variable---that is we made use of what we may call the ``Shakespeare" theorem (what's in a name?). Now, let us take $\bar x = x +\delta_\alpha x$ where $\delta_\alpha x=\alpha \Delta x$ is an infinitesimal transformation of the coordinate $x$ parametrized by a continuous parameter $\alpha$, as above. If such transformation is a (classical) symmetry, the Noether theorem implies that $S[\bar x] = S[x] +\int \dot\alpha Q$, with $\alpha$ made time-dependent and $Q$ being the associated classically-conserved charge. Then to linearized level in $\alpha$ we have
\begin{align}
F[j] = \int Dx ~J_\alpha~e^{iS[x] + i\int j x} \Big[ 1 +i\int d\tau \alpha \big( j \Delta x-\dot Q\big) \Big]
\end{align}
after partial integration---$\alpha(\tau)$ can be taken to vanish at the boundary. The jacobian term $J_\alpha$ comes from the change of integration variable, from $\bar x = x +\alpha \Delta x$ to $x$, in the path integral measure. If the  Jacobian is equal one, $F[j]$ also appears on the r.h.s, so that subtracting it from both sides leaves
\begin{align}
\int Dx ~e^{iS[x] + i\int j x} \Big[ j(\tau) \Delta x(\tau)-\dot Q(\tau)\Big] =0~,
\label{eq:Qnoether}
\end{align}
that is the quantum counterpart of the classical Noether theorem. In particular setting $j=0$ one gets
\begin{align}
\int Dx ~e^{iS[x]} \dot Q =0~,
\end{align}
that is the quantum-mechanical conservation of the charge $Q$. However, notice that~\eqref{eq:Qnoether}  contains a lot more information then its classical counterpart. In fact, one can differentiate~\eqref{eq:Qnoether} w.r.t. $j$ an arbitrary number of times and set $j=0$ in the end. This leads to an infinite tower of identities called ``Ward identities". For example if we take one derivative w.r.t. $j(\tau')$ and set $j=0$ we have
\begin{align}
\int Dx ~e^{iS[x] + i\int j x} \Big[\delta(\tau-\tau') \Delta x(\tau)-i\dot Q(\tau) x(\tau')\Big] =0
\end{align} 
and similarly for higher order differentiations. On the other hand if the Jacobian is {\it not} equal to one, then the classical symmetry does not hold at the quantum level, i.e. $\int Dx ~e^{iS[x] } \dot Q = {\cal A} \neq 0$. The quantity ${\cal A} $ is called ``anomaly". Notice that, since we are dealing with infinitesimal transformations, we always (schematically) have  $J_\alpha = 1+ i\alpha {\cal J} $. The anomaly is then precisely the path integral average of ${\cal J} $. In other words, the anomaly sits in the path integral Jacobian~\cite{Fujikawa:1980vr}. For the free particle examples  considered above it is easy to see that the Jacobian is identically one.  Let us check it for instance  for the rotations: we have
\begin{align}
J_\theta = \det \frac{\partial \bar x^{i'}(\tau')}{\partial x^i(\tau)} &=\det \big( \delta^{i' i}\delta(\tau -\tau') +\theta^{i'i}   \delta(\tau -\tau')\big) \nonumber\\
&= 1 +\tr \big(\theta^{i'i}   \delta(\tau -\tau') \big)=1 
\end{align}   
where, in the second equality we used that $\det (1+A) = 1 +\tr A +o(A^2)$ if $A$ is infinitesimal, whereas the third equality follows from the antisymmetry of $\theta$. Similarly one can show that the other three symmetries considered above give rise to a unit Jacobian. 

What described in the present section for a particle generalizes to quantum field theories. In particular let us conclude mentioning that (cancellation of) anomalies have played a crucial role in theoretical physics in the past decades to describe or predict several physical processes, the $\pi^0$ decay, the Aharonov-Bohm effect and the quark top prediction, just to name a few.

 \chapter{Lectures on the Worldline Formalism}
 \label{ch:ch2}


\section{History of the worldline formalism}

In 1948, Feynman developed the path integral approach to nonrelativistic quantum mechanics (based on
earlier work by Wentzel and Dirac). Two years later, he started his famous series of papers that laid the
foundations of relativistic quantum field theory (essentially quantum electrodynamics at the time) and
introduced Feynman diagrams. However, at the same time he also developed a representation of the  
QED S-matrix in terms of relativistic particle path integrals. It appears that he considered this approach
less promising, since he relegated the information on it to appendices of  \cite{Feynman:1950ir} and
\cite{feynman:pr84}. And indeed, no essential use was made of those path integral representations
for many years after; and even today path integrals are used in field theory mainly as integrals over fields,
not over particles. Excepting an early brilliant application by Affleck et al. \cite{afalma} to pair creation by electric fields in 1984, the potential
of this particle path integral or ``worldline'' formalism to improve on standard field theory methods, at least
for certain types of computations, was recognized  only in the early nineties through
the work of Bern and Kosower \cite{berkos} and Strassler \cite{strassler}. In these lectures, we will concentrate
on the case of the one-loop effective actions in QED and QCD, and the associated photon/gluon amplitudes,
since here the formalism has been shown to be particularly efficient, and to offer some distinct advantages over
the usual Feynman diagram approach. At the end we will shortly treat also the gravitational case, which so far has been
much less explored. This is due to a number of mathematical difficulties which arise in the construction of path integrals
in curved space,  and which have been resolved satisfactorily only recently, as discussed in Section~\ref{sec:nonrel-curved}.

\section{Scalar QED}
\label{l1: scalar qed}

\subsection{The free propagator}
\label{sec: freeprop}

We start with the free scalar propagator, that is, the Green's function for the Klein-Gordon operator equation:

\bear
D_0^{x'x} \equiv \la 0|{\rm T} \phi (x') \phi (x) |0\ra = 
\langle x' | {1\over -\square + m^2} | x\rangle \, .
\label{freeprop}
\ear
We work with euclidean conventions, defined by starting in Minkowski space with the metric $(-+++)$
and performing a Wick rotation (analytic continuation)

\bear
E=k^0 = -k_0 &\to& ik_4 \, , \non
t=x^0 = -x_0 &\to & ix_4 \, .\non
\label{wick}
\ear 
Thus the wave operator $\square$ turns into the four-dimensional Laplacian,

\bear
\square = \sum_{i=1}^4 \frac{\partial^2}{{\partial x_i}^2} \, .
\label{defwaveop}
\ear
As usual in QFT we will set $\hbar = c=1$.

We exponentiate the denominator using a Schwinger proper-time parameter $T$.
This gives

\bear
D_0^{x'x} &=& 
\langle x'| \int_0^{\infty}dT\, {\rm exp}\Bigl\lbrack - T (  -\square + m^2)\Bigr\rbrack  | x\rangle
\non
&=&
 \int_0^{\infty}dT \,\e^{-m^2T}
 \langle x' | \, {\rm exp}\Bigl\lbrack - T (  -\square )\Bigr\rbrack  | x\rangle \, .
\label{c7exponentiate}
\ear
Rather than working from scratch to transform the transition amplitude
appearing under the integral into a path integral, let us compare with the formula derived in Chapter~\ref{ch:ch1}
for the transition amplitude of the free particle in quantum mechanics:

\bear
\langle \vec x',t | \vec x,0\rangle \equiv
\langle \vec x' | \e^{-it H} |\vec x\rangle 
=
 \int_{x(0)=\vec x}^{x(t)=\vec x'} {\cal D}x(\tau) \,\e^{i\int_0^t d\tau \frac{m}{2} \dot x^2}
\label{transNR}
\ear
where $H = - \frac{1}{2m} \nabla^2 $. Using (\ref{transNR}) with the formal replacements

\bear
\nabla^2 &\to & \square \, , \non
m &\to & \half \, ,\non
\tau &\to & -i \tau \, , \non
t &\to & -i T \, .\non
\label{NRtoR}
\ear
we get

\bear
D_0^{x'x}=
\int_0^{\infty}
dT\,
\e^{-m^2T}
\int_{x(0)=x}^{x(T)=x'}
{\cal D}x(\tau)\,
e^{-\int_0^T d\tau
\kinb}
\, .
\non
\label{scalpropfreepi}
\ear\no
This is the {\it worldline path integral representation} of the relativistic propagator of a scalar
particle in euclidean spacetime from $x$ to $x'$. Note that now $ \dot x^2 = \sum_{i=1}^4 \dot x_i^2$.
The parameter $T$ for us will just be an integration variable, but as discussed
in Chapter~\ref{ch:ch1}  also has a deeper mathematical meaning related to one-dimensional diffeomorphism
invariance. In the following we shall give a brief overview of how this free-particle path integral can be calculated, and shall return to this representation of the propagator later in section 3. 

Having found this path integral, let us calculate it, as a consistency check and also to start developing our technical tools.
First, let us perform a change of variables from $x(\tau)$ to $q(\tau)$, defined by

\bear 
x^\mu(\tau) = x^\mu_{cl}(\tau)+ q^\mu(\tau) =
\Bigl[x^\mu +  \frac{\tau}{T} (x^{\prime \mu}-x^\mu) \Bigr]+ q^\mu(\tau)  \, .
\label{defq}
\ear
That is, $x_{cl}(\tau)$ is the free (straight-line) classical trajectory fulfilling the path integral boundary conditions,
$x_{cl}(0)= x$, $x_{cl}(T)= x'$,  and $q(\tau)$ is the fluctuating quantum variable around it,
fulfilling the Dirichlet boundary conditions (``DBC'' in the following)

\bear
q(0) = q(T) = 0 \, .
\label{BCDBC}
\ear
This is still very familiar from quantum mechanics. 
From (\ref{defq}) we have

\bear
\dot x^{\mu}(\tau) =   \dot q^{\mu}(\tau) + \frac{1}{T} (x^{\prime\mu}-x^\mu)\, .
\label{dotx}
\ear
Plugging this last equation into the kinetic term of the path integral, and using that
$\int_0^Td\tau \dot q(\tau)=0$ on account of the boundary conditions (\ref{BCDBC}), we find

\bear
\int_0^T d\tau \kinb = \int_0^T d\tau \fourth \dot q^2 + \frac{(x'-x)^2}{4T} \, .
\label{transkin}
\ear
Implementing this change of variables also in the path integral
(this is just a linear shift and thus does not induce a Jacobi determinant factor)
we get 

\bear
D_0^{x'x}=
\int_0^{\infty}
dT\,
\e^{-m^2T}
\e^{- \frac{(x'-x)^2}{4T}}
\int_{DBC}
{\cal D}q(\tau)\,
e^{-\int_0^T d\tau \frac{1}{4}\dot q^2} \, .
\non
\label{scalpropfreepiq}
\ear\no
The new path integral over the fluctuation variable $q(\tau)$ depends only on $T$, not on $x,x'$. 
And again we need not calculate it ourselves; using eq. (\ref{freepiqm}) of Chapter~\ref{ch:ch1} and our formal substitutions (\ref{NRtoR}), and taking into
account that our path integral has $D$ components, we get 

\bear
\int_{DBC}
{\cal D}q(\tau)\,
e^{-\int_0^T d\tau \frac{1}{4}\dot q^2}
= (4\pi T)^{-\frac{D}{2}}
\, .
\label{freedet}
\ear
Thus we have 

\bear
D_0^{x'x}=
\int_0^{\infty}
dT\,
 (4\pi T)^{-\frac{D}{2}}
\e^{-m^2T}
\e^{- \frac{(x'-x)^2}{4T}} \, .
\non
\label{scalpropfreehk}
\ear\no
This is indeed a representation of the free propagator in $x$-space, but let us now Fourier transform
it (still in $D$ dimensions) to get the more familiar momentum space representation:

\bear
D_0^{p'p} &\equiv&
\int\int dx dx' \, e^{ip\cdot x} e^{ip'\cdot x'}
D_0^{x'x}
\nonumber\\
 &=& 
 \int_0^{\infty}
dT\,
 (4\pi T)^{-\frac{D}{2}}
\e^{-m^2T}
\int\int dx dx' \, e^{ip\cdot x} e^{ip' \cdot x'}
\e^{- \frac{(x'-x)^2}{4T}}
\nonumber\\
&\stackrel{x'\to x'+x}{=}&
 \int_0^{\infty}
dT\,
 (4\pi T)^{-\frac{D}{2}}
\e^{-m^2T}
\int  dx \, \e^{i(p'+p)\cdot x}
\int dx' \,  e^{ip'\cdot x'}
\e^{- \frac{x'^2}{4T}}
\nonumber\\
&=&
(2\pi)^D \delta^D(p+p') 
\int_0^\infty dT \,
e^{- T (p'^2 +m^2) } 
\nonumber \\ 
\!\!\ &=& \!\!
(2\pi)^D \delta^D(p+p') \, {1\over p^2 + m^2} \ .
\label{scalpropfreehkx}
\ear

\subsection{Coupling to the electromagnetic field}
\label{em}

We are now ready to couple the scalar particle to an electromagnetic
field $A_{\mu}(x)$.  The worldline action was obtained in Eq.(\ref{SA}):

\bear
S[x(\tau)] = \int_0^Td\tau \Bigl(\kinb +ie\,\dot x\cdot A(x(\tau)) \Bigr) \, .
\label{Sem}
\ear
This is also what one would expect
from classical Maxwell theory (the factor $i$ comes from the analytic continuation). 
We then get the ``full'' or ``complete'' propagator $D^{x'x}[A] $ for a scalar particle, that interacts
with the background field $A$ continuously while propagating from $x$ to $x'$:

\bear
D^{x'x}[A] &=&
\int_0^{\infty}
dT\,
\e^{-m^2T}
\int_{x(0)=x}^{x(T)=x'}
{\cal D}x(\tau)\,
e^{-\int_0^T d\tau\bigl(
\kinb
+ie\,\dot x\cdot A(x(\tau))
\bigr)}
\, .
\non
\label{scalproppi}
\ear\no
Note, that in \eqref{freeprop} we started with the simplest case of the propagator of a real scalar field, but we now switched to
the complex case, to be able to couple the scalar to a Maxwell field. Formally, this does not make a difference for the path integral and we shall indeed compute this for the special case of a plane wave background in section 3. This will lead us to a compact formula for tree level scattering amplitudes. 

Similarly, in Chapter~\ref{ch:ch1} we introduced already the {\it effective action} for the scalar particle in the background field: 

\bear
\Gamma [A]
&=&
\Tintm
\int_{x(T)=x(0)}{\cal D}x(\tau)\,
e^{-\int_0^T d\tau\bigl(
\kinb
+ie\,\dot x\cdot A(x(\tau))
\bigr)}
\, .
\non
\label{scalarQEDpi}
\ear\no
Note that we now have a $dT/T$, and that the path integration is over closed loops; those trajectories can
therefore belong only to virtual particles, not to real ones. The effective action contains the quantum effects
caused by the presence of such particles in the vacuum for the background field. In particular, it causes
electrodynamics to become a nonlinear theory at the one-loop level, where photons can interact with each
other in an indirect fashion.

\subsection{Gaussian integrals}

As was already mentioned, the path integral formulas (\ref{scalproppi}) and (\ref{scalarQEDpi}) 
were found by Feynman already in 1950 \cite{Feynman:1950ir}. Techniques for their efficient calculation
were, however, developed only much later. Presently there are three different methods available, namely

\benn

\item
The analytic or ``string-inspired'' approach, based on the use of worldline Green's functions.

\item
The semi-classical approximation, based on a stationary trajectory (``worldline instanton'').

\item
A direct numerical calculation of the path integral (``worldline Monte Carlo'').

\enn

In these lectures we will treat mainly the ``string-inspired'' approach; only in chapter three will we shortly
discuss also the ``semi-classical'' approach.

In the ``string-inspired'' approach all path integrals are brought into gaussian form; usually this requires
some expansion and truncation. They are then
calculated by a formal extension of the $n$-dimensional gaussian integration formulas to infinite
dimensions. This is possible because gaussian integration involves only very crude information on operators,
namely their determinants and inverses (Green's functions). We recall that, in $n$ dimensions,

\bear
\int d^nx \,\e^{- \fourth x\cdot M \cdot x } &=& 
\frac{(4\pi)^{n/2}}{(\det M)^{1/2}} \, ,
\label{gaussfree}
\\
\frac{
\int d^nx \,\e^{-\fourth x\cdot M \cdot x + x\cdot j}
}
{
\int d^nx \,\e^{-\fourth x\cdot M \cdot x }
}
 &=& 
\,\e^{j\cdot M^{-1}\cdot j} \, ,
\label{gauss}
\ear
where the $n\times n$ matrix $M$  is assumed to be symmetric and positive definite. 

\no
Also, by multiple differentiation of the second formula with respect to the components of the
vector $j$ one gets 

\bear
\frac{
\int d^nx \, x_ix_j \,\e^{-\fourth x \cdot M \cdot x }
}
{
\int d^nx \,\e^{-\fourth x \cdot M \cdot x }
}
 &=& 
2M^{-1}_{ij} \, , \nonumber\\
\frac{
\int d^nx \, x_ix_jx_kx_l \,\e^{-\fourth x \cdot M \cdot x }
}
{
\int d^nx \,\e^{-\fourth x \cdot M \cdot x }
}
 &=& 
4\Bigl( M^{-1}_{ij}M^{-1}_{kl} +  M^{-1}_{ik}M^{-1}_{jl} +  M^{-1}_{il}M^{-1}_{jk} \Bigl) \, ,\nonumber\\
\vdots && \vdots 
\label{gaussinsert}
\ear
(an odd number of $x_i$'s gives zero by antisymmetry of the integral). Note that on the right hand sides
we always have one term for each way of grouping all the $x_i's$ into pairs; each such grouping
is called a ``Wick contraction''. In the canonical formalism the same combinatorics arises from
the canonical commutator relations. 

As will be seen, in flat space calculations these formulas can be generalized to the worldline
path integral case in a quite naive way, while in curved space there arise considerable
subtleties. Those subtleties are, to some extent, present already in nonrelativistic quantum 
mechanics. 

\subsection{The $N$-photon amplitude}
\label{Nphotscal}

We will focus on the closed-loop case in this section, since it turns out to be simpler than the
propagator one that we deal with only in section 3. Nevertheless, it should be emphasized that everything that we will do in the following
for the effective action can also be done for the propagator, as we shall see further below.

We could use (\ref{scalarQEDpi}) for a direct calculation of the effective action in a derivative
expansion (see \cite{41}), but let us instead apply it to the calculation of the one-loop $N$-photon amplitudes
in scalar QED. 
This means that we will now consider the special case where the scalar particle, while
moving along the closed trajectory in spacetime, absorbs or emits a fixed but arbitrary number $N$ of 
quanta of the background field, that is, photons of fixed momentum $k$ and polarization $\varepsilon$.
In field theory, to implement this we first specialize the background $A(x)$, which so far was
an arbitrary Maxwell field, to a sum of $N$ plane waves,

\bear
A^{\mu}(x) = \sum_{i=1}^N \varepsilon^{\mu}_i\,\e^{ik_i \cdot x} \, .
\label{Apw}
\ear
We expand the part of the exponential involving the interaction with $A$ as a power series, and take the term
of order $A^N$. This term looks like

\bear
\frac{(-ie)^N}{N!}\Bigl( \int_0^Td\tau \sum_{i=1}^N \varepsilon_i\cdot \dot x(\tau) \,\e^{ik_i\cdot x(\tau)}\Bigr)^N \, .
\label{termN}
\ear
Of these $N^N$ terms we are to take only the $N!$ ``totally mixed'' ones that involve all $N$ different polarizations and momenta.
Those are all equal by symmetry, so that the $1/N!$ just gets cancelled, and we remain with

\bear
(-ie)^N\
 \int_0^Td\tau_1 \varepsilon_1\cdot \dot x(\tau_1) \,\e^{ik_1\cdot x(\tau_1)}
 \cdots
  \int_0^Td\tau_N \varepsilon_N\cdot \dot x(\tau_N) \,\e^{ik_N\cdot x(\tau_N)} \, .
\label{termNreduced}
\ear
This can also be written as

\bear
(-ie)^N V^{\gamma}_{\rm scal}[k_1,\varepsilon_1]\cdots V^{\gamma}_{\rm scal}[k_N,\varepsilon_N]
\label{rewritetermN}
\ear
where we have introduced the {\it photon vertex operator} 

\bear
 V_{\rm scal}^{\gamma}[k,\varepsilon] &\equiv &  \int_0^Td\tau\, \varepsilon\cdot \dot x(\tau) \,\e^{ik\cdot x(\tau)} \, .
\label{defVscal}
\ear
This is the same vertex operator which is used in (open) string theory to describe the emission or absorption of
a photon by a string.
We can thus write the $N$-photon amplitude as

\bear
\Gamma_{\rm scal}(k_1,\varepsilon_1;\ldots ; k_N,\varepsilon_N) &=&
(-ie)^N 
\int_0^{\infty}
\frac{dT}{T}\,
\e^{-m^2T}
\int_{x(0)=x(T)}
{\cal D}x(\tau)\,
e^{-\int_0^T d\tau
\kinb}
\nonumber\\
&& \times
 V^{\gamma}_{\rm scal}[k_1,\varepsilon_1]\cdots V^{\gamma}_{\rm scal}[k_N,\varepsilon_N]
  \nonumber\\
\label{DNpointscal}
\ear
(where we now abbreviate $x(\tau_i)=:x_i$). 
Note that each vertex operator represents the emission or absorption of a single
photon, however the moment when this happens is arbitrary and must therefore
be integrated over.   

To perform the path integral, note that it is already of Gaussian form. Doing it for arbitrary
$N$ the way it stands would still be difficult, though, due to the factors of $\dot x_i$. 
Therefore we first use a little formal exponentiation
trick, writing

\bear
\varepsilon_i \cdot \dot x_i = \e^{\varepsilon_i \cdot \dot x_i} \big\vert_{{\rm lin}(\varepsilon_i)} \, .
\label{trick}
\ear 
Now the path integral is of the standard Gaussian form (\ref{gauss}). 
Since

\bear
\int_0^Td\tau \dot x^2 = \int_0^T d\tau x \bigl(-\frac{d^2}{d\tau^2}\bigr) x
\label{IBPkin}
\ear
(the boundary terms vanish because of the periodic boundary condition)
we have the correspondence $M \leftrightarrow - \frac{d^2}{d\tau^2}$.
Thus we now need the determinant and the inverse of this operator. 
However, we first have to solve a little technical problem:
positive definiteness does not hold for the full path integral ${\cal D}x(\tau)$,
since there is a ``zero mode'';
the path integral over closed trajectories includes the constant loops, $x(\tau) = const.$, on which
the kinetic term vanishes, corresponding to a zero eigenvalue of the matrix $M$
\footnote{No such problem arose for the propagator, since $q(0)=q(T)=0$ and
$q \equiv const.$ implies that $q \equiv 0$.}.

\no
To solve it, we define the loop center-of-mass (or average position)
by

\bear
x_0^{\mu} := \frac{1}{T}\int_0^Td\tau\,  x^{\mu}(\tau) \, .
\label{defx0}
\ear
We then separate off the integration over  $x_0$, thus reducing the 
path integral to an integral over the relative coordinate $q$:
 
\begin{eqnarray}
x^{\mu}(\tau) = x^{\mu}_0  +  
q^{\mu} (\tau ), \qquad
{\dps\int}{\cal D}x(\tau) =
{\dps\int}d^D x_0{\dps\int}{\cal D} q(\tau) \, .
\label{split}
\end{eqnarray}
It follows from (\ref{defx0}), (\ref{split}) that
the variable $q(\tau)$ obeys, in addition to periodicity, the constraint equation

\begin{eqnarray}
\int_0^T d\tau\,   q^{\mu} (\tau ) = 0 \, .
\label{BCSI}
\end{eqnarray}
The zero mode integral can be done immediately, since it factors out as
(since $\dot x = \dot q$)

\bear
{\dps\int}d^D x_0 \,\e^{i\sum_{i=1}^N k_i\cdot x_0} = (2\pi)^D \delta^D\Bigl(\sum_{i=1}^N k_i\Bigr) \, .
\label{x0int}
\ear
This is just the expected global delta function for energy-momentum conservation.

In the reduced space of the $q(\tau)$'s the operator $ M = - \frac{d^2}{d\tau^2}$
has only positive eigenvalues. In the exercises you will show that, in this space,

\bear
{\rm det} M = (4T)^D
\label{det}
\ear 
and that the Green's function of $M$ corresponding to the above treatment of the zero mode is

\bear
G_B^{\rm c}(\tau,\tau') \equiv
2\la \tau | \Bigl(\frac{d^2}{d\tau^2}\Bigr)^{-2} | \tau' \ra_{SI}
=
| \tau-\tau'| 
-{{(\tau-\tau')}^2\over T}
-{T\over 6}
\, .
\label{defGBc}
\ear
\no
Note that this Green's functions is the inverse of $\frac{1}{2}\frac{d^2}{d\tau^2}$
rather than $-\frac{d^2}{d\tau^2}$, a convention introduced in \cite{strassler}. 
Note also that it is a function of the difference $\tau - \tau'$ only; the reason is that
the boundary condition (\ref{BCSI}) does not break the translation invariance in $\tau$.
The subscript ``B'' stands for ``bosonic'' (later on we will also introduce a ``fermionic'' Green's function) and
the subscript ``SI'' stands for ``string-inspired'', since in string theory the zero mode of the worldsheet path integral
is usually fixed analogously. 
The superscript ``c'' refers to the inclusion of the constant (= coincidence limit) $-T/6$.
It turns out that, in flat space calculations, this constant is irrelevant and can be omitted. Thus in flat space
we will usually use instead

\bear
G_B(\tau,\tau') &\equiv&
| \tau-\tau'| 
-{{(\tau-\tau')}^2\over T}
\, .
\label{defGB}
\ear
\no
This replacement does not work in curved space calculations, though. 
Below we will also need the first and second derivatives of this Green's function, which
are

\bear
\dot G_B(\tau,\tau') &=& {\rm sign}(\tau-\tau') - 2 \frac{\tau-\tau'}{T} \, , \label{dotG}\\
\ddot G_B(\tau,\tau') &=& 2\delta(\tau-\tau') - \frac{2}{T} \, .\label{ddotG}
\ear
Here and in the following a ``dot'' always means a derivative with respect to the first variable;
since $G_B(\tau,\tau')$ is a function of $\tau - \tau'$, we can always rewrite $\partder{}{\tau'}=-\partder{}{\tau}$.

To use the gaussian integral formula (\ref{gauss}), we define

\bear
j(\tau) &\equiv& \sum_{i=1}^N\Bigl(i\delta(\tau-\tau_i)k_i - \delta'(\tau-\tau_i)\varepsilon_i\Bigr) \, .
\label{defj}
\ear
This enables us to rewrite (recall that $\int_0^T d\tau \delta'(\tau-\tau_i)q(\tau) = - \dot q(\tau_i)$)

\bear
\e^{\sum_{i=1}^N(i k_i\cdot q_i + \varepsilon_i\cdot \dot q_i )} 
=
\e^{\int_0^Td\tau j(\tau)\cdot q(\tau)} \, .
\label{rewriteexponent}
\ear
Now the formal application of (\ref{gauss}) yields

\bear
\frac
{
\int
{\cal D}q(\tau)\,
e^{-\int_0^T d\tau
\frac{1}{4}\dot q^2}
\,\e^{\sum_{i=1}^N\bigl(i k_i\cdot q_i + \varepsilon_i\cdot \dot q_i \bigr)} 
}
{
\int
{\cal D}q(\tau)\,
e^{-\int_0^T d\tau
\frac{1}{4}\dot q^2}
}
&=&
\exp\Bigl[-\half \int_0^Td\tau\int_0^Td\tau' G_B(\tau,\tau')j(\tau)\cdot j(\tau')\Bigr]
\nonumber\\
&& \hspace{-80pt}
= \exp\Bigg\{\sum_{i,j=1}^N\left[
 \half\G_{Bij}k_{i}\cdot k_{j}
-i\dot G_{Bij}\varepsilon_{i}\cdot k_{j}+ \half 
\ddot G_{Bij}\varepsilon_{i}\cdot\varepsilon_{j}\right]\Bigg\}
\, .
\nonumber\\
\label{intqprop}
\ear
Here we have abbreviated $G_{Bij}\equiv G_B(\tau_i,\tau_j)$, and in the second step we have used
the antisymmetry of $\dot G_{Bij}$. Note that a constant added to $G_B$ would have no effect, since
it would modify only the first term in the exponent, and by a term that vanishes by momentum conservation.
This justifies our replacement of $G_B^{\rm c}$ by $G_B$.

Finally, we need also the absolute normalization of the free path integral, which turns out to be the same as in the
DBC case:

\bear
\int
{\cal D}q(\tau)\,
e^{-\int_0^T d\tau
\frac{1}{4}\dot q^2}
= (4\pi T)^{{\color{blue} -}\frac{D}{2}} \, .
\label{SInormalize}
\ear
Putting things together, we get the famous ``Bern-Kosower master formula'':

\begin{align}
&\Gamma_{\rm scal}(k_1,\varepsilon_1;\ldots;k_N,\varepsilon_N)
=
\non
&={(-ie)}^N
{(2\pi )}^D\delta (\sum k_i)
{\dps\int_{0}^{\infty}}{dT\over T}
{(4\pi T)}^{-{D\over 2}}
e^{-m^2T}
\prod_{i=1}^N \int_0^T 
d\tau_i
\nonumber\\
&
\times
\exp\biggl\lbrace\sum_{i,j=1}^N 
\Bigl\lbrack  \half G_{Bij} k_i\cdot k_j
-i\dot G_{Bij}\varepsilon_i\cdot k_j
+\half\ddot G_{Bij}\varepsilon_i\cdot\varepsilon_j
\Bigr\rbrack\biggr\rbrace
\mid_{\rm {\rm lin}(\pol_1,\ldots,\pol_N)}
\, .
\nonumber\\
\label{scalarqedmaster}
\end{align}
\no
This formula (or rather its analogue for the QCD case, see below) was first derived by Bern and Kosower from string theory \cite{berkos},
and rederived in the present approach by Strassler \cite{strassler}.
As it stands, it represents the one-loop $N$- photon amplitude in scalar QED, but 
Bern and Kosower also derived a set of rules which allows one to construct,
starting from this master formula and by purely algebraic means,
parameter integral representations for the $N$-photon amplitudes 
with a fermion loop, as well as for the $N$-gluon amplitudes involving a
scalar, spinor or gluon loop \cite{berkos,berntasi,41}. However, there is a part of those rules
that is valid only after imposing on-shell conditions, while 
the master formula itself is still valid completely off-shell.

\subsection{The vacuum polarization}

Let us look at the simplest case of the two-point function (the photon propagator or vacuum polarization). 
For $N=2$ we get from the master formula (\ref{scalarqedmaster}), 
after expanding out the exponential,

\bear
\Gamma_{\rm scal}(k_{1},\varepsilon_{1};k_{2},\varepsilon_2)
 &=&(-ie)^{2}(2\pi)^{D}\delta(p_1+p_2)
 \int_{0}^{\infty} \frac{dT}{T}(4\pi T)^{-D/2}e^{-m^2 T}
 \nonumber\\&&\times 
 \int_{0}^{T}d\tau_{1}\int_0^Td\tau_2 \, (-i)^2 P_2 \,e^{G_{B12}k_1\cdot k_2}
\label{expand}
\ear
where

\bear
P_2 = 
\dot{G}_{B12}\varepsilon_{1}\cdot k_{2}\,\dot{G}_{B21}\varepsilon_{2}\cdot k_1
-\ddot{G}_{B12}\, \varepsilon_{1}\cdot\varepsilon_{2} \, .
\label{P2}
\ear
We could now perform the parameter integrals straight away. However, it is 
advantageous to first remove the term involving $\ddot G_{B12}$ by an IBP.
This transforms $P_2$ into $Q_2$, 

\bear
Q_2 &=&
\dot{G}_{B12}\dot{G}_{B21}
\bigl(\varepsilon_{1}\cdot k_{2}\varepsilon_{2}\cdot k_1- \varepsilon_1\cdot\varepsilon_2 k_1\cdot k_2 \bigr) \, .
\label{Q2}
\ear
We use momentum conservation to set $k_1=-k_2\equiv k$, and define

\bear
\Gamma_{\rm scal}(k_{1},\varepsilon_{1};k_{2},\varepsilon_2) \equiv (2\pi)^D\delta(p_1+p_2)\pol_1\cdot\Pi_{\rm scal}\cdot\pol_2 \, .
\label{GammatoPi}
\ear
Then we have

\bear
\Pi_{\rm scal}^{\mn}(k) 
 &=& e^2(\delta^{\mu\nu}k^2 -k^{\mu}k^{\nu})
 \int_{0}^{\infty} \frac{dT}{T}(4\pi T)^{-D/2}e^{-m^2 T}
 \nonumber\\ && \times
 \int_{0}^{T}d\tau_{1}\int_0^Td\tau_2 \dot{G}_{B12}\dot{G}_{B21}
 \,e^{G_{B12}k_1\cdot k_2}
 \, .
\label{Piscal}
\ear
Note that the IBP has had the effect to factor out the usual transversal projector
$\delta^{\mu\nu}k^2 -k^{\mu}k^{\nu}$ already at the integrand level.

We rescale to the unit circle, 
$\tau_i = Tu_i, i = 1,2$, and use the translation
invariance in $\tau$ to fix the zero to 
be at the location of the second vertex operator,
$u_2=0, u_1=u$.
We have then

\bear
G_B(\tau_1,\tau_2)&=&Tu(1-u) \, .\nonumber\\
\dot G_B(\tau_1,\tau_2)&=&1-2u \, .
\nonumber\\
\label{scaledown}
\ear\no

\bear
\Pi_{\rm scal}^{\mn}(k) 
&=&
-\frac{e^2}{(4\pi)^{\frac{D}{2}}}
(\delta^{\mu\nu}k^2 -k^{\mu}k^{\nu})
\Tint\e^{-m^2T}
T^{2-{D\over 2}}
\non&& \times
\int_0^1du
{(1-2u)}^2
\e^{-Tu(1-u)k^2} \,.
\label{Piscal1}
\ear
Using the elementary integral

\bear
\int_0^{\infty}\frac{dx}{x} \,x^{\lambda} \,\e^{-ax} = \Gamma(\lambda)a^{-\lambda}
\label{Tint}
\ear
(for $a>0$) we get finally

\bear
\Pi_{\rm scal}^{\mn}(k) 
&=&
-{e^2\over {(4\pi )}^{D\over 2}}
(\delta^{\mu\nu}k^2 -k^{\mu}k^{\nu})
\Gamma\left(2-{D\over 2}\right)
\nonumber\\&&\times
\int_0^1du
(1-2u)^2
{\Bigl[
m^2 + u(1-u)k^2
\Bigr]}^{{D\over 2}-2}
\, .
\label{Piscal2}
\ear\no
This should now be renormalized, but we need not pursue this here. 

\begin{figure}[h]
\hspace{100pt}{\centering
\includegraphics[scale=.5]{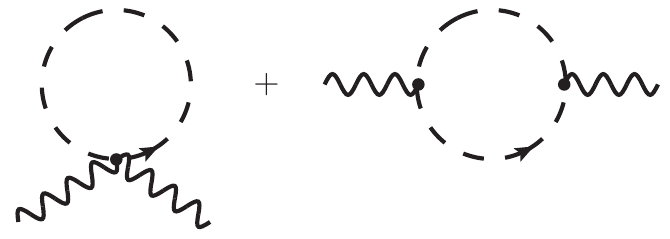}
}
\caption{Vacuum polarization diagrams in Scalar QED\label{Fig:VP}}
\label{fig1}
\end{figure}

Our result agrees, of course, with a computation of the two corresponding Feynman diagrams,
Fig.~\ref{Fig:VP}.

In this simple case our integrand before the IBP, (\ref{expand}), would have still allowed a direct comparison
with the Feynman diagram calculation; namely, the diagram involving the quartic vertex matches with the
contribution of the $\delta(\tau_1-\tau_2)$ part of the $\ddot G_{B12}$ contained in $P_2$.

Finally, it should be mentioned that, although any gaussian integral can be brought to the standard form of
(\ref{gauss}) by formal exponentiations such as (\ref{trick}), this is not always the most efficient way to
proceed. Alternatively, one can use the following set of rules for Wick contractions involving elementary
fields as well as exponentials of fields:

\begin{enumerate}

\item
The basic Wick contraction of two fields is

\bear
\langle q^{\mu}(\tau_1)q^{\nu}(\tau_2)\rangle = - G_B(\tau_1,\tau_2)\delta^{\mu\nu} \, .
\label{Wickbasic}
\ear

\item
Wick contract fields among themselves according to (\ref{gaussinsert}), e.g.,

\begin{align}
\Bigl\langle
q^{\kappa}(\tau_1)q^{\lambda}(\tau_2)
q^{\mu}(\tau_3)q^{\nu}(\tau_4)
\Bigr\rangle
&=
G_{B12}
G_{B34}
\delta^{\kappa\lambda}\delta^{\mu\nu}
+G_{B13}G_{B24}
\delta^{\kappa\mu}\delta^{\lambda\nu}
\non
&+G_{B14}
G_{B23}
\delta^{\kappa\nu}\delta^{\lambda\mu}
\, .
\label{4qcorr}
\end{align}\no

\item
Contract fields with exponentials according to

\bear
\Bigl\langle q^{\mu}(\tau_1)
\,\e^{ik\cdot q(\tau_2)}\Bigr\rangle
= i\langle q^{\mu}(\tau_1)q^{\nu}(\tau_2)
\rangle k_{\nu}
\e^{ik\cdot q(\tau_2)}
\label{wickfieldexp}
\ear\no
(the field $q^{\mu}(\tau_1)$ disappears, the exponential remains).

\item
Once all elementary fields have been eliminated, 
the contraction of the remaining
exponentials yields a universal 
factor

\bear
\Bigl\langle
\e^{ik_1\cdot q_1}
\cdots
\e^{ik_N\cdot q_N}
\Bigr\rangle 
&=&
\exp\biggl[
-\half
\sum_{i,j=1}^N
k_{i\mu}
\langle
q^{\mu}(\tau_i)
q^{\nu}(\tau_j)
\rangle
k_{j\nu}
\biggr]
\nonumber\\&=&
\exp\biggl[
\half
\sum_{i,j=1}^N
G_{Bij}
k_i
\cdot
k_j
\biggr]
\, .
\label{wickNexp}
\ear\no
\end{enumerate}
\no
Note also that, no matter how the path integral is constructed, the $\langle \rangle $ operation 
must respect linearity, so that, in particular, it commutes with taking derivatives of the fields. 


\section{Spinor QED}
\label{l2: spinor qed}

\subsection{Feynman's vs Grassmann representation}

In \cite{feynman:pr84}, Feynman presented the following generalization of the formula
(\ref{scalarQEDpi}) for the effective action to the Spinor QED case:

\bear
\Gamma_{\rm spin}[A]
&=&
-\half \Tintm
\int_{P}{\cal D}x(\tau)\,
e^{-\int_0^T d\tau\bigl(
\kinb
+ie\,\dot x\cdot A(x(\tau))
\bigr)}
{\rm Spin}[x(\tau),A]
\, .
\non
\label{spinorQEDpi}
\ear\no
Here ${\rm Spin}[x(\tau),A]$ is the ``spin factor''

\bear
{\rm Spin}[x(\tau),A] &=& {\rm tr}_{\gamma} {\cal P}
\exp\biggl[{i\frac{e}{4}\,[\gamma^{\mu},\gamma^{\nu}]
\int_0^Td\tau \, F_{\mu\nu}(x(\tau))}\biggr]
\label{defspinfactor}
\ear
where ${\rm tr}_{\gamma}$ denotes the Dirac trace, and
${\cal P}$ is the path ordering operator.
The comparison of (\ref{scalarQEDpi}) with (\ref{spinorQEDpi}) makes it clear that
the $x$-path integral, which is the same as we had for the scalar case, represents
the contribution to the effective action due to the orbital degree of freedom of the
spin $\half$ particle, and that all the spin effects are indeed due to the spin factor.
The minus sign in front of the path integral implements the Fermi statistics.

A more modern way of writing the same spin factor is 
in terms of an additional Grassmann path integral
\cite{fradkin,bermar,bdzdh,brdiho}. As we noted already in chapter 1, one has

\bear
{\rm Spin}[x(\tau),A]  &=&
\int_{A} {\cal D}\psi(\tau)
\,
\exp 
\Biggl\lbrack
-\int_0^Td\tau
\Biggl(
\half \psi\cdot \dot\psi -ie \psi^{\mu}F_{\mn}(x(\tau))\psi^{\nu}
\Biggr)
\Biggr\rbrack
\, .
\nonumber\\
\label{spinfactorgrass}
\ear
Here the path integration is over the space of anticommuting
functions antiperiodic in proper-time,
$\psi^{\mu}(\tau_1)\psi^{\nu}(\tau_2) = - \psi^{\nu}(\tau_2)\psi^{\mu}(\tau_1)$,
$\psi^{\mu}(T) = - \psi^{\mu}(0)$ (which is indicated by the subscript `A' on the
path integral). The exponential in (\ref{spinfactorgrass}) is now an ordinary one,
not a path-ordered one. 
The $\psi^{\mu}$'s effectively replace the Dirac matrices $\gamma^{\mu}$,
but are functions of the proper-time, and thus will appear in all possible orderings
after the expansion of the exponential. This fact is crucial for extending to the Spinor QED
case the above-mentioned ability of the formalism
to combine the contributions of Feynman diagrams with different orderings of
the photon legs around the loop.
Another advantage of introducing this second path integral is that, as it turns out,
there is a ``worldline'' supersymmetry between the coordinate function
$x(\tau)$ and the spin function $\psi(\tau)$ \cite{bdzdh}. 
Namely, the total worldline Lagrangian

\bear
L =  \kinb +ie\,\dot x\cdot A + \half \psi\cdot \dot\psi -ie \psi^{\mu}F_{\mn}\psi^{\nu}
\label{defLspin}
\ear
is, up to a total derivative term, invariant under the transformation

\bear
\delta x^{\mu} &=& -2 \eta\psi^{\mu} \, ,\nonumber\\
\delta \psi^{\mu} &=& \eta \dot x^{\mu}\, , \nonumber\\
\label{susy}
\ear
with a constant Grassmann parameter $\eta$.
Although this ``worldline supersymmetry'' 
is broken by the different periodicity conditions for $x$ and $\psi$, it still has
a number of useful computational consequences.

\subsection{Grassmann gauss integrals}

We need to generalize the formulas for the usual gaussian integrals to the case of
Grassmann (anticommuting) numbers.
For the treatment of QED we can restrict ourselves to real Grassmann variables.

\no
First, for a single Grassmann variable $\psi$ we define the Grassmann integration
by setting 

\bear
\int d\psi \psi = 1
\label{defintpsi}
\ear
and imposing linearity. Since $\psi^2=0$, the most general function of $\psi$ is 
of the form $f(\psi)= a + \psi b$, with constants $a$ and $b$ that could be either bosonic
or fermionic. In either case its integral will be

\bear
\int d\psi f(\psi) = b \, .
\label{psiintgen}
\ear
Note that we prefer to write $\psi b$, rather than $b \psi$, to avoid having to make a case distinction.

For two Grassmann variables $\psi_1,\psi_2$ we have the most general function
$f(\psi_1,\psi_2) = a + \psi_1 b + \psi_2 c+\psi_1\psi_2 d$, and 

\bear
\int d\psi_1 \int d\psi_2 f(\psi_1\psi_2)  = - d \, .
\label{psi1psi2int}
\ear
We can then also form a gaussian integral: let $\psi = (\psi_1,\psi_2)$ and 
$M$ a real antisymmetric matrix. Then

$$\e^{-\half \psi^T\cdot M\cdot\psi} = \e^{-M_{12}\psi_1\psi_2} = 1 - M_{12}\psi_1\psi_2$$

\no
(note that a symmetric part added to $M$ would cancel out, so that we can restrict ourselves to the
antisymmetric case from the beginning)
and

\bear
\int d\psi_1 \int d\psi_2 \,\e^{-\half \psi^T\cdot M\cdot\psi} = M_{12} \, .
\label{gg2}
\ear
On the other hand,

\bear
{\rm det}M = - M_{12}M_{21}= M_{12}^2 \, .
\label{detM}
\ear

\no
Thus we have

\bear
\int d\psi_2 \int d\psi_1 \,\e^{-\half \psi^T\cdot M\cdot\psi} = \pm ({\rm det}M)^{\half} \, .
\label{gg2fin}
\ear
It is easy to show show that this generalizes to any even dimension:
let $\psi_1,\ldots,\psi_{2n}$ be Grassmann variables and $M$ an antisymmetric
$2n\times 2n$ matrix. Then

\bear
\int d\psi_1 \cdots \int d\psi_{2n} \,\e^{-\half \psi^T\cdot M\cdot\psi} = \pm ({\rm det}M)^{\half} \, .
\label{gg2n}
\ear
Thus we have a determinant factor in the numerator, instead of the denominator as 
we had for the ordinary gauss integral (\ref{gaussfree}). 
The formula (\ref{gauss}) also generalizes to the Grassmann case, namely one finds

\begin{align}
\frac{
\int d\psi_1 \cdots \int d\psi_{2n}  \,\e^{-\half \psi^T\cdot M \cdot \psi + \psi\cdot j}
}
{
\int d\psi_1 \cdots \int d\psi_{2n} \,\e^{-\half \psi^T \cdot M \cdot \psi}
}
 &=
\,\e^{\half j\cdot M^{-1}\cdot  j} \, .
\label{gaussgrassmann}
\end{align}
By differentiation with respect to the components of $j$ one finds the Grassmann analogue of the Wick contraction
rules (\ref{gaussinsert}):

\begin{align}
\frac{
\int d\psi_1 \cdots \int d\psi_{2n} \psi_i\psi_j \,\e^{-\half \psi^T\cdot M \cdot \psi}
}
{
\int d\psi_1 \cdots \int d\psi_{2n} \,\e^{-\half \psi^T \cdot M \cdot \psi}
}
 &= 
M^{-1}_{ij} \, , \nonumber\\
\frac{
\int d\psi_1 \cdots \int d\psi_{2n} \psi_i\psi_j\psi_k\psi_l \,\e^{-\half \psi^T\cdot M \cdot \psi}
}
{
\int d\psi_1 \cdots \int d\psi_{2n} \,\e^{-\half \psi^T \cdot M \cdot \psi}
}
&= 
M^{-1}_{ij}M^{-1}_{kl} -  M^{-1}_{ik}M^{-1}_{jl} +  M^{-1}_{il}M^{-1}_{jk}  \, , \nonumber\\
\vdots\hskip1cm &\hskip1.5cm\vdots 
\label{gaussinsertgrass}
\end{align}
Note that we have now alternating signs in the permutations in the second formula in (\ref{gaussinsertgrass});
this is because now also the source $j$ must be Grassmann-valued.

Returning to our Grassmann path integral (\ref{spinfactorgrass}), we see that the matrix $M$ now corresponds to the
first derivative operator $\frac{d}{d\tau}$, acting in the space of antiperiodic functions. Thus we need its inverse, which is
very simple: one finds

\bear
G_F(\tau,\tau') \equiv
2\la \tau | \Bigl(\frac{d}{d\tau}\Bigr)^{-1} | \tau' \ra
=
{\rm sign}(\tau - \tau') 
\, .
\label{defGF}
\ear
\no
Note that in the antiperiodic case there is no zero mode problem. 
Thus we find the Wick contraction rules

\begin{align}
\langle \psi^{\mu}(\tau_1)\psi^{\nu}(\tau_2)\rangle &= \half G_F(\tau_1,\tau_2)\delta^{\mu\nu} \, ,
\nonumber\\
\Bigl\langle
\psi^{\kappa}(\tau_1)\psi^{\lambda}(\tau_2)
\psi^{\mu}(\tau_3)\psi^{\nu}(\tau_4)
\Bigr\rangle 
&=
\fourth
\biggl[
G_{F12}
G_{F34}
\delta^{\kappa\lambda}\delta^{\mu\nu}
-G_{F13}G_{F24}
\delta^{\kappa\mu}\delta^{\lambda\nu}
\non
&+G_{F14}
G_{F23}
\delta^{\kappa\nu}\delta^{\lambda\mu}
\biggr]
\, ,
\nonumber\\
\vdots\hskip1cm & \hskip1.5cm\vdots 
\label{wickgrassmann}
\end{align}
The only other information which we will need in a perturbation evaluation of (\ref{spinfactorgrass}) is the free
path integral normalization. But this was (with somewhat different conventions) already calculated in chapter 1. It is

\bear
\int_{A} {\cal D}\psi(\tau)
\,
\e^{
-\int_0^Td\tau
\half \psi\cdot \dot\psi 
}
= 2^{\frac{D}{2}} 
\, .
\label{psifreepi}
\ear
Here $D$ is any {\sl even} spacetime dimension, and we recognize the factor $2^{\frac{D}{2}}$
as the number of real degrees of freedom of a Dirac spinor in such dimension.

\subsection{The $N$-photon amplitudes}
\label{Nphotspin}

The procedure for obtaining the $N$ - photon amplitude in spinor QED from the effective action 
(\ref{spinorQEDpi}) with  (\ref{spinfactorgrass}) is completely analogous to the Scalar QED case
treated in section \ref{Nphotscal}, and we proceed\footnote{See section 3 for the analogous calculation of tree level scattering amplitudes.} straight away to the analogue of (\ref{DNpointscal}):

\begin{align}
\Gamma_{\rm spin}(k_1,\varepsilon_1;\ldots ; k_N,\varepsilon_N) &=
-\half(-ie)^N 
\int_0^{\infty}
\frac{dT}{T}\,
\e^{-m^2T}
\int_{x(0)=x(T)}
{\cal D}x(\tau)\,
e^{-\int_0^T d\tau
\kinb}
\nonumber\\
& \times
\int_{A} {\cal D}\psi(\tau)
\,
\e^{
-\int_0^Td\tau
\half \psi\cdot \dot\psi 
}
 V^{\gamma}_{\rm spin}[k_1,\varepsilon_1]\cdots V^{\gamma}_{\rm spin}[k_N,\varepsilon_N]
 \, .
  \nonumber\\
\label{DNpointspin}
\end{align}
Now $V^{\gamma}_{\rm spin}$ is the vertex operator for the emission or absorption of a photon by a spinor,

\bear
 V^{\gamma}_{\rm spin}[k,\varepsilon] &\equiv &  \int_0^Td\tau\, \Bigl[ \varepsilon\cdot \dot x(\tau)+2i \varepsilon\cdot \psi(\tau) k\cdot\psi(\tau)\Bigr] \,\e^{ik\cdot x(\tau)}
 \, ,
\label{defVspin}
\ear
the second term in brackets being the  momentum-space version of the $\psi\cdot F\cdot\psi$ - term in (\ref{spinfactorgrass}).

Now one would like to obtain a closed formula for general $N$, that is a generalization of the Bern-Kosower master formula.
This is possible, but requires a more elaborate use of the worldline supersymmetry; see section 4.2 of \cite{41} or section \ref{spinLine} in chapter 3 below. But as we will see, there is a more
efficient way to proceed. We will first revisit the case of the photon propagator.

\subsection{The vacuum polarization}

For $N=2$, (\ref{DNpointspin}) becomes
\begin{align}
\Gamma_{\rm spin}(k_1,\varepsilon_1;k_2,\varepsilon_2) &=
-\half {(-ie)}^2\Tint\e^{-m^2T}
\Dx\Dpsi
\int_0^T d\tau_1
\int_0^T d\tau_2\nonumber\\
&
 \hspace{-100pt}\times
\varepsilon_{1\mu}
\Bigl(
\dot x^{\mu}_1
+2i\psi^{\mu}_1\psi_1\cdot k_1
\Bigr)
\e^{ik_1\cdot x_1}
\varepsilon_{2\nu}
\Bigl(
\dot x^{\nu}_2
+2i\psi^{\nu}_2\psi_2\cdot k_2
\Bigr)
\e^{ik_2\cdot x_2}
{\rm e}^{-\int_0^Td\tau
\Bigl(
\kinb +\kinf
\Bigr)
}
\, .
\non
\label{spinorqed2point}
\end{align}\no 
Since the Wick contractions do not mix the $x$ and $\psi$ fields,
the calculation of ${\cal D}x$ is identical with the scalar QED calculation.
Only the calculation of ${\cal D}\psi$ is new, and amounts to
a use of the four-point Wick contraction (\ref{wickgrassmann}),

\bear
{(2i)}^2
\Bigl\langle
\psi^{\mu}_1\psi_1\cdot k_1
\psi^{\nu}_2\psi_2\cdot k_2
\Bigr\rangle
= G_{F12}^2
\Bigl[\delta^{\mu\nu}k_1\cdot k_2 -k_2^{\mu}k_1^{\nu}\Bigr]
\, .
\label{spinwick2point}
\ear\no
Adding this term to the integrand for the scalar QED case, (\ref{Piscal}) 
one finds 

\begin{align}
\Pi_{\rm spin}^{\mn}(k) 
 &=-2 e^2(\delta^{\mu\nu}k^2 -k^{\mu}k^{\nu})
 \int_{0}^{\infty} \frac{dT}{T}(4\pi T)^{-D/2}e^{-m^2 T}
 \nonumber\\ & \times
 \int_{0}^{T}d\tau_{1}\int_0^Td\tau_2 (\dot{G}_{B12}\dot{G}_{B21}-G_{F12}G_{F21})
 \,e^{G_{B12}k_1\cdot k_2}\, .
\label{Pispin}
\end{align}
Thus we see that, up to the normalization, the parameter integral for the spinor loop is obtained
from the one for the scalar loop simply by replacing, in
eq. (\ref{Piscal1}),

\bear
\dot G_{B12}\dot G_{B21} \rightarrow
\dot{G}_{B12}\dot{G}_{B21}-G_{F12}G_{F21}
\, .
\label{subs2point}
\ear\no
Proceeding as in the scalar case, one obtains the 
final parameter integral representation

\begin{align}
\Pi_{\rm spin}^{\mu\nu}(k)
&=
-8{e^2\over {(4\pi )}^{D\over 2}}
\Bigl[g^{\mu\nu}k^2-k^{\mu}k^{\nu}\Bigr]
\Gamma\bigl(2-{D\over 2}\bigl)
\non
&\times
\int_0^1du\,
u(1-u)
{\Bigl[
m^2 + u(1-u)k^2
\Bigr]
}^{{D\over 2}-2}
\label{Pispin2} 
\end{align}\no
and at this level one can easily verify the equivalence with the standard textbook
calculation of the vacuum polarization diagram in spinor QED.

\subsection{Integration-by-parts and the replacement rule}

We have made a big point out of the substitution (\ref{subs2point}) because it is actually only
the simplest instance of a general ``replacement rule'' due to Bern and Kosower \cite{berkos}.
Namely, performing the expansion of the exponential factor in (\ref{scalarqedmaster}) will yield an integrand
$\sim P_N\,e^{(\cdot)}$, where we abbreviated

\bear 
e^{(\cdot)}:= \mbox{exp}\Bigg\{\frac{1}{2}\sum_{i,j=1}^{N} G_{Bij}p_{i}\cdot p_{j}\Bigg\} \, ,
\label{abbrev}
\ear
and $P_N$ is a polynomial in $\dot G_{Bij},\ddot G_{Bij}$ and the kinematic invariants. 
It is possible to remove 
all second derivatives $\ddot{G}_{Bij}$ appearing in $P_N$ by suitable integrations-by-parts,
leading to a new integrand $\sim Q_N\,e^{(\cdot)}$ depending only on the $\dot G_{Bij}$s. 
Look in $Q_N$ for ``$\tau$-cycles'', that is, 
products of $\dot G_{Bij}$'s whose indices form a closed chain. A $\tau$-cycle can
thus be written as 
$\dot G_{Bi_1i_2} \dot G_{Bi_2i_3} \cdots \dot G_{Bi_ni_1}$
(to put it into this form may require the use of the antisymmetry of $\dot G_B$, e.g.
$\dot G_{B12}\dot G_{B12}=-\dot G_{B12}\dot G_{B21}$).
Then the integrand for the spinor
loop case can be obtained from the one for
the scalar loop simply by simultaneously replacing every $\tau$-cycle appearing in $Q_N$ by 

\begin{equation}
\dot G_{Bi_1i_2} 
\dot G_{Bi_2i_3} 
\cdots
\dot G_{Bi_ni_1}
\rightarrow 
\dot G_{Bi_1i_2} 
\dot G_{Bi_2i_3} 
\cdots
\dot G_{Bi_ni_1}
-
G_{Fi_1i_2}
G_{Fi_2i_3}
\cdots
G_{Fi_ni_1}\, ,
\nonumber\\
\label{subrule}
\end{equation}
\no
and supplying the global factor of $-2$ which we have already seen above.

This ``replacement rule'' is very convenient, since it means that we do not have to really compute the 
Grassmann Wick contractions. However, the objective of removing the $\ddot G_{Bij}$
does not fix the IBP procedure, nor the final integrand $Q_N$, and it is not at all
obvious how to proceed in a systematic way for arbitrary $N$. Moreover, our two-point
computations above suggest that the IBP procedure may also be useful for achieving
transversality at the integrand level. Thus ideally one might want to have an algorithm
for passing from $P_N$ to (some) $Q_N$ that, besides removing all
$\ddot G_{Bij}$, has also the following properties:

\benn

\item
It should maintain the permutation symmetry between the photons.

\item
It should lead to a $Q_N$ where each polarization vector $\pol_i$ is absorbed into the
corresponding field strength tensor $f_i$, defined by

\bear
f_i^{\mu\nu}&\equiv&
k_i^{\mu}\varepsilon_i^{\nu}
- \varepsilon_i^{\mu}k_i^{\nu}
\, .
\label{deff}
\ear
This assures manifest transversality at the integrand level.

\item
It should be systematic enough to be computerizable.

\enn

\no
Only very recently an algorithm has been developed that has all these properties \cite{91}.

\section{Spinor QED in a constant field}
\label{EHSpin}
In QED, an important role is played by processes in an external electromagnetic field
that can be approximated as constant in space and time. 
This is not only due to the phenomenological importance of such fields, but also
to the mathematical fact that such a field is one of a very few configurations
for which the Dirac equation can be solved in closed form. 
In the worldline formalism, the corresponding mathematical fact is that,
under the addition of the constant external field,
the worldline Green's functions (\ref{defGBc}), (\ref{defGF})  
generalize to simple trigonometric expressions in the field
\cite{41}. The resulting formalism is extremely powerful and
has already found a number of interesting applications \cite{17,18,40,50}. Here
we must be satisfied with considering just the simplest possible
constant field problem, namely the effective action in such a field itself.

\subsection{The Euler-Heisenberg Lagrangian}
\label{sec:EH}
Thus we now assume that the background field $A(x)$ has a constant field strength tensor
$F_{\mu\nu}= \partial_{\mu}A_{\nu} - \partial_{\nu}A_{\mu}$. It will be convenient
to use Fock-Schwinger gauge. This gauge is defined by fixing a ``center-point'' $x_c$, and
the gauge condition

\bear
(x-x_c)^{\mu}A_{\mu}(x) \equiv 0  \, .
\label{condFS}
\ear
The great advantage of this gauge choice is that it allows one to express derivatives of the
gauge potential $A_{\mu}$ at the point $x_c$ in terms of derivatives of
the field strength tensor at this point. Namely, it follows from (\ref{condFS}) that
the Taylor expansion of $A(x)$ at $x_c$ becomes

\begin{equation}
A_\mu(x_c+q)= \half q^\nu F_{\nu\mu}+{1\over 3}q^{\lambda}q^{\nu} \partial_\lambda F_{\nu\mu} + \ldots
\label{Aexpand}
\end{equation}
(in particular, $A(x_c )=0$).
For a constant field we have only the first term on the rhs of (\ref{Aexpand}). With the obvious
choice of $x_c=x_0$ we have along the loop (remember the definition of $x_0$ (\ref{split}))

\bear
A_{\mu} (x(\tau)) = \half q^{\nu}(\tau) F_{\nu\mu} \, .
\label{loopFS}
\ear
Thus from (\ref{spinorQEDpi}), (\ref{spinfactorgrass}) we have, for this constant field case,

\bear
\Gamma_{\rm spin}(F)
&=&
-\half \Tintm
\int d^4x_0
\int_{P}{\cal D}q(\tau)\,
e^{-\int_0^T d\tau\bigl(
\fourth \dot q^2
+\half ie\, q \cdot F \cdot \dot q
\bigr)}
\nonumber\\
&&\times 
\int_{A} {\cal D}\psi(\tau)
\,
\e^{
-\int_0^Td\tau
\bigl(
\half \psi\cdot \dot\psi -ie \psi^{\mu}F_{\mn}\psi^{\nu}
\bigr)
}
\, .
\label{constantFpi}
\ear\no
We note that the zero mode integral is empty - nothing in the integrand depends on it.
For a constant field this is expected, since by translation invariance the effective action
must contain an infinite volume factor. Factoring out this volume factor we obtain the
effective Lagrangian $\cal L$, which, as we will see, is well-defined. And this time we
will not need any expansions to get the worldline path integrals into gaussian form - they
are already gaussian!  Using our formulas (\ref{SInormalize}) and (\ref{psifreepi}) for the free
path integrals (we can now set $D=4$),
and the formulas (\ref{gaussfree}), (\ref{gg2n}) for the ordinary and Grassmann gaussian integrals,
we can right away write the integrand in terms of determinant factors: 

\bear
{\cal L}(F) 
&=&
-\half \cdot 4 \Tintm
 (4\pi T)^{-2}
\frac
{ 
\int_{P}{\cal D}q(\tau)\,
e^{-\int_0^T d\tau\bigl(
\fourth \dot q^2
+\half ie\, q \cdot F \cdot \dot q
\bigr)}
}
{
\int_{P}{\cal D}q(\tau)\,
e^{-\int_0^T d\tau
\fourth \dot q^2}
}
\nonumber\\
&&\hskip4cm\times 
\frac
{
\int_{A} {\cal D}\psi(\tau)
\,
\e^{
-\int_0^Td\tau
\bigl(
\half \psi\cdot \dot\psi -ie \psi^{\mu}F_{\mn}\psi^{\nu}
\bigr)
}
}
{
\int_{A} {\cal D}\psi(\tau)
\,
\e^{
-\int_0^Td\tau
\half \psi\cdot \dot\psi
}
}
\non
&=&
-2 \Tintm
 (4\pi T)^{-2}
\frac
{ 
{\rm Det'}^{-\half}_P\bigl(-\fourth \frac{d^2}{d\tau^2}+ \half ie F \frac{d}{d\tau} \bigr)
}
{
{\rm Det'}^{-\half}_P\bigl(-\fourth \frac{d^2}{d\tau^2} \bigr)
}
\non &&\hskip4.1cm\times
\frac
{ 
{\rm Det}^{+\half}_A\bigl(\frac{d}{d\tau}-2 ie F  \bigr)
}
{
{\rm Det}^{+\half}_A\bigl( \frac{d}{d\tau} \bigr)
}
\non
&=&
-2 \Tintm
 (4\pi T)^{-2}
{\rm Det'}^{-\half}_P\Bigl(\Eins -2 ie F \bigl(\frac{d}{d\tau}\bigr)^{-1} \Bigr)\non
&&\hskip4.1cm\times
{\rm Det}^{+\half}_A\Bigl(\Eins -2 ie F \bigl(\frac{d}{d\tau}\bigr)^{-1} \Bigr)
\, .
\non
\label{constantFL}
\ear\no
Thus we now have to calculate the determinant of the same operator,

\bear
{\cal O}(F) \equiv \Eins -2 ie F \bigl(\frac{d}{d\tau}\bigr)^{-1}
\label{defO}
\ear
acting once in the space of periodic functions (but with the zero mode eliminated, which we have 
indicated by a `prime' on the `Det' as is customary) and once in the space of antiperiodic functions
\footnote{Note that, if there was no difference in the boundary conditions, the two determinants
would cancel each other; this cancellation is related to the worldline supersymmetry
(\ref{susy}).}. As we will now see, both determinants can be calculated in the most direct
manner, by an explicit computation of the eigenvalues  of the operator and their
infinite product, and moreover these infinite products are convergent. 
This is a rather rare case for a quantum field theory computation!

This could be done using the operator ${\cal O}(F)$ as it stands, but we can use a little
trick to replace it by an operator that is diagonal in the Lorentz indices: since the
determinant of  ${\cal O}(F)$ must be a Lorentz scalar, and it is not possible to form such a scalar
with an odd number of field strength tensors $F$, it is clear that the determinant can depend
also on $e$ only through $e^2$. Thus we can write (abbreviating now $\Det {\cal O} \equiv \abs{\cal O}$)

\bear
\abs{{\cal O}(F)}^2&=&  \abs{\Eins -2 ie F \bigl(\frac{d}{d\tau}\bigr)^{-1}}  \abs{\Eins +2 ie F \bigl(\frac{d}{d\tau}\bigr)^{-1}}
\non
&=&
\abs{\Eins + 4e^2F^2  \bigl(\frac{d}{d\tau}\bigr)^{-2}} \, .
\label{abscalF}
\ear
Next, we can use the fact from classical electrodynamics that, for a generic
constant electromagnetic field, there is a Lorentz frame such that both the
electric and the magnetic field point along the $z$-axis. The euclidean field strength
tensor then takes the form 

\begin{equation}
{\bf F} =
\left(
\begin{array}{*{4}{c}}
0&B&0&0\\
-B&0&0&0\\
0&0&0&iE\\
0&0&-iE&0
\end{array}
\right)
\label{defFEucl}
\end{equation}\no
so that 

\begin{equation}
{\bf F}^2 =
\left(
\begin{array}{*{4}{c}}
-B^2&0&0&0\\
0&-B^2&0&0\\
0&0&E^2&0\\
0&0&0&E^2
\end{array}
\right)
\label{F2}
\end{equation}\no
(in Minkowski space the same relative sign between $B^2$ and $E^2$ would arise through
the raising of one index necessary in the multiplication of $F_{\mn}$ with itself).
Using (\ref{F2}) in (\ref{abscalF}), and taking the square root again, we get

\bear
\abs{{\cal O}(F)}&=&
\abs{1 + 4e^2E^2  \bigl(\frac{d}{d\tau}\bigr)^{-2}}
\abs{1 - 4e^2B^2  \bigl(\frac{d}{d\tau}\bigr)^{-2}}
\, .
\label{abscalFfin}
\ear
Thus we have managed to reduce the original matrix operator to usual (one-component)
operators. 

Next we determine the spectrum of the operator $-\frac{d^2}{d\tau^2}$ for the two
boundary conditions. Thus we need to solve the eigenvalue equation

\bear
-\frac{d^2}{d\tau^2}f(\tau) = \lambda f(\tau) \, .
\label{EVE}
\ear
In the periodic case, a basis of eigenfunctions is given by

\bear
e_n(\tau) &=& \cos(2\pi n \tau/T ), \quad n = 1,2, \ldots \non
\tilde e_n(\tau) &=& \sin(2\pi n \tau/T ),  \quad n = 1,2, \ldots \non 
\label{EFP}
\ear
with eigenvalues

\bear
\lambda_n = \frac{(2\pi n)^2}{T^2} 
\label{EVP}
\ear
and in the antiperiodic case by

\bear
p_n(\tau) &=& \cos\bigl(2\pi (n+1/2) \tau/T \bigr), \quad n = 0,1,2, \ldots \non
\tilde p_n(\tau) &=& \sin\bigl(2\pi (n+1/2) \tau/T \bigr),  \quad n = 0,1,2, \ldots \non 
\label{EFA}
\ear
with eigenvalues

\bear
\lambda_n = \frac{\bigr(2\pi (n+\frac{1}{2})\bigr)^2}{T^2} 
\, .
\label{EVA}
\ear
Using these eigenvalues in (\ref{abscalFfin}), we see that the infinite products
which we encounter are just Euler's famous infinite product representations of
the elementary trigonometric functions,

\bear
\frac{\sin x}{x} &=& \prod_{n=1}^{\infty} \Bigl(1- \frac{x^2}{(n\pi )^2}\Bigr)  \, ,\non
\frac{\sinh x}{x} &=& \prod_{n=1}^{\infty} \Bigl(1+ \frac{x^2}{(n\pi )^2}\Bigr) \, ,\non
\cos x &=& \prod_{n=0}^{\infty} \Bigl(1- \frac{x^2}{\bigl((n+1/2)\pi \bigr)^2}\Bigr) \, ,\non
\cosh x &=& \prod_{n=0}^{\infty} \Bigl(1+ \frac{x^2}{\bigl((n+1/2)\pi \bigr)^2}\Bigr) \, .\non
\nonumber\\
\label{prodform}
\ear
Combining \eqref{constantFL}, \eqref{abscalFfin}, \eqref{EVP}, \eqref{EVA} and \eqref{prodform} (here we have to take into account
that each eigenvalue occurs twice), we get

\begin{align}
{\cal L}(F) 
&=
-2 \Tintm
 (4\pi T)^{-2}
 \frac{e ET}{\sin (eET)}
  \frac{e BT}{\sinh (eBT)}
  \cos(eET)\cosh (eBT)
  \non
  &=
  -2 \Tintm
 (4\pi T)^{-2}
 \frac{e ET}{\tan (eET)}
  \frac{e BT}{\tanh (eBT)}
  \, .
\label{EHfin}
\end{align}
This is the famous Lagrangian found by Euler and Heisenberg in 1936 as one of the
first nontrivial results in quantum electrodynamics. Its existence tells us that, although in
electrodynamics there is no interaction between photons at the classical level, such
interactions do arise after quantization indirectly through the interaction of photons with
the virtual electrons and positrons in the vacuum. Moreover, it encodes this information
in a form which is convenient for the actual calculation of such processes; see, e.g., \cite{ditgie-book}.

\subsection{Schwinger pair production and worldline instantons}

The effective Lagrangian (\ref{EHfin}) contains a singularity in the $T$-integral at $T=0$. This is an ultraviolet-divergence,
and needs to be removed by renormalization (see, e.g., \cite{itzzub-book}). Moreover, except for the purely magnetic field case
this integral shows further poles at 

\bear
T_n = \frac{n\pi}{eE}, \quad n = 1,2,\ldots 
\label{polepos}
\ear
where the $\sin(eET)$ contained in the denominator  vanishes.
These poles generate an imaginary part for the effective Lagrangian. 
Restricting ourselves now to the purely electric field case, $B=0$, 
it is a simple application of complex analysis to show that

\bear
{\rm Im}{\cal L}_{\rm spin}(E) = \frac{(eE)^2}{8\pi^3}\sum_{n=1}^{\infty}{\rm exp}\biggl\lbrack -n\pi \frac{m^2}{eE}\biggr\rbrack
\, .
\label{schwinger}
\ear
Here the $n$th term in the sum is generated by the $n$th pole in (\ref{EHfin}). 

The physical interpretation of this imaginary part was already anticipated by F. Sauter in 1932: in the presence of an
electric field the vacuum becomes unstable, since a virtual electron-positron pair can, by a statistical fluctuation,
gain enough energy from the field to turn real.  However, the probability for this to happen becomes significant 
only at about $E \approx 10^{16} {\rm V}/{\rm cm}$, which is presently still out of the reach of experiment.

But let us now return to the worldline path integral for the constant field case, eq. (\ref{constantFpi}). Clearly in the presence
of an electric field component something special must happen to the path integral when $T$ takes one of the values (\ref{polepos}).
What is it?
Looking at eqs. (\ref{abscalFfin}) - (\ref{EVP}) we can see that the reason for the pole in the effective Lagrangian at $T_n$
is a zero of the operator ${\cal O}(F)$ with periodic boundary conditions (the one coming from the $x$ - path integral) due to the
eigenfunctions $e_n(\tau)$ and $\tilde e_n(\tau)$. 
Now let us return to Feynman's original worldline path integral for Spinor QED (\ref{spinorQEDpi}) and consider the classical equations of motion following from 
the coordinate worldline action, eq. (\ref{Sem}).  It is

\bear
\ddot x^{\mu} = 2ie F^{\mu\nu}\dot x_{\nu} \, .
\label{Lorentzforce}
\ear
This is, of course, simply the Lorentz force equation, only written in unfamiliar conventions and using unphysical 
(from a classical point of view) boundary conditions. 
Now in our case of a purely electric constant field, after putting $T=T_n$ we have $e_n(\tau)=\cos (2eE\tau),\tilde e_n(\tau)=\sin(2eE\tau)$,
and we can use these functions to construct the following circular solution of (\ref{Lorentzforce}):

\bear
x(\tau) =\bigl (x_1,x_2,{\cal N}e_n(\tau),{\cal N}\tilde e_n(\tau)\bigr)  \, .
\label{defwlinst}
\ear
Here $x_1,x_2$ and ${\cal N}>0$ are constants. Such a worldline trajectory that obeys both the classical field
equations and the periodic boundary conditions is called a {\it worldline instanton} \cite{afalma}. 
On this trajectory the worldline action (\ref{Sem}) gets minimized, namely it vanishes, leaving only the exponential factor
$\e^{-m^2T}$ with $T=T_n$ which reproduces the $n$th Schwinger exponential in (\ref{schwinger}). 
This suggests that Schwinger's formula (\ref{schwinger}), including the prefactors, can be obtained by a semiclassical
(stationary path) approximation of the worldline path integral, and indeed this can be made precise \cite{afalma,63,64}.
In the constant field case this approximation actually turns out to be even exact; for more general fields the method
still works as a large-mass approximation. 

Note that the electron spin does not play a role in the determination of the Schwinger exponent. In the constant field case,
the spin factor decouples from the $x$ path integral and  yields, evaluated at $T_n$, a global factor of 

\bear
4 \cos (eET_n) = 4\cos (n\pi) = 4(-1)^n \, .
\label{spinfactoreval}
\ear
In an early but highly nontrivial application of the worldline formalism \cite{afalma}, this semiclassical approximation
was, for the Scalar QED case and in the weak-field approximation, even extended to higher loop orders, yielding the following formula
for the total (summed over all loop orders) Schwinger pair creation exponential:

\bear
{\rm Im}{\cal L}_{\rm scal}^{({\rm all-loop})}(E) \stackrel{\frac{eE}{m^2}\to 0}{\approx} \frac{(eE)^2}{16\pi^3}{\rm exp}\biggl\lbrack -\pi \frac{m^2}{eE}+ \alpha \pi\biggr\rbrack
\label{schwingerallloop}
\ear
(only the first Schwinger exponential is relevant in the weak-field limit). This formula is presently still somewhat conjectural,
though, since some of the arguments of \cite{afalma} are not completely rigorous.

\section{Other gauge theories}

After having so far focused totally on the case of QED, in this last lecture let us very shortly discuss the most
important things to know about the generalization to the nonabelian case, as well as to gravitation. 

\subsection{The QCD N-gluon amplitudes}
\label{qcd}

It is easy to guess what has to be done to generalize Feynman's Scalar QED formula (\ref{scalarQEDpi}) 
to the nonabelian case, i.e. to the case where the gauge field $A_{\mu}(x)$ is a Yang-Mills field with an
arbitrary gauge group $G$, and the loop scalar transforms in some representation of that group.
We can then write 

\bear
A_{\mu}(x) = A_{\mu}^a(x)T^a
\label{decompA}
\ear
where the $T^a,a=1,\ldots ,{\rm dim}(G)$ are a basis for the Lie algebra of $G$ in the
representation of the scalar (e.g. they could be the eight Gell-Mann matrices for 
the case of $SU(3)$ and a scalar in the fundamental representation). 

Thus the $A_{\mu}(x)$ are now matrices, and the $A_{\mu}(x(\tau))$ along the loop
for different $\tau$ will in general not commute with each other. As is well known from
quantum mechanics, in such a case the ordinary exponential function has to be replaced
by a ``path-ordered exponential'' (the coupling constant $g$ in our conventions corresponds to $-e$):

\bear
{\cal P} \e^{ig\int_0^T d\tau \dot x\cdot A(x(\tau))} &\equiv&
1 + ig\int_0^T d\tau\, \dot x\cdot A(x(\tau)) \nonumber\\
&&
 +(ig)^2 \int_0^T d\tau_1\, \dot x_1\cdot A(x(\tau_1))\int_0^{\tau_1} d\tau_2\, \dot x_2\cdot A(x(\tau_2)) + \ldots
 \nonumber\\
\label{poexp}
\ear
And since the effective action must be a scalar, we will clearly also need a global color trace $\tr_c$.
Thus our nonabelian generalization of (\ref{scalarQEDpi}) becomes \cite{strassler}

\begin{equation}
\Gamma_{\rm scal}\lbrack A\rbrack   =  
{\rm tr}_c
{\dps\int_0^{\infty}}
{dT\over T}\,
e^{-m^2T}
{\dps\int}_{P} {\cal D} x
\, {\cal P}
\exp\Bigl [- \int_0^T d\tau
\Bigl (\frac{1}{4}{\dot x}^2 
- ig\, \dot x\cdot A
\Bigr )\Bigr ] 
\, . 
\label{nonabelianPI}
\end{equation}
\no
Moving on to the $N$ - gluon amplitude, it is clear that the vertex operator for a gluon should carry, in addition to a definite
momentum and a definite polarization, also a definite color assignment; thus we supplement the one for the photon (\ref{defVscal})
with one of the basis matrices $T^a$. Thus our {\it gluon vertex operator} will be

\bear
 V^{g}_{\rm scal}[k,\varepsilon,a] &\equiv & T^a  \int_0^Td\tau\, \varepsilon\cdot \dot x(\tau) \,\e^{ik\cdot x(\tau)}
 \, .
\label{defVscalna}
\ear
It is also easy to figure out what will happen to the Bern-Kosower master formula (\ref{scalarqedmaster}): due to the path ordering,
the vertex operators will have to appear in the path integral in a fixed ordering. Thus their color trace will also factor out
of the $\tau$ - integrals, leading to the nonabelian master formula 

\begin{align}
\Gamma_{\rm scal}(k_1,\varepsilon_1,a_1;\ldots;k_N,\varepsilon_N,a_N)
&=
{(ig)}^N
\tr_c (T^{a_1}T^{a_2}\cdots T^{a_N})
\nonumber\\
&
\hspace{-110pt}
\times
{\dps\int_{0}^{\infty}}{dT\over T}
{(4\pi T)}^{-{D\over 2}}
e^{-m^2T}\ 
\int_0^T d\tau_1\int_0^{\tau_1}d\tau_2 \cdots \int_0^{\tau_{N-2}}d\tau_{N-1}
\nonumber\\
&
\hspace{-110pt}
\times
\exp\biggl\lbrace\sum_{i,j=1}^N 
\Bigl\lbrack  \half G_{Bij} k_i\cdot k_j
-i\dot G_{Bij}\varepsilon_i\cdot k_j
+\half\ddot G_{Bij}\varepsilon_i\cdot\varepsilon_j
\Bigr\rbrack\biggr\rbrace
\mid_{\rm {\rm lin}(\pol_1,\ldots,\pol_N)}
\nonumber\\
\label{qcdmaster}
\end{align}
\no
(we have eliminated one integration by choosing the zero on the loop to be at the position of the $N$th
vertex operator).

Contrary to the abelian formula (\ref{scalarqedmaster}), this multi-integral does not represent the complete amplitude yet,
rather one has to still sum over all $(N-1)!$ non-cyclic permutations. A further difference to the QED case is that the one-loop
$N$-gluon amplitudes have, in contrast to the photon ones, also reducible amplitudes, which are not yet included in the
master formula. However, it is one of the remarkable features of the Bern-Kosower rules that it allows one to construct
also the reducible contributions knowing only the integrand of the master formula.

As in the photonic case, one can use the IBP procedure and the replacement rule (\ref{subrule}) to get the integrand for the
spin half loop directly from the one for the scalar loop. Moreover, for the gluon amplitudes there is also diagram
with a gluon loop, and its integrand, too, can be obtained from the scalar loop one by a rule similar to (\ref{subrule}); see
\cite{41}. 

However, in the nonabelian case there will, in general, be boundary terms, since the $\tau$ - integrals do not run over the full
loop any more. Had we restricted - unnecessarily - our integrals to run over ordered sectors in the abelian case, we would
have found that the total derivative terms added in the IBP produced boundary terms, but those would have cancelled out
between adjacent sectors (where ``adjacent sectors'' means differing only by an interchange of two neighboring vertex operators).
Here instead of cancelling those boundary terms will combine into color commutators. For example, assume that we are calculating
the four-gluon amplitude, and we are doing an IBP in the variable $\tau_3$. For the sector with the ``standard'' ordering 
$\tau_1>\tau_2>\tau_3>\tau_4$ this may produce a boundary term at the upper limit $\tau_3= \tau_2$. The same total derivative term used in
the adjacent sector with ordering $\tau_1>\tau_3>\tau_2>\tau_4$ will yield the same boundary term as a lower limit. 
The color factors of both terms will thus combine as $\tr_c (T^{a_1}[T^{a_2},T^{a_3}]T^{a_4})$.

The role of those terms is easier to analyze for the effective action than for the gluon amplitudes. The nonabelian
effective action can, in principle, be written as a series involving only Lorentz and gauge-invariant expressions 
such as $\tr_c (D_{\mu}F_{\alpha\beta}D^{\mu}F^{\alpha\beta})$, where $F$ is the full nonabelian field strength tensor

\bear
F_{\mn} \equiv F_{\mn}^a T^a = F^0_{\mn} - ig[A_{\mu}^bT^b,A_{\nu}^cT^c]
\label{defF}
\ear
where by $F^0_{\mn}$ we denote its ``abelian part'',

\bear
F^0_{\mu\nu} \equiv ( \partial_{\mu}A_{\nu}^a - \partial_{\nu}A_{\mu}^a)T^a \, .
\label{deffmn}
\ear
Each such invariant has a ``core term'' that would exist already in the abelian case; in the example, this is 
$\partial_{\mu}F^0_{\alpha\beta}\partial^{\mu}F^{0\alpha\beta}$. For the core term, the IBP goes from bulk-to-bulk,
and has a covariantizing effect in making it possible to combine vector potentials $A$ into field strength tensors
$F^0$. As one can easily check for examples (see section 4.10 of \cite{41}) the addition of the boundary terms will
complete this ``covariantization'' process by providing the commutator terms that are necessary to complete all
$F^0$s to full $F$s, and all partial derivatives to covariant ones. While this covariantization will ultimately happen
after the computation of all integrals also using any other valid computation method,
the interesting thing about the string-inspired formalism is that one can achieve manifest gauge invariance at the integrand
level, before doing any integrals. 

This property of manifest gauge invariance is somewhat less transparent in momentum space, but still very useful
for analyzing the tensor structure of the off-shell $N$-gluon amplitudes; see \cite{92} for the three-gluon case. 

\subsection{Graviton amplitudes}
\label{lgravity}

Finally, let us discuss the inclusion of gravity at least for the simplest possible case,
which is a scalar particle coupled to background gravity. 
Naively, it seems clear what to do: in the presence of a background metric
field $g_{\mn}(x)$ we should replace the free
kinetic part of the worldline Lagrangian by the geodesic one:

\bear
L_{\rm free} = \frac{\dot x^2}{4} \longrightarrow L_{\rm geo} \equiv \fourth\,\dot x^{\mu}g_{\mn}(x(\tau))\dot x^{\nu} \, .
\label{Lgeodesic}
\ear
As is well-known from General Relativity, this action yields the
classical equations of motion for a spinless particle in a background gravitational field. So we would
be tempted to write down the following formula for the one-loop effective action due to a scalar particle
in quantum gravity:

\bear
\Gamma_{\rm scal}[g] 
\stackrel{?}{=} \int_0^\infty {dT\over T }\,\e^{-m^2T} \int_{P} 
{\cal D}x
\; e^{-\int_0^T d\tau L_{\rm geo}}         
\, .        
\label{Gammagravwrong}
\ear
As is usual in Quantum Gravity, we could then introduce gravitons as small plane wave perturbations
of the metric around flat space,

\bear
g_{\mn}(x) = \delta_{\mn} + \kappa h_{\mn}(x)
\label{gtoh}
\ear
where $\kappa$ is the gravitational coupling constant and 

\bear
h_{\mn}(x) = \varepsilon_{\mn}\,\e^{ik\cdot x}
\label{defh}
\ear
with a symmetric polarization tensor $\varepsilon_{\mn}$. 
The usual perturbative evaluation of the path integral will then
yield graviton amplitudes in terms of Wick contractions with the
usual bosonic Green's function $G_B$, and each graviton presented by
a vertex operator 

\bear
V^{h}_{\rm scal}(k,\varepsilon) \stackrel{?}{=} \varepsilon_{\mn}\int_0^Td\tau \dot x^{\mu}\dot x^{\nu} \,\e^{ik\cdot x(\tau)}
\, .
\label{defVh}
\ear
However, one can immediately see that, contrary to the gauge theory case,
the Wick contractions will now lead to mathematically ill-defined
expressions, due to the $\delta(\tau -\tau')$ contained in $\ddot G_B(\tau,\tau')$.  
For example, a Wick contraction of two vertex operator will produce a term with a
$\delta^2(\tau_1-\tau_2)$, and even the one of just a single vertex operator will already
contain an ill-defined $\delta (0)$. These ill-defined terms signal UV divergences in
our one-dimensional worldline field theory. However, they are of a spurious nature.
To get rid of them we have to take the nontrivial background metric into
account not only in the Lagrangian, but also in the path integral measure. 
In general relativity the general covariance requires that
each spacetime integral should contain a factor of $\sqrt{\det{g}}$. Therefore the measure which
should be used in (\ref{Gammagravwrong}) is of the form \cite{leeyan}

\bear
{\cal D} x = Dx \prod_{ 0\leq \tau < T} \sqrt{\det g_{\mu\nu}(x(\tau))}   
\ear
where $Dx=\prod_\tau d^Dx(\tau)$ is the standard translationally 
invariant measure. 
However, in our string-inspired approach we clearly cannot use these metric
factors the way they stand; they ought to be somehow exponentiated.
A convenient way of doing this \cite{bastianelli,basvan93} is by introducing
commuting $a^\mu$ and anticommuting $b^\mu , c^\mu$ worldline ghost fields with
periodic boundary conditions 

\bear
{\cal D} x = Dx \prod_{ 0 \leq \tau < 1} \sqrt{\det g_{\mu\nu}(x(\tau))}  =
Dx \int_{PBC} { D} a { D} b { D} c \;
{\rm e}^{- S_{gh}[x,a,b,c]} 
\label{expmeasure}
\ear
where the ghost action is given by

\bear
S_{gh}[x,a,b,c]
= \int_{0}^{T} d\tau \; {1\over 4}g_{\mu\nu}(x)(a^\mu a^\nu 
+ b^\mu c^\nu) 
\, .
\label{Sghost}
\ear
Thus the final version of the path integral representation (\ref{Gammagravwrong}) is

\begin{align}
\Gamma_{\rm scal}[g] 
&= \int_0^\infty {dT\over T }\,\e^{-m^2T} \non
&\times\int_{P} 
DxDaDbDc
\,\e^{-\frac{1}{4} \int_0^T d\tau\, g_{\mn}(x(\tau))\bigl(\dot x^{\mu}(\tau)\dot x^{\nu}(\tau)
+a^{\mu}(\tau)a^{\nu}(\tau) + b^{\mu}(\tau)c^{\nu}(\tau)\bigr)}
\, .
\non
\label{Gammagravfin}
\end{align}
After the split (\ref{gtoh}) we will now get the graviton vertex operator in its final form,

\bear
V^h_{\rm scal}[k,\pol] &=& 
\pol_{\mn}\int_0^Td\tau \Bigl[\dot x^{\mu}(\tau)\dot x^{\nu}(\tau)
+a^{\mu}(\tau)a^{\nu}(\tau) + b^{\mu}(\tau)c^{\nu}(\tau)
\Bigr]
\,\e^{ik\cdot x(\tau)}
\, .
\non
\label{defVhfin}
\ear
The ghost fields have algebraic kinetic terms, so that their
Wick contractions involve only $\delta$ - functions:

\bear
\langle a^{\mu}(\tau_1)a^{\nu}(\tau_2)\rangle
&=&
2\delta(\tau_1-\tau_2)\delta^{\mn} \, ,\non
\langle b^{\mu}(\tau_1)c^{\nu}(\tau_2) \rangle
&=&
-4 \delta(\tau_1-\tau_2)\delta^{\mn} \, .\non
\label{wickrulesghostscal}
\ear
The extra terms arising from the ghost action will remove all the UV divergences.
Let us quickly check this for the self-contraction of the graviton vertex operator (\ref{defVhfin}):

\bear
\bigl\langle \dot x^{\mu}(\tau)\dot x^{\nu}(\tau)+a^{\mu}(\tau)a^{\nu}(\tau) + b^{\mu}(\tau)c^{\nu}(\tau) \bigr\rangle
& = &\Bigl( \ddot G_B(\tau,\tau) + 2 \delta(0) -4 \delta (0) \Bigr) \delta^{\mn} 
\non
&=& - \frac{2}{T} \delta^{\mn} \, .
\label{selfwick}
\ear
However, not all troubles are over yet. As it often happens in field theory, the cancellation of UV divergences
leaves behind a finite ambiguity: some of the remaining integrals require a regularization to be assigned a
definite value. This regularization dependence must then be removed by 
the addition of appropriate finite counterterms to the worldline Lagrangian.
If our one-dimensional worldline theory was an autonomous one, this would introduce some
new free parameters into the game. However our goal here is just to reproduce the same results as
would otherwise be obtained in the standard Feynman diagram approach to quantum gravity. By comparing
with some known quantities, such as conformal anomalies, one can establish \cite{Bastianelli:2006rx} that consistency with
the standard formalism can be achieved by adding, for any given regularization scheme, just three such counterterms,

\bear
L_{\rm counter} = c_1 R + c_2 g^{\mu\nu}\Gamma^\beta_{\mu\alpha} \Gamma^\alpha_{\nu\beta} + c_3 g^{\mu\nu} g^{\alpha\beta} g_{\lambda\rho}
\Gamma^\lambda_{\mu\alpha} 
\Gamma^\rho_{\nu\beta} \non
\label{Lcounter}
\ear
with the appropriate (regularization dependent) coefficients $c_i$.

Once these coefficients have been computed for a given regularization (which for the
most important regularization choices has been done) there are no more singularities or
ambiguities, and one can now go ahead and compute the effective action, or the
corresponding $N$-graviton amplitudes, as in the gauge theory case 
\footnote{To be precise, in curved space there are a few more issues related to operator ordering and
zero mode fixing, but those have been completely resolved, too \cite{Bastianelli:2006rx,bacozi1}.}.
See \cite{Bastianelli:2006rx,61,76} for  generalization to the spinor loop case and for some applications.

\section{Conclusions}

As we said at the beginning, in this short introduction to the worldline formalism
we have concentrated on the case  of gauge boson amplitudes, or the 
corresponding effective actions, since it is for those calculations that the 
formalism has been developed to the point where it offers distinct advantages
over the standard Feynman diagram approach. This does not mean that the formalism
cannot be applied to, e.g., amplitudes involving external fermions, or other couplings such
as Yukawa or axial interactions; however there is still a dearth of
convincing state-of-the-art applications. 
Moreover, we have restricted ourselves to the one-loop case (but see Chapter 3 for tree level applications),
despite the fact that at least for QED Feynman had already given a path integral construction
of the full S-matrix to all loop orders. Here the reason was lack of time rather than lack of nice examples;
the worldline formalism has been applied to a recalculation of the two-loop QED
$\beta$ function as well as of the two-loop correction to the Euler-Heisenberg Lagrangian 
(\ref{EHfin}) (these calculations are summarized in \cite{41}). This is unfortunate, since
the property of the formalism to enable one to combine various different Feynman diagrams
into a single integral representation, which we mentioned in connection with the $N$-photon amplitudes,
becomes much more interesting at higher loops. See \cite{15} and \cite{41} for a detailed
discussion of these multiloop QED issues and calculations.

\section{Problems for chapter 2}

\begin{enumerate}

\item
One can often compute, or define, the determinant of an infinite-dimensional
operator $\cal O$ using {\it $\zeta$-function regularization}, defined as follows:
assume that $\cal P$ has a discrete spectrum of real eigenvalues
$\lambda_1< \lambda_2 < \ldots $.
Define the $\zeta$-function $\zeta_{\cal O}$ of  $\cal O$ by

\bear
\zeta_{\cal O}(z)\equiv \sum_{n=1}^{\infty}\lambda_n^{-z} 
\nonumber
\ear
\noindent
where $z\in {\bf C}$. Show that {\it formally}

\bear
\Det({\cal O}) \equiv \prod_{n=1}^{\infty}\lambda_n
=
\exp\Bigl(-{d\over dz}\zeta_{\cal O}(0)\Bigr)
\, .
\nonumber
\ear
Use this method to verify (\ref{det}) (you will need some properties of the Riemann $\zeta$-function).

\item
Derive the Green's function (\ref{defGBc}), using a direct eigenfunction expansion. 

\item
Verify (\ref{intqprop}) in detail.

\item
Compute $P_3$ and $Q_3$.

\item
Verify that the first of the Scalar QED diagrams of fig. \ref{fig1} 
matches with the contribution of the $\delta(\tau_1-\tau_2)$ part of the $\ddot G_{B12}$ contained in $P_2$.

\item
Verify that (\ref{defLspin}) is invariant under (\ref{susy}) up to a total derivative term
(you will need the Bianchi identity for the field strength tensor).

\item
Show (\ref{gg2n}).

\item
Verify (\ref{gaussgrassmann}).

\item
Show (\ref{defGF}).

\item
Show (\ref{psifreepi}) using $\zeta$-function regularization
(you will need some properties of the Hurwitz $\zeta$-function).

\item
Verify (\ref{Pispin}).

\item
Using (\ref{condFS}), obtain the first two terms on the rhs of (\ref{Aexpand}).

\item
Derive (\ref{schwinger}) from (\ref{EHfin}) (circumvent the poles displacing the integration contour above).

\item
Verify (\ref{Lorentzforce}).

\item
Verify that (\ref{defwlinst}) fulfills (\ref{Lorentzforce}).

\item
Verify (\ref{spinfactoreval}).

\item
Show (\ref{expmeasure}), (\ref{Sghost}) using our formal rules for bosonic and
fermionic (grassmann) gaussian integrations. 

\item
Verify, that in the Wick contraction of two graviton vertex operators all terms
involving $\delta(0)$ or $\delta^2(\tau_1 - \tau_2)$ cancel out after the inclusion of
the terms from the ghosts. 

\end{enumerate}

\chapter{Tree-level processes in the first quantised approach}
Both the previous sections and, for the majority part, the historical development of the worldline formalism to QFT focussed on its application to one- or multi-loop amplitudes. Firstly the effective action is an extremely important object to study since it gives quantum corrections to a classical Lagrangian and, as we have seen above, contains the full information on one-loop scattering amplitudes. Secondly, effects such as the need for renormalisation appear when considering processes at one-loop order and higher and it is clear that new tools are sorely needed to make such higher order calculations manageable. From a calculational perspective, the loop processes discussed earlier have certain advantages -- for example, they are scalar quantities in coordinate and spin space, involving (functional) traces over such degrees of freedom; moreover, the effective action is gauge invariant (up to total derivatives), again due to this trace over the positions of the closed worldlines that provide the functional determinant from integrating out matter degrees of freedom. 

\begin{figure}
	\centering
	\def\svgwidth{0.7 \columnwidth}
	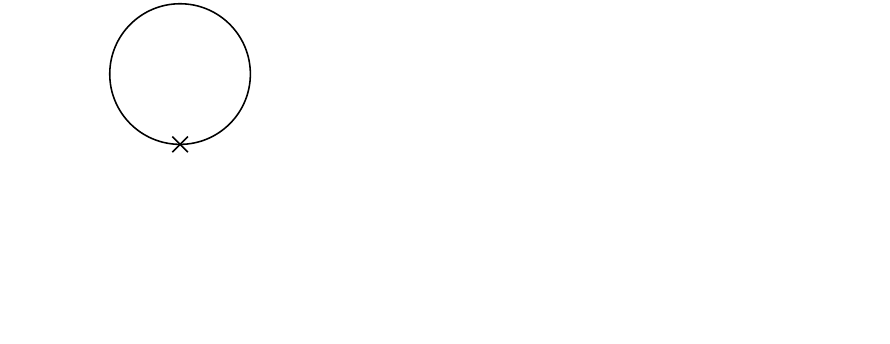
	\caption{Going from the closed loop, where the effective action requires a trace over spatial and spinor indices, to the open line, where the kernel is a function of the endpoints and the spin states at each end, or in momentum space is a function of momentum flowing into the line and the spin states. The top line represents scalar QED whilst the bottom line is for spinor QED.}
	\label{figLoopLine}
\end{figure}

At tree level, however, the fundamental object of interest shall be a path integral representation of the propagator between two spatial points, $x$ and $x^{\prime}$. Thus we will be describing a functional integral over all possible open worldlines with fixed endpoints. As a consequence, the propagator will initially be a function of these endpoints; furthermore for a fermionic propagator it will also depend upon the spin states, $\alpha$ and $\beta$, at these endpoints (see Fig \ref{figLoopLine}). Dressing this propagator with an arbitrary number of photons will provide access to tree level scattering amplitudes -- i.e. amplitudes with external scalar or spinor states. We will eventually Fourier transform to momentum space whereby the propagator will become dependent upon the inflowing momentum, $p$, as well as the photon momenta going into the line. Since there are no longer traces over the endpoints of the line or the spin indices, the propagator also becomes a gauge covariant (rather than invariant) object. From a calculational perspective, the change in topology of the worldlines requires different boundary conditions on the path integral, worldline Green function and etc. (see also Problem 3.1).

Our motivations for studying tree level processes using worldline techniques include a more efficient calculation of on-shell tree-level scattering amplitudes and their gauge structure -- in particular for a more tractable approach to determining high order corrections to $g$-$2$ -- and (off-shell) particle self energies and their tensor decomposition, or to study the so-called LKF transformations that give the behaviour correlation functions under a change in the photon gauge parameter. Quantum propagation in an electromagnetic background is also a natural problem to consider in first quantisation, given the clear benefits discussed in earlier sections. In any case, the propagators are needed for a complete description of the QED S-Matrix and one would be justified in feeling that the worldline formalism would be incomplete were it unable to be adapted to tree level. In this section we begin as usual with scalar QED followed by spinor QED in vacuum, before extending the results to the case of a constant electromagnetic background. We will demonstrate how to derive the particle self energies and linear Compton amplitudes and cross sections in these cases. Indeed, for the spinor propagator it is only very recently that the worldline formalism has been adapted to provide an alternative to the standard approach that maintains the same advantages present at one loop order. 

\section{Scalar QED}
For the worldline representation of the propagator of a scalar field coupled to an Abelian gauge potential $A^{\mu}$, we can again refer to Feynman's initial construction \cite{Feynman:1950ir} that was briefly mentioned in section 2.2.2. Denoting this scalar propagator between fixed points $x$ and $x'$ by $\mathfrak{D}^{x^{\prime}x}$, we search for a first quantised representation of the transition amplitude (note that based on our convention the starting and ending points of the particle in configuration (momentum) space are $x (p)$ and $x'(p')$ accordingly)
\begin{equation}
	\mathfrak{D}^{x'x} = \Big< x' \Big |\left( -D^{2} + m^{2} \right)^{-1} \Big| x\Big>,
	\label{Dxxp}
\end{equation}
where the covariant derivative is defined by $D_{\mu} = \partial_{\mu} + ieA_{\mu}$. This is of course a Green function for the Klein-Gordon operator with minimal coupling to an electromagnetic field. The inverse of the operator is exponentiated with a (convergent) Schwinger proper time integral to get (working in Euclidean space)
\begin{align}
	\mathfrak{D}^{x'x} &= \int_{0}^{\infty}dT\, \Big< x' \Big|\e^{-T \left( -D^{2} + m^{2} \right)} \Big| x\Big>\,.
\end{align}
The matrix element, considered now as a quantum transition amplitude with Hamiltonian $H = -D^{2} + m^{2}$, can be represented as in (\ref{scalproppi}), as a path integral over point particle trajectories that travel from $x$ to $x'$ in time $T$ whilst interacting with a background field:
\begin{equation}
	\mathfrak{D}^{x^{\prime}x}[A] = \int_{0}^{\infty} dT\, \e^{-m^{2}T} \int_{x(0) = x}^{x(T) = x^{\prime}}\hspace{-1em}\mathscr{D}x(\tau)\, \e^{-\int_{0}^{T}d\tau \left[\frac{\dot{x}^{2}}{4} + i e \dot{x} \cdot A(x(\tau))\right]}\,.
	\label{Dxx}
\end{equation} 
We will eventually be interested in extracting scattering amplitudes which are most easily analysed in momentum space, so it will be useful to study the Fourier transform of this object with respect to the endpoints\footnote{Our convention is that all momenta are inflowing which is why we use the same sign in the Fourier exponent for each endpoint}
\begin{equation}
	\mathfrak{D}^{p^{\prime}p}[A] := \int d^{D}x' \int d^{D}x\,  \mathfrak{D}^{x^{\prime}x}[A] \, \e^{ i p^{\prime} \cdot x^{\prime}+ip\cdot x}.
	\label{Dpp}
\end{equation}

\label{Photon dressed propagator}
We will later see that the functional form of the propagator ensures that we get momentum conservation for $\mathfrak{D}^{p^{\prime}p}[A]$. Before proceeding it is important to highlight the different measure on the proper time integral (c.f. $\Gamma[A]$ defined in (\ref{scalarQEDpi})). This can be understood as following from the fact that there is no longer a translational symmetry with respect to points along a closed loop so it is not necessary to divide by $\frac{1}{T}$ to compensate for over-counting (a more formal way of deriving this measure starts from a locally translational invariant one-dimensional supergravity involving an einbein $e(\tau)$; the local symmetry allows the einbein to be gauge fixed to a constant, $\hat{e}(\tau) = T$, but this remains as a modulus to be integrated over. The Faddeev-Popov determinant associated to the gauge fixing gives a measure $\frac{dT}{T}$ on the loop and $dT$ on the line -- the interested reader can find full details in Appendix B of \cite{MP2}).

\subsection{Scattering amplitudes}
As in the case of the one-loop processes discussed in the previous sections, to study photon scattering off the scalar line we will set
\begin{equation}
	A_{\mu}[x(\tau)] = \bar{A}_{\mu}[x(\tau)] + \sum_{i = 1}^{N}\ep_{\mu i} \e^{i k_{i} \cdot x(\tau_i)}	
\end{equation}
where $\bar{A}$ is a background field that we shall take either to vanish (vacuum processes) or be a constant field. It has been seen in chapter 2 that this background gives access to photon scattering amplitudes in momentum space. The only difference here is that we will now be able to study photon scattering off external scalar legs. As early as 1996, Daikouji et. al., starting from field theory, found a parameter integral expression for $N$-photon scattering amplitudes in vacuum \cite{Daik}, whose essential form we shall now re-derive from (\ref{Dxx}). These details were first worked out in \cite{line2, LineNA}, whose presentation we follow. So we set $\bar{A}_{\mu} = 0$ and as in section 2.2.4 we expand the exponent of (\ref{Dxx}) to order $e^{N}$ with $N$ distinct polarisation vectors. Repeating the steps of equations (\ref{termN}) to (\ref{trick}) we arrive at a similar path integral with vertex operator insertions (see (\ref{defVscal}) for the vertex):
\begin{equation}
	\mathfrak{D}^{x^{\prime}x}[A] = (-ie)^{N}\int_{0}^{\infty} dT\, \e^{-m^{2}T} \int_{x(0) = x}^{x(T) = x'} \hspace{-1em}\mathscr{D}x(\tau)\, \e^{-\int_{0}^{T}d\tau \, \frac{\dot{x}^{2}}{4}} \prod_{i = 1}^{N} V_{\scal}^{\gamma}[k_{i}, \ep_{i}]\,.
	\label{Dxxk}
\end{equation}
Now comes the first effect of the open line boundary conditions. In the closed-loop case we noted that there was a zero mode that was separated from the path integral by expanding the trajectory about the loop centre of mass, $x_{0}$; this led to the string inspired boundary conditions of equation (\ref{BCSI}). Here it is also convenient to absorb the boundary conditions by expanding the trajectory about some fixed path from $x$ to $x^{\prime}$. We could do this with any such path, but a useful choice is the straight line between the endpoints\footnote{A single zero eigenfunction but a set of measure zero.}, so that we set
\begin{equation}
	x(\tau) = x+ (x^{\prime}-x)\frac{\tau}{T} + q(\tau) := \hat{x}(\tau) + q(\tau)\,,
\end{equation}
where $q(\tau)$ represents the deviation from the straight line path as illustrated in Fig. \ref{figDefQ}. Then we find that the path integral over $x(\tau)$ becomes one over $q(\tau)$ with \textit{Dirichlet boundary conditions}, $q(0) = 0 = q(T)$:
\begin{figure}
\centering
\def\svgwidth{0.45 \columnwidth}
\begingroup%
  \makeatletter%
  \providecommand\color[2][]{%
    \errmessage{(Inkscape) Color is used for the text in Inkscape, but the package 'color.sty' is not loaded}%
    \renewcommand\color[2][]{}%
  }%
  \providecommand\transparent[1]{%
    \errmessage{(Inkscape) Transparency is used (non-zero) for the text in Inkscape, but the package 'transparent.sty' is not loaded}%
    \renewcommand\transparent[1]{}%
  }%
  \providecommand\rotatebox[2]{#2}%
  \newcommand*\fsize{\dimexpr\f@size pt\relax}%
  \newcommand*\lineheight[1]{\fontsize{\fsize}{#1\fsize}\selectfont}%
  \ifx\svgwidth\undefined%
    \setlength{\unitlength}{76.55236672bp}%
    \ifx\svgscale\undefined%
      \relax%
    \else%
      \setlength{\unitlength}{\unitlength * \real{\svgscale}}%
    \fi%
  \else%
    \setlength{\unitlength}{\svgwidth}%
  \fi%
  \global\let\svgwidth\undefined%
  \global\let\svgscale\undefined%
  \makeatother%
  \begin{picture}(1,0.69606526)%
    \lineheight{1}%
    \setlength\tabcolsep{0pt}%
    \put(0,0){\includegraphics[width=\unitlength,page=1]{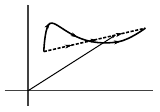}}%
    \put(0.22556743,0.32354){\makebox(0,0)[lt]{\lineheight{1.25}\smash{\begin{tabular}[t]{l}$x$\end{tabular}}}}%
    \put(0.89523432,0.44296898){\makebox(0,0)[lt]{\lineheight{1.25}\smash{\begin{tabular}[t]{l}$x^{\prime}$\end{tabular}}}}%
    \put(0.6823826,0.52517956){\makebox(0,0)[lt]{\lineheight{1.25}\smash{\begin{tabular}[t]{l}$\hat{x}(\tau)$\end{tabular}}}}%
    \put(0.67403647,0.35164132){\makebox(0,0)[lt]{\lineheight{1.25}\smash{\begin{tabular}[t]{l}$x(\tau)$\end{tabular}}}}%
    \put(0,0){\includegraphics[width=\unitlength,page=2]{Qdef.pdf}}%
    \put(0.32543784,0.57226737){\makebox(0,0)[lt]{\lineheight{1.25}\smash{\begin{tabular}[t]{l}$q(\tau)$\end{tabular}}}}%
  \end{picture}%
\endgroup%

	\caption{The decomposition of open trajectories in terms of a deviation $q(\tau)$ about the straight line path (dashed line, denoted $\hat{x}(\tau)$) between the endpoints: $x(\tau) = \hat{x}(\tau) + q(\tau)$.}
	\label{figDefQ}
\end{figure}
\begin{equation}
	\int_{x(0) = x}^{x(T) = x'} \hspace{-1em}\mathscr{D}x(\tau) \, \Omega[x(\tau)] \longrightarrow \int_{q(0) = 0}^{q(T) = 0}  \hspace{-1em}\mathscr{D}q(\tau) \, \Omega[\hat{x}(\tau) + q(\tau)]\,,
	\label{xqline}
\end{equation}
for any functional of trajectories $\Omega$. In this case, at least in vacuum, the kinetic term in the action is positive definite over the space of open trajectories due to the Dirichlet boundary conditions so there is no zero mode problem (note that $q(0) = 0 = q(T)$ ensures that a general linear function of $\tau$ would need to be identically zero).

In the particular case of (\ref{Dxxk}), the decomposition (\ref{xqline}) means that the kinetic term becomes
\begin{equation}
	-\int_{0}^{T} d\tau\, \frac{\dot{x}^{2}}{4} = -\frac{(x^{\prime}-x)^{2}}{4T} - \int_{0}^{T}d\tau\, \frac{\dot{q}^{2}}{4}\,,
\end{equation}
which is left to the reader to verify through appropriate integration by parts. Moreover, using the exponentiation trick, (\ref{trick}), of the polarisation vectors in the vertex operators allows them to be written as
\begin{equation}
	V_{\scal}^{\gamma}[k, \ep] \longrightarrow \int_{0}^{T}d\tau\, \e^{i k \cdot \hat{x}(\tau) + \frac{\ep}{T} \cdot (x^{\prime}-x) + i k \cdot q(\tau) + \ep \cdot \dot{q} }\bigg|_{\textrm{lin } \ep}.
\end{equation}
The $q$-independent terms can be taken out of the path integral, whilst the pieces linear in $q$ can be absorbed into the action as in section 2.2.4 by a coupling to the current $j(\tau)$ in equation (\ref{defj}). With these ingredients, it is a simple exercise to show that the $N$-photon dressed propagator becomes
\begin{align}
	&\mathfrak{D}^{x'x}[k_{1}, \ep_{1}; \ldots; k_{N}, \ep_{N}] = (-ie)^{N}\int_{0}^{T}dT\, \e^{-m^{2}T} \e^{-\frac{(x^{\prime}-x)^{2}}{4T}}\non
	&\times \prod_{i = 1}^{N} \int_{0}^{T}d\tau_{i} \, \e^{\sum_{i = 1}^{N} \left[ i k_{i} \cdot \hat{x}(\tau) + \frac{\ep}{T} \cdot (x^{\prime}-x)\right] } \int_{q(0) = 0}^{q(T) = 0} \hspace{-1em}\mathscr{D}q(\tau) \, \e^{-\int_{0}^{T}d\tau\, \big[ \frac{\dot{q}^{2}}{4} - ie j(\tau) \cdot q(\tau) \big]}\bigg|_{\linep}.
	\label{Dxxq}
\end{align}
Here, as before, at the end of the computation of the path integral we must take the terms that are multi-linear in the polarisation vectors. 

The path integral over $q(\tau)$ can be calculated with the help of two ingredients; the overall normalisation and the worldline Green function for Dirichlet boundary conditions that are respectively (note the functional determinant evaluates to the same $T$-dependent factor as in the one-loop case) 
\begin{align}
	\int_{q(0) = 0}^{q(T) = 0} \hspace{-1em}\mathscr{D}q(\tau) \, \e^{-\int_{0}^{T}d\tau\,  \frac{\dot{q}^{2}}{4}} = \textrm{Det}\left[- \frac{1}{4}\frac{d^{2}}{d\tau^{2}}\right]_{\DBC} = \Big(4\pi T\Big)^{-\frac{D}{2}} \,,
	\label{Detline}
	\end{align}
with the following open-line Green function 
	\begin{align}
	\left<q^{\mu}(\tau)q^{\nu}(\tau^{\prime})\right> = -2\delta^{\mu\nu}\Delta(\tau, \tau^{\prime}) \,,\label{Gline}
\end{align}
where it is left as an exercise to verify that
\begin{equation}
	\Delta(\tau, \tau^{\prime}) := \frac{1}{2}\left|\tau - \tau^{\prime}\right| - \frac{1}{2}(\tau + \tau^{\prime}) + \frac{\tau \tau^{\prime}}{T}
	\label{Delta}
\end{equation}
satisfies the defining equation for a Green function and vanishes when either point is on the boundary. Here the second two terms are solutions of the homogeneous equation defining the Green function that impose Dirichlet boundary conditions. To understand the origin of (\ref{Detline}) and (\ref{Delta}) it is useful to note that a basis of eigenfunctions of the second derivative operator satisfying Dirichlet boundary conditions is given by
\begin{equation}
	f(\tau) = \sqrt{\frac{2}{T}}\sin\big(\frac{n \pi \tau}{T}\big), \quad \textrm{with eigenvalues} \quad \lambda_{n} = \frac{n^{2}\pi^{2}}{T^{2}}.
\end{equation}
Then with $\zeta$-function regularisation \cite{HawkingZ} (introduced in section 1.2.2 and discussed as a problem of chapter 2) we define the functional determinant by
\begin{equation}
\hspace{-2em}	 \textrm{Det}\left[- \frac{1}{4}\frac{d^{2}}{d\tau^{2}}\right]_{\DBC} := \left(\prod_{n = 1}^{\infty} \frac{n^{2}\pi^{2}}{4T^{2}}\right)^{-\frac{D}{2}} = \exp\left[-\frac{D}{2}\frac{d}{dz} \left(\frac{\pi}{2T}\right)^{-2z}\zeta(2z) \bigg|_{z = 0}\right] \longrightarrow \Big(4\pi T\Big)^{-\frac{D}{2}}.
\end{equation}
Moreover, using the spectral decomposition of the Green function we may equally write
\begin{equation}
	\Delta(\tau, \tau^{\prime}) = \frac{T}{2\pi^{2}}\sum_{n \neq 0}^{\pm \infty} \frac{1}{n^{2}}\left[\e^{\frac{i n \pi (\tau + \tau^{\prime})}{T}} - \e^{\frac{i n \pi (\tau - \tau^{\prime})}{T}}\right].
\end{equation}
Finally we note that the breaking of translational symmetry by the boundary conditions means that the new Green function, $\Delta(\tau, \tau')$, is no longer a function of $\tau - \tau^{\prime}$ (as indeed is clear in the spectral decomposition above). In particular the coincidence limit (i.e. $\Delta(\tau, \tau) = \tau(\frac{\tau}{T} - 1)$) no longer vanishes and we must also now distinguish between derivatives with respect to the first and second argument. A useful notation for the latter is to indicate derivatives with respect to the first (second) argument with a $\bullet$ to the left (right) of $\Delta$ so that:
\begin{align}
	\ddel(\tau, \tau^{\prime}) &= \frac{\tau^{\prime}}{T} + \frac{1}{2} \sigma(\tau - \tau^{\prime}) - \frac{1}{2}\,, \\
	\deld(\tau,\tau')&=\frac{\tau}{T} - \frac{1}{2} \sigma(\tau - \tau^{\prime}) - \frac{1}{2}\,,\\
	\ddeld(\tau, \tau^{\prime}) &= -\delta(\tau - \tau^{\prime}) + \frac{1}{T}\,,
\end{align}
and so forth. Finally we record the coincidence limit of the first derivatives for later, $\ddel(\tau,\tau)=\deld(\tau,\tau)=\frac{\tau}{T}-\frac{1}{2}$. 

\begin{figure}[h]
	\centering
\includegraphics[width=5in]{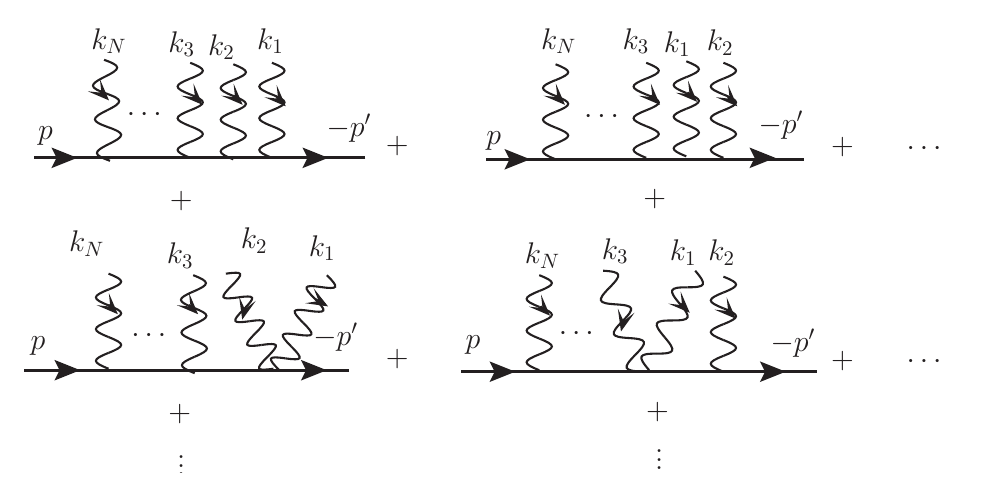}
	\caption{Multi-photon Compton scattering diagram in scalar QED (we do not distinguish the propagator of the matter field between scalar or spinor QED, choosing to indicate both with a solid line). The seagull vertices in the second row are once again produced by the $\delta$-function in the second derivative of the open-line Green function $\Delta$.}
	\label{fig-multiphoton}
\end{figure}

Armed with these results, we complete the square in the exponent of (\ref{Dxxq}) and repeating the steps between (\ref{defj}) and (\ref{scalarqedmaster}) the path integral can be computed to give
\begin{align}
&\mathfrak{D}^{x'x}[k_{1}, \ep_{1}; \ldots; k_{N}, \ep_{N}] = (-ie)^{N}\int_{0}^{T}dT\, (4\pi T)^{-\frac{D}{2}} \e^{-m^{2}T} \e^{-\frac{(x^{\prime}-x)^{2}}{4T}} \non
	&\times\prod_{i = 1}^{N} \int_{0}^{T}d\tau_{i} \, \e^{\sum_{i = 1}^{N} \left[ i k_{i} \cdot (x+ (x' - x)\frac{\tau_{i}}{T}) + \frac{\ep}{T} \cdot (x' - x)\right] }\, \e^{\sum_{i, j = 1}^{N}\big[\Delta_{ij}  k_{i} \cdot k_{j}  - 2i \ddel_{ij}\ep_{i} \cdot k_{j} - \ddeld_{ij}\ep_{i}\cdot \ep_{j} \big]}\bigg|_{\linep}\,,
		\label{DxxkFull}
\end{align}
where we denote $\Delta_{ij} := \Delta(\tau_{i}, \tau_{j})$. As mentioned above, it is often more convenient to work in momentum space, so we transform (\ref{DxxkFull}) according to (\ref{Dpp}). To do this, it is useful to change variables to $\xm := x^{\prime}-x$ and $\xp := \frac{1}{2}(x + x^{\prime})$, respectively the relative displacement of the endpoints and the midpoint of the line between them (this choice of variables lead to unit Jacobian). Then picking out the $\xp$ dependence from the exponent of (\ref{DxxkFull}) and noting that $\e^{ip \cdot x + i p^{\prime} \cdot x^{\prime}} = \e^{i \xp \cdot (p + p^{\prime}) + \frac{i}{2} \xm \cdot (p^{\prime}-p)}$ leads to an $\xp$ integral
\begin{equation}
	\int d^{D}\xp \, \e^{i \xp \cdot \big(p + p^{\prime} + \sum_{i = 1}^{N}k_{i}\big)} = (2\pi)^{D}\delta^{D}\big(p + p^{\prime} + K\big)
\end{equation}
where we denote $K := \sum_{i = 1}^{N}k_{i}$. Using this momentum conserving $\delta$-function simplifies the exponent of the Gaussian $\xm$ integral and the reader can verify that the result can be written as a ``Master Formula'' in a similar way to closed-loop amplitudes (c.f. (\ref{scalarqedmaster}))
\begin{align}
&\mathfrak{D}^{p'p}[k_{1}, \ep_{1}; \ldots; k_{N}, \ep_{N}] =(2\pi)^{D} \delta^{D}\big(p + p^{\prime} + K\big)(-ie)^{N}\int_{0}^{\infty}dT\, \e^{-m^{2}T}\non
&\hspace{0.5cm}\times \prod_{i = 1}^{N} \int_{0}^{T}d\tau_{i} \,  \e^{-T\big(p' + \frac{1}{T}\sum_{i = 1}^{N}(k_{i} \tau_{i} - i \ep_{i})\big)^{2}} \,\e^{\sum_{i, j = 1}^{N}\big[\Delta_{ij}  k_{i} \cdot k_{j}  - 2i \ddel_{ij}\ep_{i} \cdot k_{j} - \ddeld_{ij}\ep_{i}\cdot \ep_{j} \big]}\bigg|_{\linep}\,.
		\label{DppkFull}
\end{align}
Note that this is really only a function of $p$ and the $N$ independent $k_{i}$ and polarisation vectors\footnote{An equivalent form for (\ref{DppkFull}) that is sometimes useful is 
\begin{align}
&(2\pi)^{D} \delta^{D}\big(p + p^{\prime} + K\big)(-ie)^{N}\int_{0}^{T}dT\, \e^{-(p'^{2} + m^{2})T}\non
	 &\times\prod_{i = 1}^{N} \int_{0}^{T}d\tau_{i} \,  \e^{-2p'\cdot \sum_{i = 1}(k_{i} \tau_{i} - i \ep_{i})+ \sum_{i, j = 1}^{N}\big[\Upsilon_{ij}  k_{i} \cdot k_{j}  - 2i \dup_{ij}\ep_{i} \cdot k_{j} - \dupd_{ij}\ep_{i}\cdot \ep_{j} \big]}\bigg|_{\linep}
	\label{DppkFullUps}
\end{align} 
where $\Upsilon(\tau, \tau^{\prime}) := \frac{1}{2}\left|\tau - \tau^{\prime}\right| - \frac{1}{2}(\tau + \tau^{\prime})$ coincides with the Green function for the second derivative with mixed boundary conditions $q(0) = 0 = \dot{q}(T)$.}. See Fig. \ref{fig-multiphoton} for a diagrammatic representation of all Feynman diagrams generated by the master formula above -- note that as is usual in worldline calculations this generating function automatically takes into account the sum over diagrams related by exchange of external particle states. In future calculations the momentum conserving factor will often be understood implicitly and omitted. Exercise 3.4 will put this into the compact form originally given in \cite{Daik,line2}.

\subsubsection{Applications}
Now we will analyse the dressed propagators for some simple cases. Firstly, consider the $N = 0$ momentum space propagator:  (\ref{DppkFull}) gives 
\begin{equation}
	\mathfrak{D}^{p'p} = (2\pi)^{D}\delta^{D}\big(p + p^{\prime}\big) \int_{0}^{\infty}dT\, \e^{-(p'^{2} + m^{2})T} = \frac{(2\pi)^{D}\delta^{D}\big(p + p^{\prime}\big)}{p'^{2} + m^{2}}.
	\label{D0p}
\end{equation}
Of course one may just as well derive a position space proper time representation of this result using (\ref{DxxkFull}) which leads to the position space propagator related by inverse Fourier transform to (\ref{D0p}) -- see also the discussion about (\ref{scalpropfreehk} - \ref{scalpropfreehkx}). Note that this has provided us with the \textit{untruncated} vacuum propagator, so that when it comes to calculating scattering amplitudes we will need to amputate the external scalar legs by multiplying according to the LSZ prescription to convert the propagator into an amplitude
\begin{equation}
	i\hat{\mathfrak{D}}^{p'p} := (p^{\prime2} + m^{2})\mathfrak{D}^{p'p}(p^{2} + m^{2})\,.
	\label{Damp}
\end{equation}
Secondly, it is important to point out that all of our manipulations, and consequently the vacuum propagator above, are valid off-shell, both for the scalar legs and the external photons for $N > 0$. For this reason, they can be used as the building blocks for more complicated processes with multiple propagators -- that can even be internal lines -- ``sewn'' together. We will use this trick for the photons when we calculate the scalar self-energy below following \cite{line2}, but for an explicit calculation where scalar legs appear as virtual internal lines see \cite{BastLadder}. Thirdly, a key advantage of the worldline formalism is that at all steps we maintained the permutation symmetry under exchange of external photon legs, which symmetry is therefore retained by (\ref{DppkFull}). We will also discuss below how this can be exploited to maintain manifest gauge invariance throughout the calculation. This is just as in the one-loop case and is in direct contrast to the standard approach that would require us to calculate diagrams corresponding to distinct orderings separately and add their contributions together to restore the gauge invariance in the sum. 

Turning now to $N > 0$, as in the one-loop case the seagull diagrams (see figure \ref{ScalarQEDV}) of scalar QED are contained in the final term of the exponent of (\ref{DppkFull}), where the $\delta(\tau_{i} - \tau_{j})$ contained in $\ddeld$ inserts the external photons at coincident points along the line. However, for $N = 1$ there is no such term, since it is quadratic in $\varepsilon_{1}$, and expanding (\ref{DppkFull}) to $\mathcal{O}(\ep_{1})$ gives (we strip off the momentum conserving $\delta$-function)
\begin{align}
	\mathfrak{D}_{1}^{p'p}[k, \ep] &= e\ep \cdot \left(2p' + k\right) \int_{0}^{\infty}dT \, \e^{-T(p'^{2} + m^{2})}\int_{0}^{T}d\tau  \e^{\tau \left(p^{\prime 2}-p^2\right)}  \\
	&= \frac{e \ep \cdot \left(2p' + k\right)}{\left(p^{2} + m^{2}\right)\left(p^{\prime 2} + m^{2}\right)}\,.
	\label{Dpp1}
\end{align} 
Amputating the external legs according to (\ref{Damp}) one finds the expression for the scalar-photon-scalar vertex (see the left hand vertex of figure \ref{ScalarQEDV}).

Going to higher $N$, we now consider three traditional field theory calculations in turn: the linear Compton amplitude, the scalar self energy and a brief discussion of one loop corrections to the above scalar-photon vertex, see \cite{line2}. As a by-product, the Compton calculation will allow us to recover the 4-point vertex present in scalar QED. 

For the Compton amplitude we set $N = 2$ and expand (\ref{DppkFull}) to $\mathcal{O}(\ep_{1}\ep_{2})$, finding ($\Delta p := p^{\prime}-p$)
\begin{align}
\mathfrak{D}_{2}^{p^{\prime}p} &=- 2e^{2}\varepsilon_{1\mu}\varepsilon_{2\nu}\int_{0}^{\infty}dT\int_{0}^{T}\!d\tau_1\int_{0}^{T}d\tau_{2}\Big[\delta^{\mu\nu}\delta_{12} -\frac{1}{2}\left(\Delta p+\sigma_{21}k_{2}\right)^{\mu}\left(\Delta p+ \sigma_{12}k_{1}\right)^{\nu}\Big] \non
&\hspace{2cm}\times \e^{-T\left(p'^{2} + m^{2}\right)+\left|\tau_{1} - \tau_{2}\right|k_{1}\cdot k_{2} -  \Delta p\cdot \left(\tau_{1}k_{1} + \tau_{2}k_{2} \right)}\,,
	 \label{Dpp2}
\end{align}
where we shorten $\delta_{ij} := \delta(\tau_{i} - \tau_{j})$. It is left as an exercise to show that
\begin{align}
&\int_0^{\infty}dT\,\e^{-(m^2+p'^2)T}\int_0^Td\tau_1\int_0^{\tau_1}d\tau_2 \,\e^{k_1\cdot k_2 (\tau_1-\tau_2) - (p-p')\cdot (\tau_1k_1+\tau_2 k_2)}\non
&= \frac{1}{(m^2+p'^2)[m^2+(p'+k_1)^2](m^2+p^2)} 
 \label{int1}\\
 \non
&\int_0^{\infty}dT\,\e^{-(m^2+p'^2)T}\int_0^Td\tau_1\int_0^Td\tau_2 \, \delta(\tau_1-\tau_2)\,\e^{k_1\cdot k_2 \left|\tau_1-\tau_2\right| - (p-p')\cdot (\tau_1k_1+\tau_2 k_2)}\non
&=  \frac{1}{(m^2+p'^2)(m^2+p^2)}
 \label{int2}
\end{align}
that are sufficient to compute the parameter integrals in (\ref{Dpp2}). Clearly the results of (\ref{int1}) and (\ref{int2}) correspond to the familiar Feynman denominators of the scalar propagators of the Feynman diagrams for the amplitude in the standard approach (see the exercises of this section). Indeed, application of (\ref{Damp}) to truncate the external scalar legs leads to
\begin{align}
	i\hat{\mathfrak{D}}^{p'p}_{2}&= e^2\bigg\{\Big(\ep_{1} \cdot \Delta p \, \ep_{2} \cdot \Delta p - \ep_{1}\cdot k_{2} \ep_{2} \cdot k_{1}\Big)\Big(\frac{1}{\left(p'+k_{1}\right)^{2} + m^{2}}+\frac{1}{\left(p'+k_{2}\right)^{2} + m^{2}}\Big) \nonumber \\
	&\hspace{0.9cm}+ \Big( \ep_{1} \cdot \Delta p\,  \ep_{2} \cdot k_{1} - \ep_{1} \cdot k_{2}\ep_{2} \cdot \Delta p\Big)
	\Big(\frac{1}{\left(p'+k_{1}\right)^{2} + m^{2}}-\frac{1}{\left(p'+k_{2}\right)^{2} + m^{2}}\Big)
\non	&\hspace{2cm}-2\ep_{1} \cdot \ep_{2} \bigg\}.
	\label{D2amp}
\end{align}
Here the last term arises from the $\delta_{12}$ in (\ref{Dpp2}) and gives the contribution from the seagull diagram, thereby reproducing the scalar-scalar-photon-photon vertex of the scalar field theory. Now we use momentum conservation to simplify the $p$ and $p^{\prime}$ dependence and take the external photons and scalar legs on-shell, so that $k_{1}^{2} = 0 = k_{2}^{2}$ and $\ep_{1} \cdot k_{1} = 0 = k_{2} \cdot \ep_{2}$, whilst $p^{2} = -m^{2} = p^{\prime 2}$. These relations allow the scalar Compton amplitude (\ref{D2amp}) to be written as
\begin{equation}
i\hat{\mathfrak{D}}^{p'p}_{2}	=- 2e^{2}\ep_{1\mu}\left[\frac{p'^{\mu}p^{\nu}}{p'\cdot k_{1}}+ \frac{p'^{\nu}p^{\mu}}{p'\cdot k_{2}} + \eta^{\mu\nu}\right]\ep_{2\nu}.
	\label{scalarN2}
\end{equation}
This familiar result, valid in an arbitrary Lorentz frame, can be found in any standard textbook on quantum field theory. Squaring, summing over polarisations and going to the cross section (not forgetting the required two-particle phase space function!) are now standard steps. In particular since (\ref{scalarN2}) is transversal one may use 
\begin{equation}
	\sum_{\lambda} \ep_{\mu}^{\lambda \star}\ep_{\nu}^{\lambda} \rightarrow \eta^{\mu\nu}.
\end{equation}
Another option is to first choose a convenient gauge for the photon polarisation vectors (although in this case one must then compute the sum over polarisations by hand). From this point on the worldline formalism does not supply anything new, so we do not carry out these calculations. However in the spinor case elaborated below we will find that the first quantised approach can provide additional simplifications to the determination of the cross section. We shall discuss these in more detail below, where we also flesh out the steps required to determine the cross section.

\begin{figure}
	\centering
	\includegraphics[width=0.5\textwidth]{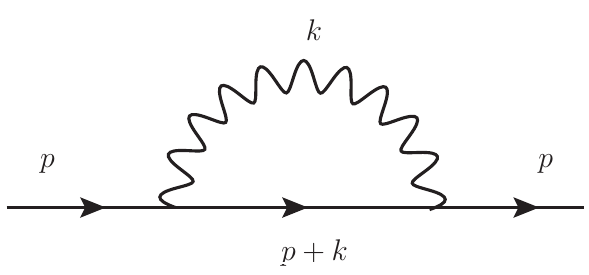}
	\caption{The standard irreducible Feynman diagram for the scalar self-energy at one-loop order.}
	\label{figSE}
\end{figure}

By now it should be clear that for general, fixed $N$, the un-amputated dressed propagator can be written in terms of polynomials analogous to the $P_{N}$ introduced in the previous section of these notes:
\begin{align}
&\mathfrak{D}^{p'p}[k_{1}, \ep_{1}; \ldots; k_{N}, \ep_{N}] =(2\pi)^{D} \delta^{D}\big(p + p^{\prime} + K\big)(-ie)^{N}(-i)^{N}\int_{0}^{\infty}dT\, \e^{-m^{2}T}\non
&\hspace{0.5cm}\times \prod_{i = 1}^{N} \int_{0}^{T}d\tau_{i} \, \bar{P}_{N} \e^{-T\big(p' + \frac{1}{T}\sum_{i = 1}^{N}k_{i} \tau_{i}\big)^{2}+\sum_{i, j = 1}^{N}\Delta_{ij}  k_{i} \cdot k_{j} } \,,
		\label{DppkFullP}
\end{align}
where we denoted the result of expanding the exponent to multi-linear order in the polarisation vectors according to
\begin{align}
&&\e^{-T\big(p' + \frac{1}{T}\sum_{i = 1}^{N}(k_{i} \tau_{i}-i\varepsilon_{i})\big)^{2}} \,\e^{\sum_{i, j = 1}^{N}\big[\Delta_{ij}  k_{i} \cdot k_{j}  - 2i \ddel_{ij}\ep_{i} \cdot k_{j} - \ddeld_{ij}\ep_{i}\cdot \ep_{j} \big]}\Big|_{\textrm{lin }\varepsilon_{1}\ldots\varepsilon_{N}}\non
&&:= (-i)^{N}\bar{P}_{N} \e^{-T\big(p' + \frac{1}{T}\sum_{i = 1}^{N}k_{i} \tau_{i}\big)^{2} + \sum_{i,j=1}^{N}\Delta_{ij}k_{i}\cdot k_{j}}\,.
\label{eqDefPbar}
\end{align}
As an illustrative example, it is easy to check that the above calculation for $N=1$ photons revealed the first non-trivial polynomial (clearly $\bar{P}_{0} = 1$)
\begin{align}
	\bar{P}_{1} &= 2\ddel_{11} \varepsilon_{1}\cdot k_{1} - 2\varepsilon_{1}\cdot (p' + k_{1})\cdot \frac{\tau_{1}}{T} \nonumber \\
	&= -\varepsilon_{1}\cdot (2p' + k_{1})\,.
\end{align}

\subsubsection{Loop corrections}
For the scalar self energy, indicated in Figure \ref{figSE}, we may take advantage of the fact that our Master Formula is valid with the external photons off-shell. The diagram we need can in fact be obtained by using the $N = 2$ photon formula if we identify $k_{1} = k = -k_{2}$ and sew the legs together with a photon propagator. In momentum space this is achieved by the replacement\footnote{Here we use Feynman gauge for the virtual photon, but we can work in an arbitrary covariant gauge by modifying the sewing procedure to $\ep_{1\mu}\ep_{2\nu} \rightarrow \int \frac{d^{D}k}{(2\pi)^{D}}\, \Big[\frac{\eta_{\mu\nu}}{k^{2}} - (1 - \xi)\frac{k_{\mu}k_{\nu}}{(k^{2})^{2}}\Big]$. The eager student may like to repeat the calculation including the gauge dependent part and compare with \cite{line2}.}  
\begin{equation}
	\ep_{1\mu}\ep_{2\nu} \rightarrow \int \frac{d^{D}k}{(2\pi)^{D}}\, \frac{\eta_{\mu\nu}}{k^{2}}.
	\label{sewEp}
\end{equation}
This means that (\ref{Dpp2}) can be reused and applying (\ref{sewEp}) it can be simplified to (upon sewing we earn $p = -p^{\prime}$ from the momentum conserving $\delta$-function)
\begin{align}
\Sigma_{\scal} &= e^{2}\int_{0}^{\infty} dT\ \e^{-(p'^{2} + m^{2}) T} \int_{0}^{T}d\tau_{1}\int_{0}^{\tau_{1}}d\tau_{2} \non
&\hspace{1cm}\times\int \frac{d^{D} k}{(2\pi)^{D}} \frac{1}{k^{2}}\left[ (2p '+ k)^{2} - 2D \delta_{12}\right] \e^{-(\tau_{1} - \tau_{2})\left(k^{2} + 2 k \cdot p'\right)}.
\end{align}
Note that to avoid overcounting we have chosen a fixed ordering of the insertion of the photon $\tau_{2} \leqslant \tau_{1}$ that fixes the momentum flow in the Feynman diagram. Recalling that the self-energy diverges in $D = 4$ we have for the moment kept the dimension arbitrary. In such dimensional regularisation the contribution from the seagull diagram vanishes when the $k$-integral is done. This leaves us with the first term in square brackets. We use
\begin{equation}
	\int_{0}^{T}d\tau_{1}\int_{0}^{\tau_{1}}d\tau_{2}\,  \e^{-(\tau_{1} - \tau_{2})\left(k^{2} + 2 k \cdot p'\right)} = \frac{T}{k^{2} + 2 k\cdot p'} - \frac{1 - \e^{-T \left(k^{2} + 2 k \cdot p'\right)}}{\left(k^{2} + 2 k \cdot p'\right)^{2}}
\end{equation}
and combine it with the exponent $\e^{-T(p'^{2} + m^{2})}$. The resulting proper time integral yields the simple product of Feynman denominators
\begin{align}
 \frac{1}{\left(p'^{2} + m^{2}\right)\left((p'+k)^{2} + m^{2}\right)\left(p'^{2} + m^{2}\right)}\,.
\end{align}
Of course the first and last factor on the denominator come from the external legs that have not yet been amputated, corresponding as before to these scalar propagators. Putting this back into our calculation for $\Sigma$ gives
\begin{equation}
	\Sigma_{\scal} = \frac{e^{2}}{\left(p'^{2} + m^{2}\right)^{2}}\int \frac{d^{D}k}{(2 \pi)^{D}} \frac{\left(2p' + k\right)^{2}}{k^{2}\left[(p'+k)^{2} + m^{2}\right]}.
\end{equation}
This is in agreement with calculation based on the scalar QED Feynman rules, as the studious student will be glad to verify. An alternative way of arriving at this result is from (\ref{D2amp}), putting $k_{1} = k= -k_{2}$, $p^{\prime} =-p$, sewing the polarisation tensors and integrating over $q$ as above. Consulting the list of integrals in the appendix and truncating the external legs provides
\begin{equation}
	\widehat{\Sigma}_{\scal} = \frac{e^{2}(m^{2})^{\frac{D}{2}-2}}{(4\pi)^{\frac{D}{2}}} \Gamma\left(1 - \frac{D}{2}\right)\left[m^{2} - 2(m^{2} - p'^{2})\,_2F_{1}\big(2-\frac{D}{2},1;\frac{D}{2};-\frac{p'^2}{m^2}\big)\right].
\end{equation}
The divergence as $D \rightarrow 4$ is contained in the leading $\Gamma$-function that has simple poles at the negative integers\footnote{The Hypergeometric function in $\widehat{\Sigma}_{\textrm{scal}}$ has expansion about $D = 4$ that takes the form $_2F_{1}\big(2-\frac{D}{2},1;\frac{D}{2};-\frac{p'^2}{m^2}\big) = 1 + \mathcal{O}(D-4)$.}. For the result in an arbitrary covariant gauge see \cite{line2}. The above result can now be used for one-loop mass and field strength renormalisation according to standard analysis of the propagator. 

\begin{figure}
	\centering
	\includegraphics[width=0.8\textwidth]{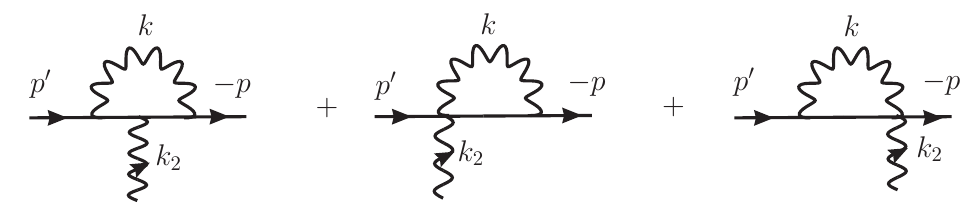}
	\caption{The three relevant diagrams for the scalar-photon vertex correction. These are sufficient for the present purpose, but restricting the integration regions to this ordering does break gauge invariance.}
	\label{figVert}
\end{figure}
At $N=3$ it is possible to study non-linear Compton scattering or, as we shall do here for simplicity, one loop corrections to the scalar-photon vertex as in Figure \ref{figVert}. Here we shall be very brief and leave the detailed calculations to the interested reader. The worldline representation of these diagrams comes from $\mathfrak{D}^{p'p}_{3}[k_{1}, \ep_{1}; k_{2}, \ep_{2}; k_{3}, \ep_{3}]$ where photon $1$ and $3$, say, are sewn in the same way as for the self energy calculation above. After suitable manipulation one finds
\begin{align}
\hspace{-2em}	&D_{3}^{p'p}[k, \ep_{1}; k_{2}, \ep_{2}; -k, \ep_{3}]\bigg|_{\textrm{sewn}} = -e^{3}\int \frac{d^{D}k}{(2\pi)^{D}} \frac{1}{k^{2}}\int_{0}^{\infty}dT\, \e^{-(p'^{2} + m^{2})T} \int_{0}^{T}d\tau_{1}\int_{0}^{\tau_{1}}d\tau_{2}\int_{0}^{\tau_{2}}d\tau_{3} \nonumber \\
\hspace{-2em}	&\times \Big[(2p' - k)\cdot (2p + k)\, \ep_{2}\cdot \left(k_{2} + 2(p' - k)\right) - 2\ep_{2} \cdot (2p+ k)\delta_{12} + 2\ep_{2} \cdot (2p' - k)\delta_{23}\Big]\nonumber \\
\hspace{-2em} & \times \e^{-(k^{2} - 2p'\cdot k)\tau_{1} - (k_{2}^{2} + 2 k_{2} \cdot (p' - k))\tau_{2} - (-k^{2} + 2k \cdot (p' + k_{2}) )\tau_{3}}.
	\label{Dpp3}
\end{align}
The attentive reader will note that the ordering of the photon insertions has been fixed -- this is because the three terms in square brackets represent the three diagrams in Figure \ref{figVert} in order, as can be seen by examining the $\delta$-functions of the second and third term that generate the seagull diagrams as usual. The important point is that the exponent is linear in each parameter $\tau_{i}$ so that the three integrals can easily be computed. Just to give a flavour of the calculations we focus on the first diagram on the left in figure \ref{figVert}. The parameter and proper time integrals yield, following truncation,
\begin{equation}
	-e^{3} \int \frac{d^{D}k}{(2\pi)^{D}} \frac{ \ep_{2}\cdot [k_{2} + 2(p' - k)]}{k^{2}}\frac{(2p' - k)\cdot (2p + k)}{\left[(p'-k)^{2} + m^{2}\right]\left[(p + k)^{2} + m^{2}\right]}\,,
	\label{Verta}
\end{equation} 
which can now be expressed in terms of the hypergeometric functions given in the appendix. Computing the other contributions leads to a particularly convenient tensor decomposition of the vertex (and the self energy). Specifically the contribution (\ref{Verta}) to the vertex can be written in Feynman gauge (in $D$-dimensions and where $\Gamma_{a} = \Gamma_{a}^{\mu} \ep_{2\mu}$) as 
\begin{align}
	\hspace{-2em}\Gamma_{a}^{\mu} = -\frac{e^{3}}{(2\pi)^{D}}\left[ -\Delta p^{\mu} K^{(0)} + 2K_{\mu}^{(1)} + 2\Delta p^{\nu} \left( \Delta p^{\mu} J_{\nu}^{(1)} - 2J_{\mu\nu}^{(2)} \right) +4 p\cdot p^{\prime}\left( \Delta p^{\mu} J^{(0)} - 2J_{\mu}^{(1)} \right) \right]
\end{align}
where as earlier $\Delta p = p'-p$ and the integrals are defined in the appendix. For further details and calculation of the other contributions the reader is referred to \cite{line2}. Beyond the vertex calculation we have reviewed, the authors also investigated the LKF transformations associated to a gauge transformation of individual photons attached to the open line.

In the special case that the external photons are assumed to have vanishingly small energy, the interesting physical information is contained in part of the amplitude that is multi-linear not only in the polarisation vectors but also in the photon momenta. This limit is particularly amenable to a worldline treatment and we shall return to this point when we discuss the scalar propagator in a constant electromagnetic background field in section \ref{ScalField}.

\subsection{Gauge invariance}
\label{secGauge}
As hinted at above, we can exploit the form of the vertex operator to arrive at a manifestly transverse form of the amplitude for on-shell processes. As is of course well known, transversality of the on-shell amplitude\footnote{Here we return to the more traditional notation for the amplitude, which we recall is defined in Minkowski space, writing it as $\mathcal{M}$ for both scalar and spinor QED. In the present case we have $\mathcal{M}^{p'p}_{N} = \mathfrak{D}^{p'p}_{N}$. In the main text we extract the polarisation vectors from $\mathcal{M}$ -- by convention we take those to the left(right) of the tensor $\mathcal{M}_{\mu\cdots \rho \,\alpha \cdots \zeta}$ to be associated to incoming (outgoing) photons.} decomposed as $\mathcal{M} := \varepsilon_{\mu}\ldots \varepsilon_{\rho}\mathcal{M}^{\mu\cdots \rho \alpha \cdots \zeta}\varepsilon_{\alpha}\ldots \varepsilon_{\zeta}$ is guaranteed by the Ward identity which encodes the gauge symmetry of the theory. 

At the level of the on-shell amplitude the Ward identity is realised by the statement that $\mathcal{M}$ should remain invariant under the transformation $\varepsilon_{\mu}(k_{i}) \rightarrow \varepsilon_{\mu}(k_{i}) + \lambda_{i}k_{i\mu}$, or in other words we can add any longitudinal part to the photon polarisation vectors without changing the prediction for the amplitude. As a consequence we learn that, on-shell, we must have $k_{i\xi}\mathcal{M}^{\mu\cdots \xi\cdots \zeta} = 0$ for any $i \in 1, \ldots N$. It is common knowledge that in the standard approach individual Feynman diagrams do not (necessarily) obey this relation, but it is rather the sum of diagrams that recovers this result. Since in the worldline formalism our Master Formulae take into account the permutations of external legs we expect them to satisfy this result already at the level of the parameter integrals -- here we show this is the case and describe briefly how to make this gauge invariance manifest.

We first consider gauge transforming the vertex operator for a photon inserted either along an open line or (returning to the previous chapter) closed loop:
\begin{align}
	V_{\textrm{scal}}^{\gamma}[k, \varepsilon + \lambda k] &= \int_{0}^{T} d\tau \big[\varepsilon \ \cdot \dot{x} + \lambda k \cdot \dot{x}(\tau)\big]\e^{i k \cdot x(\tau)}\nonumber \\
&= \int_{0}^{T} d\tau \Big[\varepsilon \cdot \dot{x} - i \lambda\frac{d}{d\tau}\Big]\e^{i k \cdot x(\tau)}\,.
\label{eqVgauge}
\end{align}
Around the loop the total derivative trivially integrates to zero. For the open lines discussed in this section it evaluates to a function of the endpoints (boundary terms):
\begin{equation}
	\Delta V_{\textrm{scal}}^{\gamma}[k, \varepsilon] = i\lambda\big(\e^{i k \cdot x} - \e^{ik  \cdot x'}\big)\,.
	\label{eqDeltaV}
\end{equation}
Despite this, the additional terms will not survive the amputation indicated in (\ref{Damp}) and as such will not contribute in the LSZ formula. To understand this, consider the effect of the exponents in (\ref{eqDeltaV}) when passing to momentum space; the Fourier exponents will be shifted by $p^{\prime}\rightarrow p^{\prime}+ k$ and $p\rightarrow p- k$ which will change the location of the poles in $p^{2}$ or $p^{\prime 2}$, moving them away from $-m^{2}$ so that when we evaluate (\ref{Damp}) for the variation we will get a sum of terms involving variations in amplitude
\begin{equation}
	\hspace{-0.75em}i \Delta \hat{\mathfrak{D}}^{p'p} = i\lambda \Big[\frac{p^{\prime 2} + m^2}{(p^{\prime} + k)^2 + m^2}\delta_{+}\mathcal{M}^{p'p}(p^2 + m^2) - (p^{\prime 2} + m^2)\delta_{-}\mathcal{M}^{p'p}\frac{p^2 + m^2}{(p-k)^2+m^2}\Big]\,,
\end{equation}
whose on-shell limit vanishes. Thus we have proved that the Master Formulae will be transversal when used to generate on-shell amputated amplitudes.

\subsubsection{Manifest transversalilty}
We can, in fact, take advantage of the above results to write Master Formulae in a manifestly transverse way, that will be equal to our earlier versions up to terms that will not survive in the on-shell amplitudes after amputation. To do this, recall that the Maxwell field strength tensor, $F_{\mu\nu}$ is invariant under a gauge transformation $A_{\mu}(x) \rightarrow A_{\mu}(x) - \partial_{\mu}\alpha(x)$. For our photons represented by a sum of plane waves, with field strength tensor $F_{\mu\nu}(x) = \sum_{i=1}^{N}if_{i\mu\nu}\e^{i k_{i} \cdot x}$ with $f_{i\mu\nu} = k_{i\mu}\varepsilon_{i\nu} - k_{i\nu}\varepsilon_{i\mu}$ this is realised by the invariance of $f_{i}$ under $\varepsilon_{\mu}(k_{i}) \rightarrow \varepsilon_{i}(k_{i})$, making the individual photon field strength tensors transverse.

To write the amplitude explicitly in terms of the field strength tensors we choose $\lambda$ in (\ref{eqVgauge}) as $\lambda = - \frac{r\cdot \varepsilon}{r\cdot k}$, where $r$ is an arbitrary ``reference vector'' that only has to satisfy $r\cdot k \neq 0$. With this choice we arrive at a new vertex operator
\begin{equation}
	V_{\textrm{scal}}^{\gamma}[k, \varepsilon ; r] := \int_{0}^{T} d\tau \, \frac{r\cdot f \cdot \dot{x}}{r\cdot k} \e^{i k \cdot x(\tau)}\,,
\end{equation}
that is written entirely in terms of the photon's field strength $f$. This vertex operator is equally valid for use in the Master Formula for the aim of producing \textit{on-shell} scattering amplitudes and corresponds to the simple replacement $\varepsilon_{i} \rightarrow \frac{r_{i}\cdot f_{i}}{r_{i}\cdot k_{i}}$. The resulting scattering amplitudes must ultimately turn out to be independent of the choice of the reference after amputation and going on-shell. The replacement can be made in the one-loop formulae (\ref{scalarqedmaster}) or (the scalar part of) (\ref{DNpointspin}), or in the tree-level result (\ref{DppkFull}) which turns into
\begin{align}
\hspace{-1.5em}&\mathfrak{D}^{pp'}[k_{1}, \ep_{1}; \ldots; k_{N}, \ep_{N}] \simeq (2\pi)^{D} \delta^{D}\big(p + p^{\prime} + K\big)(-ie)^{N}\int_{0}^{\infty}dT\, \e^{-m^{2}T}\non
\hspace{-1.5em}&\hspace{0.5cm}\times \prod_{i = 1}^{N} \int_{0}^{T}d\tau_{i} \,  \e^{-T\big(p' + \frac{1}{T}\sum_{i = 1}(k_{i} \tau_{i} - i \frac{r_{i}\cdot f_{i}}{r_{i}\cdot k_{i}})\big)^{2}} \,\e^{\sum_{i, j = 1}^{N}\big[\Delta_{ij}  k_{i} \cdot k_{j}  - 2i \ddel_{ij}\frac{r_{i}\cdot f_{i}\cdot k_{j}}{r_{i}\cdot k_{i}}  + \ddeld_{ij}\frac{r_{i}\cdot f_{i} \cdot f_{j}\cdot r_{j}}{r_{i}\cdot k_{i} r_{j}\cdot k_{j}}\big]}\bigg|_{\textrm{lin } f_{1}\ldots f_{N}}\,.
		\label{DppkFullf}
\end{align}
The symbol $\simeq$ indicate equivalence on-shell, even though the actual functions are not really equal. To make practical use of this formula, a ``good'' choice of the reference vectors is needed. For $N > 1$, choosing $r_{i} = p^{\prime}$ has proven to be useful in reducing the complexity of the resulting parameter integrals. On the loop, the natural choice would be a symmetric identification $r_{i} = k_{i+1}$ modulo $N$ or similar. 

Here we restrict our attention to the simplest example of recovering the vertex for $N = 1$. On-shell, the Bern-Kosower exponent doesn't contribute ($k^2 = 0$ and $f \cdot k = 0$) so expanding to linear order in $\varepsilon$ provides
\begin{align}
	&2 \frac{r \cdot f \cdot p^{\prime}}{r \cdot k} \int_{0}^{T}dT \int_{0}^{T}d\tau \, \e^{-T(p^{\prime 2} + m^{2})}\nonumber\\
	=& \frac{1}{p^{\prime 2} + m^2}2 \frac{r \cdot f \cdot p^{\prime}}{r \cdot k} \frac{1}{p^2 + m^2}\,.
\end{align}
Then expanding the field strength tensor, the quotient evaluates to $2\varepsilon \cdot p^{\prime}$ thanks to the on-shell relation\footnote{In fact this is why we cannot fix $r = p^{\prime}$ or $r = p$ in this case.} $p^{\prime} \cdot k = 0$. This agrees with (\ref{Dpp1}) on mass-shell. The above choices for the reference vectors at higher $N$ are explored in the problems for this section. 

\section{Spinor QED}
\label{spinLine}
Moving on from scalar QED, here we consider the open fermion line which presents considerably more mathematical complexity in its description. Most of this stems from the spin structure, since we must specify spinor indices $\alpha$ and $\beta$ at either end of the line. In fact for some time it was thought that this additional structure would preclude a first quantised representation of the dressed fermion propagator sufficiently simple to offer a viable alternative to the standard approach. However, here we will show, following the results presented in \cite{SpinProp, SpinProp2}, that there is a suitable worldline representation of this object that retains the calculational efficiency seen at higher loop order\footnote{For an alternative approach using coherent states see \cite{Olindo}.}. Before beginning the construction it is important to clarify a point of notation: for the $\gamma$-matrices required for the spin degrees of freedom of a Dirac field we use the convention $\{\gamma^{\mu}, \gamma^{\nu}\} = -2\eta^{\mu\nu}$ with metric signature $(-, +, +, +)$ in Minkowski space. In the Euclidean space for the path integral quantisation we use $\{\gamma^{\mu}, \gamma^{\nu}\} = -2\delta^{\mu\nu}$.

The matrix element to be computed is given in configuration space by the transition amplitude between the endpoints of the line analogous to (\ref{Dxxp}) 
\begin{equation}
	S_{\beta\alpha}^{x^{\prime}x} := \big<x', \beta \big|\big[-i\slashed{\partial} + e \As + m\big]^{-1}\big| x, \alpha\big>
\end{equation}
which represents the propagation of a spin $\frac{1}{2}$ particle from spin state $\alpha$ at fixed point $x$ to spin state $\beta$ at fixed point $x'$ whilst interacting with an electromagnetic field $A^\mu$. This is clearly a Green function for the Dirac operator $\hat{\ps} + e\As + m$. We apply a useful trick related to squaring the Dirac operator to rewrite this as 
\begin{equation}
	S_{\beta\alpha}^{x^{\prime}x}  =\left(-\hat{\ps} - e \As(x') + m\right) \big<x', \beta \big|\big[-\left(\hat{\ps} + e \As\right)^{2} + m^{2}\big]^{-1}\big|x, \alpha\big>.
	\label{Ssecond}
\end{equation}
Noting now that $$-\left(\hat{\ps} + e \As\right)^{2} + m^{2} = \left(\hat{p} + e A\right)^{2} + m^{2} + \frac{ie}{4}\big[\gamma^{\mu}, \gamma^{\nu}\big]F_{\mu\nu}\,,$$ it becomes clear that the matrix element takes the form of the propagator for a scalar particle (with dynamics given by the Klein-Gordon equation) interacting additionally with a vector valued potential representing the spin coupling to the electromagnetic field\footnote{The knowledgeable student may recognise this as representing a transition between the first- ad second-order formalism of spinor QED -- this forms a key part of \cite{Morgan}.}. Then following the same steps as in scalar QED above, we are able to express the propagator in the form
\begin{align}
	S_{\beta\alpha}^{x^{\prime}x} &= \Big[m \bone + i \gamma \cdot \big(\partial'+ ieA(x')\big)\Big]_{\beta \rho}K_{\rho \alpha}^{x^{\prime}x} \,,\label{Sk}
	\end{align}
	where $\partial'\equiv \frac{\partial}{\partial x'}$ and the kernel is given by  
	\begin{align}
	K_{\rho \alpha}^{x^{\prime}x} &= \int_{0}^{\infty} dT\, \e^{-m^{2}T} \int_{x(0) = x}^{x(T) = x'}\hspace{-1em}\mathscr{D}x(\tau) \, \e^{-\int_{0}^{T}d\tau \left[\frac{\dot{x}^{2}}{4} + i e \dot{x}\cdot A(x(\tau))\right]}\mathcal{S}_{\rho \alpha}[x(\tau), A[x(\tau)]]\,.
	\label{Korig}
\end{align}
Now the spin factor, $\mathcal{S}[x, A(x)]$, introduced in equations (\ref{spinorQEDpi}) and (\ref{defspinfactor}), enters with spinor indices attached, associated to the spin of the fermion at either end of the line. This matrix-valued path ordered exponential has a path integral representation over Grassmann variables, as in the one-loop case, only with an appropriate change in boundary conditions for open lines. This is not the only way to deal with the matrix-valued potential (one may simply work with path ordered exponentiated line integrals, or alterntively make use a coherent state basis in the path integral) but it has been proven to be convenient in worldline calculations. 

The explicit form of this path integral\footnote{The path integral is normalised as in (\ref{psifreepi}), such that $$\int_{\xi(0) + \xi(T) = \eta}\hspace{-1.5em}\mathscr{D}\xi(\tau) \, \e^{-\int_{0}^{T}d\tau \frac{1}{2}\xi \cdot \dot{\xi}} = 2^{\frac{D}{2}}$$ counts the degrees of freedom of a spinor field in dimension $D$.} can be shown to be as follows, recalling that $\mathscr{P}$ indicates path ordering as described in previous sections of these notes, 

\begin{align}
\hspace{-1.5em} \mathcal{S}_{\rho \alpha}[x(\tau), A(x(\tau))] :&=	\mathscr{P}\left\{ \e^{-\frac{ie}{2}\int_{0}^{T}d\tau\, \gamma^{\mu} F_{\mu\nu}(x(\tau)) \gamma^{\nu}} \right\}_{\rho \alpha} \non
\hspace{-1.5em} &= 2^{-\frac{D}{2}} \textrm{symb}^{-1}\left\{ \int_{\xi(0) + \xi(T) = 2\eta}\hspace{-2.5em}\mathscr{D}\xi(\tau) \, \e^{-\int_{0}^{T}d\tau \left[\frac{1}{2}\xi \cdot \dot{\xi} - ie \xi \cdot F(x(\tau)) \cdot \xi \right] - \frac{1}{2} \xi(T) \cdot \xi(0)}\right\}_{\rho\alpha}.
 \label{Sxi}
\end{align}
Note that the boundary conditions on $\xi$ have been modified from the earlier anti-periodic conditions in the case of the loop to depend now upon a Grassmann parameter $\eta$ (and the boundary contribution $ - \frac{1}{2} \xi(T) \cdot \xi(0)$ is then required both for a consistent variational principle and to maintain a worldline supersymmetry along with the bosonic coordinates that we describe below). As expected, the closed loop result can be recovered through taking the trace. For open lines, however, the path integral above becomes	 a function of $\eta$ which generates the Dirac matrix structure through the so-called \textit{symbol map}.

This symbol map has appeared at various points in the historic literature on first quantised representations of open lines (see, for example, \cite{FradkinGitman}). Its function is to generate the $\gamma$-matrix structure of the propagator by converting between fully anti-symmetrised products\footnote{The square brackets around $n$ indices indicate anti-symmetrisation with a combinatorical factor of $\frac{1}{n!}$, or $$\gamma^{[\alpha_1}\gamma^{\alpha_2}\ldots \gamma^{\alpha_n]}\equiv\frac{1}{n!}\sum_{\pi\in S_n}{\rm sign}(\pi)\,\gamma^{\alpha_{\pi(1)}}\gamma^{\alpha_{\pi(2)}}\cdots \gamma^{\alpha_{\pi(n)}}.$$} of $n$ $\gamma$ matrices and $n$ Grassmann vectors as:
\begin{equation}
	\textrm{symb}\Big\{\gamma^{[\alpha}\gamma^{\beta}\ldots \gamma^{\rho]}\Big\} := (-i\sqrt{2})^n\eta^{\alpha}\eta^{\beta}\ldots\eta^{\rho}\,.
\end{equation}
Note that the right hand side of this equation vanishes for a product of more than $D$ $\eta$s whilst examination of the left hand side shows that the inverse symbol map in (\ref{Sxi}) expresses the result in the Clifford basis of the Dirac algebra; in 4 dimensions this is spanned by $\{\bone, \gamma^{\mu}, \frac{1}{2}\sigma^{\mu\nu}, \gamma^{\mu}\gamma_{5}, \gamma_{5}\}$, in which we use $\sigma^{\mu\nu} = \frac{1}{2}[\gamma^{\mu}, \gamma^{\nu}]$. The proof of (\ref{Sxi}) is beyond the scope of these notes -- see \cite{SpinProp} -- but exercise (3.8) asks you to prove a simpler relation of a similar form. 

To get practice with the use of this symbol map, it is instructive to consider the following result (that arises for the fermion propagator in a constant electromagnetic background where the anti-symmetric matrix $\mathcal{Z}_{\mu\nu}$ would be proportional to the background field strength tensor) in $D = 4$:
\begin{align}
	&\textrm{symb}^{-1}\left\{ \e^{\eta \cdot \mathcal{Z} \cdot \eta} \right\} = \textrm{symb}^{-1}\left\{1 + \eta^{\mu}\eta^{\nu}\mathcal{Z}_{\mu\nu} + \frac{1}{2}\eta^{\mu}\eta^{\nu}\eta^{\alpha}\eta^{\beta}\mathcal{Z}_{\mu\nu}\mathcal{Z}_{\alpha \beta} \right\} \nonumber \\
&= 1 - \frac{1}{2} \times \frac{1}{2!} \big[\gamma^{\mu}, \gamma^{\nu}\big]\mathcal{Z}_{\mu\nu} + \frac{1}{2\times 4 \times4!} \sum_{\sigma \in S_{4}} (-1)^{\sigma} \gamma^{\sigma(\mu)}\gamma^{\sigma(\nu)} \gamma^{\sigma(\alpha)}\gamma^{\sigma(\beta)}\mathcal{Z}_{\mu\nu}\mathcal{Z}_{\alpha\beta} \nonumber \\
	&= 1 -\frac{1}{2} \sigma^{\mu\nu}\mathcal{Z}_{\mu\nu} - \frac{i}{8}\gamma_{5} \epsilon^{\mu\nu\alpha\beta}\mathcal{Z}_{\mu\nu}\mathcal{Z}_{\alpha\beta} \nonumber \\
&= 1 - \frac{1}{4}\big[\gamma^{\mu}, \gamma^{\nu}\big]\mathcal{Z}_{\mu\nu}  + \frac{i}{4} \gamma_{5}\tr(\mathcal{Z} \cdot \widetilde{\mathcal{Z}})\,,
	\label{symbZ}
\end{align}
where the dual of a matrix is defined by $\widetilde{\mathcal{Z}}^{\mu\nu} := \frac{1}{2}\epsilon^{\mu\nu\alpha\beta}\mathcal{Z}_{\alpha\beta}$. 

Now, looking again at the path integral (\ref{Sxi}) it is convenient to shift the integration variable so as to homogenise the boundary conditions: setting $\xi^{\mu}(\tau) := \psi^{\mu}(\tau) + \eta^{\mu}$ we get an equivalent path integral that is more similar to the closed loop case,
\begin{align}
 \hspace{-0.5em}\mathcal{S}_{\rho \alpha}[x(\tau), A(x(\tau))] &=	\mathscr{P}\left\{ \e^{-\frac{ie}{2}\int_{0}^{T}d\tau\, \gamma^{\mu} F_{\mu\nu}(x(\tau)) \gamma^{\nu}} \right\}_{\rho \alpha}\nonumber \\ 
\hspace{-0.5em} &= 2^{-\frac{D}{2}} \textrm{symb}^{-1}\left\{ \oint_{AP} \hspace{-0.5em}\mathscr{D}\psi(\tau) \, \e^{-\int_{0}^{T}d\tau \left[\frac{1}{2}\psi \cdot \dot{\psi} - ie (\psi + \eta) \cdot F(x(\tau)) \cdot (\psi +  \eta) \right]} \right\}_{\rho\alpha}.
 \label{Spsi2}
\end{align}
Note also that this has removed the awkward boundary contribution and provides anti-periodic (AP) boundary conditions on $\psi$, at the expense of introducing explicit dependence upon $\eta$ in the action. With this change we finally arrive at a worldline action for the path integral representation of the kernel, $K$, similar to that used in the loop case in equation (\ref{defLspin}),
\begin{equation}
	\hspace{-0.75em}S[x, \psi; A] = \int_{0}^{T}d\tau \left[\frac{\dot{x}^{2}}{4} + \frac{1}{2} \psi \cdot \dot{\psi} + i e \dot{x} \cdot A(x(\tau)) - ie (\psi + \eta)\cdot F(x(\tau)) \cdot (\psi + \eta)\right].
	\label{Sline}
\end{equation}
Before proceeding we note that, as in the closed-loop, one of the benefits of a path integral version of the open line spin factor is that (\ref{Sline}) is invariant under a global supersymmetry generated by a Grassmann parameter $\varrho$ (c.f. (\ref{susy}) above)
\begin{align}
	\delta_{\varrho}x^{\mu} &= -2\varrho \psi^{\mu} \nonumber \\
	\delta_{\varrho}\psi^{\mu} &= \varrho \dot{x}^{\mu}.
	\label{susyline}
\end{align}
Despite being broken by the different boundary conditions on these variables the supersymmetry nonetheless provides useful calculational simplifications and will eventually motivate a superspace formulation of the path integral.

\subsection{Master formula}
To extract scattering amplitudes in vacuum we again specialise the gauge field to a sum of plane waves:
\begin{equation}
	A_{\mu}(x(\tau)) = \sum_{i = 1}^{N}\ep_{i\mu}\, \e^{i k_{i} \cdot x(\tau)}\, ; \qquad F_{\mu\nu}(x(\tau)) = i \sum_{i = 1}^{N} f_{i\mu\nu} \, \e^{i k_{i} \cdot x(\tau)},
\end{equation}
where we reuse the photon field strength tensor introduced in section \ref{secGauge}, $f_{i\mu\nu} := \left(k_{i\mu}\ep_{i\nu} - \ep_{i\mu}k_{i\nu}\right)$. As is now familiar, we insert this into the worldline action (\ref{Sline}) and then expand to $\mathcal{O}(e^{N})$, taking the $N!$ terms with distinct polarisations. Doing so produces the spinor equivalent of (\ref{Dxxk}),
\begin{align}
K^{x^{\prime}x}[A] &= (-ie)^{N}2^{-\frac{D}{2}}\int_{0}^{\infty} dT\, \e^{-m^{2}T} \int_{x(0) = x}^{x(T) = x'}\hspace{-1.5em}\mathscr{D}x(\tau)\, \e^{-\int_{0}^{T}d\tau \, \frac{\dot{x}^{2}}{4}} \non
	&\times \textrm{symb}^{-1}\left\{ \oint_{AP} \mathscr{D}\psi(\tau) \, \e^{-\int_{0}^{T}d\tau\, \frac{1}{2} \psi \cdot \dot{\psi}}\prod_{i = 1}^{N} V_{\spin}^{\gamma}[k_{i}, \ep_{i}; \eta]\right\}\,,
	\label{KPI}
\end{align}
with vertex operator insertions
\begin{equation}
	V_{\spin}^{\gamma}[k, \ep; \eta] := \int_{0}^{T}d\tau \left[ \ep \cdot \dot{x}(\tau) + 2i \ep \cdot (\psi + \eta)\, k \cdot (\psi +\eta)\right] \e^{i k \cdot x(\tau)}.
	\label{eqVspineta}
\end{equation}
As in the scalar case, we absorb the open line boundary conditions on $x(\tau)$ by setting $x(\tau) = \hat{x}(\tau) + q(\tau)$, implying Dirichlet boundary conditions on $q(\tau)$. With these variables, and applying the usual exponentiation trick, the vertex operator becomes
\begin{equation}
	V_{\eta}^{x^{\prime}x}[k, \ep] := \int_{0}^{T}d\tau \, \e^{i k \cdot \hat{x}(\tau) + i k \cdot q(\tau) + \ep \cdot \left(\frac{\xm}{T} + \dot{q}\right) + 2i \ep \cdot (\psi + \eta)\, k \cdot (\psi + \eta)}\bigg|_{\textrm{lin} \ep}
\end{equation}
so that we must calculate the path integral
\begin{align}
 K^{x^{\prime}x}[A] &= (-ie)^{N}2^{-\frac{D}{2}}\int_{0}^{\infty}\! dT\, \e^{-m^{2}T} \e^{-\frac{\xm^{2}}{4T}} \int_{\DBC} \mathscr{D}q(\tau)\, \e^{-\int_{0}^{T}d\tau \, \frac{\dot{q}^{2}}{4}}\non
 &\times \textrm{symb}^{-1}\left\{ \oint_{AP}\mathscr{D}\psi(\tau) \, \e^{-\int_{0}^{T}d\tau\, \frac{1}{2} \psi \cdot \dot{\psi}}\prod_{i = 1}^{N} V_{\eta}^{x^{\prime}x}[k_{i}, \ep_{i}]\right\}\bigg|_{\linep}.
	\label{KDBC}
\end{align}
Note that as in the scalar case the exponent of the vertex operators contains a part linear in $q$ that can be combined with the bosonic kinetic term through completing the square. However there is now an additional piece quadratic in $\psi$. In principle this could be absorbed into the fermionic kinetic term, but it would lead to a complicated worldline Green function and a functional determinant dependent upon the polarisation vectors. One choice is to forgo the exponentiation of the fermionic part of the vertex operator, and subsequently to apply Wick's theorem for contractions of the Grassmann fields according to (\ref{defGF} -- \ref{wickgrassmann}):
\begin{equation}
	\left<\psi^{\mu}(\tau) \psi^{\nu}(\tau^{\prime})\right>  = \frac{1}{2}G_{F}(\tau, \tau^{\prime}) = \frac{1}{2}\,{\rm sign}(\tau - \tau^{\prime}).
\end{equation}
Below we shall do precisely this in such a way as to arrive at an appealing ``spin-orbit decomposition.'' First, however, we describe an alternative, which is to find a way to linearise the interaction terms. We do this by appealing to the supersymmetry (\ref{susyline}) which allows us to frame the worldline theory in superspace and obtain a Master Formula similar to the scalar result. 

\subsubsection{Superspace}
The idea of superspace is to make the supersymmetry (\ref{susyline}) manifest. We introduce an additional Grassmann coordinate $\theta$, to pair along with the familiar bosonic coordinate $\tau$, so that our parameter domain is modified to $\tau \rightarrow \tau | \theta$. This requires us also to introduce a new derivative -- a covariant derivative -- defined by $D := \partial_{\theta} - \theta \partial_{\tau}$. With this machinery we can combine our fields $q^{\mu}(\tau)$ and $\psi^{\mu}(\tau)$ into a \textit{superfield},
\begin{equation}
	Q^{\mu}(\tau) := q^{\mu}(\tau) + \sqrt{2}\theta \psi^{\mu}(\tau).
\end{equation} 
Functions of superfields have a finite expansion in $\theta$ due to $\theta^{2} := 0$. To write the worldline action as a superspace integral we must also define integration with respect to Grassmann variables. Our convention is 
\begin{equation}
	\int d\theta \, 1 := 0\, ; \qquad \int d\theta\, \theta := 1\,.
\end{equation}
It is a useful exercise to convince oneself that the supersymmetry transformation in (\ref{susyline}) can be compactly written as $\delta_{\varrho}Q^{\mu} = -\sqrt{2}D P_{\varrho}^{\mu}$ where we defined the (Grassmann odd) superfield $P_{\varrho}^{\mu} = \varrho q^{\mu} + \sqrt{2} \theta (\varrho \psi^{\mu})$.

Now we can write the kinetic terms and the vertex operator as (left as an exercise to show) 
\begin{align}
	&\int_{0}^{T}d\tau \, \left[\frac{\dot{q}^{2}}{4} + \frac{1}{2}\psi \cdot \dot{\psi}\right] = -\frac{1}{4}\int_{0}^{T}d\tau \int d\theta \, Q D^{3}Q \label{Ssuper}\\
	\non
	&V_{\eta}^{x^{\prime}x}[k, \ep] = \int_{0}^{T}d\tau \int d\theta \, e^{i k \cdot \big(Q(\tau)+\hat{x}(\tau) +\sqrt{2} \theta \eta\big) + \ep \cdot \big( DQ + \sqrt{2}\eta - \theta \frac{\xm}{T}\big)}\bigg|_{\textrm{lin } \ep}. \label{Vsuper}
\end{align}
There is one subtlety in the second equality: to get the correct signs one has to suppose that now the polarisation vector anti-commutes with the variable $\theta$ (and also with polarisation vectors associated to other vertex operators): after expanding to linear order in $\ep$ one must arrange for all $\ep$ to be to the left of all $\eta$, following which the polarisations can be demoted back to ordinary vectors. Here we make two vital points about the superspace formulation:
\begin{itemize}
	\item The exponent of the vertex operator is now linear in the superfield, $Q(\tau)$, and as such can be shifted into the quadratic kinetic term of (\ref{Ssuper}) by completing the square. 
	\item The supersymmetry is manifest: as you will show in the exercises, a product of superfields remains a superfield with the same transformation properties under supersymmetry, (\ref{susyline}). 
	\item Moreover, an integral of a superfield, $\mathbb{X}$, over the whole superspace, $\int d\tau \int d\theta\, \mathbb{X}$ just picks out the term proportional to $\theta$ in $\mathbb{X}$ and integrates it with respect to $\tau$; but this term transforms as a total derivative (see (\ref{susyline}))and so the net variation of such an integral is a pure boundary term\footnote{Were the boundary conditions equal on $x$ and $\psi$ this would even vanish on the loop, but as we have already pointed out, in both cases the boundary conditions break the supersymmetry; nonetheless, the superspace formalism provides a powerful calculational bookkeeping principle.}.
\end{itemize}
Using these ideas in (\ref{KDBC}) allows for the path integral to be cast as a super-path integral where the interaction with the photons is finally linear in the superfield:
\begin{align}
K^{x^{\prime}x}[A] &= (-ie)^{N}2^{-\frac{D}{2}}\int_{0}^{\infty} dT\, \e^{-m^{2}T} \e^{-\frac{\xm^{2}}{4T}} \int_{x(0) = x}^{x(T) = x'}\hspace{-1em}\mathscr{D}Q(\tau, \theta)\, \e^{-\int_{0}^{T}d\tau d\theta \, [-\frac{1}{4} QD^{3}Q]}\,\non
&\hspace{2cm}\times\textrm{symb}^{-1}\left\{ \prod_{i = 1}^{N} V_{\eta}^{x'x}[k_{i}, \ep_{i}]\right\}\,,
\end{align}
where the vertex insertions give exponents linear in $Q$. 

To compute the path integral we define a super-current generalisation of $j^{\mu}(\tau)$ as $J^{\mu}(\tau, \theta) := \sum_{i = 1}^{N}\left(i k_{i} - \ep_{i}D\right)\delta(\tau - \tau_{i})$ and a super-vector combining the boundary variables, $X(\tau, \theta) := \hat{x}(\tau) + \sqrt{2}\,\theta\, \eta$. Then noting that $\ep \cdot \frac{\xm}{T} = \ep \cdot \dot{\hat{x}}$, the path integral is equivalent to (\ref{Dxxq}) with the replacements 
\bear
&&\int d\tau \rightarrow \int d\tau \int d\theta; \qquad \frac{d}{d\tau_{i}} \rightarrow D_{i}; \qquad q(\tau) \rightarrow Q(\tau, \theta); \non
&&\quad j^{\mu}(\tau) \rightarrow J^{\mu}(\tau, \theta); \qquad \hat{x}(\tau) \rightarrow \hat{X}(\tau, \theta).
\ear
Completing the square in the superfield the only element missing for a direct generalisation of (\ref{DxxkFull}) is the worldline super-Green function. This is given by
\begin{align}
	\left<Q^{\mu}(\tau, \theta)Q^{\nu}(\tau^{\prime}, \theta^{\prime})\right> &:= -2\delta^{\mu\nu}\hat{\Delta}(\tau, \theta; \tau^{\prime}, \theta^{\prime}) \\
	\hat{\Delta}(\tau, \theta; \tau^{\prime}, \theta^{\prime}) &:= \Delta(\tau, \tau^{\prime}) + \frac{1}{2}\theta \theta^{\prime}G_{F}(\tau, \tau^{\prime}).
	\label{Gsuper}
\end{align}
Exercise 3.11 asks for verification of the following properties of the super-Green function
\begin{align}
	D_{i}\hat{\Delta}_{ij} &= -\theta_{i}\ddel_{ij} + \frac{1}{2}\theta_{j}G_{Fij} \label{DiG}\\
	D_{j}\hat{\Delta}_{ij} &= -\theta_{j}\deld_{ij} -\frac{1}{2}\theta_{i}G_{Fij} \label{DjG}\\
	D_{i}D_{j}\hat{\Delta}_{ij} &= -\frac{1}{2}G_{Fij} + \theta_{i}\theta_{j}\ddeld_{ij}\label{DijG}\\
	D^{3}_{i}\hat{\Delta}_{ij} &= \delta(\theta_{i} - \theta_{j})\delta(\tau_{i} - \tau_{j}). \label{D3G}
\end{align}
With these properties it is straightforward to see that the path integral yields
\begin{align}
K^{x'x}[k_{1}, \ep_{1}; \ldots; k_{N}, \ep_{N}] &=(-ie)^{N}\textrm{symb}^{-1}\bigg\{\int_{0}^{T}dT\, (4\pi T)^{-\frac{D}{2}} \e^{-m^{2}T} \e^{-\frac{( x^{\prime}-x)^{2}}{4T}}\non
&\times \prod_{i = 1}^{N} \int_{0}^{T}d\tau_{i} d\theta_{i} \e^{\sum_{i = 1}^{N} \left[ i k_{i} \cdot x +  \frac{x'-x}{T}  (i\tau_{i}k_{i} + \theta_{i}\ep_{i}) -\sqrt{2}\eta\cdot (\ep_{i} + i \theta_{i}k_{i}) \right] } \nonumber \\
&\times \e^{\sum_{i, j = 1}^{N}\big[\hat{\Delta}_{ij}  k_{i} \cdot k_{j}  + 2i D_{i}\hat{\Delta}_{ij}\ep_{i} \cdot k_{j} + D_{i}D_{j}\hat{\Delta}_{ij}\ep_{i}\cdot \ep_{j} \big]}\bigg|_{\linep} \bigg\}\, .
		\label{KxxkFull}
\end{align}
This represents the master formula for $N$-photon scattering amplitudes at tree level in spinor QED in configuration space. To avoid sign errors, we recall that due to our exponentiation of the vertex operators requiring temporary promotion of the polarisations $\varepsilon_{i}$ to Grassmann-valued vectors, after expansion they must be anti-commuted to one side of all other Grassmann variables before being restored to conventional commuting vectors.

\subsubsection{Scattering amplitudes}
For the calculation of scattering amplitudes it is convenient to switch to momentum space. We do this as in the scalar case. Generalising (\ref{Dpp}) to the spinor case and appealing to (\ref{Sk}) we must find (as before we take all momenta to be ingoing)
\begin{align}
	S_{N}^{p'p} := \int d^{D}x \int d^{D}x^{\prime}\, \e^{ i (p^{\prime} \cdot x^{\prime}+ p \cdot x)} \bigg[ i \gamma \cdot \Big(\partial' + ie \sum_{i = 1}^{N}\ep_{i} \e^{i k_{i} \cdot x}\Big) + m\bigg]K_{N}^{x^{\prime}x}\bigg|_{\linep}.
\end{align}
Changing variables to $x_{\pm}$ and considering the two ways of selecting the terms multi-linear in the polarisation vectors it is straightforward to see that
\begin{equation}
S_{N}^{p'p}= \left[\ps' + m\right]K_{N}^{p^{\prime}p} - e\sum_{i = 1}^{N}\s\varepsilon_{i} K_{N-1}^{p'+k_{i}, p}\,,
\label{Sppk}
\end{equation}
where $K_{N-1}^{p'+k_{i}, p} = K_{N-1}^{p'+k_{i}, p}[k_{1}, \ep_{1}; \ldots ; k_{i-1}, \ep_{i-1}; k_{i+1}, \ep_{i+1}; \ldots ; k_{N}, \ep_{N}]$ is a function of the $N-1$ photon momenta and polarisations excluding the $i^{\textrm{th}}$. For example, with $N = 2$, $K_{1}^{p'+k_{1}, p}[k_{2}; \ep_{2}]$ will be a function of the momentum and polarisation of the second photon, with the ingoing momentum shifted by $k_{1}$. Here $K_{N}^{p^{\prime}p}$ is the Fourier space version of (\ref{KxxkFull}),
\begin{align}
	K^{p'p}&= (-ie)^{N}\textrm{symb}^{-1}\,\int_{0}^{\infty}dT\, \e^{-m^{2}T} \prod_{i = 1}^{N} \int_{0}^{T}d\tau_{i} d\theta_{i} \e^{{\rm Exp}}\Big|_{\linep}\,,
		\label{KppkFull}
\end{align}
where 
\bear
{\rm Exp}&=&-T \Big(p' + \frac{1}{T} \sum_{i = 1}^{N} \left(  k_{i} \tau_{i} - i \theta_{i}\ep_{i} \right)\Big)^{2} - \sqrt{2}\eta \cdot \sum_{i=1}^{N} \left(\ep_{i} + i \theta_{i}k_{i}\right) \non		
&& +\sum_{i, j = 1}^{N}\big[\hat{\Delta}_{ij}  k_{i} \cdot k_{j}  + 2i D_{i}\hat{\Delta}_{ij}\ep_{i} \cdot k_{j} + D_{i}D_{j}\hat{\Delta}_{ij}\ep_{i}\cdot \ep_{j} \big]\,.
\ear
In these equations we have neglected the momentum conserving $\delta$-function that arises from the integral over $\xp$, leaving it implicit as in the scalar case. It is important to highlight that as in the scalar case our quantisation procedure is valid off-shell, both for external fermions and the attached photons. Permutation symmetry of these photons has also been preserved, meaning that we continue to sum over all Feynman diagrams related by a re-ordering of the photon legs. Thus the representation (\ref{Sppk}) achieves the aim of a first quantised version of the propagator that retains the calculational efficiencies of the worldline approach that have been seen at one-loop order, particularly with respect to manifest gauge invariance. 

\subsubsection{Spin-Orbit decomposition}
The particular form of the worldline vertex operator, (\ref{eqVspineta}) or its one-loop counterpart with $\eta = 0$, that neatly separates orbital interactions ($\varepsilon \cdot \dot{x}$, identical in scalar QED) from spin interactions (terms involving $\psi \cdot F \cdot \psi$) motivates a novel decomposition of scattering amplitudes. This method of breaking down the amplitude into processes with a fixed number of spin and orbital interaction vertices\footnote{We should be clear that this separation is not, of course, Lorentz invariant, but the decomposition is valid in any fixed reference frame.} will be called the \textit{spin-orbit decomposition} and it considerably unclear how one would arrive at this way of writing the amplitudes starting in the standard formalism (although some hints can be glimpsed in \cite{Morgan}). 

We thus consider the decompositions of the effective action or propagator depending on the process under study,
\begin{align}\label{eqGammaNS}
	\Gamma_{N} &= \sum_{S=0}^{N} \Gamma_{NS}\\
	K_{N} &= \sum_{S=0}^{N} K_{NS}\,,
	\label{eqKNS}
\end{align}
where on the RHS the summands denote the contributions to the kernel from $S$ spin interactions and $N-S$ orbital interactions in the relevant vertex operators. It is immediately obvious that $\Gamma_{N0}$ and $K_{N0}$ coincide with the scalar results.
	
Using the fact that the scalar and spinor path integrals do not mix, we can deal with the spin and orbital parts of the vertex insertions separately. For $S$ spin interactions we require expectation values formed from the second term in the vertex operator, viz.,
\begin{equation}
	W_{\eta}(1\ldots S) := i^{S}\Big\langle V_{\eta}[k_{1}, \varepsilon_{1}]\cdots V_{\eta}[k_{S}, \varepsilon_{S}]\Big\rangle \,,\qquad V_{\eta}[k, \varepsilon] := -i(\psi + \eta)\cdot f \cdot (\psi + \eta)\,.
\end{equation}
In \cite{SpinProp}, a closed formula for the case $\eta = 0$ is given in terms of a Pfaffian determinant along with a more complicated result for the general case; here we restrict ourselves to some simple examples in $D=4$ which the dedicated reader will enjoy verifying:
\begin{align}
\hspace{-1.5em}	W_{\eta}(1) &= \eta \cdot f_{1} \cdot \eta \nonumber \\
\hspace{-1.5em}	W_{\eta}(12) &= -\frac{1}{2}G_{F12}G_{F21}\tr(f_{1}\cdot f_{2}) + 2G_{F12}\eta \cdot f_{1} \cdot f_{2} \cdot \eta  + \eta \cdot f_{1} \cdot \eta \, \eta \cdot f_{2} \cdot \eta \nonumber \\
\hspace{-1.5em}	W_{\eta}(123) &= -G_{F12}G_{F23}G_{F31}\tr(f_{1} \cdot f_{2} \cdot f_{3}) - \frac{1}{2}\Big[G_{F12}G_{F21}\tr (f_{1}\cdot f_{2})\eta \cdot f_{3} \cdot \eta + \textrm{2 perms} \Big] \nonumber \\
\hspace{-1.5em}	&\,\,\hphantom{=} + 2\Big[G_{F12}G_{F23}\eta \cdot f_{1}\cdot f_{2}\cdot f_{3}\cdot \eta +  \textrm{2 perms} \Big] \nonumber \\
 \hspace{-1.5em}	&\,\,\hphantom{=}+ 2\Big[G_{F12}\eta \cdot f_{1} \cdot f_{2}\cdot \eta\, \eta \cdot f_{3} \cdot \eta + \textrm{2 perms} \Big]\,,
 \label{eqW123}
	\end{align}
where the permutations simple cycle the variables. Of course, putting $\eta = 0$ provides the results for the closed loop. We also note that in contrast to the superspace formalism, here the polarisation vectors remain ordinary commuting variables throughout.

Continuing with the orbital contribution, then, the contribution of $N-S$ orbital interactions takes a form similar to the scalar case,
\begin{equation}
	\Big\langle V[\varepsilon_{S+1}]\cdots V[\varepsilon_{N}] \e^{i \sum_{i = 1}^{N} k_{i} \cdot x_{i}}\Big\rangle \,, \qquad V[ \varepsilon] := \varepsilon \cdot \dot{x}
\end{equation}
(note the sum in the exponent runs over \textit{all} photons). We can evaluate this in analogy to scalar QED with some minor modifications to (\ref{DppkFull}) to take into account the absence of the $S$ photons we assigned to the spin interaction. Using the 4-vectors $b = p' +  \frac{1}{T}\sum_{i = 1}^{N}(k_{i}\tau_{i} - i\varepsilon_{i})$ and $b_{0} \equiv b\big|_{\textrm{all }\varepsilon_{i} = 0}$ we define polynomials $\bar{P}_{NS}^{\{1\dots S\}}$ by (compare to (\ref{eqDefPbar}))
\begin{align}
&&\e^{-Tb^2} \,\e^{\sum_{i, j = 1}^{N}\big[\Delta_{ij}  k_{i} \cdot k_{j}  - 2i \ddel_{ij}\ep_{i} \cdot k_{j} - \ddeld_{ij}\ep_{i}\cdot \ep_{j} \big]}\Big|_{\varepsilon_{1}=\cdots=\varepsilon_{S} = 0}\Big|_{\varepsilon_{S+1}\ldots\varepsilon_{N}}\non
&&:= (-i)^{N-S}\bar{P}_{NS}^{\{1\ldots S\}} \e^{-Tb_{0}^{2} + \sum_{i,j=1}^{N}\Delta_{ij}k_{i}\cdot k_{j}}\,.
\label{eqDefPBarS}
\end{align}
The ``double bar'' notation on the LHS indicates that we first remove all terms with $\varepsilon_{1}$ to $\varepsilon_{S}$ from the exponent and then expand the resulting exponential to multi-linear order in the remaining variables. It should be clear that $\bar{P}_{N0}^{\{\}} = \bar{P}_{N}$ from the scalar theory defined in (\ref{eqDefPbar}) and that $\bar{P}_{NN}^{\{1\ldots N\}} = 1$ when all $N$ photons are assigned to the spin interaction. 

Making use of this notation we can form the fixed $N$ and $S$ contributions to (\ref{eqGammaNS}) and (\ref{eqKNS}) by summing over the different ways to assign $S$ of the photons to the spin interaction part of the vertex
\begin{align}
	\Gamma_{NS} &= \sum_{\{i_{1} i_{2}\ldots i_{S}\}}\Gamma_{NS}^{\{i_{1} i_{2}\ldots i_{S}\}}\\
	K_{NS} &= \sum_{\{i_{1} i_{2}\ldots i_{S}\}}K_{NS}^{\{i_{1} i_{2}\ldots i_{S}\}}
\end{align}
where the sums run over all choices of $S$ of the $N$ variables, with the contributions on the RHS are defined by
\begin{align}
\hspace{-5.5em}	\Gamma_{NS}^{\{i_{1} i_{2}\ldots i_{S}\}} = (-ie)^{N}(-i)^{N}  \int_{0}^{\infty}\frac{dT}{T} \e^{-m^{2}T}\prod_{i=1}^{N} \int_{0}^{T}d\tau_{i}\, W_{0}(i_{1} i_{2}\ldots i_{S})P_{NS}^{\{i_{1} i_{2}\ldots i_{S}\}} \e^{\sum_{i,j=1}^{N} G_{Bij}k_{i}\cdot k_{j}}\,,
\end{align}
and
\begin{align}
\hspace{-5.5em}	K_{NS}^{\{i_{1} i_{2}\ldots i_{S}\}} = (-ie)^{N}(-i)^{N} \symb^{-1}\Big\{ \int_{0}^{\infty}dT\, \e^{-m^{2}T}\prod_{i=1}^{N} \int_{0}^{T}d\tau_{i}\,  W_{\eta}(i_{1} i_{2}\ldots i_{S})\bar{P}_{NS}^{\{i_{1} i_{2}\ldots i_{S}\}} \e^{-T b_{0}^{2} + \sum_{i,j=1}^{N} \Delta_{ij}k_{i}\cdot k_{j}}\Big\}\,,
\end{align}
respectively.
\subsubsection{Applications}
To illustrate the use of the momentum space representation of the dressed fermion propagator we give its explicit form for $N = 0, 1, 2$, deriving the fermion-photon-fermion vertex and investigating linear Compton scattering. First note that the momentum space propagator is untruncated with respect to the external fermions. To see this set $N = 0$ which provides
\begin{equation}
	K_{0}^{p^{\prime}p} = \textrm{symb}^{-1}\left\{ \int_{0}^{\infty}dT\, \e^{-T(p'^{2} + m^{2})} \right\} = \frac{\bone}{p'^{2} + m^{2}}.
\end{equation}
Then (\ref{Sppk}) implies
\begin{equation}
	S_{0}^{p^{\prime}p} = (\ps' + m)K_{0}^{p^{\prime}p} = \frac{\ps' + m}{p'^{2} + m^{2}} = \frac{1}{-\pps + m} = \frac{1}{\ps + m}.
\end{equation}
Therefore, as is familiar from the scalar case, we must amputate external legs to determine scattering amplitudes according to
\begin{equation}
	\hat{S}_{N}^{p^{\prime}p} := (-\ps' + m)S_{N}^{p^{\prime}p}(\ps + m).
	\label{Shat}
\end{equation}
The bare vertex is found for $N = 1$. In the superspace formalism, $K_{1}^{p^{\prime}p}$ is found by expanding the Master Formula to linear order in $\ep$ and integrating over $\theta$, thus picking out the piece linear in that variable too:
\begin{align}
	K_{1}^{p^{\prime}p}[k, \ep] &= (-ie) \,\textrm{symb}^{-1}\bigg\{  \int_{0}^{\infty} dT\, \e^{-T(p'^{2} + m^{2})} \int_{0}^{T}d\tau \int d\theta\non
	&\hspace{3cm}\times \left[i\theta \ep \cdot \Delta p - \sqrt{2} \eta \cdot \ep \right] \e^{i \theta \sqrt{2}\eta \cdot k - \tau \Delta p \cdot k} \bigg\} \non
	&= e\, \textrm{symb}^{-1} \Big\{  \ep \cdot (2p' + k) - 2 \eta \cdot \ep\, \eta \cdot k \Big\}\non
	&\hspace{3cm}\times\int_{0}^{\infty}  dT\, \e^{-T(p'^{2} + m^{2})} \int_{0}^{T}d\tau\, \e^{\tau (p'^{2} - p^2)}\,,\non
	\label{eqK1MF}
\end{align}
where in the second equality was expanded the remaining exponent to linear order in $\theta$ and then computed the integral over that variable (recall here that $\varepsilon$ remains  Grassmann vector until moved to the right of all $\eta$). 

In contrast, the spin orbit decomposition arrives at the same result via the expansion 
\begin{align}
	K_{1} &= K_{10} + K_{11} \nonumber\\
	&= -e\,\symb^{-1}\Big\{ \int_{0}^{T}dT\, \e^{-m^2 T} \int_{0}^{T} d\tau \big[ \bar{P}_{1} + W_{\eta}(1)\big] \e^{-T(p' + k\frac{\tau}{T})^2 - \tau(1-\frac{\tau}{T})k^2}\Big\}\nonumber\\
	&=-e\,\symb^{-1}\Big\{ \int_{0}^{T}dT\, \e^{-(p'^2 + m^2) T} \int_{0}^{T} d\tau \big[-\varepsilon \cdot (2p' + k) + \eta \cdot f \cdot \eta \big]\e^{\tau(p'^2 -p^2)}\Big\}\,.
\end{align}
The proper time integral was calculated for (\ref{Dpp1}). The inverse symbol map can then be applied to yield (first ensuring that all polarisation vectors in (\ref{eqK1MF}) are moved to the left of the $\eta$) 
\begin{equation}
	K_{1}^{p^{\prime}p} = e \frac{\ep \cdot (2p' + k)\bone + \frac{1}{2}(\ks \s\ep - \s\ep \ks)}{(p'^{2} + m^{2})(p^{2} + m^{2})}=e\frac{\s\ep(\slashed{p}-m)-(\slashed{p}'-m)\s\ep}{(p'^2+m^2)(p^2+m^2)}.
	\label{Kpp1}
\end{equation}
As mentioned above, the symbol map has provided the answer written in the Clifford basis. The kernel has evaluated to a scalar piece proportional to the identity in Dirac space familiar from (\ref{Dpp1}), plus an additional piece coming from the spin coupling to the photon. This decomposition is made manifest in the spin orbit decomposition and is described in detail in \cite{SpinProp}.

To form the familiar amplitude, $S_{1}^{p'p}$, we must also add the subleading term in (\ref{Sppk}), given by
\begin{equation}
	-e \slashed{\ep} K_{0}^{p' + k, p} = -e\frac{\slashed\ep}{\ps + m} = -e\slashed{\ep} \frac{-\ps + m}{p^{2} + m^{2}}\,.
\end{equation} 
Then using momentum conservation and amputating as per (\ref{Shat}) it is left as an exercise to demonstrate that this reproduces the Dirac vertex, familiar in the Feynman diagrams of the standard formalism. That is, firstly we have
\begin{equation}
	\hat{S}_{1}^{p'p} = (\pps + m)K_{1}^{p'p} -e \slashed\ep K_{0}^{p' + k, p} = \frac{(\pps + m)\slashed\ep(\ps - m)}{(p^2 + m^2)(p'^2 + m^2)}\,,
 	\label{sVertnohat}
\end{equation}
which provides
\begin{equation}
	\hat{S}_{1}^{p'p}  = -\slashed\ep
 	\label{sVert}
\end{equation}
as required. 

At order $N = 2$ we can access both the one-loop contribution to the fermion self energy and the linear Compton cross section. So to begin with we expand the Master Formula for $K_{2}$ to order $\varepsilon_{1}\varepsilon_{2}$ which gives
\begin{eqnarray}
 \hspace{-2.5em}K_{2}^{p' p} (k_1,\varepsilon_1; k_2, \varepsilon_2)\!\! &\! = \!&\! \!- e^2 \text{symb}^{-1}\bigg\{ \int_0^\infty dT \int_0^T d\tau_1 d\tau_2 
\varepsilon_{1\mu}\varepsilon_{2\nu} \int \int d\theta_1 d\theta_2\non
\hspace{-2.5em}&&\times\bigg[
\dfrac{1}{2}\eta^\mu \eta^\nu + \sigma_{12}\delta^{\mu\nu} + i \theta_1 \Big( \dfrac{\eta^\mu}{\sqrt{2}}\sigma_{21}k_1^\nu +
\dfrac{\eta^\nu}{\sqrt{2}}\sigma_{12}k_1^\mu - \dfrac{\eta^\nu}{\sqrt{2}}(p'-p)^\mu \Big) \non
\hspace{-2.5em}&&-i\theta_2\Big( \dfrac{\eta^\nu}{\sqrt{2}}\sigma_{12}k_2^\mu + \dfrac{\eta^\mu}{\sqrt{2}}\sigma_{21}k_2^\nu - \dfrac{\eta^\mu}{\sqrt{2}}(p'-p)^\nu \Big) \nonumber \\
\hspace{-2.5em}&& +\theta_1\theta_2\Big( 2\delta^{\mu \nu} \delta_{12} - (p'-p)^\mu(p'-p)^\nu
+\sigma_{12}k_2^\mu(p'-p)^\nu+\sigma_{21}(p'-p)^\mu k_1^\nu \Big)\bigg] \nonumber \\
\hspace{-2.5em}&& \times \, \e^{  -T(p'^2+m^2) + |\tau_1-\tau_2|k_1 \cdot k_2 - (p'-p)\cdot (\tau_1 k_{1} + \tau_2 k_{2}) + i(\theta_1k_{1}+\theta_2k_{2})\cdot \frac{\eta}{\sqrt{2}} 
+ \theta_1\theta_2\sigma_{12} k_1 \cdot k_2 }\bigg\}\non
\end{eqnarray}
so that integrating over the Grassmann variables yields 
\begin{align}
\hspace{-5em}K_{2}^{p'p} (k_1,\varepsilon_1; k_2, \varepsilon_2) &= - e^2 \text{symb}^{-1}\bigg\{ \int_0^\infty dT \int_0^T d\tau_1 d\tau_2\,  
\varepsilon_{1\mu}\varepsilon_{2\nu} \nonumber \\
&\times\bigg[ \Big( \dfrac{1}{2}\eta^\mu\eta^\nu + \sigma_{12}\delta^{\mu\nu}\Big)
\Big( \sigma_{12} k_1 \cdot k_2 + \dfrac{1}{2} k_1 \cdot \eta k_2 \cdot \eta \Big)
+2\delta^{\mu\nu}\delta_{12} - (p'-p)^\mu(p'-p)^\nu \nonumber \\
& + \sigma_{12}k_2^\mu(p'-p)^\nu + \sigma_{21}(p'-p)^\mu k_1^\nu
-\dfrac{1}{2}\eta^\nu \eta^\rho \sigma_{12} k_2^\mu (p+p^\prime)_\rho 
-\dfrac{1}{2}\eta^\mu \eta^\rho \sigma_{21} k_1^\nu (p+p^\prime)_\rho  \nonumber \\
& -\dfrac{1}{2}\Big( \eta^\mu \eta^\rho (p'-p)^\nu k_{1\rho} 
+ \eta^\nu\eta^\rho(p'-p)^\mu k_{2\rho} \Big) \bigg]
e^{-T(p'^2+m^2)+|\tau_1-\tau_2|k_1 \cdot k_2 -(p'-p)\cdot (\tau_1k_{1}+\tau_2k_{2})}\bigg\} . 
\label{K2Exp}
\end{align}
This is the starting point of the first quantised representation of the Compton amplitude, where we shall see that the first quantised representation of the propagator provides considerable simplification and an easy way to extract the gauge invariant information that survives in the on-shell cross-section. So carrying out the parameter integrals and applying the inverse symbol map we get to
\begin{align}
\hspace{-4em}K_2^{p'p} & = \frac{e^2}{(m^2+p^2)(m^2+p'^2)} 
\biggl \lbrace 
-2\ep_{1}\cdot\ep_{2} 
+ 
\biggl\lbrack \frac{1}{m^2+(p'+k_1)^2}
\Bigl( 
\ep_{1}\cdot(p'-p-k_2)\ep_{2}\cdot(p'-p+k_1) 
\nonumber\\
\hspace{-4em} &
+\ep_{1}\cdot k_2\ep_{2}\cdot k_1 
-k_1\cdot k_2 \ep_{1}\cdot\ep_{2} 
+\half \ep_{1}\cdot\ep_{2} [\slash k_1,\slash k_2] + \half k_1\cdot k_2 [\slash\ep_{1},\slash \ep_{2}] 
\nonumber\\
\hspace{-4em}&
+\half \ep_1\cdot k_2 [\slash\ep_2,\slash k_1] 
-\half [\slash\ep_1,\slash k_2]\ep_2\cdot k_1
-\half\ep_1\cdot(p'-p-k_{2})[\slash\ep_2,\slash k_2] \non
\hspace{-4em}&-\half [\slash\ep_1,\slash k_1]\ep_2\cdot(p'-p+k_{1})
+i\gamma_5 \varepsilon(\ep_1,\ep_2,k_1,k_2)
\Bigr)
+ (1\leftrightarrow 2)
\biggr\rbrack
\biggr\rbrace
\label{K2fin}  
\end{align}
where $\varepsilon(\ep_1,\ep_2,k_1,k_2) := \epsilon^{\mu\nu\alpha\beta}\ep_{1\mu}\ep_{2\nu}k_{1\alpha}k_{2\beta}$. Once again the result comes in the Clifford basis; if desired one may translate back to the standard Dirac basis to verify the amplitude is as expected:
\begin{align}
\hspace{-4em} K_2^{p'p} & =  \frac{e^2}{(m^2+p^2)(m^2+p'^2)} 
\biggl\lbrace 
 \frac{1}{m^2+(p'+k_1)^2}
 \Bigl\lbrack
- \slash\varepsilon_1(\slash p'+\slash k_1 + m)\slash\varepsilon_2 (\slash p-m) - (\slash p' - m)  \slash\varepsilon_1\slash\varepsilon_2 (\slash p-m)
 \nonumber\\
 \hspace{-4em} & \hspace{185pt}
 +  (\slash p' - m)  \slash\varepsilon_1(\slash p'+\slash k_1 - m)\slash\varepsilon_2 
 \Bigr\rbrack
+ (1\leftrightarrow 2)
\biggr\rbrace
 \label{K2alt}
 \end{align}
which is a straightforward exercise in Dirac algebra. Using (\ref{K2alt}) in (\ref{Shat}) and subtracting also the contribution from the subleading term gives
 \begin{align}
\hspace{-4em} S_2^{p'p} &=e^2
\frac{({\slash p'}+m)}{(p^2+m^2)(p'^2+m^2)} 
\biggl\lbrace 
 \frac{1}{(p'+k_1)^2+m^2}
 \Bigl\lbrack
- \slash\varepsilon_1(\slash p'+\slash k_1 + m)\slash\varepsilon_2 (\slash p-m)\non
&\hspace{2cm} - (\slash p' - m)  \slash\varepsilon_1\slash\varepsilon_2 (\slash p-m)
 +  (\slash p' - m)  \slash\varepsilon_1(\slash p'+\slash k_1 - m)\slash\varepsilon_2 
 \Bigr\rbrack
+ (1\leftrightarrow 2)
\biggr\rbrace
\nonumber\\
\hspace{-4em}& - e^2\biggl\lbrace \slash\varepsilon_1 \frac{\slash \varepsilon_2 ({\slash {p}}-m) - (\slash p'+k_1-m) \slash\varepsilon_2}{[(p'+k_1)^2+m^2](p^{2}+m^2)}+ (1\leftrightarrow 2)
\biggr\rbrace
\label{S2expl}\non
\end{align}
which after a little more algebra leads to 
\begin{align}
\hat S_2^{p'p} = (-{\pps}+m) S_2^{p'p} (\ps +m) =
e^2\Big[ \slash\varepsilon_{1} \frac{\pps + \ks_{1} + m}{\left(p' + k_{1}\right)^{2} + m^{2}}\slash \varepsilon_{2} + \slash\varepsilon_{2} \frac{\pps + \ks_{2} + m}{\left(p' + k_{2}\right)^{2} + m^{2}}\slash\varepsilon_{1} \Big]\,.
\label{S2final}
\end{align}
It is a quick task to check that this is what comes out of the standard Feynman diagram calculation of the Compton amplitude from the Dirac field, which verifies that the worldline formalism gives the same result (note that as yet we have not gone on-shell). Of course the attentive student may ask about the use of the spin-orbit decomposition; we hope they will verify that the $N=2$ kernel is reproduced in this fashion according\footnote{For completion we provide the formulae \begin{align}
	\bar{P}_{21}^{\{1\}} &= 2\big( \ddel_{21}\varepsilon_{2}\cdot k_{1} + \ddel_{22}\varepsilon_{2}\cdot k_{2} \big) - 2\varepsilon_{2}\cdot b_{0} \\
	\bar{P}_{21}^{\{2\}} &= 2\big( \ddel_{11}\varepsilon_{1}\cdot k_{1} + \ddel_{12}\varepsilon_{1}\cdot k_{2} \big) - 2\varepsilon_{1}\cdot b_{0}
\end{align}}
\begin{align}
	K_{2}^{p'p} = \symb^{-1} \int_{0}^{\infty}dT\, &\e^{-(p'^{2} + m^{2})T}\int_{0}^{T}d\tau_{1}\int_{0}^{T}d\tau_{2}\e^{k_{1}\cdot k_{2}|\tau_{1} - \tau_{2}| - (p'-p)\cdot (\tau_{1}k_{1} + \tau_{2}k_{2})} \nonumber\\
	\times & \Big[ \bar{P}_{2} + W_{\eta}(1)\bar{P}_{21}^{\{1\}}+ W_{\eta}(2)\bar{P}_{21}^{\{2\}} + W_{\eta}(12) \Big]
	\label{eqSON2P}
\end{align}

All this has ultimately been in the spirit of showing that the worldline formalism reproduces the standard, Feynman diagram approach. However, since the worldline calculation naturally provides its result in the Clifford basis, it is worth asking whether we can gain some simplifications in our formalism for a direct calculation of the on-shell cross section without the effort of translating to the standard Dirac basis. In the following section we shall return to this question to show that it is indeed the case that, on-shell, the worldline formalism can streamline the calculation of the spin summed cross section substantially, after a short deviation to discuss loop corrections. 

\subsubsection{Electron self energy diagram}

First, however, we continue at the level of the amplitude, since similar to the scalar case, one can use the above results with off-shell external states to determine the one-loop correction to the fermion propagator (or the electron self-energy), see Fig. (\ref{figSE}q). One could return to (\ref{K2Exp}), before the parameter and proper time integrals have been done. As before, to sew in Feynman gauge we set $k_{1} = -q = k_{2}$, which forces $p' = -p$, and effect the replacement $\varepsilon_{1\mu}\varepsilon_{2\nu} \rightarrow \frac{\delta_{\mu\nu}}{q^{2}}$ following which we integrate over the loop momentum $q^{\mu}$. Once again fixing the order of the photon legs provides a simplified integrand
\begin{align}
\hspace{-3em}-  \text{symb}^{-1}\bigg\{ \int \frac{d^Dq}{(2\pi)^{D}} \int_0^\infty dT \int_0^T d\tau_1 \int_{0}^{\tau_{1}}d\tau_2\,  
\frac{e^2}{q^{2}}   \bigg[&- D q^2 
+2D\delta_{12} - 4p'^\mu p'^\nu
 +4 \sigma_{12}q^\mu p'^\nu \bigg] \nonumber \\
 &\times e^{-T(p'^2+m^2)-|\tau_1-\tau_2|q^2 +2p^{\prime} \cdot q (\tau_1-\tau_2)}\bigg\}.
 \label{Ksewold}
\end{align}
We leave the remaining calculation of the parameter, proper-time and momentum integrals as an exercise, however, since the more efficient way to do this calculation is by using the existing result for $N=2$ in equation (\ref{K2fin}).
We make the familiar sewing in Feynman gauge, directly in (\ref{K2fin}) and then integrate over the loop momentum $q^{\mu}$. 
 
Doing this it is easy to check that $\varepsilon(\varepsilon_{1}, \varepsilon_{2}, k_{1}, k_{2})\big|_{k_{1} = q = -k_{2}} = 0$ by anti-symmetry and that the coefficient of the commutator $[\gamma^{\alpha},\gamma^{\beta}]$ is
\begin{align}
	B_{2\,\alpha \beta}\big|_{k_{1} = q = -k_{2}} \propto Dq^{\alpha}q^{\beta} + q^{2}\delta^{\alpha \beta}.
\end{align}
Multiplying this symmetric tensor into the commutator of the Dirac matrices clearly gives a result that vanishes. This leaves only the part proportional to the identity in spinor space. Carrying out the sewing we can split this into its ``scalar'' (that are also present in the case of scalar QED) and ``spin'' parts that take the following form:
\begin{align}
	\hspace{-2em}\textrm{scal}\big|_{k_{1} = q = -k_{2}} &\rightarrow -\frac{2}{q^{2}}\frac{1}{p^{\prime 2} + m^{2}}\left(D - \frac{\left(2p^{\prime} + q\right)^{2}}{(p^{\prime} + q)^{2} + m^{2}} \right)\frac{1}{p^{\prime 2} + m^{2}}\, ,\nonumber\\
	\hspace{-2em}\textrm{spin}\big|_{k_{1} = q = -k_{2}} &\rightarrow \frac{2 (D - 1)}{\big[p^{\prime 2} + m^{2}\big]\big[(p^{\prime} + q)^{2} + m^{2}\big]\big[p^{\prime 2} + m^{2}\big]}\,, \nonumber\\
\end{align}
where we have taken advantage of the freedom to change the variable of integration $q \rightarrow -q$ to simplify the results. In fact the spin contribution turns out to be independent of the covariant gauge, since this part of the interaction involves the part of the worldline vertex depending only on the field strength tensor. Putting these together the surviving contribution to the self energy is (we use $d^{D}\bar{q} := \frac{d^{D}q}{(2\pi)^{D}}$)
\begin{equation}
K^{pp'}_{(2,{\rm sew})}=2e^2\int \frac{d^{D}\bar{q}}{q^{2}} \frac{1}{p^{\prime 2} + m^{2}}\left[ \frac{(2p^{\prime} + q)^{2} + (D - 1)q^{2}}{(p^{\prime} + q)^{2} + m^{2}} - D\right]\frac{1}{p^{\prime 2} + m^{2}}.
\label{K2sewn}
\end{equation}
The interested reader will derive this from (\ref{Ksewold}) to show that both methods agree. Now, the very last term corresponds to the diagrams with the seagull vertex and this vanishes in dimensional regularisation so it could therefore be dropped. The remaining terms come from the two other spinor-photon-spinor vertices in the second order formalism of spinor QED. 

To proceed we add to this the subleading terms in the decomposition. From the $N = 1$ result, (\ref{Kpp1}), we apply the sewing procedure to $-\s\varepsilon_{1}K_{1}$ and $-\s\varepsilon_{2}K_{1}$ and find that they double up as
\begin{equation}
	-\slash{\varepsilon}_{1} K_{(1, \textrm{sew})}^{p^{\prime} + k_{1}, p}(k_{2}, \varepsilon_{2})-\slash{\varepsilon}_{2} K_{(1, \textrm{sew})}^{p^{\prime} + k_{2}, p}(k_{1}, \varepsilon_{1}) =-2 \int \frac{d^{D}\bar{q}}{q^{2}}\Big[\frac{(2 \pps - (D-2)\slash{q})}{(p^{\prime 2} + m^{2})((p^{\prime} + q)^{ 2} + m^{2})}\Big].
	\label{SubleadingIntK}
\end{equation}
Now here we have over-counted the two orderings of the endpoints of the sewn photon (essentially swapping $q$ with -$q$) so as we now finally construct the sewn propagator we divide by two, using conservation of momentum and the freedom to change variables between $\pm q$ to find
\begin{equation}
\Sigma(\ps)=	 \int d^Dq \dfrac{e^2}{q^2(m^2+p^2)}\left[ \frac{\slashed{p}' + m}{p^{2} + m^{2}}\left(
 \dfrac{(D-1) q^2 + (2p + q)^{2}}{(p+q)^2+m^2} -D\right) +  \dfrac{ 2\slashed{p} -(D-2) \slashed{q}}{(p+q)^2+m^2}\right].
 \end{equation}
Amputation in accordance with (\ref{Shat}) provides
\begin{equation}
\hspace{-4em}\hat{\Sigma}(\ps) =	\int d^Dq \dfrac{e^2}{q^2(m^2+p^2)}\left[\left(
\dfrac{(D-1) q^2 + (2p + q)^{2}}{(p+q)^2+m^2}-D\right)(\slashed{p} + m) - (\slashed{p}' - m) \dfrac{ 2\slashed{p} -(D-2) \slashed{q}}{(p+q)^2+m^2}  (\slashed{p} + m)\right].
\label{Sigmaspin}
\end{equation}
Combining the terms together, a little algebra turns the result into
\begin{align}
	\Sigma(\ps) &= -\int d^Dq \dfrac{e^2}{q^2}\frac{(D-2)(\slashed{p} + \slashed{q}) + Dm}{((p+q)^2+m^2)} \nonumber\\
	&=-\int d^Dq \dfrac{e^2}{q^2}\gamma^{\mu}\frac{\slashed{p} + \slashed{q} -m}{((p+q)^2+m^2)}\gamma_{\mu} \nonumber \\
	&=\int d^Dq \dfrac{e^2}{q^2}\gamma^{\mu}\frac{1}{\slashed{p} + \slashed{q} + m}\gamma_{\mu}.
\end{align}
This is of course just the (unregulated) self energy $\Sigma(\slashed{p})$ of a Dirac fermion, as is easy to verify using the familiar Feynman rules of the Dirac field. 

Taking advantage to discuss our result in a little more detail, we observe that from (\ref{Sigmaspin}) one may easily decompose the self energy into the structure
\begin{equation}
	\Sigma(\ps) := \alpha(p^{2})\pps + \beta(p^{2})\bone
\end{equation}
using the integrals in the appendix. Specifically, in Feynman gauge this leads to (with dimensional regularisation in which the seagull contribution vanishes)
\begin{align}
	\alpha(p^{2}) &= \frac{e^{2}}{2p^{2}}\frac{(m^{2})^{\frac{D}{2} - 2}}{(4\pi)^{\frac{D}{2}}} (D-2)\Gamma\left(1 - \frac{D}{2}\right)\left[ (m^{2} - p^{2}) {}_{2}F_{1} \left( 2 - \frac{D}{2}, 1, \frac{D}{2}; -\frac{p^{2}}{m^{2}}\right)- m^{2} \right]\\
	\beta(p^{2}) &= e^{2}m \frac{(m^{2})^{\frac{D}{2} - 2}}{(4\pi)^{\frac{D}{2}}} D \Gamma\left(1 - \frac{D}{2}\right){}_{2}F_{1} \left( 2 - \frac{D}{2}, 1, \frac{D}{2}; -\frac{p^{2}}{m^{2}}\right).
\end{align}
This can also be achieved in an arbitrary gauge for comparison to \cite{Andrei}. The advantages of the worldline formalism for this tensor decomposition will be even greater in the non-Abelian case -- the reader may see, for instance at one-loop order, the form factor decomposition of off-shell three- and four-gluon vertices \cite{92,4-gluon}.

\subsection{On-shell amplitudes and cross sections}
Now we return to the Compton scattering amplitude to show how cross sections can be calculated in the worldline scheme. Here we shall see some significant simplifications for on-shell particles. Firstly, we must recall that in contrast to the scalar amplitudes in spinor QED, aside from amputation we also need to include the spinor wave functions associated to the external spinor legs so that the scattering matrix element becomes
\begin{equation}
i{\cal M}_{(N)s^{\prime}s}^{ p' p}	
	= \bar{u}_{s'}(-p')\hat{S}_N^{p'p}u_{s}(p)
	\label{defTau}
\end{equation}
where the spinor wave functions are normalised such as to give the completeness relation when summing over spin states
\begin{equation}
	\sum_{s}u_{s}(q) \bar{u}_{s}(q) = (-\slashed{q} + m \bone)\,.
	\label{completeU}
\end{equation}
Referring back to (\ref{defTau}) the spin summed cross section comes from
\begin{equation}
	\textrm{phase-space } \times \frac{1}{2}\sum_{s^{\prime}s}\big|{\cal M}_{(N)s^{\prime}s}^{ p^{\prime}p}	\big|^{2}
\end{equation}
where the two particle phase space function can be found in any standard reference book. Unlike in the standard approach, it turns out that the first quantised formalism allows us to separate completely the dependence on photon polarisations and fermion spin. This point is given significant weight in the presentation of \cite{SpinProp2}. Here, for brevity we focus on the unpolarised cross-section, summing (or averaging as appropriate) first over electron spins and later over photon polarisations.

In this spirit then, we first note that the completeness relation (\ref{completeU}) simplifies the spin summed modulus square of the amplitude (we average over initial spins):
\begin{equation}
	\frac{1}{2}\sum_{ss^{\prime}}\big|{\cal M}_{(N)s^{\prime}s}^{p^{\prime}p}	\big|^{2} = \frac{1}{2}\tr_{\gamma}\left[(\pps + m)\hat{S}^{p^{\prime}p}_{N}(-\ps + m) \gamma^{0} \hat{S}_{N}^{p^{\prime}p \dagger}\gamma^{0}\right].
	\label{T2}
\end{equation}
Recalling that $\gamma^{\mu \dagger} = \gamma^{0}\gamma^{\mu}\gamma^{0}$, it is easy to see that $\gamma^{0} \hat{S}_{N}^{p p^{\prime} \dagger}\gamma^{0} = \hat{S}_{N}^{p p^{\prime} \textrm{ rev}}$, where on the right hand side, in $\hat{S}_{N}^{p p^{\prime} \textrm{ rev}}$, all $\gamma$ matrices in $\hat{S}_{N}^{p p^{\prime}}$ have their order reversed. To take advantage of this, it is instructive to note that the identity used to rewrite the propagator in the second order formalism, (\ref{Ssecond}) could just have easily been written with the Dirac operator involving $p$ on the right. This would give an equivalent to (\ref{Sppk}) of the form
\begin{equation}
		S_{N}^{p'p} = K_{N}^{p'p}(-\ps + m) - e\sum_{i = 1}^{N} K_{N-1}^{p^{\prime}, p+k_{i}}\s\varepsilon_{i}
		\label{Sppk2}
\end{equation}
which completely reorganises the arrangement of the terms making up the propagator. With this rewriting, we get an alternative version of $\hat{S}_{N}^{p p^{\prime} \textrm{ rev}}$, which takes the form
\begin{equation}
	S_{N}^{ p \prime p  \textrm{ rev}} = (-\ps + m)K_{N}^{p^{\prime}p \textrm{ rev}} - e\sum_{i = 1}^{N}\s\varepsilon_{i} K_{N-1}^{ p',p + k_{i} \textrm{ rev}}.
\end{equation}
Substituting this into (\ref{T2}) leads to
\begin{equation}
	\frac{1}{2}\sum_{ss^{\prime}}\big|{\cal M}_{(N)s^{\prime}s}^{p^{\prime} p}	\big|^{2} = \frac{e^{2N}}{2}\tr_{\gamma}\left[\mathfrak{K}_{N}^{ p^{\prime}p}\mathfrak{K}_{N}^{p^{\prime}p \textrm{ rev}}\right] + \textrm{subleading},
	\label{TrKKs}
\end{equation}
where we have defined 
\begin{equation}
	K_{N}^{p^{\prime}p} := (-ie)^{N}\frac{\mathfrak{K}_{N}^{p^{\prime}p}}{(p^{2} + m^{2})(p^{\prime 2} + m^{2})}.
\end{equation}
In fact, using this notation, we can be more explicit with the pole structure even at the level of amplitude. Indeed, recall that the LSZ formula gives scattering amplitudes as the residues of poles when the external legs are on-shell. Since $p^2 + m^2 = (\ps + m)(-\ps + m)$ and the Dirac equation for the spinors, 
\begin{equation}
	\bar{u}_{s'}(-p')(-\pps + m) = 0 = (\ps + m)u_{s}(p)\,,
\end{equation}
applies on-shell, we can examine how the zeros from the on-shell conditions for the spinors cancel against the mass-shell poles:
\bear
\mathcal{M}_{(N)s's}^{p'p} \supset&& (-ie)^{N}\bar{u}_{s'}(-p') (-{\slashed p'}+m) (\slashed p' +m)\frac{\mathfrak{K}_N^{p^{\prime}p}}{\left(p'^{2} + m^{2}\right)\left(p^2 + m^{2}\right)}
 (\slashed p +m) u_{s}(p) 
\non
&& =
(-ie)^{N} \bar{u}_{s'}(-p') \frac{\mathfrak{K}_N^{p^{\prime}p}}{-\slashed p + m} u_{s}(p)  \non
&& =
(-ie)^{N}  \bar{u}_{s'}(-p') \frac{\mathfrak{K}_N^{p^{\prime}p}}{2m} u_{s}(p) \, .
 \label{Knopole}
\ear

The subleading terms in (\ref{TrKKs}) are more complicated, but it is already clear that the momentum shifts from $p' \rightarrow p' + k_{i}$ and/or $p \rightarrow p+k_{i}$ move the poles away from the mass-shell (c.f the discussion in section \ref{secGauge}). Thus, given that all of the subleading terms in (\ref{TrKKs}) involve at least one kernel with such a shifted momentum, and that they are sandwiched between on-shell spinors, these factors do not provide the correct pole structure to form part of the on-shell amplitude! So for the on-shell amplitudes (\ref{Knopole}) becomes exact (we use $\textrm{OS}$ to indicate equality On-Shell),
\begin{equation}
	\mathcal{M}_{(N)s's}^{p'p} \overset{\textrm{OS}}{=} (-ie)^{N}  \bar{u}_{s'}(-p') \frac{\mathfrak{K}_N^{p^{\prime}p}}{2m} u_{s}(p) \, .
	\label{eqMOS}
\end{equation}
This is a remarkable result and, as we shall see below, is related to how the gauge information of the amplitude is organised between leading and subleading terms. It comes out very easily in the worldline formalism and states that only the leading contributions to the fermion propagator can contribute to scattering amplitudes on-shell. We must stress, however, that the subleading terms certainly \textit{do} contribute off-shell, as we have seen that when they entered the calculation of the self energy above. 

\subsubsection{Overcounting the cross-section}
Continuing with the spin-summed cross-section, we are tempted to conclude that the subleading terms \textit{should not} contribute to (\ref{TrKKs}) either. As such we \textit{expect} (\ref{TrKKs}) to become
\begin{equation}
	\frac{1}{2}\sum_{ss^{\prime}}\big|{\cal M}_{(N)s^{\prime}s}^{p^{\prime} p}	\big|^{2} \overset{?}{=} \frac{e^{2N}}{2}\tr_{\gamma}\left[\mathfrak{K}_{N}^{p^{\prime}p}\mathfrak{K}_{N}^{p^{\prime} p \textrm{ rev}}\right].
	\label{TrKKOver}
\end{equation}
It turns out, however, that there is a subtle overcounting in (\ref{TrKKOver}), related precisely to the subleading terms. The problem centres on the fact that when we sum over spins, the completeness relation, (\ref{completeU}), replaces spinors with Dirac matrices (in fact, projectors onto the on-shell subspaces of the Dirac operator), which allows the subleading contributions hidden in the leading term to resurge. One way to see this (but see also the following section) is to make a direct comparison to the standard formalism, which we present in Appendix \ref{appFactor2}. As we explain there, the final result is that the subleading terms imply an overcounting by a simple factor of 2: thus we should replace (\ref{TrKKOver}) with the corrected relation
\begin{equation}
	\frac{1}{2}\sum_{ss^{\prime}}\big|{\cal M}_{(N)s^{\prime}s}^{p^{\prime} p}	\big|^{2} \rightarrow \frac{e^{2N}}{4}\tr_{\gamma}\left[\mathfrak{K}_{N}^{p^{\prime}p}\mathfrak{K}_{N}^{p^{\prime} p \textrm{ rev}}\right].
	\label{TrKK}
\end{equation}

With this machinery in place, we can finally compute cross-sections. If we further decompose the leading term as
\begin{equation}
	\mathfrak{K}_{N}^{p'p} = A_N\bone + B_{N\mu\nu}\frac{1}{2}[\gamma^{\mu},\gamma^{\nu}] -i C_N\gamma_{5}\,,
\end{equation}
then we only need one further result for the traces of products of Dirac matrices in our basis of the Clifford algebra,
\begin{eqnarray}
&&\frac{1}{4}\textrm{tr}_{\gamma}\Big[ \big(A\bone + B_{\mu\nu}\frac{1}{2}[\gamma^{\mu}, \gamma^{\nu}] -iC\gamma_{5} \big)\big(A^{\star}\bone - B_{\mu\nu}^{\star}\frac{1}{2}[\gamma^{\mu}, \gamma^{\nu}] + iC^{\star}\gamma_{5} \big) \Big]\non
&& \hspace{2cm}= |A|^{2} + 2B_{\mu\nu}B^{\star \mu\nu} - |C|^{2}
	\label{trCliff}
\end{eqnarray}
which show that the basis vectors do not mix ($\star$ denotes complex conjugation of the coefficients). Using this in (\ref{TrKK}) we get
\begin{equation}
	\frac{1}{2}\sum_{ss^{\prime}}\big|{\cal M}_{(N)s^{\prime}s}^{p^{\prime} p}	\big|^{2} = e^{2N}\Big[|A|^{2} + 2B_{\mu\nu}B^{\star \mu\nu} - |C|^{2}\Big]\,,
	\label{eqM2final}
\end{equation}
diagonal in the coefficients of the kernel's decomposition in the Clifford algebra. If this heuristic derivation has appeared dubious, in the following section we explain how it can be derived starting from (\ref{eqMOS}) by exploiting relations between the coefficients. Before this, we return to complete the calculation of the Compton cross-section.

\subsubsection{Compton scattering}
Returning to the specific case of $N=2$ Compton scattering, we first note that by using momentum conservation and the on-shell conditions, (\ref{K2fin}) can be written in a manifestly transversal way. Of course given our lessons in the scalar case, presented in section \ref{secGauge}, this is to be expected (we formalise this below). So, combining the denominators of the two different orderings inside the square brackets yields
\bear
\mathfrak{K}_{2}^{p'p} &=& A_2\bone + B_{2\mu\nu}\frac{1}{2}[\gamma^{\mu},\gamma^{\nu}] -i C_2\gamma_{5} 
\label{K2gothicf}
\ear
where (we denote the dual of a rank two tensor by $\widetilde{T}^{\mu\nu} := \frac{1}{2}\epsilon^{\mu\nu\alpha\beta}T_{\alpha \beta}$)

\bear
A _2&=& 
 -\frac{8p'\cdot f_1\cdot f_2\cdot p'+p'\cdot (k_1+k_2){\rm tr}(f_1\cdot f_2)}{4p'\cdot k_1  p'\cdot k_2}
 \nonumber\\
 B_2^{\mu\nu} &=& \frac{1}{2}  \Bigl(\frac{1}{2p'\cdot k_1}- \frac{1}{2p'\cdot k_2}\Bigr) [f_1,f_2]^{\mu\nu}
 -\frac{1}{2}   \frac{f_1^{\mu\nu}p'\cdot  f_2 \cdot k_1+ f_2^{\mu\nu}p'\cdot f_1\cdot k_2 }{p'\cdot k_1  p'\cdot k_2}
  \nonumber\\
C_2 &=& -\frac{1}{4}\Bigl(\frac{1}{2p'\cdot k_1} + \frac{1}{2p'\cdot k_2}\Bigr)\epsilon_{\alpha \beta \gamma \delta}f_{1}^{\alpha \beta}f_{2}^{\gamma \delta} =  \frac{1}{2}\Bigl(\frac{1}{2p'\cdot k_1} + \frac{1}{2p'\cdot k_2}\Bigr) \tr\left( f_{1}\widetilde{f}_{2} \right) \, . \nonumber\\
\label{abc}
\ear
We can insert this directly into (\ref{eqM2final}) and finally sum / average also over photon polarisations using the replacement, valid because each term in our amplitude is (manifestly!) transverse:
\bear
\sum_{\lambda = \pm} \ep_{i\lambda}^{\mu}\ep_{i\lambda}^{\nu\ast} \to \eta^{\mn}.
\label{sumpol}
\ear
At the level of the field strength tensors this amounts to
\begin{equation}
	\sum_{\lambda = \pm}f_{i\lambda}^{\mu\nu}f_{i\lambda}^{\alpha \beta \ast} \to k_{i}^{\mu}k_{i}^{\alpha} \eta^{\nu \beta} - k_{i}^{\mu}k_{i}^{\beta}\eta^{\nu\alpha} - k_{i}^{\nu}k_{i}^{\alpha}\eta^{\mu\beta} + k_{i}^{\beta}k_{i}^{\nu}\eta^{\mu\alpha}\,.
\end{equation}
For the mod-squared coefficients we require simple tensor manipulations, the proofs of which we omit for brevity. 

For $|C|^{2}$, anti-symmetry of the Levi-Civita tensor leads to the result (assuming on-shell photons in $D=4$)
\begin{align}
	\sum_{\lambda, \lambda' = \pm} \tr(f_{1\lambda}\cdot \widetilde{f}_{2\lambda'})\tr(f^{\ast}_{1\lambda}\cdot \widetilde{f}^{\ast}_{2\lambda'}) &\to 4\epsilon_{\mu\nu\alpha  \beta}\epsilon^{\mu\nu}{}_{ab}k_{1}^{\alpha}k_{2}^{\beta}k_{1}^{a}k_{2}^{b}\nonumber \\
	&\to -8 (k_{1}\cdot k_{2})^{2}\,.
\end{align}
The calculations of the other coefficients are a little longer. If we define a tensor
\begin{equation}
	F^{\mu\nu\alpha\beta} := \sum_{\lambda, \lambda' = \pm} (f_{1\lambda}\cdot f_{2\lambda'})^{\mu\nu} (f^{\ast}_{1\lambda}\cdot f^{\ast}_{2\lambda'})^{\alpha\beta} \,,
\end{equation}
then the results we need to form $|A|^{2}$ are (on-shell, $D=4$),
\begin{align}
	F^{\mu}{}_{\mu}{}^{\alpha}{}_{\alpha} &= 8(k_{1}\cdot k_{2})^{2}\\
	F^{\mu}{}_{\mu}{}^{\alpha\beta}p'_{\alpha}p'_{\beta} &= -2m^{2}(k_{1}\cdot k_{2})^{2} \\
	p'_{\mu}p'_{\nu}F^{\mu\nu\alpha\beta}p'_{\alpha}p'_{\beta} &= 2 (p'\cdot k_{1})^{2}(p'\cdot k_{2})^{2} +2m^{2}k_{1}\cdot k_{2} p'\cdot k_{1} p'\cdot k_{2} + m^{4}(k_{1}\cdot k_{2})^{2}\,.
\end{align}
Similarly, for $B_{\mu\nu}B^{\mu\nu\star}$ the following non-vanishing structures appear:
\begin{align}
\hspace{-2em}	F^{\mu\nu}{}_{\mu\nu} &= 4(k_{1}\cdot k_{2})^{2} = F^{\mu\nu}{}_{\nu\mu}\\
\hspace{-2em}	\sum_{\lambda, \lambda' = \pm} \tr(f_{1\lambda}\cdot f_{2\lambda'}^{\ast}) p'\cdot f_{1\lambda}^{\ast}\cdot k_{2} p'\cdot f_{2\lambda '}\cdot k_{1} &= 4(k_{1}\cdot k_{2})^{2}p'\cdot k_{1} p'\cdot k_{2} + 2m^{2}(k_{1}\cdot k_{2})^{3} \nonumber \\ 
\hspace{-2em}	&= \sum_{\lambda, \lambda' = \pm} \tr(f_{2\lambda'}\cdot f_{1\lambda}^{\ast}) p'\cdot f_{1\lambda}\cdot k_{2} p'\cdot f_{2\lambda '}^{\ast}\cdot k_{1} 
\end{align}
where we continue working on-shell in $D=4$. An alternative approach is to work out the fixed polarisation amplitudes and then compute the sum manually -- this is worked out using the spinor-helicity formalism in \cite{SpinProp2} to which we direct readers interested in polarised amplitudes.

It is a straightforward exercise to use these results and some on-shell kinematic relations to show that this leads to the correct spin / polarisation summed / averaged cross section that is derived in the standard formalism in any introductory text book on quantum field theory:
\begin{equation}
	2e^{4}\Big[m^{4} \left(\frac{1}{p' \cdot k_{1}} + \frac{1}{ p' \cdot k_{2}}\right)^{2} -2m^{2}\left(\frac{1}{ p' \cdot k_{1}} + \frac{1}{ p' \cdot k_{2}}\right)- \frac{p' \cdot k_{1}}{ p' \cdot k_{2}} - \frac{p' \cdot k_{2}}{ p' \cdot k_{1}} \Big]\,.
\end{equation}
Full details of the worldline representation of this amplitude, including for fully polarised electron and/or photon spins and various curious relationships between the scalar and spinor results, can be found by the interested reader in \cite{SpinProp2}.

The experienced student who recalls the standard approach to calculating the Compton cross-section, with its myriad traces of products of up to 8 $\gamma$-matrices will be able to appreciate some of the efficiencies of the approach presented here, and the relatively simple formulas we have arrived at that present the cross-section as a sum of scalar contributions.

\subsection{The coefficients $A$, $B_{\mu\nu}$ and $C$}
We pause for a moment to present some additional information regarding the coefficients $A$, $B_{\mu\nu}$ and $C$ defined in (\ref{K2gothicf}). In particular we will give formulae that generate manifestly transverse versions of them and will note some important relations between them for fixed $N$ (there also exist recursion relations between the coefficients for different values of $N$ that are given in \cite{SpinProp2}, that we will not discuss here).

\subsubsection{Manifest transversality}
In direct analogy to the scalar case -- see section \ref{secGauge} -- we expect to achieve a rewriting of the amplitude in terms of photon field strength tensors, $f_{i}$. Since  (\ref{eqMOS}) teaches us that the on-shell amplitude boils down to the leading kernel, $\mathfrak{K}_{N}$, the linear independence of the Clifford algebra basis tells us that in fact each component should \textit{individually} turn out to be manifestly transverse (as we found for the $N=2$ case in (\ref{abc}), for example).

We formalise this by combining our previous rewriting of the scalar kernel in (\ref{DppkFullf}) with the spin-orbit decomposition. We first introduce some notation, defining polynomials $\bar{R}_{N}$ from (\ref{DppkFullf}) analogous to the $\bar{P}_{N}$ already introduced. So we will write the result of expanding the exponential in the second line of (\ref{DppkFullf}) as 
\begin{equation}
	(-i)^{N}\bar{R}_{N}\, {\rm e}^{ -Tp'^2 + \frac{1}{2}\sum_{i,j=1}^N  \vert \tau_i - \tau_i \vert k_i\cdot k_j + (p-p') \cdot \sum_{i=1}^N k_i \tau_i}
\end{equation}
so that (dropping the obvious momentum conserving $\delta$-function)
\begin{align}
	\hspace{-1em}D_N^{p'p}(k_1,\veps_1,r_1;\cdots;k_N,\veps_N,r_N) = &(-e)^N \int_0^\infty dT\, {\rm e}^{-m^2T} \nonumber \\
\hspace{-1em} &\times \int_0^T\prod_{i=1}^{N} d\tau_{i} \bar R_N\, {\rm e}^{ -Tp'^2 + \frac{1}{2}\sum_{i,j=1}^N  \vert \tau_i - \tau_i \vert k_i\cdot k_j + (p-p') \cdot \sum_{i=1}^N k_i \tau_i}. \label{Master_es_onshell}
\end{align}
If the student worked out the details of the scalar case they will agree that the first three polynomials defined in this way are
\begin{align}
\begin{split}
\hspace{-2.5em}\bar R_1 &=
\frac{r_1\cdot f_1\cdot (p-p')}{r_1\cdot k_1}
\, ,
\\
\hspace{-1.5em}\bar R_2 &= \Big[ \frac{r_1\cdot f_1\cdot (p-p')}{r_1\cdot k_1}  + \sigma_{12}\frac{r_1\cdot f_1\cdot k_{2}}{r_1\cdot k_1}\Big]\Big[ \frac{r_2\cdot f_2\cdot (p-p')}{r_2\cdot k_2}  + \sigma_{21}\frac{r_2\cdot f_2\cdot k_{1}}{r_2\cdot k_2}\Big]\\
\hspace{-1.5em}&+ 2\delta(\tau_1-\tau_2)
{r_1\cdot f_1\cdot f_2\cdot r_2\over r_1\cdot k_1 \, r_2\cdot k_2}\,, \\
\hspace{-1.5em}\bar R_3 &=\Big[ \frac{r_1\cdot f_1\cdot (p-p')}{r_1\cdot k_1}  +\sum_{j\neq 1}^{3} \sigma_{1j}\frac{r_1\cdot f_1\cdot k_{j}}{r_1\cdot k_1}\Big]\Big[ \frac{r_2\cdot f_2\cdot (p-p')}{r_2\cdot k_2}  +\sum_{j\neq 2}^{3} \sigma_{2j}\frac{r_2\cdot f_2\cdot k_{j}}{r_2\cdot k_2}\Big]\\
\hspace{-11.5em}&\times \Big[ \frac{r_3\cdot f_3\cdot (p-p')}{r_3\cdot k_3}  +\sum_{j\neq 3}^{3} \sigma_{3j}\frac{r_3\cdot f_3\cdot k_{j}}{r_3\cdot k_3}\Big]  \\
\hspace{-2.5em}&+ 2\delta_{12} 
{r_1\cdot f_1\cdot f_2\cdot r_2\over r_1\cdot k_1 \, r_2\cdot k_2}
\Big[ \frac{r_3\cdot f_3\cdot (p-p')}{r_3\cdot k_3}  +\sum_{j\neq 3}^{3} \sigma_{3j}\frac{r_3\cdot f_3\cdot k_{j}}{r_3\cdot k_3}\Big] \\
\hspace{-1.5em}&+ 2\delta_{13} 
{r_1\cdot f_1\cdot f_3\cdot r_3\over r_1\cdot k_1 \, r_3\cdot k_3}
\Big[ \frac{r_2\cdot f_2\cdot (p-p')}{r_2\cdot k_2}  +\sum_{j\neq 2}^{3} \sigma_{2j}\frac{r_2\cdot f_2\cdot k_{j}}{r_2\cdot k_2}\Big] \\
\hspace{-1.5em}&+ 2\delta_{23} 
{r_2\cdot f_2\cdot f_3\cdot r_3\over r_2\cdot k_2 \, r_3\cdot k_3}
\Big[ \frac{r_1\cdot f_1\cdot (p-p')}{r_1\cdot k_1}  +\sum_{j\neq 1}^{3} \sigma_{1j}\frac{r_1\cdot f_1\cdot k_{j}}{r_1\cdot k_1}\Big]\,, \\
\end{split}
\label{barR1R2}
\end{align}
and the pattern should be clear from here. 

Returning to the spinor case, we recall the spin-orbit decomposition which writes the kernel as
\begin{equation}
	K_{N} = \sum_{S=0}^{N}K_{NS}\,.
\end{equation}
Moreover, the $K_{NS}$ are written in terms of the functions $W_{\eta}(S)$ and polynomials $\bar{P}_{NS}^{\{i_{1}\ldots i_{S}\}}$. Now the spin interaction (that produces the $W_{\eta}(S)$) is already written in terms of the field strength tensors, $f_{i}$, and are thus manifestly transverse (c.f (\ref{eqW123})). In contrast, the orbital interaction (producing the $\bar{P}_{NS}^{\{i_{1}\ldots i_{S}\}}$) is, as it stands, written in terms of the photon polarisation vectors, so our only task is to transform these polynomials into some equivalent ones that depend instead on their field strength tensors.

As should be clear, we can only expect to be able to do this for on-shell amplitudes, and in fact we already have everything we need to achieve this rewriting. We simply replace the exponent in (\ref{eqDefPBarS}) with our manifestly transverse version introduced in (\ref{DppkFullf}). The expansion according to (\ref{eqDefPBarS}) then defines new polynomials, $\bar{R}_{NS}^{\{i_{1}\ldots i_{S}\}}$ that will be written entirely in terms of the field strength tensors:
\begin{align}
	&\e^{-T\big(p' + \frac{1}{T}\sum_{i = 1}(k_{i} \tau_{i} - i \frac{r_{i}\cdot f_{i}}{r_{i}\cdot k_{i}})\big)^{2}} \,\e^{\sum_{i, j = 1}^{N}\big[\Delta_{ij}  k_{i} \cdot k_{j}  - 2i \ddel_{ij}\frac{r_{i}\cdot f_{i}\cdot k_{j}}{r_{i}\cdot k_{i}}  + \ddeld_{ij}\frac{r_{i}\cdot f_{i} \cdot f_{j}\cdot r_{j}}{r_{i}\cdot k_{i} r_{j}\cdot k_{j}}\big]}\bigg|_{f_{i_1}=\ldots =f_{i_S}=0}\bigg|_{\textrm{lin } f_{i_{S+1}}\ldots f_{i_{N}}} \nonumber \\
	 &:= (-i)^{N}\bar{R}_{NS}^{\{i_{1}\ldots i_{S}\}}\e^{-T\big(p' + \frac{1}{T}\sum_{i = 1}k_{i} \tau_{i}\big)^{2} + \sum_{i, j = 1}^{N}\Delta_{ij}  k_{i} \cdot k_{j}}\,.
\end{align}
As in the original spin-orbit decomposition, we immediately have $\bar{R}_{N0}^{\{\}} = \bar{R}_{N}$ and $\bar{R}_{NN}^{\{1\ldots N\}} = 1$. For the polynomials relevant in the Compton amplitude, appearing when we make the replacement $\bar{P}_{NS}^{\{i_{1}\ldots i_{S}\}} \rightarrow \bar{R}_{NS}^{\{i_{1}\ldots i_{S}\}}$ in (\ref{eqSON2P}),  we ask the interested reader to confirm
\begin{align}
	\bar{R}_{21}^{\{1\}} &= \frac{r_{2}\cdot f_{2}\cdot \big(p-p' + \sigma_{21}k_{1}\big)}{r_{2}\cdot k_{2}}\,,\\
	\bar{R}_{21}^{\{2\}} &= \frac{r_{1}\cdot f_{1}\cdot \big(p-p' + \sigma_{12}k_{2}\big)}{r_{1}\cdot k_{1}}\,.
\end{align}
Using these expressions it is now a mechanical exercise to reproduce the coefficients $A$, $B_{\mu\nu}$ and $C$ in (\ref{abc}) without having to work out how to combine the polarisation vectors into transverse field strength tensors. To do this, one can set each of the reference vectors $r_{i} = p'$, for example. 

\subsubsection{Relations between coefficients}
For on-shell amplitudes we sandwich the kernel between spinors -- recall (\ref{eqMOS}) -- and here we will uncover some useful relations that hold between the coefficients in $A$, $B_{\mu\nu}$ and $C$ in the on-shell case. We do this by appealing again to the ``inverted'' version of the propagator introduced in (\ref{Sppk2}) in which the Dirac operator appears on the right hand side of the kernels. Equating this expression with our usual one for $S_{N}^{p'p}$,
\begin{equation}
	(\pps + m)K_{N}^{p'p} - e\sum_{i=1}^{N}\slashed{\varepsilon}_{i}K_{N-1}^{p'+k_{i}, p} \overset{!}{=}K_{N}^{p'p}(-\ps + m) - e\sum_{i=1}^{N}K_{N-1}^{p', p+k_{i}}\slashed{\varepsilon}_{i}\,,
\end{equation}
and taking the external momenta on-shell leads to
\bear
 \slashed p' (A_N \bone + B_{N\alpha\beta}\sigma^{\alpha\beta} -i C_N\gamma_{5})=
 (A_N \bone + B_{N\alpha\beta}\sigma^{\alpha\beta} -i C_N\gamma_{5})(-\slashed p) \, .
\ear
If we then decompose the resulting product back onto the standard basis of the Clifford algebra and compare linear and cubic terms we obtain the relations
\bear
A_N(p+p')^{\alpha} &=& 2B_{N\,\,\nu}^{\,\, \,\,\alpha}(p-p')^{\nu}\,, \label{idAB1}\\
C_N(p-p')^{\alpha} &=& -2\widetilde B_{N\,\,\nu}^{\,\, \,\,\alpha}(p+p')^{\nu} \,,\label{idCB1}
\ear
that can be turned into scalar identities by contracting with external momenta:
\bear
A_N(m^2-p\cdot p') &=& 2p^\mu B_{N\mu\nu} p'^\nu \label{idAB2}\,,\\
C_N(m^2+p\cdot p')&=& 2p^\mu  \widetilde B_{N\mu\nu}p'^\nu \, .\label{idCB2}
\ear
It is a useful exercise to verify that these hold for the simple amplitudes at order $N=1$ and $N=2$ derived above. We see, then, that one only really needs to determine the coefficient $B_{\mu\nu}$ in order to determine the full information of on-shell amplitudes.

Armed with this, we can give a clearer derivation of the final formula for the spin-summed cross-section, (\ref{eqM2final}), starting from the on-shell matrix element, (\ref{eqMOS}). From this starting point the poles of the leading term have already been cancelled and we can discard the subleading terms for not containing the required pole structure. Then using the completeness relations for the spinors leads us directly to
\bear
\hspace{-2em}\langle |\mathcal{M}^{p^{\prime} p}_N|^2 \rangle &=& \dfrac{e^{2N}}{2 m^2} \Big\{ \left(p'\cdot p + m^2\right) \left( |A_N|^2 + 2 B_N^{\alpha\beta}B^\ast_{N\alpha \beta} \right) + \left(p' \cdot p - m^2\right) |C_N|^2 \nonumber \\
\hspace{-2em}&& \hspace{30pt} -2 \Re \left[ 2 p' \cdot B_N \cdot p \ A_N^\ast - 4 p' \cdot B_N \cdot B_N^\ast \cdot p 
- 2p' \cdot \tilde B_{N}\cdot p \ C_N^\ast \right] \Big\}.\non
\label{suboptimal}
\ear
Now \eqref{idAB1} provides
\bear
4 (p-p')\cdot B^{\ast}\cdot B\cdot (p-p') = -(p+p')^2 |A|^2 = 2(m^2-p\cdot p')|A|^2 \;,
\ear
whilst  \eqref{idCB1} gives
\bear
4(p+p')\cdot \tilde B^{\ast}\cdot \tilde B \cdot (p+p')
=
-(p-p')^2 |C|^2 
= 2(m^2+p\cdot p') |C|^2  \;.
\ear
If we decompose the second term on the second line of(\ref{suboptimal}) into parts symmetric and anti-symmetric in $p'$ and $p$, we can use the above relations along with \eqref{idAB2} and \eqref{idCB2} to turn its RHS into the more compact formula
\begin{equation}
	\frac{1}{2}\sum_{ss^{\prime}}\big|{\cal M}_{(N)s^{\prime}s}^{p^{\prime} p}	\big|^{2} = e^{2N}\Big[|A|^{2} + 2B_{\mu\nu}B^{\star \mu\nu} - |C|^{2}\Big]\,,
	\label{eqM2final2}
\end{equation}
this time with no subtlety around any awkward factors of $2$. For recursion relations between the coefficients at different orders, the reader is directed to the appendix of \cite{SpinProp2}.

\section{Propagators in a constant background}
\label{ScalField}
In section \ref{EHSpin} the one loop effective action in a constant electromagnetic background field was considered and the Euler-Heisenberg Lagrangian derived. The key ingredients were the use of Fock-Schwinger gauge (\ref{loopFS}) and the modification of the functional determinant entering the path integral -- see (\ref{abscalFfin}). Now we will give similar results for the scalar and spinor propagators in a constant background. We will be brief and omit many details in order to focus on the structure of the resulting representation of the propagator. The full calculation can be found in \cite{LineEM}.

\subsubsection{Scalar propagator}
Introducing a constant background electromagnetic field, it continues to be advantageous to use Fock-Schwinger gauge, only this time centered at one of the endpoints of the line:
\begin{equation}
	\bar{A}^{\mu}(x(\tau)) = -\frac{1}{2}F^{\mu\nu}(x(\tau) - x)_{\nu}.
\end{equation}
Since the uniform background does not break translational invariance we still expect the propagator depend only on the separation of the endpoints. Indeed, using the Fock-Schwinger representation of the gauge potential and expanding about the straight line path the generalisation of (\ref{Dxxq}) for $N = 0$ photons is found to be
\begin{equation}
	\mathfrak{D}^{x'x}[F] = \int_{0}^{\infty} dT\, \e^{-m^{2}T} \e^{-\frac{\xm^{2}}{4T}} \int_{DBC}\hspace{-0.5em} \mathscr{D}q(\tau)\, \e^{-\int_{0}^{T}d\tau \big[ \frac{1}{4}q \cdot \big(- \frac{d^{2}}{d\tau^{2}} + 2ie F \frac{d}{d\tau}\big)\cdot q - \frac{ie}{T}\xm \cdot F \cdot q \big]}.
\end{equation}
Clearly the exponent remains quadratic in $q$, so that by completing the square we can compute the path integral with knowledge only of the overall normalisation and the worldline Green function in the constant background.

The functional determinant is unchanged from the closed loop case, given in (\ref{abscalFfin}). On the other hand, the worldline Green function must obey open line boundary conditions. In exercise 3.3 you showed that in vacuum $\Delta(\tau, \tau^{\prime}) = \frac{1}{2}\left[G_{B}(\tau, \tau^{\prime}) - G_{B}(\tau, 0) - G_{B}(0, \tau^{\prime}) + G_{B}(0,0)\right]$. It turns out that this relationship holds also in the presence of the background field, just replacing vacuum Green functions with their constant field counterparts:
\begin{equation}
	\delC(\tau, \tau^{\prime}) = \frac{1}{2}\left[\mathcal{G}_{B}(\tau, \tau^{\prime})- \mathcal{G}_{B}(\tau, 0) - \mathcal{G}_{B}(0, \tau^{\prime}) + \mathcal{G}_{B}(0, 0)\right]
\end{equation}
where $\mathcal{G}_{B}(\tau, \tau^{\prime})$ is the constant field worldline Green function with string inspired boundary conditions generalising $G_{B}(\tau, \tau^{\prime})$:
\begin{eqnarray}
{{\cal G}_B}^{\mu\nu}(\tau,\tau') &=& 2\Big\langle \tau \Big| \Big( \frac{d^{2}}{d\tau^{2}} - 2ie F \frac{d}{d\tau} \Big)^{-1} \Big| \tau' \Big\rangle^{\mu\nu} \non
&=&
\left[{T\over 2{\cal Z}^2}
\biggl({{\cal Z}\over{{\rm sin}({\cal Z})}}
\,{\rm e}^{-i{\cal Z}\dot G_B(\tau,\tau') }
+ i{\cal Z}\dot G_B(\tau,\tau')  - 1\biggr)\right]^{\mu\nu} \, .
\label{GBF}
\end{eqnarray}
where $\mathcal{Z} = eFT$. This is now a non-trivial matrix in Lorentz space. Using this allows us to complete the square in the path integral and with the additional result
\begin{equation}
	\int_{0}^{T}d\tau \int_{0}^{T}d\tau^{\prime}\ \delC(\tau, \tau^{\prime}) = \frac{T^{3}}{4\mathcal{Z}}\left(\cot (\mathcal{Z} ) - \frac{1}{\mathcal{Z}}\right)
\end{equation}
we get
\begin{equation}
	\mathfrak{D}^{x'x}[F] = \int_{0}^{T}dT\, \e^{-m^{2}T} (4 \pi T)^{-\frac{D}{2}}\textrm{det}^{-\frac{1}{2}}\left[\frac{\sin(\mathcal{Z})}{\mathcal{Z}}\right]\e^{-\frac{1}{4T} \xm \cdot \mathcal{Z} \cot (\mathcal{Z} )\cdot \xm}.
	\label{DxF}
\end{equation}
Fourier transforming to momentum space, it is a simple exercise to demonstrate that 
\begin{equation}
		\mathfrak{D}^{p'p}[F] = \int_{0}^{T}dT\, \e^{-m^{2}T} \frac{\e^{-T p' \cdot \frac{\tan (\mathcal{Z})}{\mathcal{Z}}\cdot p'}}{\textrm{det}^{\frac{1}{2}}\left[\cos (\mathcal{Z})\right]}.
		\label{DpF}
\end{equation}
In \cite{LineEM} this result is extended to the dressed propagator where $N$-photon tree level scattering amplitudes can be derived. 

Note that this result is valid in Fock-Schwinger gauge for the background field; however, it is well known how to transform this to an arbitrary gauge \cite{ditgie-book}. Define the difference of holonomies $\e^{-i e \varphi(x',x)}$ where we denote ${\varphi(x', x) := \int_{x}^{x'} (\bar{A} - \hat{\bar{A}}) \cdot dx}$ where $\hat{\bar{A}}$ is the gauge potential in some reference gauge (such as Fock-Schwinger above) and $\bar{A}$ is related to $\hat{\bar{A}}$ by a gauge transformation with gauge function $\Lambda(x)$. Then $\varphi(x', x) = \Lambda(x') - \Lambda(x)$, independent of the path chosen between the endpoints, and (\ref{DxF}) changes under a gauge transformation as 
\begin{align}
	\mathfrak{D}^{x'x}[F]  &= \e^{-i e \varphi(x', x)}\hat{\mathfrak{D}}^{x'x}[F] \nonumber \\
	 &= \e^{-i e \varphi(x', x)} \int_{0}^{T}dT\, \e^{-m^{2}T} (4 \pi T)^{-\frac{D}{2}}\textrm{det}^{-\frac{1}{2}}\left[\frac{\mathcal{Z}}{\sin(\mathcal{Z})}\right]\e^{-\frac{1}{4T} \xm \cdot \mathcal{Z} \cot (\mathcal{Z} )\cdot \xm}.
	\label{DxFg}
\end{align}
where $\hat{\mathfrak{D}}^{x'x}[F]$ is of course the kernel in the Fock-Schwinger reference gauge, (\ref{DxF}). 

As an example application (or verification) of the result (\ref{DpF}), we consider the tree-level scattering of photons in the low energy limit. This would correspond to taking the part of the amplitude that is not only multi-linear in the polarisations (as in the full master formula (\ref{DppkFull}) above) but also multi-linear in the photon momenta $k_{1}, k_{2}, \ldots k_{N}$. This is familiar from the expansion of the one-loop Euler-Heisenberg Lagrangian mentioned in section \ref{sec:EH}, which likewise contains information about the one-loop low energy photon amplitudes. To understand this, one notes that a plane wave photon in the low energy limit looks to be constant at any finite scale since its phase changes so slowly with spatial position (derivative corrections can be included if one makes a derivative expansion of the effective action in the background). Here, for illustration, we focus on the $2$- and $4$-photon amplitudes in the low energy limit.

To do this we follow the following prescription (see \cite{Itz} for the four-photon case and \cite{LowE} for the general case):
\begin{enumerate}
	\item Expand the Euler-Heisenberg Lagrangian or, as in our case, the constant background propagator to $\mathcal{O}(F^{N})$, where $N$ is the number of low energy photons under study.
	\item Replace $F$ by $F_{\textrm{tot}} = \sum_{i = 1}^{N}f_{i}$, with a linearised field strength tensor $f_{i}^{\mu\nu} = k_{i}^{\mu}\ep_{i}^{\nu} - k_{i}^{\nu}\ep_{i}^{\mu}$ for each low energy photon.
	\item From the expression at $\mathcal{O}(F^{N})$ from point 1 retain only the parts that are multi-linear in the $f_{i}$ of point 2. This gives the amplitude for $N$-photons in the limit of vanishingly small energies.
\end{enumerate}
It is worth stating that the algorithm can be further refined to fixed helicity photons to assist the analysis of low energy photons where $K$ have helicity $+$ and $L = N-K$ have helicity $-$, as explored at one- and two-loop order in \cite{LowE} and \cite{LowERed}. 

The first step of this process yields in momentum space, to $\mathcal{O}(F^{4})$, 
\begin{align}
\hspace{-3em}\mathfrak{D}^{p'p}[F] =  \int_{0}^{T}dT\, \e^{-(p'^{2} + m^{2})T}\Big[1 &+ \frac{e^{2}T^{2}}{4}\tr(F^{2}) - \frac{e^{2}T^{3}}{3}p' \cdot F^{2} \cdot p' \nonumber \\
	&+ \frac{e^{4}T^{4}}{24}\tr(F^{4}) + \frac{e^{4}T^{4}}{32}\tr(F^{2})^{2} - \frac{2e^{4}T^{5}}{15}p' \cdot F^{4}\cdot p' + \frac{e^{4}T^{6}}{18}(p' \cdot F^{2} \cdot p')^{2} \nonumber \\
	&- \frac{e^{4}T^{5}}{12}\tr(F^{2})p' \cdot F^{2} \cdot p' + \cdots\Big].
	\label{DpExp}
\end{align}
Carrying out the proper time integral leads to the vacuum propagator plus a series of terms representing (in Fock-Schwinger gauge) interactions with photons:
\begin{align}
\hspace{-3em}	\mathfrak{D}^{p'p}[F] = \frac{1}{p'^{2} + m^{2}} &+ \frac{e^{2}}{2}\frac{\tr(F^{2})}{(p'^{2} + m^{2})^{3}} - 2e^{2} \frac{p' \cdot F^{2} \cdot p'}{(p'^{2} + m^{2})^{4}} \nonumber \\
	&+e^{4}\frac{\tr(F^{4})}{(p'^{2} + m^{2})^{5}} + \frac{3}{4}e^{4} \frac{\tr(F^{2})^{2}}{(p'^{2} + m^{2})^{5}} -16e^{4}\frac{p' \cdot F^{4} \cdot p'}{(p'^{2} + m^{2})^{6}} + 40e^{4}\frac{ (p' \cdot F^{2} \cdot p')^{2}}{(p'^{2} + m^{2})^{7}} \nonumber \\
	&- 10e^{4}\frac{\tr(F^{2})p' \cdot F^{2} \cdot p'}{(p'^{2} +m^{2})^{6}}  + \cdots
	\label{DPexp4}
\end{align}
In the next step we replace $F$ by $f_{1} + f_{2}$ for the terms at order $F^{2}$ and by $f_{1} + f_{2} + f_{3} + f_{4}$ at quartic order in $F$. We also amputate the external legs which requires multiplying all terms except for the first one by $(p'^{2} + m^{2})^{2}$. The final step is to keep hold of the parts linear in the field strengths of each photon. This requires the structures:
\begin{align}	\label{Flin2}
	\tr(F^{2}) &\longrightarrow 2 \tr(f_{1} \cdot f_{2})\\
	p' \cdot F^{2} \cdot p' &\longrightarrow 2p'\cdot f_{1} \cdot f_{2} \cdot p'
\end{align}
for the terms at quadratic order (the factors of two come from the cyclicity of the trace), whilst for the quartic terms we need the replacements
\begin{align}
\hspace{-3em}		\tr(F^{4}) &\longrightarrow 8\tr(f_{1} \cdot f_{2} \cdot f_{3} \cdot f_{4}) + 8\tr(f_{1} \cdot f_{2} \cdot f_{4} \cdot f_{3}) + \perms \\
\hspace{-3em}		\tr(F^{2})^{2} &\longrightarrow 8\tr(f_{1} \cdot f_{2}) \tr(f_{3} \cdot f_{4}) + 8\tr(f_{1} \cdot f_{3})\tr(f_{2} \cdot f_{4}) + \perms \\
\hspace{-3em}		p'\cdot F^{4} \cdot p' &\longrightarrow 2p' \cdot f_{1} \cdot f_{2} \cdot f_{3} \cdot f_{4} \cdot p' + 2 p' \cdot f_{1} \cdot f_{2} \cdot f_{4} \cdot f_{3} \cdot p' + \perms \\
\hspace{-3em}		(p' \cdot F^{2} \cdot p')^{2} & \longrightarrow 8p'\cdot f_{1} \cdot f_{2} \cdot p'\, \, p'\cdot f_{3} \cdot f_{4}\cdot p' + 8p'\cdot f_{1} \cdot f_{3} \cdot p' \,\,p'\cdot f_{2} \cdot f_{4}\cdot p' + \perms\\
\hspace{-3em}		\tr(F^{2})p' \cdot F^{2} \cdot p' &\longrightarrow 4\tr(f_{1} \cdot f_{2})p'\cdot f_{3} \cdot f_{4}\cdot p' + 4\tr(f_{1} \cdot f_{3})p'\cdot f_{2} \cdot f_{4}\cdot p'  +\perms
\label{Flin4}
\end{align}
where the sum over permutations involves additional terms with orderings of the field strength tensors that are inequivalent under the trace or contractions with momentum. Aside from the vacuum propagator, (\ref{D0p}), we also get (after amputation) the two-photon amplitude
\begin{equation}
	\mathfrak{D}^{p'p}[f_{1}, f_{2}] = e^{2}\frac{\tr(f_{1} \cdot f_{2})}{(p'^{2} + m^{2})} - 4e^{2} \frac{p' \cdot f_{1} \cdot f_{2} \cdot p'}{(p'^{2} + m^{2})^{2}}
	\label{2photonp}
\end{equation}
and the four-photon amplitude
\begin{align}
\hspace{-4em}	\mathfrak{D}^{p'p}[f_{1}, f_{2}, f_{3}, f_{4}] &= 8e^{4}\frac{\tr(f_{1} \cdot f_{2} \cdot f_{3} \cdot f_{4})}{(p'^{2} + m^{2})^{3}} + 6 e^{4} \frac{\tr(f_{1} \cdot f_{2}) \tr(f_{3} \cdot f_{4})}{(p'^{2} + m^{2})^{3}} -32 e^{4}\frac{p' \cdot f_{1} \cdot f_{2} \cdot f_{3} \cdot f_{4} \cdot p'}{(p'^{2} + m^{2})^{4}}  \nonumber \\
	\hspace{-4em}&+ 320e^{4}\frac{ p'\cdot f_{1} \cdot f_{2} \cdot p'\, \, p'\cdot f_{3} \cdot f_{4}\cdot p'}{(p'^{2} + m^{2})^{5}} - 40e^{4}\frac{\tr(f_{1} \cdot f_{2})p'\cdot f_{3} \cdot f_{4}\cdot p' }{(p'^{2} +m^{2})^{6}}   +\perms.
	\label{4photonp}
\end{align}
For the two photon case, we point out (leaving the proof as an exercise) that the second term in (\ref{2photonp}) reproduces the Thomson cross-section corresponding to the low energy limit of Compton scattering, whilst the first term provides a subleading contribution suppressed by $\frac{\omega}{m}$.

Of course one would like to verify that these results are in agreement with a straightforward worldline calculation of the photon amplitudes in the low energy limit, which we now turn to.

As described above, in the low energy limit we should expand the worldline vertex, $V^{\gamma}_{\textrm{scal}}$, to linear order in momentum. In this case, writing $\tilde{x}(\tau) = x(\tau) - x = \xm \frac{\tau}{T} + q(\tau)$, it takes the following form after an integration by parts (note we begin in position space):
\begin{equation}
	\lim_{k\rightarrow 0} V^{\gamma}_{\textrm{scal}}[k, \varepsilon] \simeq \e^{i k \cdot x'}\int_{0}^{T}d\tau \,\frac{i}{2} \tilde{x}\cdot f \cdot \dot{\tilde{x}},
	\label{limkV}
\end{equation} 
where we discard the boundary term because in Fock-Schwinger gauge $\varepsilon \cdot \xm = 0$ and retained the leading exponential because it will eventually provide us with momentum conservation. Expanding $\tilde{x}$ and integrating by parts again we find the low energy vertex in Fock-Schwinger gauge\footnote{In the closed loop case the equivalent vertex involves only the first term in the integrand and once replaces $\e^{ik \cdot x}$ with $\e^{i k \cdot x_{0}}$.}
\begin{equation}
	V^{0}_{\textrm{scal}}[f] = \frac{i}{2}\e^{i k \cdot x} \int_{0}^{T}d\tau \, \left[q \cdot f \cdot \dot{q} - \frac{2}{T}\xm \cdot f \cdot q\right].
	\label{defV0}
\end{equation}
Next we consider the two photon amplitude in this limit
\begin{equation}
e^{2}\int_{0}^{\infty} dT\, \e^{-m^{2}T} (4 \pi T)^{-\frac{D}{2}}\e^{i (k_{1} + k_{2}) \cdot x} \e^{-\frac{\xm^{2}}{4T}} \big<V^{0}_{\textrm{scal}}[f_{1}]V^{0}_{\textrm{scal}}[f_{2}]\big>.
\end{equation}
Using (\ref{defV0}) one must compute the Wick contraction of up to four fields according to (\ref{Gline}). However, the anti-symmetry of the field strength tensors mean that many of these will not contribute. Moreover one requires an even number of fields for a non-zero result so that the expectation value above evaluates to	
\begin{equation}
	-f_{1\mu\nu}f_{2\alpha\beta} \int_{0}^{T}d\tau_{1} \int_{0}^{T}d\tau_{2} \left[ \left(\delta^{\mu\beta}\delta^{\nu\alpha} - \delta^{\mu\alpha}\delta^{\nu\beta}\right)\ddel_{12}\ddel_{21} - \frac{2}{T^{2}}\xm^{\mu} \xm^{\alpha} \delta^{\nu\beta}\del_{12}\right]
	\label{Wick2V0}
\end{equation}
where we have used the result $\deld_{12} = \ddel_{21}$ and integrated by parts to remove second derivatives of the Green function. The parameter integrals are easily evaluated to
\begin{align}
\label{intDelta12}
	\int_{0}^{T}\int_{0}^{T} d\tau_{1}d\tau_{2} \del_{12} &= -\frac{T^{3}}{12}\\
	\int_{0}^{T}\int_{0}^{T} d\tau_{1}d\tau_{2} \ddel_{12}\ddel_{21} &= -\frac{T^{2}}{12}.
\label{intDDelta1221}
\end{align}
Thence carrying out the contractions in the expectation value we find that the two photon amplitude evaluates to
\begin{equation}
	\int_{0}^{\infty}dT \,\e^{-m^{2}T} (4 \pi T)^{-\frac{D}{2}} \e^{-\frac{\xm^{2}}{4T}}\e^{i (k_{1} + k_{2}) \cdot x} \left[\frac{e^{2}T^{2}}{6} \tr(f_{1} \cdot f_{2}) + \frac{e^{2}T}{6}\xm \cdot f_{1} \cdot f_{2} \cdot \xm\right].
	\label{F2x}
\end{equation}
We leave it as an exercise to check that this is exactly what arises if one expands (\ref{DxF}) to $\mathcal{O}(F^{2})$ and applies the above algorithm for low energy photon amplitudes in position space. 

The remaining step is to Fourier transform to momentum space in line with (\ref{Dpp}). As usual the $\xp$ integral gives momentum conservation (which is why $\e^{i k \cdot x}$ was stripped from the low energy vertex before expansion). The one non-trivial integral becomes
\begin{equation}
	\int d^{D}\xm \e^{-\frac{\xm^{2}}{4T} + i p' \cdot \xm}\left[\frac{e^{2}T^{2}}{6} \tr(f_{1} \cdot f_{2}) + \frac{e^{2}T}{6}\xm \cdot f_{1} \cdot f_{2} \cdot \xm\right]
\end{equation}
which after completing the square in the exponent and shifting the integration variable evaluates to 
\begin{equation}
	(4\pi T)^{\frac{D}{2}} \left[3 T^{2} \tr(f_{1} \cdot f_{2}) - 4T^{3}p'\cdot f_{1} \cdot f_{2} \cdot p'\right]\e^{-Tp'^{2}}.
	\label{Wick4V0}
\end{equation}
Putting this together, then, we get the proper time representation of the amplitude in momentum space (omitting the momentum conserving $\delta$ function)\footnote{The analogous result at one-loop order retains only the first term involving the trace, albeit with a different coefficient and measure on the proper time integral.}
\begin{equation}
	\int dT \, \e^{-T(p'^{2} + m^{2})}\left[\frac{e^{2}T^{2}}{3}\tr(f_{1} \cdot f_{2}) - \frac{2 e^{2}T^{3}}{3} p' \cdot f_{1} \cdot f_{2} \cdot p'\right].
\end{equation}
This is in agreement with (\ref{DpExp}) after applying steps 2 and 3 of the algorithm above, so that a final integration over proper time leads to the two photon amplitude (\ref{2photonp}). We invite the reader to verify the expectation value of a product of four low energy vertices, 
\begin{equation}
\hspace{-3em}	e^{4}\int_{0}^{\infty} dT\, \e^{-m^{2}T} (4 \pi T)^{-\frac{D}{2}}\e^{i (\sum_{i = 1}^{4}k_{i}) \cdot x} \e^{-\frac{\xm^{2}}{4T}} \Big<V^{0}_{\textrm{scal}}[f_{1}]V^{0}_{\textrm{scal}}[f_{2}]V^{0}_{\textrm{scal}}[f_{3}]V^{0}_{\textrm{scal}}[f_{4}]\Big>
\end{equation}
yields the $\mathcal{O}(F^{4})$ expansion of (\ref{DxF}) in position space and reproduces (\ref{4photonp}) after Fourier transforming to momentum space. 

\subsubsection{Spinor propagator}
For the spinor case we will be briefer, just outlining the principal results, initially presented in (\cite{GK2}). The application to low energy amplitudes will be left to the reader. 

The generalisation of the propagator, (\ref{Sk}), and its kernel, (\ref{Korig}), to the constant background case is again facilitated by Fock-Schwinger gauge in which they become
\begin{align}
	S^{x'x}(F)  &= \Big[m + i \gamma \cdot \Big(\frac{\partial}{\partial x'} - \frac{ie}{2}F \cdot x_-\Big)\Big]\,
	K^{x'x}(F) 
	\label{D0photon}
\end{align}
where now the ``kernel'' function is
\begin{align}
	K^{x'x}(F)& =\int_0^\infty \e^{-m^2 T} dT\int_{x(0)=x}^{x(T)=x'}\hspace{-2em}\mathscr{D}x\, \e^{-\int_0^Td\tau \big ({1\over 4}{\dot x}^2 
+ ieA\cdot\dot x \big)   }\nonumber\\
& \quad  \times 2^{-\frac{D}{2}} {\rm symb}^{-1}\biggl\{\oint_{AP}\mathscr{D}\psi\, \e^{-\int_0^Td\tau \big [ {1\over
2}\psi\cdot\dot\psi
 - ie\, (\psi+\eta)\cdot F\cdot (\psi + \eta)
\big]} \biggr\} \,.
\end{align}
Of course the scalar path integral on the first line is familiar from above; the new ingredient is the spin path integral that follows. Note that the interaction with the background is quadratic in the variable $\psi$ so this term can be absorbed into the kinetic term by completing the square. Again, the path integral normalisation is replaced by the same  matrix determinant as at one-loop order discussed in the previous chapter. Moreover, since the boundary conditions on $\psi$ are homogeneous, the relevant Green function is just the one-loop Green function in a constant background:
\begin{align}
	\mathcal{G}_{F}(\tau, \tau') &= 2\Big\langle \tau \Big| \big(\frac{d}{d\tau} - 2ie F\big)^{-1} \Big| \tau'\Big\rangle^{\mu\nu} \nonumber \\
	&= \Big[G_{F}(\tau, \tau') \frac{\e^{-i\Zz \dot{G}_{B}(\tau, \tau')}}{\cos(\Zz)}\Big]^{\mu\nu}\,.
\end{align}
With this Green function and normalisation we can complete the square in the Grassmann path integral to arrive at
\begin{align}
\hspace{-2em}K^{x'x}(F)&=\int_0^{\infty}
dT\,
\e^{-m^2T}
\e^{-\frac{x_-^2}{4T}}
\int_{\textrm{DBC}}
\mathscr{D}q\,
\e^{-\int_0^Td\tau\big[\frac{1}{4}q\cdot \big(-\frac{d^2}{d\tau^2}+2ieF\frac{d}{d\tau}\big)\cdot q-\frac{ie}{T}x_-\cdot F \cdot q\big]}
\nonumber\\
\hspace{-2em}&  \times 2^{-\frac{D}{2}}
{\rm symb}^{-1}\Big\{
\oint_{AP} \mathscr{D}\psi
\, \e^
{-\int_0^Td\tau\,
\bigl[\half\psi\cdot \big(\frac{d}{d\tau}-2ieF\big)\cdot\psi-2ie\psi\cdot F\cdot \eta-ie\eta\cdot F\cdot \eta\bigl]
}\Big\} \nonumber \\
\hspace{-2em}&= \int_{0}^{\infty} dT \e^{-m^{2}T} (4\pi T)^{-\frac{D}{2}} \detZs 
	\e^{-\frac{1}{4T} \xm \cdot \Zz \cdot \cot (\Zz) \cdot \xm}  \textrm{\rm symb}^{-1}\Big\{\e^{i \eta \cdot \tan (\Zz) \cdot \eta}\Big\} \, ,
\label{K0photon}
\end{align}
so that
\begin{align}
S^{x'x}(F)  = \Big[m + i \gamma \cdot \Big(\frac{\partial}{\partial x'} - \frac{ie}{2}F \cdot x_-\Big)\Big]& \int_{0}^{\infty} dT \e^{-m^{2}T} (4\pi T)^{-\frac{D}{2}} \detZs  \nonumber\\
	&\times\e^{-\frac{1}{4T} \xm \cdot \Zz \cdot \cot (\Zz) \cdot \xm}  \textrm{\rm symb}^{-1}\Big\{\e^{i \eta \cdot \tan (\Zz) \cdot \eta}\Big\}\\
	= \int_{0}^{\infty} dT \e^{-m^{2}T} (4\pi T)^{-\frac{D}{2}} \detZs &\Big[m - \frac{ie}{2} \gamma \cdot F\cdot \Big( \cot (\Zz) + i\bone\Big) \cdot x_-\Big] \nonumber\\
	&\times\e^{-\frac{1}{4T} \xm \cdot \Zz \cdot \cot (\Zz) \cdot \xm}  \textrm{\rm symb}^{-1}\Big\{\e^{i \eta \cdot \tan (\Zz) \cdot \eta}\Big\}
\end{align}
This is the spinor analogy to (\ref{DxF}) and is trivially transformed to momentum space:
\begin{equation}
	\hspace{-1em}S^{p'p}(F) = \int_{0}^{\infty}dT\big[m- (\pps + i \tan(\Zz)\cdot p')\big] \e^{-T(m^2 + p' \cdot \frac{\tan(\Zz)}{\Zz}\cdot p')}\symb^{-1}\Big\{ \e^{i\eta\cdot \tan(\Zz)\cdot \eta}\Big\}\,.
\end{equation}
Although it goes beyond the scope of these lectures, it is fairly straightforward to show that the symbol map representation of the spin coupling to the background field is in agreement with previous first quantised expressions (such as those presented in \cite{ditgie-book}), which is achieved by establishing the correspondence
\begin{equation}
	\symb^{-1}\Big\{ \e^{i\eta\cdot \tan(\Zz)\cdot \eta}\Big\}= \det{}^{-\frac{1}{2}}\big[\cos (\Zz)\big] \e^{\frac{i}{2} \sigma \cdot \Zz}\,.
\end{equation}
Rather than continue to expand this in powers of the field strength to examine low energy amplitudes we leave such a task to students who wish to explore further.

\section{Outlook}
In this final chapter we have reported recent developments adapting the first quantised approach to processes with external scalar and spinor legs. In doing this we extended the original ``bare'' propagator given by Feynman to derive master formulae for non-linear Compton scattering and gave an equally efficient version for the case of external fermion legs. The applications thus far may seem straightforward, since they are largely verifications of known results that can be derived using standard techniques. On the other hand, we anticipate that the worldline approach will show significant advantages with an increasing number of photons, especially considering the versions of the master Formulae that produce manifestly gauge invariant on-shell amplitudes; furthermore, standard the master formulae are valid off-shell and, through the sewing procedure, can be used as building blocks of higher loop processes. Finally we have given a proper time representation of the propagators in a constant field and discussed their relation to low energy photon amplitudes (in the scalar case), in an obvious generalisation of Euler-Heisenberg Lagrangians. 

There remain a great deal of open questions for study, both for worldline applications and multi-loop order as discussed in Chapter 2. It is worth saying that the bare and one-photon propagators have already been constructed and used along with (\ref{DxF}) in \cite{GK1, GK2, GK4}, along with their one-loop counterparts, to explore and extend the recent discovery of one-particle \textit{reducible} contributions to the particle self energies in a constant background and the two-loop Euler-Heisenberg Lagrangian \cite{GKA, GKB}. Here the first quantised approach allowed for a particularly elegant proof involving only manipulations under the worldline path integral, that would be completely obscured in the usual Feynman diagram approach. More recently the $N$-photon amplitudes in a plane wave background have been worked out using the worldline formalism in \cite{NPhotonPlane}; the inclusion of a graviton emitted or absorbed by the photon-dressed line was treated in \cite{Ahmadiniaz:2019ppj}.

From the three chapters of material that has been briefly presented in these notes, we hope that it is clear that this line of research can offer a viable alternative to the study of quantum fields. One illustrative example would be the recent application (after these lectures were finished) to study the so-called LKF gauge transformations that shed light on the non-perturbative structure of correlation functions in QED \cite{line2, LKFT1, LKFT2}. We hope it can be used to throw fresh light on existing problems and perhaps to motivate study of new problems in the years to come.

If the reader is interested in finding out more, they are directed to the original papers of Feynman \cite{Feynman:1950ir, feynman:pr84}, the classical review \cite{41} and a recent update \cite{UsReport}.

\subsection*{Acknowledgements}
JPE is grateful to Victor Banda, Guopeng Xu, C\'{e}sar Mata, Anabel Trejo, Misha Lopez and Predrag Cvitanovic for helpful comments on these notes (and for pointing out various typos!) and acknowledges support from CONACyT and PRODEP. NA acknowledges support from the organisers of the Helmholtz-DIAS International Summer School: {\it Quantum Field Theory at
the Limits: from Strong Fields to Heavy Quarks (2019}), especially David Blaschke, and also Ralf Sch\"utzhold for making it happen.  

\subsection*{Problems for chapter 3}
\begin{itemize}
	\item \textbf{3.1} Convince yourself that indeed the topology of closed and open trajectories is distinct.  \textit{Hint:} What happens if you remove one point, not an endpoint, in either case; what topologically invariant property distinguishes these from one another?
	\item \textbf{3.2} Fill in the gaps to derive equation (\ref{Dxxq}) by using (\ref{xqline}) in  (\ref{Dxxk}).
	\item \textbf{3.3} Show that $\Delta(\tau, \tau^{\prime})$ is an appropriate Green function for the second derivative operator on the interval $[0, T]$ with Dirichlet boundary conditions. Furthermore, demonstrate that $$\Delta(\tau, \tau^{\prime}) = \frac{1}{2}\left[G_{B}(\tau, \tau^{\prime}) - G_{B}(\tau, 0) - G_{B}(0, \tau^{\prime}) + G_{B}(0,0)\right]$$ and find the inverse transformation that expresses the one-loop Green function in terms of $\Delta$.
	\item \textbf{3.4} Show that by defining $\kappa_{0} := p$, $\kappa_{i} := k_{i}$, $\kappa_{N+1} := p^{\prime}$ along with $t_{0} := T$, $t_{i} := \tau_{i}$, $t_{N+1} := 0$ and $\varepsilon_{0} := 0$, $\varepsilon_{i} := \ep_{i}$, $\varepsilon_{N+1} := 0$ for $i \in \{1, \ldots N\}$ then the momentum space propagator (\ref{DppkFull}) can be written analogously to the one-loop effective action (2.43) as $\mathfrak{D}^{pp'}[k_{1}, \ep_{1}; \ldots; k_{N}, \ep_{N}] =$
	\begin{eqnarray}
	&&(2\pi)^{D} \delta^{D}\bigg(\sum_{i = 0}^{N+1}\kappa_{i}\bigg)(-ie)^{N}\int_{0}^{T}dT\, \e^{-m^{2}T} \prod_{i = 1}^{N} \int_{0}^{T}d\tau_{i} \, \nonumber\\
		&&\times \e^{ \sum_{i, j = 0}^{N+1}\big[\Delta^{\infty}_{ij}  \kappa_{i} \cdot \kappa_{j}  - 2i \dot{\Delta}^{\infty}_{ij}\varepsilon_{i} \cdot \kappa_{j} + \ddot{\Delta}^{\infty}_{ij}\varepsilon_{i}\cdot \varepsilon_{j} \big]}\bigg|_{\linep}
		\end{eqnarray}
	where $\Delta^{\infty}(\tau, \tau') = \abs{\tau - \tau'}$; this is the result originally found in \cite{Daik}.
	\item \textbf{3.5} Verify (\ref{int1}) and (\ref{int2}). Show also that $\int_{0}^{T}d\tau_{1} \int_{\tau_{1}}^{T} d\tau_{2}\, f(\tau_{1}, \tau_{2}) = \int_{0}^{T}d\tau_{1} \int_{0}^{\tau_{1}} d\tau_{2}\, f(\tau_{2}, \tau_{1})$ for any function $f$. Explain why swapping $\tau_{1}$ and $\tau_{2}$ in the exponent of (\ref{Dpp2}) is tantamount to exchanging $k_{1} \leftrightarrow k_{2}$.
	\item \textbf{3.6} Draw the Feynman diagrams corresponding to Compton scattering in scalar QED. Use the rules given in (\ref{ScalarQEDV}) to write down the amplitude associated to each diagram and compare the propagators to (\ref{int1}) and (\ref{int2}).
	\item \textbf{3.7} Derive (\ref{Dpp3}) from (\ref{DppkFull}) by sewing photons $1$ and $3$. 
	\item \textbf{3.8} Using the manifestly gauge invariant Master Formula, (\ref{DppkFullf}), determine the $N=2$ on-shell Compton amplitude fixing $r_{1} = p^{\prime} = r_{2}$. You should find the compact result
	\begin{equation}
		i\hat{\mathfrak{D}}_{2}^{p'p} \propto \frac{p' \cdot f_{1} \cdot f_{2} \cdot p'}{p'\cdot k_{1} p'\cdot k_{2}}\,,
	\end{equation}
	and it is a simple (and important) exercise to show that this coincides with (\ref{D2amp}). \\
	This should illustrate the point that it is much more convenient to generate the amplitude already in terms of the field strength tensors instead of starting with polarisation vectors and trying to massage them into products of the $f_{i}$.
	\item \textbf{3.9} Show that $\left(\ps + e \As\right)^{2} + m^{2} = -\left(p + e A\right)^{2} + m^{2} - \frac{e}{2}\sigma^{\mu\nu}F_{\mu\nu}$ where $\frac{i}{2}\sigma^{\mu\nu}:= \frac{i}{4}\big[\gamma^{\mu}, \gamma^{\nu}\big]$ are generators of the Lorentz group in the spin $\frac{1}{2}$ representation.
	\item \textbf{3.10*} Consider the path integral over Grassmann functions $\psi^{\mu}(\tau)$ dependent upon an arbitrary time-dependent Grassmann vector $\theta^{\mu}(\tau)$ 
	\begin{equation}
		\textrm{symb}^{-1}\left\{ \oint \mathscr{D}\psi(\tau) \, \e^{-\int_{0}^{T}d\tau \left[\frac{1}{2} \psi \cdot \dot{\psi} - e\theta(\tau) \cdot \left(\psi  + \eta\right)\right]}\right\}.
	\end{equation}
	By completing the square in the fermionic path integral and then expanding the $\eta$-dependent piece, show that this is equal to ($\hat{\gamma}^{\mu} = \frac{i}{\sqrt{2}}\gamma^{\mu}$)
	\begin{equation}
	2^\frac{D}{2}	\e^{-e\int_{0}^{T}d\tau \, \theta(\tau) \cdot \hat{\gamma} - \frac{e^{2}}{4} \int_{0}^{T}d\tau \int_{0}^{T}d\tau^{\prime} \, \theta_{\mu}(\tau) \sigma(\tau - \tau^{\prime}) \theta^{\mu}(\tau^{\prime})}.
	\end{equation}
	By differentiating with respect to $T$ (you will have to expand the first exponential) show that this expression satisfies the differential equation of a path ordered exponentiated line integral, so that you have proven an identity used in \cite{FradkinGitman}
	\begin{equation}
		\textrm{symb}^{-1}\left\{ \oint \mathscr{D}\psi(\tau) \, \e^{-\int_{0}^{T}d\tau \left[\frac{1}{2} \psi \cdot \dot{\psi} - e\theta(\tau) \cdot \left(\psi  + \eta \right)\right]}\right\} = 2^{\frac{D}{2}}\mathscr{P}\left\{ \e^{-e\int_{0}^{T}d\tau\, \theta(\tau) \cdot \hat{\gamma}} \right\}.
	\end{equation}
	\item \textbf{3.11} Show (\ref{Spsi2}) using the previous exercise and consequently prove that the worldline action is indeed (\ref{Sline}). 
	\item \textbf{3.12} Calculate the superspace integrals of equations (\ref{Ssuper}) and (\ref{Vsuper}) to verify they give the correct expressions. 
	\begin{itemize}
		\item Explain why the product of two superfields can be written in superfield form; demonstrate that the transformations of the bosonic and fermionic part of the product superfield implied by (\ref{susyline}) take an identical form to (\ref{susyline}). 
		\item Carry out the $\theta$ integral of an arbitrary superfield and demonstrate that its transformation under supersymmetry yields a total derivative; now put the three parts of the question together to argue that an integral over superspace of a function of superfields is (almost!) invariant under supersymmetry. 
	\end{itemize}
	\item \textbf{3.13} Verify (\ref{DiG} - \ref{D3G}), the latter being the defining equation of the super-Green function. \textit{Hint:} How does $(\theta^{\prime} - \theta)$ act under a Grassmann integral against an arbitrary superfield,  $\int d\theta^{\prime} \, (\theta^{\prime} - \theta) \mathbb{X}(\tau, \theta^{\prime})$? Show, by expanding the superfields in their components, that the Green function definition is consistent with the Wick contractions of the bosonic and fermionic pieces separately. 
	\item \textbf{3.14} Derive (\ref{Sppk}).
	\item \textbf{3.15*} Show that (\ref{sVert}) follows from (\ref{Kpp1}).
	\item \textbf{3.16} Calculate the diagrams that contribute to the one loop fermion self energy using the standard Feynman approach to second order spinor QED and verify you get (\ref{K2sewn}). Apply sewing also to the subleading terms to verify (\ref{SubleadingIntK}).
	\item \textbf{3.17} Prove (\ref{trCliff}).
	\item \textbf{3.18} Demonstrate that $\mathcal{G}_{B}(\tau, \tau^{\prime})$ is a Green function for the differential operator $\big(- \frac{d^{2}}{d\tau^{2}} + 2ie F \frac{d}{d\tau}\big)$ with the string inspired boundary conditions seen at one-loop. Show also that $\delC(\tau, \tau^{\prime})$ is indeed a Green function for the same operator now satisfying Dirichlet boundary conditions. 
	\item \textbf{3.19} Derive (\ref{DxF}) and subsequently (\ref{DpF}).
	\item \textbf{3.20} By expanding the constant field propagator in momentum space and hence carrying out the proper time integrals verify (\ref{DPexp4}). Follow the algorithm to derive the replacements (\ref{Flin2} - \ref{Flin4}) so as to find the two-and four-photon amplitudes in Fock-Schwinger gauge.
	\item \textbf{3.21} Expand the vertex operator $V^{\gamma}_{\textrm{scal}}$ to linear order in $k$ and integrate by parts to derive (\ref{limkV}). Hence verify (\ref{defV0}). What would the low energy vertex be in the spinor case?
	\item \textbf{3.22} Compute the necessary contractions of fields in $\big<V^{0}_{\textrm{scal}}[f_{1}]V^{0}_{\textrm{scal}}[f_{2}]\big>$ to arrive at (\ref{Wick2V0}). Compute the integrals (\ref{intDelta12}) and (\ref{intDDelta1221}).
	\item \textbf{3.23} Expand $\mathfrak{D}^{x'x}[F]$ of (\ref{DxF}) to $\mathcal{O}(F^{4})$. At quadratic order you should find (\ref{F2x}).
	\item \textbf{3.24} Compute the contractions (involving products of up to $8$ fields!) of (\ref{Wick4V0}). Subsequently carry out the parameter integrations over $\tau_{1}, \ldots, \tau_{4}$ and verify the resulting expression agrees with the result of the previous problem. Compute the Fourier transformation of the result and compare to (\ref{DpExp}) before doing the proper time integral and after to (\ref{4photonp}).
\end{itemize}

\section{Appendix: Momentum Integrals}
In section 3 various momentum integrals appear that can be expressed in terms of hypergeometric functions. To this end, define a hypergeometric function
\begin{equation}
	_{2}F_{1}(a, b, c; z) \equiv  \sum_{n=0}^{\infty} \frac{(a)_{n}(b)_{n}}{(c)_{n}}\frac{z^{n}}{n!}
\end{equation}
with the rising factorial
\begin{equation}
	(q)_{n} \equiv q(q+1)...(q+n-1)
\end{equation}
subject to $(q)_{0} = 1$. We can then express some elementary momentum integrals in terms of this function:
\begin{align}
\hspace{-4em}I_1&=\int \frac{d^Dq}{(2\pi)^D}\frac{1}{m^2+(p-q)^2}=\frac{(m^2)^{\frac{D}{2}-1}}{(4\pi)^{\frac{D}{2}}}\Gamma\Big(1-\frac{D}{2}\Big)\\
\hspace{-4em}I_2&=\int \frac{d^Dq}{(2\pi)^D}\frac{1}{q^2[m^2+(p-q)^2]}=-\frac{(m^2)^{\frac{D}{2}-2}}{(4\pi)^{\frac{D}{2}}}\Gamma\Big(1-\frac{D}{2}\Big)\,_2F_{1}(2-\frac{D}{2},1;\frac{D}{2};-\frac{p^2}{m^2})\\
\hspace{-4em}I_3^\mu&=\int \frac{d^Dq}{(2\pi)^D}\frac{q^\mu}{q^2[m^2+(p-q)^2]}=-\frac{1}{2}\frac{\Gamma\Big(1-\frac{D}{2}\Big)(m^2)^{\frac{D}{2}-2}p^\mu}{(4\pi)^\frac{D}{2}}\Big\{(1+\frac{m^2}{p^2})\,_2F_{1}\big(2-\frac{D}{2},1;\frac{D}{2};-\frac{p^2}{m^2}\big)-\frac{m^2}{p^2}\Big\}\\
\hspace{-4em}J_1&=\int\frac{d^Dq}{(2\pi)^D}\frac{1}{q^4[m^2+(p-q)^2]}=\frac{(m^2)^{\frac{D}{2}-3}}{(4\pi)^{\frac{D}{2}}}\Gamma\Big(1-\frac{D}{2}\Big)\,_2F_1(3-\frac{D}{2},2,\frac{D}{2},-\frac{p^2}{m^2})\\
\hspace{-4em}J_2^\mu&=\int\frac{d^Dq}{(2\pi)^D}\frac{q^\mu}{q^4[m^2+(p-q)^2]}=-\frac{1}{2}\frac{(m^2)^{\frac{D}{2}-2}p^\mu}{(4\pi)^{\frac{D}{2}}p^2}\Gamma\Big(1-\frac{D}{2}\Big)\Big\{\,_2F_1(2-\frac{D}{2},1,\frac{D}{2},-\frac{p^2}{m^2})\\
\hspace{-4em}&\hspace{9cm}-\frac{m^2+p^2}{m^2}\,_2F_1(3-\frac{D}{2},2,\frac{D}{2},-\frac{p^2}{m^2})\Big\}
\end{align}
For the calculation of the one-loop vertex in scalar QED we define the following integrals; one can express them in terms of various special functions but in fact we are content to use them here as a ``basis'' for the decomposition of the vertex.
\begin{align}
	K^{(0)}(p, p') &= \int d^{D}q\, \frac{1} {\left[ m^{2} + (p-q)^{2} \right] \left[ m^{2} + (q+p'^{2})^{2}\right]} \\
	K^{(0)}_{\mu}(p, p') &= \int d^{D}q\, \frac{q_{\mu}} {\left[ m^{2} + (p-q)^{2} \right] \left[ m^{2} + (q+p'^{2})^{2}\right]}\\
	J^{(0)}(p, p') &= \int d^{D}q\, \frac{1} {q^{2}\left[ m^{2} + (p-q)^{2} \right] \left[ m^{2} + (q+p'^{2})^{2}\right]}\\
	J^{(0)}_{\mu}(p, p') &= \int d^{D}q\, \frac{q_{\mu}} {q^{2}\left[ m^{2} + (p-q)^{2} \right] \left[ m^{2} + (q+p'^{2})^{2}\right]}\\
	J^{(0)}_{\mu\nu}(p, p') &= \int d^{D}q\, \frac{q_{\mu}q_{\nu}} {q^{2}\left[ m^{2} + (p-q)^{2} \right] \left[ m^{2} + (q+p'^{2})^{2}\right]}.
\end{align}
These integrals arise often in quantum field theory calculations and have been studied in detail.

\section{Appendix: Spin summed cross-section}
\label{appFactor2}
In the main text we claimed that in the spin-summed cross section that our explicit omission of the subleading terms of the propagator leads to an overcounting in the spin-summed cross section. In this Appendix we guide the motivated student to demonstrate this by comparing the contributions in the leading term to the standard formalism. It will therefore read as an extended exercise.

We are motivated by the cancellation rendered by the subleading parts in $S_{1}$ ((\ref{sVert})) and $S_{2}$ ((\ref{S2expl}) and (\ref{S2final})) to leave behind the familiar (first order) Feynman diagram result. The starting point is to observe the general decomposition (where clear from context we omit momentum labels of various quantities)
\begin{equation}
	K_{N} = K_{N}^{F} + K_{N}^{S}\, ; \qquad (\pps + m)K_{N}^{F} = S_{N}\, ,
\end{equation}
where we will demand that $K_{N}^{F}$ produces the desired final amplitude. 
\begin{itemize}
	\item Begin by showing that if we ask for
	\begin{equation}
	\hat{S}_{N} = (p^{\prime 2} + m^{2})K_{N}^{F}(\ps + m) \, ,
\end{equation} 
	then we must have $ K_{N}^{F} = \frac{\hat{S}_{N}(-\ps + m)}{(p^{2} + m^{2})(p^{\prime 2} + m^{2})}$.
\end{itemize} 
This allows us to express the physical part of $K_{N}$ in terms of the known Feynman amplitude. 

The remaining part of the kernel, $K_{N}^{S}$, is the piece that must cancel with the subleading part of $S$. But then $(\pps + m)K_{N}^{S} = \sum_{i=1}^{N}\slashed{\varepsilon}_{i}K_{N-1}$:
\begin{itemize}
	\item Show that this is solved by
\begin{equation}
	K_{N}^{S} = -\sum_{i=1}^{N}\frac{(\pps - m)\slashed{\varepsilon}_{i}K_{N-1}}{(p^{\prime 2} + m^{2})}.
\end{equation}
\end{itemize} 
Hence we have concluded that we can rewrite the kernel as
\begin{equation}
\hspace{-1em}(-ie)^{N} \mathfrak{K}_{N}^{p'p} = (p^{2} + m^{2})(p^{\prime 2} + m^{2})	K_{N}^{p'p} = \hat{S}_{N}^{p'p}(-\ps + m) - (p^{2} + m^{2})\sum_{i=1}^{N}(\pps - m)\slashed{\varepsilon}_{i}K_{N-1}^{p'+k_{i}, p}\, ,
\label{KFS}
\end{equation}
which manifests the leading term that contributes to on-shell amplitudes and the subleading information that should vanish on-shell.

Using this we will go on to form the spin-summed cross-section, for which we will need a second version of the above decomposition.
\begin{itemize}
	\item Show the similar result arising from the ``reversed kernel:''
	\begin{equation}
		(p^{2} + m^{2})(p^{\prime 2} + m^{2})	K_{N}^{p'p} = (\pps + m)\hat{S}_{N}^{p'p} + (p^{\prime 2} + m^{2})\sum_{i=1}^{N}K_{N-1}^{p', p+k_{i}}\slashed{\varepsilon}_{i}(\ps + m)\, ,
	\end{equation}
\end{itemize}  It is the ``subleading'' part (second terms in the decompositions) that we have uncovered hidden in this leading term that is able to resurge when we apply the on-shell completeness relation for the spinors, as we now show. 

The idea is to substitute this decomposition (and its reversed version) into (\ref{TrKKOver}). By construction, the leading terms (the first term on the RHS of (\ref{KFS})) produces the correct Feynman diagram expression. If this were all (in other words, if the other contributions vanished on-shell) then we would be done. However, it is simple, if labourious, to convince oneself of the following:
\begin{enumerate}
	\item The decomposition of $K_{N}$ can be iterated in the RHS of (\ref{KFS}) and its reversed version, right down to $K_{0}$.
	\item Almost all ``cross-terms'' between $S_{N}$ in (\ref{KFS}) and the subleading contributions vanish on-shell (providing zeros rather than poles).
	\item The ``diagonal-terms'' that arise in the final iteration\footnote{To see the subleading terms contribute it is not necessary to effect the iteration; however, without this it is hard to show that they simple double up the contribution of the leading term.} (at the level of $K_{0}$) survive on shell; moreover, \textit{they give a second contribution equal to that of the leading term}.
\end{enumerate}
Since we already identified the contribution to the amplitude in the truly leading terms we  see that in arriving at (\ref{TrKKOver}) we effectively double counted by explicitly ignoring the subtraction of the subleading terms from the leading term.
\begin{itemize}
	\item Convince yourself that this can happen because of subtle order of limits in applying the completeness relation (that replaces the spinors with projectors) and taking the amplitude on-shell.
\end{itemize}
The final conclusion, then, is that we must replace (\ref{TrKKOver}) with 
\begin{equation}
	\frac{1}{2}\sum_{ss^{\prime}}\big|{\cal M}_{(N)s^{\prime}s}^{p^{\prime} p}	\big|^{2} \rightarrow \frac{e^{2N}}{4}\tr_{\gamma}\left[\mathfrak{K}_{N}^{p^{\prime}p}\mathfrak{K}_{N}^{p^{\prime} p \textrm{ rev}}\right].
	\label{TrKKap}
\end{equation}


\begin{thebibliography}{99}

\bibitem{Feynman:1965}
R.~P.~Feynman and A.~R.~Hibbs, ``Quantum Mechanics and Path Integrals'', New York: McGraw-Hill (1965).

\bibitem{Chaichian:2001cz}
  M.~Chaichian and A.~Demichev,
  ``Path integrals in physics. Vol. 1: Stochastic processes and quantum mechanics'',
  Bristol, UK: IOP (2001).

\bibitem{Schulman:1981vu}
  L.~S.~Schulman,
  ``Techniques And Applications Of Path Integration'',
 Wiley (1981). 
  
\bibitem{Polchinski:1985zf}
  J.~Polchinski,
 ``Evaluation of the One Loop String Path Integral'',
  Commun.\ Math.\ Phys.\  {\bf 104} (1986) 37.\\
  E.~Kiritsis,``Introduction to superstring theory'',
 Leuven notes in mathematical and theoretical physics. B9, arXiv: 9709062 [hep-th].


\bibitem{Bastianelli:2006rx}
  F.~Bastianelli and P.~van Nieuwenhuizen,
  ``Path integrals and anomalies in curved space'',
  Cambridge University Press (2006).

\bibitem{18}
  M.~Reuter, M.~G.~Schmidt and C.~Schubert,
  ``Constant external fields in gauge theory and the spin 0, 1/2, 1 path integrals'',
  Annals Phys.\  {\bf 259}, 313 (1997), arXiv: 9610191 [hep-th].
  
\bibitem{41}
  C.~Schubert,
  ``Perturbative quantum field theory in the string inspired formalism'',
  Phys.\ Rept.\  {\bf 355}, 73 (2001), arXiv: 0101036 [hep-th].

\bibitem{Bastianelli:2008nm}
  F.~Bastianelli, O.~Corradini and E.~Latini,
  ``Spinning particles and higher spin fields on (A)dS backgrounds'',
  JHEP {\bf 0811} (2008) 054, arXiv: 0810.0188 [hep-th];\\
  O.~Corradini,
  ``Half-integer Higher Spin Fields in (A)dS from Spinning Particle Models'',
  JHEP {\bf 1009} (2010) 113, arXiv: 1006.4452 [hep-th].

\bibitem{15}
  M.~G.~Schmidt and C.~Schubert,
  ``Multiloop calculations in the string inspired formalism: The single spinor loop in QED'',
  Phys.\ Rev.\ D {\bf 53}, 2150 (1996), arXiv: 9410100 [hep-th].

\bibitem{Dai:2008bh}
  P.~Dai, Y.~-t.~Huang and W.~Siegel,
  ``Worldgraph Approach to Yang-Mills Amplitudes from N=2 Spinning Particle'',
  JHEP {\bf 0810} (2008) 027, arXiv: 0807.0391 [hep-th].
  
\bibitem{Feynman:1950ir}
  R.~P.~Feynman,
  ``Mathematical formulation of the quantum theory of electromagnetic interaction'',
  Phys.\ Rev.\  {\bf 80}, 440 (1950).


\bibitem{Bastianelli:2012bn}
  F.~Bastianelli, R.~Bonezzi, O.~Corradini and E.~Latini,
  ``Effective action for higher spin fields on (A)dS backgrounds'',
 JHEP {\bf 1212} (2012) 113, arXiv: 1210.4649 [hep-th].
  
\bibitem{Bastianelli:2007pv}
  F.~Bastianelli, O.~Corradini and E.~Latini,
  ``Higher spin fields from a worldline perspective'',
  JHEP {\bf 0702} (2007) 072, arXiv: 0701055 [hep-th].

\bibitem{Sorokin:2004ie}
D.~Sorokin,
``Introduction to the classical theory of higher spins'',
AIP Conf. Proc. {\bf 767}, 172 (2005), arXiv: 0405069 [hep-th].

\bibitem{Bastianelli:2005rc}
  F.~Bastianelli,
  ``Path integrals in curved space and the worldline formalism'',
8th International Conference on Path Integrals from Quantum Information to Cosmology,
6-10 Jun 2005, Prague, arXiv: 0508205 [hep-th].
  
\bibitem{Bastianelli:2015tha}
  F.~Bastianelli, R.~Bonezzi, O.~Corradini and E.~Latini,
  ``Spinning particles and higher spin field equations'',
  arXiv: 1504.02683 [hep-th].

\bibitem{Gies:2003cv}
  H.~Gies, K.~Langfeld and L.~Moyaerts,
  ``Casimir effect on the worldline'',
  JHEP {\bf 0306} (2003) 018, arXiv: 0303264 [hep-th].
  
\bibitem{Gopakumar:2003ns}
  R.~Gopakumar,
  ``From free fields to AdS'',
  Phys.\ Rev.\ D {\bf 70}, 025009 (2004), arXiv: 0308184 [hep-th].
  
  \bibitem{61}
F. Bastianelli and C. Schubert,
``One-loop photon-graviton mixing in an electromagnetic field: Part 1",
JHEP {\bf 0502} (2005) 069, arXiv: 0412095 [gr-qc].  

\bibitem{Gies:2005sb}
  H.~Gies, J.~Sanchez-Guillen and R.~A.~Vazquez,
  ``Quantum effective actions from nonperturbative worldline dynamics'',
  JHEP {\bf 0508} (2005) 067, arXiv: 0505275 [hep-th].

\bibitem{Hollowood:2007kt}
  T.~J.~Hollowood and G.~M.~Shore,
  ``Causality and Micro-Causality in Curved Spacetime'',
  Phys.\ Lett.\ B {\bf 655}, 67 (2007), arXiv: 0707.2302 [hep-th].

  
\bibitem{Bastianelli:2008vh}
  F.~Bastianelli, O.~Corradini, P.~A.~G.~Pisani and C.~Schubert,
  ``Scalar heat kernel with boundary in the worldline formalism'',
  JHEP {\bf 0810} (2008) 095, arXiv: 0809.0652 [hep-th].
  
\bibitem{Bonezzi:2012vr}
  R.~Bonezzi, O.~Corradini, S.~A.~Franchino Vinas and P.~A.~G.~Pisani,
  ``Worldline approach to noncommutative field theory'',
  J.\ Phys.\ A {\bf 45}, 405401 (2012), arXiv: 1204.1013 [hep-th];\\
  N.~Ahmadiniaz, O.~Corradini, D.~D’Ascanio, S.~Estrada-Jiménez and P.~Pisani,
  ``Noncommutative U(1) gauge theory from a worldline perspective,''
  JHEP {\bf 1511} (2015) 069,
  arXiv: 1507.07033 [hep-th].

\bibitem{Negele:1988vy}
  J.~W.~Negele and H.~Orland,
  ``Quantum Many Particle Systems'', Frontiers in Physics, 68, Addison-Wesley (1988).
  
\bibitem{Fujikawa:1980vr}
  K.~Fujikawa,
  ``Comment on Chiral and Conformal Anomalies'',
  Phys.\ Rev.\ Lett.\  {\bf 44}, 1733 (1980).


\bibitem{feynman:pr84}
R. P. Feynman, 
``An Operator calculus having applications in quantum electrodynamics'', 
Phys. Rev. {\bf 84}, 108 (1951).

\bibitem{afalma}
I. K. Affleck, O. Alvarez and N. S. Manton, 
 ``Pair Production At Strong Coupling In Weak External Fields'',
Nucl. Phys. {\bf B 197}, 509 (1982).


\bibitem{berkos}
Z. Bern and D. A. Kosower, 
``Efficient calculation of one loop QCD amplitudes'', 
Phys. Rev. Lett. {\bf 66}, 1669 (1991);\\ 
``The computation of loop amplitudes in gauge theory'',  
Nucl. Phys. B {\bf 379}, 451 (1992).

\bibitem{strassler}
M. J. Strassler, 
``Field theory without Feynman diagrams: One-loop effective actions'', 
Nucl. Phys. B {\bf385}, 145 (1992), arXiv: 9205205 [hep-ph].


\bibitem{berntasi}
Z. Bern, 
``String-based perturbative methods for gauge theories'',
TASI Lectures, Boulder TASI 92, 471, arXiv: 9304249 [hep-ph].

\bibitem{91}
N. Ahmadiniaz, C. Schubert and V.M. Villanueva, 
``String-inspired representations of photon/gluon amplitudes'',
JHEP {\bf 1301} (2013) 312, arXiv: 1211.1821 [hep-th].

\bibitem{fradkin}
E. S. Fradkin, 
``Application of functional methods in quantum field theory and quantum statistics (II)'',
Nucl. Phys. {\bf 76}, 588 (1966).

\bibitem{bermar}
F. A. Berezin, M. S. Marinov, 
``Particle spin dynamics as the Grassmann variant of classical mechanics'',
Ann. Phys. (N.Y.) {\bf 104}, 336 (1977).

\bibitem{bdzdh}
L. Brink, S. Deser, B. Zumino, P. Di Vecchia, P. Howe,
``Local Supersymmetry for Spinning Particles'', 
Phys. Lett. B {\bf 64}, 435 (1976).

\bibitem{brdiho}
L. Brink, P. Di Vecchia and  P. Howe, 
``A Lagrangian Formulation of the Classical and Quantum Dynamics of Spinning Particles'', 
Nucl. Phys. B {\bf 118}, 76 (1977). 

\bibitem{17}
S.L. Adler and C. Schubert, 
``Photon splitting in a strong magnetic field: Recalculation and comparison with previous calculations'',
Phys. Rev. Lett. {\bf 77}, 1695 (1996), arXiv: 9605035 [hep-th].

\bibitem{40} 
C. Schubert,
``Vacuum polarisation tensors in constant electromagnetic fields: Part I'', 
Nucl. Phys. B {\bf 585}, 407 (2000), arXiv: 0001288 [hep-ph].

\bibitem{50} 
G. V. Dunne and C. Schubert, 
``Closed-form two-loop Euler-Heisenberg Lagrangian in a self-dual background'', 
Phys. Lett. B {\bf 526}, 55 (2002), arXiv: 0111134 [hep-th].

\bibitem{ditgie-book}
W. Dittrich and H. Gies, 
``Probing the Quantum Vacuum. Perturbative Effective Action Approach in Quantum Electrodynamics and its Application'',
Springer Tracts Mod. Phys. {\bf 166}, 1 (2000).
 
 \bibitem{itzzub-book}
C. Itzykson, J. Zuber, ``Quantum Field Theory'', McGraw-Hill (1985).

\bibitem{63}
G. V. Dunne and C. Schubert,
``Worldline instantons and pair production in inhomogeneous fields'', 
Phys. Rev. D {\bf 72}, 105004 (2005), arXiv: 0507174 [hep-th]. 

\bibitem{64}
G. V. Dunne, Q.-h. Wang, H. Gies and C. Schubert,  
``Worldline instantons and the fluctuation prefactor'',
Phys. Rev. D {\bf  73}, 065028 (2006),  arXiv: 0602176 [hep-th].

\bibitem{92}
N. Ahmadiniaz and C. Schubert, 
``Covariant representation of the Ball-Chiu vertex'',
Nucl. Phys. B {\bf 869}, 417 (2013), arXiv: 1210.2331 [hep-ph].

\bibitem{4-gluon}
N. Ahmadiniaz and C. Schubert,
``QCD gluon vertices from the string-inspired formalism'',
 International Journal of Modern Physics E. {\bf25}, 1642004 (2016), arXiv: 1811.10780 [hep-th].

\bibitem{leeyan} 
T.~D.~Lee and C.~N.~Yang,
``Theory Of Charged Vector Mesons Interacting With The  Electromagnetic Field,''
Phys.\ Rev.\  {\bf 128} (1962) 885.

\bibitem{bastianelli}
F. Bastianelli, 
``The Path integral for a particle in curved spaces and Weyl anomalies'', 
Nucl. Phys. B {\bf 376}, 113 (1992), arXiv: 911203 [hep-th].

\bibitem{basvan93}
F. Bastianelli and P. van Nieuwenhuizen, 
``Trace anomalies from quantum mechanics'', 
Nucl. Phys. B {\bf 389}, 53 (1993), arXiv: 9208059 [hep-th].

\bibitem{bacozi1}
F.~Bastianelli, O.~Corradini and A.~Zirotti,
``BRST treatment of zero modes for the worldline formalism in curved space'', 
JHEP {\bf 0401} (2004) 023, arXiv: 0312064 [hep-th].  


\bibitem{76}
F. Bastianelli, J. M. D\'avila and C. Schubert, 
``Gravitational corrections to the Euler-Heisenberg Lagrangian'',
JHEP {\bf 0903} (2009) 086, arXiv: 0812.4849 [hep-th].


\bibitem{MP2}
J.~P. Edwards and P.~Mansfield, ``{Delta-function Interactions for the Bosonic
  and Spinning Strings and the Generation of Abelian Gauge Theory},'' JHEP {\bf 01} (2015) 127, arXiv: 1410.3288 [hep-th]

\bibitem{Daik}
K.~Daikouji, M.~Shino, and Y.~Sumino, ``{Bern-Kosower rule for scalar QED},''
Phys. Rev. D {\bf 53}, 4598 (1996), arXiv: 9508377 [hep-th].

\bibitem{line2}
N.~Ahmadiniaz, A.~Bashir, and C.~Schubert, ``{Multiphoton amplitudes and
  generalized Landau-Khalatnikov-Fradkin transformation in scalar QED},'' 
  Phys. Rev. D {\bf93}, 045023 (2016), arXiv: 1511.05087 [hep-ph].

\bibitem{LineNA}
N.~Ahmadiniaz, F.~Bastianelli, and O.~Corradini, ``{Dressed scalar propagator
  in a non-Abelian background from the worldline formalism}'', Phys.
  Rev. D {\bf93}, 025035 (2016), arXiv: 1508.05144 ,\\
  Addendum: Phys. Rev. D {\bf 93}, 049904 (2016).

\bibitem{HawkingZ}
S.~W. Hawking, ``{Zeta Function Regularization of Path Integrals in Curved
  Space-Time}'', Commun. Math. Phys. {\bf 55}, 133 (1977).

\bibitem{BastLadder}
F.~Bastianelli, A.~Huet, C.~Schubert, R.~Thakur, and A.~Weber, ``{Integral
  representations combining ladders and crossed-ladders}'', JHEP
  {\bf07} (2014) 066, arXiv: 1405.7770 [hep-ph].

\bibitem{SpinProp}
N.~Ahmadiniaz, V. M. Banda Guzm\'an, F.~Bastianelli, O.~Corradini, J.~Edwards, and C.~Schubert,
  ``Worldline master formulas for the dressed electron propagator, parts 1: off-shell amplitudes", JHEP {\bf 49} (2020) 018, arXiv: 2004.01391 [hep-th].

\bibitem{SpinProp2}
N.~Ahmadiniaz, V.~M.~B.~Guzm\'an, F.~Bastianelli, O.~Corradini, J.~P.~Edwards and C.~Schubert,
	``Worldline master formulas for the dressed electron propagator, part 2: On-shell amplitudes'', 
arXiv:2107.00199 [hep-th].

\bibitem{Olindo}
O.~Corradini and G.~D.~Esposti,
``Dressed Dirac propagator from a locally supersymmetric N=1 spinning particle'',
Nucl. Phys. B \textbf{970}, 115498 (2021), 
 arXiv: 2008.03114 [hep-th].

\bibitem{Morgan}
A.~G.~Morgan,
``Second order fermions in gauge theories,''
Phys. Lett. B \textbf{351}, 249 (1995),
 arXiv: 9502230 [hep-ph].

\bibitem{FradkinGitman}
E.~S.~Fradkin and D.~M.~Gitman,
``Path integral representation for the relativistic particle propagators and BFV quantization'', 
Phys. Rev. D \textbf{44}, 3230 (1991), 
\bibitem{Andrei}
A.~I. Davydychev, P.~Osland, and L.~Saks, ``{Quark gluon vertex in arbitrary
  gauge and dimension}'', Phys. Rev. D {\bf 63}, 014022 (2001), arXiv: 0008171 [hep-ph].

\bibitem{LineEM}
A.~Ahmad, N.~Ahmadiniaz, O.~Corradini, S.~P. Kim, and C.~Schubert, ``{Master
  formulas for the dressed scalar propagator in a constant field}'',  Nucl.
  Phys. B {\bf 919}, 9 (2017), arXiv: 1612.02944 [hep-ph].



\bibitem{Itz}
C. Itzykson and J. B. Zuber, ``Quantum Field Theory'', McGraw-Hill (1980).

\bibitem{LowE}
  L.~C.~Martin, C.~Schubert and V.~M.~Villanueva Sandoval, ``On the low-energy limit of the QED N photon amplitudes'',
  Nucl.\ Phys.\ B {\bf 668}, 335 (2003), 
  arXiv: 0301022 [hep-th].

\bibitem{LowERed}
  J.~P.~Edwards, A.~Huet and C.~Schubert,
  ``On the low-energy limit of the QED N-photon amplitudes: part 2'', 
  Nucl.\ Phys. B {\bf 935}, 198 (2018), 
  arXiv: 1807.10697 [hep-th].
  
\bibitem{GK1}
  J.~P.~Edwards and C.~Schubert,
  ``One-particle reducible contribution to the one-loop scalar propagator in a constant field'',
  Nucl.\ Phys.\ B {\bf 923}, 339 (2017),
  arXiv: 1704.00482 [hep-th].

\bibitem{GK2}
  N.~Ahmadiniaz, F.~Bastianelli, O.~Corradini, J.~P.~Edwards and C.~Schubert,
  ``One-particle reducible contribution to the one-loop spinor propagator in a constant field'',
  Nucl.\ Phys.\ B {\bf 924}, 377 (2017), 
  arXiv: 1704.05040 [hep-th].  

  \bibitem{GK4}
  N.~Ahmadiniaz, J.~P.~Edwards and A.~Ilderton,
  ``Reducible contributions to quantum electrodynamics in external fields'', JHEP {\bf05} (2019) 38, 
  arXiv: 1901.09416 [hep-th].
  
\bibitem{GKA}
  H.~Gies and F.~Karbstein,
  ``An Addendum to the Heisenberg-Euler effective action beyond one loop",
  JHEP {\bf 03} (2017) 108, 
  arXiv: 1612.07251 [hep-th].

\bibitem{GKB}
  F.~Karbstein,
  ``Tadpole diagrams in constant electromagnetic fields,''
  JHEP {\bf 10} (2017) 075,
  arXiv: 1709.03819 [hep-th].

\bibitem{NPhotonPlane}
J.~P.~Edwards and C.~Schubert,
``N-photon amplitudes in a plane-wave background'', 
Phys. Lett. B \textbf{822}, 136696 (2021), 
arXiv: 2105.08173 [hep-th].

\bibitem{Ahmadiniaz:2019ppj}
N.~Ahmadiniaz, F.~M.~Balli, O.~Corradini, J.~M.~D\'avila and C.~Schubert,
``Compton-like scattering of a scalar particle with $N$ photons and one graviton'', 
Nucl. Phys. B \textbf{950}, 114877 (2020), 
arXiv: 1908.03425 [hep-th].

\bibitem{LKFT1}
J.~Nicasio, J.~P.~Edwards, C.~Schubert and N.~Ahmadiniaz,
``Non-perturbative gauge transformations of arbitrary fermion correlation functions in quantum electrodynamics'', arXiv: 2010.04160 [hep-th].

\bibitem{LKFT2}
N.~Ahmadiniaz, J.~P.~Edwards, J.~Nicasio and C.~Schubert,
``Generalized Landau-Khalatnikov-Fradkin transformations for arbitrary N-point fermion correlators'', 
Phys. Rev. D \textbf{104}, 025014 (2021), 
arXiv: 2012.10536 [hep-th].

\bibitem{UsReport}
J.~P.~Edwards and C.~Schubert,
``Quantum mechanical path integrals in the first quantised approach to quantum field theory'', arXiv: 1912.10004 [hep-th].
\end{thebibliography}
\end{document}